\documentclass[runningheads]{llncs}
\usepackage{graphicx}
\usepackage{tcolorbox}
\usepackage{multirow}
\usepackage{mathtools}

\usepackage[linesnumbered, ruled, vlined]{algorithm2e}

\SetCommentSty{mycommfont}
\usepackage{amssymb,amsmath}
\usepackage{subcaption}
\usepackage{hyperref}
\hypersetup{
    colorlinks,
    citecolor=blue,
    filecolor=black,
    linkcolor=red,
    urlcolor=green
}
\usepackage{float}

\begin{document}
%
\title{Towards Regret Free Slot Allocation in Billboard Advertisement
\thanks{The work of Suman Banerjee is supported by the Seed Grant (SGT-100047) provided by the Indian Institute of Technology Jammu, India. A very preliminary version of this paper has appeared as a \emph{student abstract} in \textbf{37th AAAI Conference on Artificial Intelligence (AAAI 2023)\cite{ali2023efficient}.}}
}
%
%
\author{Dildar Ali \and Suman Banerjee  \and Yamuna Prasad }
\authorrunning{Ali et al.}
\institute{Department of Computer Science and Engineering, \\ Indian Institute of Technology Jammu, Jammu \& Kashmir 181221, India. \\
\email{\{2021rcs2009,suman.banerjee,yamuna.prasad\} @iitjammu.ac.in}\\
}

\maketitle              
\begin{abstract}
Creating and maximizing influence among the customers is one of the central goals of an advertiser, and hence, remains an active area of research in recent times. In this advertisement technique, the advertisers approach an influence provider for a specific number of views of their content on a payment basis. Now, if the influence provider can provide the required number of views or more, he will receive the full, else a partial payment. In the context of an influence provider, it is a loss for him if he offers more or less views. This is formalized as `Regret', and naturally, in the context of the influence provider, the goal will be to minimize this quantity. In this paper, we solve this problem in the context of billboard advertisement and pose it as a discrete optimization problem. We propose four efficient solution approaches for this problem and analyze them to understand their time and space complexity. We implement all the solution methodologies with real-life datasets and compare the obtained results with the existing solution approaches from the literature. We observe that the proposed solutions lead to less regret while taking less computational time.   
\keywords{ Out-of-home Advertisement \and Regret Minimization \and Influence Provider \and Advertisers \and Discrete Optimization.}
\end{abstract}

\section{Introduction} \label{Sec:Intro}
In recent times, Outdoor advertising has emerged as a $29$ billion-dollar industry, and every year it is increasing by $3 \%$ to $4 \%$ \footnote{\url{https://www.marketing-interactive.com/ooh-advertising-spend-to-soar-to-us33-billion-by-2021/}}. Several outdoor advertisement approaches exist, such as billboards, public transportation systems, etc. and at least $74 \%$ of its growth comes from billboard advertisements. Billboard advertisement has emerged as an effective approach for the out-of-home advertisement technique. In this advertisement technique, display boards (e.g., billboards) are placed in popular locations (e.g., Mall, Park, Cafeteria, Side of Highways, etc.), and an E-Commerce company displays advertisement content (i.e., image, video, animation, etc.) to ingrain it to the people's mind. It is reported that $80 \%$ of the people look at the display content while driving\footnote{\url{http://www.runningboards.com.au/outdoor/relocatable-billboards}}. From a recent market survey, it is known that 
in billboard advertisement, the investment of return is $65\%$ more compared to other advertisement techniques \footnote{\url{https://topmediadvertising.co.uk/billboard-advertising-statistics/}}. Also, a recent study by Zhang et al. \cite{zhang2019optimizing} shows that more than $50 \%$ of travelers are impressed by at least five billboards on each trip. Recently, billboards have become digital and are available to the e-commerce house slot-wise.

\par Now, it is an essential question of how to quantify the influence of a billboard slot or a given set of billboard slots. In the consumer behavior literature, it has been reported that if a traveler sees an advertisement several times then it is highly unlikely that they will take any action \cite{feder1985adoption,greene2003econometric,lee2014examining,sierzchula2014influence,train2009discrete}. As an example if the advertisement is for a product then the action may be as follows: the traveler may purchase the item or influence others regarding the product. Also, some evidence shows an $S$-like function should measure the repetition of any advertisement content \cite{campbell2003brand,malaviya2007moderating,palda1965measurement,taylor2009once}.
 The influence of billboard slots is less when the number of views on that billboard slots is less, and the influence will increase dramatically upon seeing more. However, when the same advertisement content is shown more often, the effect of additional impressions will decline as the extra information diminishes. Also, the marketing literature has reported that a person sees at least five billboards in a trip \cite{zhang2019optimizing}. It is feasible for an influence provider to rent multiple billboard slots (possibly in different billboards) to ingrain the content in the traveler's mind. In many existing studies, the basic triggering model has been used to compute the influence \cite{ali2022influential,ali2023influential,lotfi2017multi,wang2019efficiently,zhang2018trajectory}. So, one important point to highlight here is that to compute the influence of a billboard slot (or a set of billboard slots) we need the information about the traveler's location (which is called as trajectory database) and information about the billboards (e.g., location, cost, etc. and this is known as billboard database).

\par In billboard advertisement technique, the billboards belong to an influence provider, and several advertisers approach the influence provider to obtain the required amount of influence on a payment basis. The influence provider allocates billboard slots to the advertisers based on the payment and influence demand. The payment model follows: ``The E-Commerce house will pay the quoted price if the quoted influence demand (or more) is satisfied, else a partial payment on a pro-rata basis will be made". Now, if the influence provider allocates billboard slots that lead to more influence than the quoted influence, this is a loss for the influence provider. The reason here is that, first of all, there is no extra incentive for the extra influence. Secondly, the billboard slots that cause the extra influence may be allocated to one of the advertisers whose quoted influence is unsatisfied. As the quoted influence of that advertiser is not satisfied, this advertiser will not make the full payment to the influence provider. This loss is formally quantified in terms of regret. This regret can be of two types:
\begin{itemize}
\item The first kind of regret is formalized as \emph{excessive regret} when an extra influence is provided to the advertiser rather than the quoted influence.
\item The second kind of regret is formalized as \emph{unsatisfied regret} when less influence is provided.
\end{itemize}
   Several studies in the literature in the context of billboard advertisement techniques exist. In most of the studies, the problem that has been considered is locating the influential locations to place the billboards or finding out a limited number of influential billboard slots \cite{lotfi2017multi,wang2019efficiently,zhang2018trajectory,zhang2020towards,zhang2019optimizing}. However, in practice, there exist multiple advertisers. We are surprised to see a very limited amount of literature exists in the multiple advertiser setting. In particular, in the presence of multiple advertisers, the problem becomes more challenging. Next, we summarize our contributions to this regret minimization problem. 

\paragraph{\textbf{Our Contribution}} In this paper, we make the following contributions:
\begin{itemize}
\item We study the \textsc{Regret Free Slot Allocation in Billboard Advertisement} Problem in multi-advertiser settings for which the literature is very limited.
\item We propose four efficient heuristic solutions: Effective Allocation Policy, Effective Advertiser Driven One-by-One Exchange Policy, Effective Billboard Driven One-by-One Exchange Policy and Effective Billboard Driven One-by-Two Exchange Policy.
\item We perform an extensive set of experiments in real-world datasets and compare the performance of the proposed allocation strategies with the available methods in the literature. 
\end{itemize}

\paragraph{\textbf{Organization of the Paper}} 
The subsequent sections of the paper are structured in the following manner:  Firstly, in Section \ref{Sec:Related_work}, we explore the existing literature related to our problem. Section \ref{Sec:BPD} provides an overview of the context and formally defines the problem. Section \ref{Sec:PS} discusses the proposed solution approaches for solving the problem. Section \ref{Sec:Experimental_Details} comprehensively describes the experimental setup and the required parameters. Section \ref{Sec:Experimental_Evaluations}  discusses the observations of the proposed approaches over different datasets and parameter settings. Lastly, Section \ref{Sec:CFD} serves as the final part of our study and provides suggestions for future research.

\section{Related Work}\label{Sec:Related_work}\label{Sec:Related_Work}
As mentioned previously, our study comes under the umbrella of outdoor advertisement. Our literature survey is categorized into four major categories: (1) Influential Site Selection, (2) Trajectory Driven Influence Maximization, (3) Influential Billboard Selection, and (4) Regret Minimization. First, we start with influential site selection problems. 

\subsection{Influential Site Selection}
In the past few years, influential site or zone selection problems have gained the attention of researchers due to their vast application areas. In this direction, Cabello et al. \cite{cabello2006reverse,CABELLO201099} introduce facility location problems to find the most influential place or sites and introduce three optimization problems. They propose an approach based on Euclidean distance to address the issue of locating all data points in a dataset with a given query point as their closest neighbor. Next, Stanoi et al. studied the problem of finding the RNN, and they showed how the finding of the influential set can be reduced to the nearest neighbor and range queries; they proposed a solution in which the R-Tree is called multiple times, and this is the drawback of their work. To overcome this drawback, Xia et al. \cite{10.5555/1083592.1083701} introduce a problem of finding the top-$t$ most influential sites from a given spatial region. They propose a new metric to prune less influential areas and an efficient algorithm to solve the problem. Cabello et al. \cite{CABELLO201099} find the location in which placing a service site guarantees the visit of a maximum number of customers. However, a customer can visit $k$ places for multiple product requirements. So, Zhou et al.\cite{zhou2011maxfirst} extended the MaxBRNN problem to the MaxBRKNN and proposed an efficient solution, the `Maxfirst', by partitioning regions into different quadrants to solve this. Next, Wang et al. \cite{wang2016pinocchio} studied a more generalized problem where the goal is to place several new facilities to maximize a score. This score can be defined as profit or influence given by the placed object. They proposed a solution methodology called PRIME-LS, which considers mobility and probability of movement factors into consideration.
Further, they improved the proposed solution approach by incorporating two different optimization strategies. Later, Huang et al. \cite{10.1145/2063576.2063971} studied the problem of the most influential location selection problem and introduced a new type of facility query. They proposed two pruning techniques to reduce the search space region and efficiently process the query. An extensive set of experiments with real and synthetic datasets shows the efficiency of their proposed solutions.

\subsection{Trajectory Driven Influence Maximization}
Due to the recent advancement of wireless internet and handheld mobile devices, tracking of locations of moving objects become easier. Hence, several trajectory datasets are publicly available \cite{zheng2015trajectory,wang2020big}. 
 Zhang et al. \cite{zhang2017automatic} studied the problem of transit advertisement where the goal is to select Top-$k$ bus routes such that the targeted influence is maximized. They proposed a framework that captures the mobility pattern of the passengers and the characteristics of bus stops. Guo et al.\cite{7676319} studied the influence maximization problem using trajectory databases. They identify the $k$ most optimal trajectories for associating with a specific advertisement to maximize the anticipated influence among a large audience. They propose three upper-bound estimation techniques to speed up the processing time. They also presented a greedy approach that guarantees an approximation ratio of $(1 - \frac{1}{e})$ and a cluster-based technique that further improves the greedy approach. Next, Wang et al. \cite{8118111} studied a new query to find the capacity and route planning. Their main goal is to find an optimal route from the starting point to the endpoint to attract the maximum number of passengers according to the predefined travel distance. In social network analysis, Cai et al. \cite{7524471} studied the problem of influence maximization in social networks and used mobile crowdsource data from location-based social network services. They proposed a network and propagation models for social networks and considered the physical world.
Further, Cai et al. \cite{cai2022survey} study a survey on influence maximization in online(e.g., social network) and offline(e.g., billboard) advertisements. They analyze the existing location-driven influence maximization works and classify them into different categories. Next, Zhang et al. \cite{10.1145/3394486.3403327} studied the problem of finding a set of locations within a budget constraint such that it achieves the maximum user reach. They proposed a deterministic algorithm over a simple greedy approach and a learning-based variant, `NN-Sower', to improve efficiency further. Their exhaustive experiments over real-world datasets show the effectiveness and efficiency of the proposed methods.

\subsection{Influential Billboard Selection}
Recently, several studies have been done in the context of trajectory-driven billboard advertisements. Zhang et al. \cite{zhang2020towards} studied the trajectory\mbox{-}driven influential billboard placement problem where a set of billboards, location, and cost are given. The goal here is to choose a subset of the billboards within the budget that influence the largest number of trajectories. Later, Wang et al. \cite{wang2022data} studied the problem of the  Targeted Outdoor Advertising Recommendation (TOAR) problem considering the facts of user\mbox{-}profiles and advertisement topics. Their main contribution was a targeted influence model that characterizes the advertising influence spread along with user mobility.
Based on the divide and conquer approach, they developed two solution strategies. Implementation with real-world datasets shows the efficiency and effectiveness of the proposed solution approaches. Later, Liu et al. \cite{liu2016smartadp} performed visual data analytics of large-scale taxi trajectories to place billboards. Wang et al. \cite{wang2019efficiently} studied the problem of selecting $k$ influential billboards using trajectory data to empower audience-targeted billboard advertising. They proposed a noble method to quantify the influence by integrating traffic conditions, speed, etc., and semantic advertisement topics. As this problem is a large-scale combinatorial optimization problem, they employed the popular divide-and-conquer approach and proposed a utility-driven optimal searching approach. Zhang et al. \cite{zhang2019optimizing} studied the problem of optimizing impression count where given a billboard database, trajectory database, and budget, the goal is to select a subset of $k$ billboard slots within the budget that has the maximum influence. Zahradka et al. \cite{zahradka2021price} did a case study analysis of the cost of billboard advertising in the different regions of the Czech Republic.
Recently, Ali et al. \cite{ali2022influential,ali2023influential,ali2024influential} studied the problem of finding Top-$k$ influential billboard slots and proposed a submodularity graph-based pruning technique to prune less influential billboard slots. Further, they extended their work for the same problem and proposed three spatial clustering-based solution methodologies. An extensive set of experiments shows the effectiveness of their proposed approaches. 

\subsection{Regret Minimization}
There are several studies on regret minimization in database query optimization. The $k$-regret queries, which handle decision-making based on several factors, were studied \cite{10.14778/1920841.1920980,6816699,10.1145/3183713.3196903} as a step towards database query optimization. Their approach aims to get around the problems with both the top-$k$ query and skyline queries. The top-$k$ query needs an exact utility function from the user, and the skyline query can't control the size of the result. The purpose of a $k$-regret query is to choose $k$ tuples from a specified database so that, regardless of the exact utility function used, the utility of the user's preferred tuple among these $k$ selections is only slightly less than the usefulness of their preferred tuple in the entire database. Next, Xie et al. \cite{10.1145/3299869.3300068} represent a simple framework for interactive regret minimization and find the most favorable tuple in the database. They proposed two solutions with asymptotically optimal interactions with 2-dimensional space and two with theoretical guarantee in a d-dimensional space where each dimension corresponds to an attribute in a tuple in the database. Later, Aslay et al. \cite{10.14778/2752939.2752950} introduce a new domain of allocating social network users to the advertiser to minimize the total regret of an influence provider. They have proposed a greedy approach with an upper bound on the regret. Their real-world datasets experimentation shows the scalability and quality of the solutions compared to the baseline methods. There are few studies in the context of billboard advertising that consider the minimization of regret that is caused by providing influence to the advertiser. In this direction, Zhang et al. \cite{zhang2021minimizing} studied the problem of regret minimization in billboard advertisements and proposed several solution methodologies. Experimentation with real-world datasets showed the efficiency of the proposed solution methodologies. Further, Ali et al. \cite{ali2023efficient} introduce a problem of regret minimization of an influence provider in the billboard advertisement. They present an efficient algorithm to minimize the regret under a multi-advertiser setting.

\section{Background and Problem Definition} \label{Sec:BPD}
In this section, we describe the background and formally define our problem. Consider $\ell$ number of billboards $\mathcal{B}=\{b_1, b_2, \ldots, b_{\ell}\}$ are placed at different locations of a city. Each billboard $b_i \in \mathcal{B}$ runs for the duration $[T_1,T_2]$. As mentioned previously, digital billboards are leased by the E-Commerce house for a fixed time slot. For simplicity, we assume that all the billboards' slots have the same duration, and $\Delta$ denotes it. We call each $\Delta$ duration as a billboard slot. Now, we formally state what a billboard slot is in Definition \ref{Def:Slot}.

\begin{definition}[Billboard Slot] \label{Def:Slot}
A billboard slot is defined as a tuple of the from $(b_i,[t,t+\Delta])$ where $b_i \in \mathcal{B}$ and $t \in \{T_{1}, T_{1}+\Delta+1, T_{1}+2\Delta+1, \ldots, \frac{T_2-T_1}{\Delta}-\Delta+1 \}$.
\end{definition}
Given a set of billboards $\mathcal{B}$ and the working duration $[T_1, T_2]$, we denote the set of all the billboard slots as $\mathbb{BS}$. Given a subset of billboard slots, we aim to calculate its influence. First, we have defined trajectory and billboard databases, which are stated in Definition \ref{Def:Trajectory} and \ref{Def:Billboard}, respectively.

\begin{definition}[Trajectory Database] \label{Def:Trajectory}
A trajectory $\mathcal{D}$ contains location information about a number of persons and each tuple of the following form: $<\texttt{P\_id}, \ \texttt{L\_id}, \texttt{time\_stamp}>$ which signifies that person with the id $\texttt{P\_id}$ was there at the location with the id $\texttt{L\_id}$ at the  
 $\texttt{time\_stamp}$.
\end{definition}

\begin{definition}[Billboard Database] \label{Def:Billboard}
A billboard database contains billboard information, including their location, slot duration, cost, and so on. So, each tuple of this database will be of the following form: $<\texttt{B\_id}, \ \texttt{L\_id}, \texttt{Slot\_Duration},$ $ \texttt{Slot\_Cost}>$. 
\end{definition}

Given the information about trajectory and billboard database, for a given subset of billboard slots $\mathcal{S} \subseteq \mathbb{BS}$, we denote its influence as $I(\mathcal{S})$ and can be computed in different ways. In this study, we consider the influence is happening by the rule of triggering model \cite{kempe2003maximizing}.
\begin{definition} [Influence of Billboard Slots]\label{Def:4}
Given a subset of billboard slots $\mathcal{S}$, a trajectory database $\mathcal{D}$, and the influence of $\mathcal{S}$ under the triggering model can be computed using the following equation 
\begin{equation} \label{Eq:1}
I(\mathcal{S})= \underset{t_j \in \mathcal{D}}{\sum} 1- \underset{bs_i \in \mathbb{BS}}{\prod} \ (1-Pr(bs_i,t_j))
\end{equation}
where $Pr(bs_i,t_j)$ denotes the influence probability of the billboard slot $bs_i$ on the trajectory $t_j$.
\end{definition}
In this study, we assume that for all $bs_{i} \in \mathbb{BS}$ and $t_{j}\in \mathcal{D}$, the value of $Pr(bs_i,t_j)$ is known. However, there are standard methods to calculate this value \cite{ali2022influential,zhang2020towards,zhang2021minimizing}.

Consider there are $n$ advertiser $\mathcal{A}=\{a_1, a_2, \ldots, a_n\}$ and one influence provider $\mathcal{X}$. Each advertiser $a_i \in \mathcal{A}$ submits a campaign proposal to $\mathcal{X}$ as follows: ``The advertiser $a_i$ want an influence of amount $\sigma_{i}$ based on payment $u_i$." The influence provider $\mathcal{X}$ has the advertiser database $\mathbb{A}$ whose content are the tuples of the form $(a_{i}, \sigma_{i}, u_{i})$ for all $i=1, \ldots,n$. Now, for any advertiser, the payment rule is as follows: if the influence provided to $a_i \in \mathcal{A}$ is more than or equal to $\sigma_{i}$, then the payment of amount $u_i$ will be made by $a_i$ else a partial payment (e.g., pro-rata basis which will be lesser than the $u_i$). Now, there are a few important points to highlight.

\begin{itemize}
\item The first one is for any advertiser, if the influence provided by $\mathcal{X}$ is less than what is expected, then this is a loss for $\mathcal{X}$ because a partial payment will be received. We call this regret as the \emph{Unsatisfied Regret}.
\item On the other hand, if $\mathcal{X}$ provides more influence than expected, that also leads to a loss for $\mathcal{X}$. The reason is as follows. Assume that for an advertiser $a_i$, the billboard slots that are allocated is $\mathcal{S}_{i}$ by the influence provider $\mathcal{X}$. Now, if $I(\mathcal{S}_{i})> \sigma_{i}$, then $\mathcal{X}$ does not get any extra incentive for the excessive influence amount $(I(\mathcal{S}_{i}) - \sigma_{i})$. Also, a subset of billboard slots from $\mathcal{S}_{i}$ may be allocated to the other advertisers, which may help them reach the required influence. We call this kind of regret the \emph{Excessive Regret}. 
\end{itemize}
Combining the above-mentioned two cases, we define the regret model in Definition \ref{Def:5}.

\begin{definition}[The Regret Model] \label{Def:5}
For any advertiser $a_i \in \mathcal{A}$ if the allocated billboard slots $\mathcal{S}_{i}$ by $\mathcal{X}$ then the associated regret with this allocation is denoted by $\mathcal{R}(\mathcal{S}_{i})$ can be given by the following conditional equation:
\[
    \mathcal{R}(\mathcal{S}_{i})= 
\begin{cases}
    u_{i} \cdot (1- \gamma \cdot \frac{I(\mathcal{S}_{i})}{\sigma_{i}}),& \text{if } \sigma_{i} > I(\mathcal{S}_{i})\\
    u_{i} \cdot \frac{I(\mathcal{S}_{i})- \sigma_{i}}{\sigma_{i}},              & \text{otherwise}
\end{cases}
\]
\end{definition}
Here, the fraction $\frac{I(\mathcal{S}_{i})}{\sigma_{i}}$ is part of the satisfied influence, and the $\gamma$ is a parameter that is called the penalty ratio due to the unsatisfied demand. The developed solution methodologies are independent of the choice of $\gamma$, which has been explained further in Section \ref{Sec:Experimental_Details} of this paper. So, from the influence provider's point of view, the goal is to find a billboard slot allocation plan to minimize the overall regret. Consider $\mathcal{Y}=\{\mathcal{S}_{1}, \mathcal{S}_{2}, \ldots, \mathcal{S}_{n}\}$ is an allocation of billboard slots to the advertisers. Now, we define the total regret associated with this allocation.

\begin{definition} [Total Regret Associated with an allocation]\label{Def:6}
Given an allocation of the billboard slots $\mathcal{Y}=\{\mathcal{S}_{1}, \mathcal{S}_{2}, \ldots, \mathcal{S}_{n}\}$ the total regret associated with the allocation $\mathcal{Y}$ is denoted by $\mathcal{R}(\mathcal{Y})$ defined as the sum of the regret associated with the individual advertisers.
\begin{equation} \label{Eq:2_Total Regret}
\mathcal{R}(\mathcal{Y})= \underset{\mathcal{S}_{i} \in \mathcal{Y}}{\sum} \ \mathcal{R}(\mathcal{S}_{i})
\end{equation}   
\end{definition}

We denote an optimal allocation of the billboard slots by $\mathcal{Y}^{OPT}$. From the influence providers' point of view, the goal will be to find an optimal allocation, i.e., an allocation that minimizes the total regret. We call this problem the \textsc{Regret Minimization in Billboard Advertisement} Problem. We formally state this problem in Definition \ref{Def:7}.

\begin{definition}[Regret Minimization in Billboard Advertisement Problem] \label{Def:7}
Given the billboard slot information $\mathbb{BS}$, a trajectory database $\mathcal{D}$, and the advertiser database $\mathbb{A}$, the goal of this problem is to create an allocation $\mathcal{Y}=\{\mathcal{S}_{1}, \mathcal{S}_{2}, \ldots, \mathcal{S}_{n}\}$ in which the set of billboard slots $\mathcal{S}_{i}$ are allocated to the advertiser $a_{i}$ such that the total regret is minimized and for one billboard slot is allotted to only one advertiser. Mathematically, this problem can be posed as follows:

\begin{equation}
\mathcal{Y}^{OPT} = \underset{\mathcal{Y}_{i} \in \mathcal{L}(\mathcal{Y})}{argmin} \ \mathcal{R}(\mathcal{Y}_{i})
\end{equation}
Here, the constraint is for all $\mathcal{S}_{i}$, $\mathcal{S}_{j}$ and they need to be disjoint i.e. $\mathcal{S}_{i} \cap \mathcal{S}_{j} = \emptyset$.
\end{definition}
From the computational point of view, this problem can be written as follows:

\begin{center}
\begin{tcolorbox}[title=\textsc{Regret Minimization in Billboard Advertisement} Problem, width=12.5cm] 
\textbf{Input:} The Set of Billboard Slots $\mathbb{BS}$, The Influence Function $I()$, The Trajectory Database $\mathcal{D}$, Advertiser Database $\mathbb{A}$.

\textbf{Problem:} Find out an optimal allocation $\mathcal{Y}^{OPT}=\{\mathcal{S}_{1}, \mathcal{S}_{2}, \ldots, \mathcal{S}_{n}\}$ of the billboard slots that minimizes the overall regret.
\end{tcolorbox}
\end{center}

\begin{definition}[Effective Split of Billboard Slots]\label{Def:8}
Given the billboard slot information $\mathbb{BS}^{'}$ and $\mathbb{BS}^{''}$, a split is said to be an effective split if an allocation $\mathcal{Y}$ from $\mathcal{BS}^{''}_{3}$ where $ \mathcal{BS}^{''}_{3}\subseteq \mathcal{BS}^{''}$ and $\mathcal{Y}=\{\mathcal{S}_{1}, \mathcal{S}_{2}, \ldots, \mathcal{S}_{n}\}$  in which the set of billboard slots $\mathcal{S}_{i}$ are allocated to the advertiser $a_{i}$ such that the total regret $\mathcal{R(Y)}$ is minimized. Mathematically, it can be written as follows:
\begin{equation}
 \mathcal{{Y}^{ES}} = \underset{\mathcal{Y} \in \mathcal{BS}^{'} \cup \mathcal{BS}^{''}_{3}} {argmin} \ \mathcal{R(Y)} 
\end{equation}
\end{definition}

Here, $\mathcal{BS}^{''}$ is divided into three parts in the tree fashion. At first, $\mathcal{BS}^{''}$ is divided into two parts where the first one $\mathcal{BS}^{''}_{1}$, consists [(1/3)* length of $\mathcal{BS}^{''}$]  and the second one $\mathcal{BS}^{''}_{f}$, consists [(2/3)* length of $\mathcal{BS}^{''}$] many number of billboard slots. Again $\mathcal{BS}^{''}_{f}$ is divided into two parts in the same way named $\mathcal{BS}^{''}_{2}$, and $\mathcal{BS}^{''}_{3}$. Now, we can say that $\mathcal{BS}^{''}_{1}$ contains maximum influential, $\mathcal{BS}^{''}_{2}$ is medium influential, and $\mathcal{BS}^{''}_{3}$ is less influential billboard slots as they are already sorted as per influence in descending order. This $\mathcal{BS}^{''}_{3}$ is further used in Algorithm \ref{Algo: Algorithm_Heuristic_3} and Algorithm \ref{Algo: Algorithm_Heuristic_4}.

The following hardness of approximation result of the regret minimization problem in the context of billboard advertisement has been mentioned in \cite{zhang2021minimizing}.
\begin{theorem}
    The Regret Minimization Problem in Billboard Advertisement is \textsf{NP-hard} and \textsf{NP-hard} to approximate within any constant factor.
\end{theorem}

This negative result motivates the development of efficient heuristic solutions for this problem. There are a few heuristic solutions proposed by Zhang et al. \cite{zhang2021minimizing}. This paper describes four heuristic solutions that dominate the existing solution methodologies in the literature.

\section{Proposed Solution Approaches} \label{Sec:PS}
Considering hardness, the Regret Minimization problem says that efficient algorithms w.r.t to optimal regret do not exist unless $P = NP$. In this section, we propose four solution methodologies. The first is the Effective Allocation of the billboard slots to the advertisers. Now, we describe this solution methodology.

\subsection{Effective Allocation Policy (EA)}
In this section, we discuss the Effective Allocation policy of the billboard slots to the advertisers. At first, we sort the advertisers based on budget over demand and subsequently compute the influence of each billboard slot. Next, sorted slots based on `payment per unit influence' are allocated simultaneously as long as they reach the advertiser's influence demand. It will ensure that all ideal billboard slots should not be assigned to the few most budget-effective advertisers, and low-budget-effective advertisers are still unsatisfied or only get less influential
billboard slots. During allocation, when a billboard slot satisfies the condition $(demand < supply)$, then we will not allocate that billboard slot to the advertiser. 
\par Before further describing this procedure, we define the marginal regret of a billboard slot in Definition \ref{Def:12}.

\begin{definition} [Marginal Regret of a Billoard Slot] \label{Def:12}
Given a set of billboard slots $\mathcal{S}$ and a billboard slot $b \in \mathbb{BS} \setminus \mathcal{S}$, the marginal regret of this slot is defined as the difference between regret value when $b$ is not included and when it is included. Mathematically, this has been mentioned in Equation \ref{Eq:6}.

\begin{equation} \label{Eq:6}
\mathcal{R}(b|\mathcal{S})=  \mathcal{R}(\mathcal{S}) - \mathcal{R}(\mathcal{S} \cup \{b\})
\end{equation}
\end{definition}

 Consider, $\mathcal{S}_{i}^{'}$ is the set of billboard slots allocated so far and $\mathcal{S}^{''}$ = $\mathbb{BS} \setminus \mathcal{S}_{i}^{'}$ be the remaining billboard slots. Now, we compute the value of $I(\mathcal{S}_{i}^{'})$ using Equation \ref{Eq:1} and from this it is easy to observe that $I(\mathcal{S}_{i}^{'}) \leq \underset{s \in \mathcal{S}_{i}^{'}}{\sum} \ I(s)$. 

 So, we keep on putting the billboard slots into $\mathcal{S}_{i}^{'}$ based on maximum marginal regret using Equation \ref{Eq:7} to obtain $\mathcal{S}_{i}^{'}$ such that $I(\mathcal{S}_{i}^{'}) \geq \sigma_{i}$. As long as unallocated billboard slots are available, we repeat this process for all the advertisers one by one. Now, we consider the ratio of the marginal regret and the individual influence of the billboard slot.
 
 \begin{equation}\label{Eq:7}
 b^{*} \longleftarrow \underset{b \in \mathbb{BS} \setminus \mathcal{S}^{'}_{i}}{argmax} \frac{ \mathcal{R}(\mathcal{S}) - \mathcal{R}(\mathcal{S} \cup \{b\})}{I(b)}
 \end{equation}


\begin{algorithm}[h!]
\scriptsize
\SetAlgoLined
\KwData{Trajectory Database $\mathcal{D}$, Billboard Slot Information $\mathbb{BS}$, Advertiser Database $\mathbb{A}$, and the Influence Function $I()$.}
\KwResult{  An allocation $\mathcal{Y}$ of the billboard slots that minimizes the total regret}
$\text{initialize} \ \mathcal{Y} \leftarrow \{\mathcal{S}_{1}^{'}, \mathcal{S}_{2}^{'},\ldots,\mathcal{S}_{|\mathcal{A}|}^{'}\} \ \text{of empty lists}$\; 
 \For{$\text{All }a_i \in \mathcal{A}$}{
 $\mathcal{Z}_{i} \longleftarrow \frac{u_i}{\sigma_{i}} $\;
 }
 $\mathcal{A} \longleftarrow \text{Sort the advertisers based on descending order in } \mathcal{Z}$\;
 \For{$\text{All }bs_i \in \mathbb{BS}$}{
 $\text{Calculate } I(bs_i) \text{ using Equation No. }\ref{Eq:1}$\;
 }
 $\mathbb{BS} \longleftarrow \text{Sort the billboard slots based on the individual influence value}$\;

 $c \longleftarrow 1$\;
 \For{$i=1 \text{ to }|\mathcal{A}|$}{ 
 $\mathcal{S}_{i}^{'} \longleftarrow \emptyset$; $I(\mathcal{S}_{i}^{'}) \longleftarrow 0 $\;
 \While{$I(\mathcal{S}_{i}^{'}) < \sigma_{i}$ and $\mathbb{|BS|} \neq \emptyset$}{
 $\mathcal{S}_{i}^{'} \longleftarrow \mathcal{S}_{i}^{'} \bigcup \mathbb{BS}[c]$\;
 $I(\mathcal{S}_{i}^{'}) \longleftarrow I(\mathcal{S}_{i}^{'}) + I(\mathbb{BS}[c])$\;
 $\mathbb{BS} \longleftarrow \mathbb{BS} \setminus (\mathbb{BS}[c])$\;
 $c \longleftarrow c+1$
 
 }
 $\text{Calculate the value of }I(\mathcal{S}_{i}^{'}) \text{ using Equation No. }\ref{Eq:1}$\;
 $\sigma^{'}_{i} \longleftarrow \sigma_{i} - I(\mathcal{S}_{i}^{'})$\;
}

\For{$i=1 \text{ to }|\mathcal{A}|$}{   
 \While{$I(\mathcal{S}^{'}_{i}) < \sigma^{'}_{i}$ and  $\mathbb{|BS|} \neq \emptyset$}{
$b^{*} \longleftarrow \underset{b \in \mathbb{BS} \setminus \mathcal{S}^{'}_{i}}{argmax} \frac{ \mathcal{R}(\mathcal{S}) - \mathcal{R}(\mathcal{S} \cup \{b\})}{I(b)}$\;
$\mathcal{S}^{'}_{i} \longleftarrow \mathcal{S}^{'}_{i} \cup \{b\}$\; $\mathbb{BS} \longleftarrow \mathbb{BS} \setminus \{b\}$
}}

$\text{Return} \ \mathcal{Y} \leftarrow \{\mathcal{S}_{1}^{'}, \mathcal{S}_{2}^{'},\ldots,\mathcal{S}_{|\mathcal{A}|}^{'}\}$\\
\caption{ Effective Allocation Policy of the Billboard Slots for the Regret Minimization Problem}
 \label{Algo: Algorithm_Heuristic_1}
\end{algorithm}

\paragraph{\textbf{Complexity Analysis.}} We analyze this algorithm to understand its time and space requirements. In Line No. $1$ initializing all the $\mathcal{S}_{i}$ together will take $\mathcal{O}(n)$ time. Now, in Line No. $2$ to $4$ for all the advertisers, computing their payment per unit influence will take $\mathcal{O}(n)$ time. Sorting the advertisers based on this value will take $\mathcal{O}(n \log n)$ time in Line no $5$. It is easy to observe from Equation No. \ref{Eq:1} that for any billboard slot $bs_i \in \mathbb{BS}$ computing the value of $I(bs_i)$ will take $\mathcal{O}(t)$ time where $t$ is the number of tuples in the trajectory database. Assume that the number of billboard slots in the database $\mathbb{BS}$ is $\ell$. Hence, the time requirement from Line No. $6$ to $8$ will be $\mathcal{O}(\ell \cdot t)$. Sorting the billboard slots based on the individual influence value will take $\mathcal{O}(\ell \log \ell)$ time. Now, it is easy to observe that the \texttt{for loop} at Line No. $11$ will iterate $\mathcal{O}(n)$ times. The initialization statements at Line No. $12$ will take $\mathcal{O}(1)$ time. Now, it is easy to observe that in the worst case, the \texttt{while loop} at Line No. $13$ will execute at most the number of billboard slots in $\mathbb{BS}$ which is of $\mathcal{O}(\ell)$. All the statements from Line No. $14$ to $17$ will take $\mathcal{O}(1)$ time. As the size of $\mathcal{S}^{'}_{i}$ can be $\mathcal{O}(\ell)$ in the worst case, so computing the influence of the billboard slots in Line No. $19$ using Equation No. \ref{Eq:1} will take $\mathcal{O}(\ell \cdot t)$ time. Statement in Line No. $20$ will take $\mathcal{O}(1)$ time to execute. So, the total time required between Line No. $11$ to $21$ is $\mathcal{O}(n \cdot \ell \cdot t )$. It is clear that \texttt{For loop} in Line No. $22$ will take $\mathcal{O}(n)$ time and the \texttt{while loop} of Line No. $23$ will also iterate $\mathcal{O}(\ell \cdot t)$ times as to calculate regret we need to calculate influence for $\ell$ number of slots and all the remaining statements in Line no. $25$ and $26$ will take $\mathcal{O}(1)$  time. Hence, the total time required between Line No. $22$ to $28$ is $\mathcal{O}(n \cdot \ell \cdot t )$. So, the total time requirement by Algorithm \ref{Algo: Algorithm_Heuristic_1} will be of $\mathcal{O}(n + n \log n + t \cdot \ell + \ell \log \ell + n\cdot \ell \cdot t)=\mathcal{O}(n \log n + \ell \log \ell + n .\ell. t)$. 

The extra space consumed by Algorithm \ref{Algo: Algorithm_Heuristic_1} to store the lists $\mathcal{S}_{i}$ and $\mathcal{S}^{'}_{i}$, the list $\mathcal{Z}$, and the additional column in the billboard database to store the individual influence. It is easy to observe that in the worst case, the size of $\mathcal{S}^{'}_{i}$ will be of  $\mathcal{O}(\ell)$. Hence, the space requirement for storing these lists will be $\mathcal{O}(n \cdot \ell)$. The space requirement to store the list $\mathcal{Z}$ and the extra column in the billboard database will be of $\mathcal{O}(n)$ and $\mathcal{O}(\ell)$, respectively. Hence, the total space requirement will be of $\mathcal{O}(n \cdot \ell + n + \ell)=\mathcal{O}(n \cdot \ell)$. So, Theorem \ref{Th:4} holds. 

\begin{theorem} \label{Th:4}
Time and space requirement by Algorithm \ref{Algo: Algorithm_Heuristic_1} is of $\mathcal{O}(n \log n + \ell \log \ell + n .\ell. t)$ and $\mathcal{O}(n \cdot \ell)$, respectively.
\end{theorem}

\paragraph{\textbf{An Illustrative Example.}} Assume an influence provider owns ten slots, $\mathbb{BS}=\{ b_1, b_2,$ $ \ldots, b_{10}\}$ with corresponding individual influence for each slot as reported in Table \ref{ETable:1}. Three advertisers, $\mathcal{A}= \{a_1, a_2, a_3\}$ approach to the influence provider for their required influence demand $I$ and corresponding budget $L$ as reported in Table \ref{ETable:2}. To satisfy the advertisers, the influence provider allocates a set of billboard slots using initial allotment, and the results are represented in Table \ref{ETable:3}. From Table \ref{ETable:3}, it is observed that all the advertisers are satisfied, and possible allocations are $a_{1}=\{b_{4}\}$, $a_{2}=\{b_{1}, b_{3}\}$, $a_{3}=\{b_{2}, b_{6}, b_{8}\}$. But, in practice, for any subset of billboard slots, their aggregated influence will always be less than or equal to their individual influence sum. Therefore, the influence provided to the advertisers will be less than the required, and the remaining required influence will be supplied using marginal regret, as represented in definition \ref{Def:12}. Finally, all the advertisers are satisfied, and the allocation of the slot is $a_{1}=\{b_{4}\}$, $a_{2}=\{b_{1}, b_{3}, b_{7}\}$, $a_{3}=\{b_{2}, b_{5}, b_{6}, b_{8}, b_{9}\}$. From Table \ref{ETable:4}, it is clear that some excessive regret (ER) occurs in the case of advertisers $a_{2}$ and $a_{3}$. Still, no unsatisfied regret (UR) occurs as billboard providers have sufficient slots.

\begin{table}
\begin{subtable}[c]{0.5\textwidth}
\begin{center}
    \begin{tabular}{| c | c | c | c | c | c | c | c | c | c | c |}
    \hline
    $\mathcal{BS}_{i}$ & $b_{1}$ & $b_{2}$ & $b_{3}$ & $b_{4}$ & $b_{5}$ & $b_{6}     $ & $b_{7}$ & $b_{8}$ & $b_{9}$ & $b_{10}$ \\ \hline
    $I(b_{i})$ & 4 & 6 & 5 & 7 & 3 & 2 & 3 & 2 & 3 & 3 \\ \hline
    \end{tabular}
    \subcaption{\label{ETable:1} Billboard Influence.}
\end{center}
\end{subtable}
\begin{subtable}[c]{0.5\textwidth}
\begin{center}
    \begin{tabular}{ | c | c | c | c |}
    \hline
    $\mathcal{A}$ & $a_{1}$ & $a_{2}$ & $a_{3}$ \\ \hline
    $\mathcal{BS}_{i}$ & $b_{4}$ & $b_{1},b_{3}$ & $b_{2},b_{6},b_{8}$ \\ \hline
    I($\mathcal{BS}_{i}$)- $I_{i}$ & 0 & 0 & 0 \\ \hline
    Satisfied & Yes & Yes & Yes \\ \hline
    \end{tabular}
    \subcaption{\label{ETable:3}Initial Allotment.}
\end{center}
\end{subtable}

\begin{subtable}[c]{0.5\textwidth}
\begin{center}
\begin{tabular}{ | c | c | c | c |}
\hline
    $\mathcal{A}$ & $a_{1}$ & $a_{2}$ & $a_{3}$ \\ \hline
    $I_{i}$ & 7 & 9 & 10 \\ \hline
    $L_{i}$ & \$12 & \$14 & \$15 \\ \hline
    \end{tabular}
    \subcaption{\label{ETable:2} Advertiser Information.}
\end{center}
\end{subtable}
\begin{subtable}[c]{0.5\textwidth}
\begin{center}
    \begin{tabular}{ | c | c | c | c | c |}
    \hline
    $\mathcal{A}$ & $a_{1}$ & $a_{2}$ & $a_{3}$ \\ \hline
    $\mathcal{BS}_{i}$ & $b_{4}$ & $b_{1},b_{3},b_{7}$ & $b_{2},b_{5},b_{6},b_{8},b_{9}$ \\ \hline
    I($\mathcal{BS}_{i}$)- $I_{i}$ & 0 & 0.3 & 0.65 \\ \hline
    Satisfied & Yes & Yes & Yes \\ \hline
    Regret & ER: No & ER: Yes & ER: Yes \\ \hline
    \end{tabular}
    \subcaption{\label{ETable:4}Final Allotment.}
\end{center}
\end{subtable}
\end{table}

\subsection{Effective Advertiser Driven One-by-One Exchange Policy (EAOE)}
The generation of this local search policy consists of three steps. First, we assign billboard slots to each advertiser using an effective allocation policy described in Algorithm \ref{Algo: Algorithm_Heuristic_1}. Next, we calculate the regret for each advertiser. After that, we perform a one-by-one exchange policy to explore its neighborhood search space by exchanging billboard slots assigned to one advertiser with the set of billboard slots assigned to other advertisers. This approach iteratively selects two advertisers from the advertiser set $\mathcal{A}$ and checks if exchanging assigned billboard slots leads to less regret than the previous allocation, then exchanging billboard slots between them occurs. We can also say that by taking advertisers one by one where $ \ a_{i}  \in \mathcal{A}$ for all $i = 1,2,...,n.$ and for each advertiser $ \ a_{j} \in \mathcal{A} \setminus a_{i},$ if there exists {$\ S_{x} \in a_{i} \ and \ S_{y} \in a_{j}$ such that exchange $[S_{x}, S_{y}]$ will satisfy $Regret^{New}(a_{i}) < Regret^{Old}(a_{i})$} then we exchange billboard slots between $ \ a_{i}$ and $ \ a_{j}$. Otherwise, no exchange occurs, i.e., allocations remain the same. This exchange policy terminates its executions when there are no improvements in regret minimization after exchanging slots between advertisers. We called this local search an Advertiser Driven One-by-One Exchange policy and presented it in Algorithm \ref{Algo: Algorithm_Heuristic_2}.

\begin{algorithm}[h!]
\scriptsize
\SetAlgoLined
\KwData{Trajectory Database $\mathcal{D}$, Billboard Slot Information $\mathbb{BS}$, Advertiser Database $\mathbb{A}$, and the Influence Function $I()$.}
\KwResult{An allocation $\mathcal{Y}$ of the billboard slots that minimizes the total regret}
$\text{Initialize} \ \mathcal{Y} \leftarrow \{\mathcal{S}_{1}, \mathcal{S}_{2},\ldots,\mathcal{S}_{|\mathcal{A}|}\} \ \text{of empty lists}$\; 
$\mathcal{Y}^{New} \leftarrow \text{EffectiveAllocationPolicy}(\mathcal{D}, \mathbb{BS},\mathbb{A}, \mathcal{Y})$\;
\While {True}{
$\mathcal{Y}^{Old} \leftarrow \mathcal{Y}^{New}$
 \For {each $a_{i} \in \mathcal{A}$} {
 \For {each $a_{j} \in \mathcal{A} \setminus \{a_{i}\}$} {
 \If{$ \exists ~ S_{x} \in a_{i},~ S_{y} \in a_{j}~ \text{such that exchange}~ [S_{x}, S_{y}]~ will ~ minimize ~\mathcal{R}(\mathcal{Y}^{Old})$}{
 $a_{i} \leftarrow a_{i} \cup \{S_{y}\}$\;
 $a_{j} \leftarrow a_{j} \cup \{S_{x}\}$\;
 $a_{i} \leftarrow a_{i} \setminus \{S_{x}\}$\;
 $a_{j} \leftarrow a_{j} \setminus \{S_{y}\}$\;
 }}}
\If{ $\mathcal{R}(\mathcal{Y}^{Old}) < \mathcal{R}(\mathcal{Y}^{New}) $}{
$\mathcal{Y}^{New} \leftarrow \mathcal{Y}^{Old}$
}
\Else
{
 Exit;
}}
return $ \mathcal{Y}^{New}$;
\caption{Effective Advertiser driven One-by-One Billboard Slot Exchange Policy for the Regret Minimization Problem}
 \label{Algo: Algorithm_Heuristic_2}
\end{algorithm}

\paragraph{\textbf{Complexity Analysis.}}
Now, we analyze the time and space requirement of Algorithm \ref{Algo: Algorithm_Heuristic_2}. First, in Line No. $1$ initializing $\mathcal{Y}$ will take $\mathcal{O}(n)$. In-Line No. $2$, the `EA' approach will take $\mathcal{O}(n \log n + \ell \log \ell + n .\ell. t)$ time in total as shown in Algorithm \ref{Algo: Algorithm_Heuristic_1}. Next, in Line No $4$ copying $\mathcal{Y}^{Old}$ to $\mathcal{Y}^{New}$ will take $\mathcal{O}(\ell)$ time in the worst case. In-Line No. $5$ \texttt{For Loop} will execute for $\mathcal{O}(n)$ time and \texttt{For Loop} at Line No. $6$ will execute for $(n-1)$ times as total $n$ number of advertisers are there. At Line No. $7$ in the worst case, exchanging slots between $a_{i}$ with $a_{j}$ will take $\mathcal{O}(\ell)$ time and to calculate regret of $\mathcal{Y}^{Old}$ we need to compute the influence of $\ell$ number of slots that will take $\mathcal{O}(n^{2}.\ell.t)$ time where $t$ is the number of tuple in the trajectory database. In Line No. $8$ to $11$ will take $\mathcal{O}(n^{2})$ time to execute. Hence, from Line No. $4$ to $13$ will take $\mathcal{O}(n^{2}.\ell.t + n^{2}.\ell +n^{2})$ time. In Line No. $15$, regret will be calculated twice, and it will take $\mathcal{O}(2.n^{2}.\ell.t)$ time. Finally, in Line No. $16$ copying from $\mathcal{Y}^{Old}$ to $\mathcal{Y}^{New}$ will take $\mathcal{O}(n)$ time. Therefore, Algorithm \ref{Algo: Algorithm_Heuristic_2} will take total $\mathcal{O}(n^{2}.\ell.t + n^{2}.\ell +n^{2} + 2.n^{2}.\ell.t + n^{2})$ i.e., $\mathcal{O}(n^{2}.\ell.t)$ time to execute one iteration of the \texttt{While Loop} in Line no $3$.

Now, the additional space requirement for Algorithm \ref{Algo: Algorithm_Heuristic_2} is for $\mathcal{Y}^{New}$ and $\mathcal{Y}^{Old}$ will be $\mathcal{O}(n)$ and $\mathcal{O}(n)$ respectively. In Line No. $2$ effective allocation will take extra $\mathcal{O}(n \cdot \ell)$ Hence, total space requirement of Algorithm \ref{Algo: Algorithm_Heuristic_2} will be $\mathcal{O}(n + n + n \cdot \ell)$ i.e., $\mathcal{O}(n \cdot \ell)$.

\subsection{Effective Billboard Driven One-by-One Exchange Policy (EBOE)}
In this section, we describe the Effective Billboard Driven One-by-One Exchange policy. When an advertiser holds a large no of billboard slots, the Advertiser Driven One-by-One Exchange policy can escape a local minimum, and exchanging already allocated billboard slots among advertisers could miss a better solution. To overcome the drawback of the Advertiser Driven One-by-One Exchange policy, we introduce a billboard slots allocation policy where instead of exchanging already allocated slots to the advertiser, we exchange already allocated slots with remaining unallocated slots as shown in Algorithm \ref{Algo: Algorithm_Heuristic_3}.

First, we allocate the billboard slots to the advertisers using the Effective Allocation policy shown in Algorithm \ref{Algo: Algorithm_Heuristic_1}. After Effective Allocation, the remaining unallocated billboard slots are divided into three clusters using the binary tree fashion. This tree method divides the total billboard slots into two clusters. In the first cluster $(C_1)$ assigned first $\lfloor$ (1/3)*length of total unallocated billboard slots $\rfloor$ many billboard slots from top to bottom and other cluster $(C_1^{'})$ holds remaining $\lceil$ (2/3)*length of total billboard slots $\rceil$ many billboard slots. Again, the remaining billboard slots in $C_1^{'}$ are divided into two clusters, the first cluster $(C_2)$ assigned $\lfloor$ (1/3)*length of remaining total billboard slots $\rfloor$ many billboard slots from top to bottom and another cluster $(C_3)$ holds remaining billboard slots. Here, one thing needs to be observed as billboard slots are already sorted based on their influence: cluster $C_1$ holds maximum influential, cluster $C_2$ holds medium influential, and cluster $C_3$ holds low influential billboard slots. In Algorithm \ref{Algo: Algorithm_Heuristic_3}, $C_3$ is denoted as $\mathbb{BS}_{3}^{''}$ and only $\mathbb{BS}_{3}^{''}$ is used in One-by-One exchange policy as most influential billboard slots are already allocated during the Effective allocation. Next, we calculate the regret for each advertiser. After that, we perform a one-by-one exchange policy to explore its neighborhood search space by exchanging billboard slots assigned to an advertiser with one billboard slot from the remaining unallocated billboard slots. This approach iteratively selects one advertiser from the advertiser set $\mathcal{A}$ and checks if exchanging an assigned billboard slot with the remaining unallocated slot leads to less regret than the previous allocation, then exchanging billboard slots performs. We can say that for each $ \ a_{i}  \in \mathcal{A}$ and $\ b_{j} \in \mathbb{BS}_{3}^{''}$ if there exists $ \ b_{i} \in a_{i} \ and \ b_{j} \in \mathbb{BS}_{3}^{''}$ such that exchange $[b_{i},b_{j}]$ will satisfy $Regret^{New}(a_{i}) < Regret^{Old}(a_{i})$ then exchange billboard slots between them. This local search terminates when no further improvement in terms of regret minimization occurs. Hence, all advertisers should get a combination of the most influential and less influential billboard slots after the exchange operation.


\begin{algorithm}[h!]
\scriptsize
\SetAlgoLined
\KwData{Trajectory Database $\mathcal{D}$, Billboard Slot Information $\mathbb{BS}$, Advertiser Database $\mathcal{A}$, and the Influence Function $I()$.}
\KwResult{  An allocation $\mathcal{Y}$ of the billboard slots that minimizes the total regret}
$\text{initialize} \ \mathcal{Y} \leftarrow \{\mathcal{S}_{1}, \mathcal{S}_{2},\ldots,\mathcal{S}_{|\mathcal{A}|}\} \ \text{of empty lists}$\; 
$\mathcal{Y}^{New} \leftarrow EffectiveAllocationPolicy(\mathcal{D}, \mathbb{BS},\mathcal{A}, \mathcal{Y})$\\
$\mathcal{S}^{'} \leftarrow$ Assigned Billboard Slots in Allocation $\mathcal{Y}$\;
$\mathbb{BS}^{'} = \mathbb{BS} \setminus \mathcal{S}^{'}$\;
\For{$i = 1 \text{ to } 2$}{
$C_{i} \leftarrow \lceil(1/3)*len(\mathbb{BS}^{'})\rceil$ no. of Billboard Slots from $\mathbb{BS}^{'}$ in descending order\;
$\mathbb{BS}^{'} \leftarrow \mathbb{BS}^{'} \setminus C_{i}$\;
}
$\mathbb{BS}_{3}^{''} \leftarrow \mathbb{BS}^{'}$\;
\While{TRUE}{
$\mathcal{Y}^{Old} \leftarrow \mathcal{Y}^{New}$\;
\For{$each \ a_{i}  \in \mathcal{A}$}{
\For{$each \ b_{j} \in \mathbb{BS}_{3}^{''}$}{
\If {$\exists \ b_{i} \in a_{i} \ and \ b_{j} \in \mathbb{BS}_{3}^{''}$ such that exchange $[b_{i},b_{j}] will ~ minimize ~\mathcal{R}(\mathcal{Y}^{Old})$}{ 
$\mathbb{BS}_{3}^{''} \leftarrow \mathbb{BS}_{3}^{''} \cup \{b_i\}$ \;
$a_i \leftarrow a_i \cup \{b_j\}$ \;

$\mathbb{BS}_{3}^{''} \leftarrow \mathbb{BS}_{3}^{''} \setminus \{b_j\}$ \;
$a_i \leftarrow a_i \setminus \{b_i\}$
}}}
\If{ $\mathcal{R}(\mathcal{Y}^{Old}) < \mathcal{R}(\mathcal{Y}^{New}) $}{
$\mathcal{Y}^{New} \leftarrow \mathcal{Y}^{Old}$
}
\Else
{
 Exit;
}}
return $ \mathcal{Y}^{New}$;
\caption{ Effective Billboard Driven One-by-One Exchange Policy for the Regret Minimization Problem}
 \label{Algo: Algorithm_Heuristic_3}
\end{algorithm}	

\paragraph{\textbf{Complexity Analysis.}}
Now, we analyze the time and space requirement of Algorithm \ref{Algo: Algorithm_Heuristic_3}. First, in Line No. $1$ initializing $\mathcal{Y}$ will take $\mathcal{O}(n)$. In-Line No. $2$, the `EA' approach will take $\mathcal{O}(n \log n + \ell \log \ell + n .\ell. t)$ time in total as shown in Algorithm \ref{Algo: Algorithm_Heuristic_1}. Next, copying assigned billboard slots to $S^{'}$ from $\mathcal{Y}$ will take $\mathcal{O}(n.\ell)$ time as each advertiser can hold $\ell$ many billboard slots in the worst case. In Line $4$ removing already allotted slots $S^{'}$ from the $\mathbb{BS}$ will take $\mathcal{O}(1)$. Now, \texttt{for loop} in Line No. $5$ to $8$ will take $\mathcal{O}(1)$ time. Next, in Line No $11$ copying $\mathcal{Y}^{Old}$ to $\mathcal{Y}^{New}$ will take $\mathcal{O}(\ell)$ time in the worst case. In-Line No. $12$ \texttt{for loop} will execute for $\mathcal{O}(n)$ time and \texttt{for loop} at Line No. $13$ will execute for $\mathcal{O}(\ell)$ times as total $\ell$ number of billboard slots are there in worst case.  At Line No. $14$ in the worst case, exchanging slots between $a_{i}$ with $\mathbb{BS}^{'}$ will take $\mathcal{O}(\ell)$ time and to calculate regret of $\mathcal{Y}^{Old}$ we need to compute the influence of $\ell$ number of slots that will take $\mathcal{O}(n .\ell^{2}.t)$ time where $t$ is the number of tuple in the trajectory database. In Line No. $15$ to $18$ will take $\mathcal{O}(n. \ell)$ time to execute. Hence, from Line No. $12$ to $21$ will take $\mathcal{O}(n.\ell^{2}.t + n.\ell)$ time. In Line No. $22$ will take constant time while Line No. $23$ copying from $\mathcal{Y}^{Old}$ to $\mathcal{Y}^{New}$ will take $\mathcal{O}(n)$ time. Therefore, Algorithm \ref{Algo: Algorithm_Heuristic_3} will take total $\mathcal{O}(n \log n + \ell \log \ell + n .\ell. t + n.\ell + \ell + n + n .\ell^{2}.t + \ell + \ell +\ell +n.\ell^{2}.t + n.\ell + n)$ i.e., $\mathcal{O}(n \log n + \ell \log \ell+n.\ell^{2}.t)$ time to execute one iteration of the \texttt{While Loop} in Line no $10$.

Now, the additional space requirement for Algorithm \ref{Algo: Algorithm_Heuristic_3} is for $S^{'}$, $\mathbb{BS}^{'}$, $C_{i}$, $\mathcal{Y}^{New}$ and $\mathcal{Y}^{Old}$ will be $\mathcal{O}(n)$, $\mathcal{O}(\ell)$, $\mathcal{O}(\ell)$, $\mathcal{O}(n)$ and $\mathcal{O}(n)$ respectively. Additionally, in Line No. $2$ effective allocation space requirement is $\mathcal{O}(n \cdot \ell)$ Hence, total space requirement of Algorithm \ref{Algo: Algorithm_Heuristic_3} will be $\mathcal{O}(n +\ell + \ell + n + n +n \cdot \ell)$ i.e., $\mathcal{O}(n \cdot \ell)$.

\subsection{Effective Billboard Driven One-by-Two Exchange Policy (EBTE)}
In this section, we describe the Effective Billboard Driven One-by-Two Exchange Policy. In advertiser-driven local search \cite{zhang2021minimizing}, releasing all billboards from an advertiser is a coarse-grained optimization. Also, this can escape a local minimum when an advertiser holds many billboard slots. Now, to overcome this drawback, we propose a one-by-two exchange policy where we exchange one billboard slot from an advertiser with two billboard slots from remaining unallocated billboard slots instead of exchanging two sets of billboard slots at a time as proposed in advertiser-driven local search \cite{zhang2021minimizing}. In particular, given an efficient allocation policy, we try to find a neighborhood search space around efficient allocation such that total regret should be minimized, as shown in Algorithm \ref{Algo: Algorithm_Heuristic_4}. Initially, we allocate billboard slots to the advertiser using an effective allocation policy. Next, we split the remaining unallocated billboard slots into three different clusters $C_1$, $C_2$, and $C_3$ as shown in Algorithm \ref{Algo: Algorithm_Heuristic_4}. Now, for each advertiser $ \ a_{i}  \in \mathcal{A}$ and for each set of two billboard slots $\ (b_{x}, b_{y}) \in \mathbb{BS}_{3}^{''}$ if there exists $ \ b_{i} \in a_{i} \ and \ (b_{x},b_{y}) \in \mathbb{BS}_{3}^{''}$ such that exchange $[b_{i},(b_{x}, b_{y})]$ will minimize the total regret then exchange billboard slots between them. Finally, if an individual advertiser's regret is minimized, then total regret is also minimized. This Billboard Driven One-by-Two Exchange Policy will terminate its execution when no improvement can be achieved from further moves.

\begin{algorithm}[h!]
\scriptsize
\SetAlgoLined
\KwData{Trajectory Database $\mathcal{D}$, Billboard Slot Information $\mathbb{BS}$, Advertiser Database $\mathcal{A}$, and the Influence Function $I()$.}
\KwResult{  An allocation $\mathcal{Y}$ of the billboard slots that minimizes the total regret}
$\text{initialize} \ \mathcal{Y} \leftarrow \{\mathcal{S}_{1}, \mathcal{S}_{2},\ldots,\mathcal{S}_{|\mathcal{A}|}\} \ \text{of empty lists}$\; 
$\mathcal{Y}^{New} \leftarrow EffectiveAllocationPolicy(\mathcal{D}, \mathbb{BS},\mathcal{A}, \mathcal{Y})$\\
$\mathcal{S}^{'} \leftarrow$ Assigned Billboard Slots in Allocation $\mathcal{Y}$\;
$\mathbb{BS}^{'} = \mathbb{BS} \setminus \mathcal{S}^{'}$\;
\For{$i = 1 \text{ to } 2$}{
$C_{i} \leftarrow \lceil(1/3)*len(\mathbb{BS}^{'})\rceil$ no. of Billboard Slots from $\mathbb{BS}^{'}$ in descending order\;
$\mathbb{BS}^{'} \leftarrow \mathbb{BS}^{'} \setminus C_{i}$\;
}
$\mathbb{BS}_{3}^{''} \leftarrow \mathbb{BS}^{'}$\;
\While{TRUE}{
$\mathcal{Y}^{Old} \leftarrow \mathcal{Y}^{New}$\;
\For{$each \ a_{i}  \in \mathcal{A}$}{
\For{$each \ \{b_{x}, b_{y}\} \in \mathbb{BS}_{3}^{''}$}{
\If {$\exists \ b_{i} \in a_{i} \ and \ \{b_{x},b_{y}\} \in \mathbb{BS}_{3}^{''}$ such that exchange $[b_{i},\{b_{x}, b_{y}\}]$ $will ~ minimize ~\mathcal{R}(\mathcal{Y}^{Old})$}{ 
$\mathbb{BS}_{3}^{''} \leftarrow \mathbb{BS}_{3}^{''} \cup \{b_i\}$ \;
$a_i \leftarrow a_i \cup \{b_x,b_y\}$ \;

$\mathbb{BS}_{3}^{''} \leftarrow \mathbb{BS}_{3}^{''} \setminus \{b_x,b_y\}$ \;
$a_i \leftarrow a_i \setminus \{b_i\} $
}}}
\If{ $\mathcal{R}(\mathcal{Y}^{Old}) < \mathcal{R}(\mathcal{Y}^{New}) $}{
$\mathcal{Y}^{New} \leftarrow \mathcal{Y}^{Old}$
}
\Else
{
 Exit;
}}
return $ \mathcal{Y}^{New}$;
 \caption{ Effective Billboard Driven One-by-Two Exchange Policy for the Regret Minimization Problem}
 \label{Algo: Algorithm_Heuristic_4}
\end{algorithm}

\paragraph{\textbf{Complexity Analysis.}}
Now, we analyze the time and space requirement of Algorithm \ref{Algo: Algorithm_Heuristic_4}. First, in Line No. $1$ initializing $\mathcal{Y}$ will take $\mathcal{O}(n)$. In-Line No. $2$, the `EA' approach will take $\mathcal{O}(n \log n + \ell \log \ell + n .\ell. t)$ time in total as shown in Algorithm \ref{Algo: Algorithm_Heuristic_1}. Next, copying assigned billboard slots to $S^{'}$ from $\mathcal{Y}$ will take $\mathcal{O}(n.\ell)$ time as each advertiser can hold $\ell$ many billboard slots in the worst case. In Line $4$ removing already allotted slots $S^{'}$ from the $\mathbb{BS}$ will take $\mathcal{O}(1)$. Now, \texttt{for loop} in Line No. $5$ to $8$ will take $\mathcal{O}(1)$ time. Next, in Line No $11$ copying $\mathcal{Y}^{Old}$ to $\mathcal{Y}^{New}$ will take $\mathcal{O}(\ell)$ time in the worst case. In-Line No. $12$ \texttt{for loop} will execute for $\mathcal{O}(n)$ time and \texttt{for loop} at Line No. $13$ will execute for $\mathcal{O}(\ell)$ times as total $\ell$ number of billboard slots are there in worst case.  At Line No. $14$ in the worst case, exchanging slots between $a_{i}$ with $\mathbb{BS}^{'}$ will take $\mathcal{O}(\ell)$ time and to calculate regret of $\mathcal{Y}^{Old}$ we need to compute the influence of $\ell$ number of slots that will take $\mathcal{O}(n .\ell^{2}.t)$ time where $t$ is the number of tuple in the trajectory database. In Line No. $15$ to $18$ will take $\mathcal{O}(n. \ell)$ time to execute. Hence, from Line No. $12$ to $21$ will take $\mathcal{O}(n.\ell^{2}.t + n.\ell)$ time. In Line No. $22$ will take constant time while Line No. $23$ copying from $\mathcal{Y}^{Old}$ to $\mathcal{Y}^{New}$ will take $\mathcal{O}(n)$ time. Therefore, Algorithm \ref{Algo: Algorithm_Heuristic_4} will take total $\mathcal{O}(n \log n + \ell \log \ell + n .\ell. t + n.\ell + \ell + n + n .\ell^{2}.t + \ell + \ell +\ell +n.\ell^{2}.t + n.\ell + n)$ i.e., $\mathcal{O}(n \log n + \ell \log \ell+n.\ell^{2}.t)$ time to execute one iteration of the \texttt{While Loop} in Line no $9$.

Now, the additional space requirement for Algorithm \ref{Algo: Algorithm_Heuristic_4} is for $S^{'}$, $\mathbb{BS}^{'}$, $C_{i}$, $\mathcal{Y}^{New}$ and $\mathcal{Y}^{Old}$ will be $\mathcal{O}(n)$, $\mathcal{O}(\ell)$, $\mathcal{O}(\ell)$, $\mathcal{O}(n)$ and $\mathcal{O}(n)$ respectively. In-Line No. $2$ to execute effective allocation policy will take $\mathcal{O}(n \cdot \ell)$ amount of additional space. Hence, total space requirement of Algorithm \ref{Algo: Algorithm_Heuristic_4} will be $\mathcal{O}(n +\ell + \ell + n + n + n \cdot \ell)$ i.e., $\mathcal{O}(n \cdot \ell)$.

\section{Experimental Details} \label{Sec:Experimental_Details}
This section describes the experimental evaluation of the proposed solution approaches. Initially, we start by describing the datasets.

\subsection{Description of the Datasets}
In this section, we describe the experimental evaluation of the proposed solution methodology. Initially, we start by describing the datasets.\par
The datasets used in our study have also been used by existing studies as well \cite{zhang2020towards}. The trajectory data is obtained from two real-world datasets. The first one is the TLC trip record dataset \footnote{\url{http://www.nyc.gov/html/tlc/html/about/trip_record_data.shtml.}} for NYC, and the second one is the Foursquare check-in dataset \footnote{\url{https://sites.google.com/site/yangdingqi/home.}} Both these datasets contain the records of green taxi trips from Jan 2013 to Sep 2016 and different location types such as Mall, Beach, Bank, and so on. From these trajectory datasets, we separate the tuples corresponding to the following 5 locations: `Beach', `Mall', `Bank', `Park', and `Airport', and create five different datasets. Next, the billboard dataset provided by LAMAR \footnote{\url{http://www.lamar.com/InventoryBrowser.}} is also divided into five different datasets according to location types: `Beach', `Mall', `Bank', `Park', and `Airport'. The basic statistics of this dataset have been listed in Table \ref{table:Dataset}, and here, `\# Rows1' and `\# Rows2' denote the size of the billboard and trajectory, respectively.

Next, we describe different parameter settings along with the experimental details.

\begin{table} [h!]
    \centering
    \caption{Basis Statistics of the Datasets}
    \begin{tabular}{ || c c c c || }
    \hline
    Location Type & \# Rows1 & \# Rows2  & \# Billboard Slots\\ [0.5 ex]
    \hline \hline
    Beach & 76 & 575 & 21888 \\
    Mall & 86 & 1186 & 24768 \\
    Bank & 671 & 2232 &  193248 \\
    Airport & 313 & 2852 & 90144 \\
    Park & 536& 4804 &  154368 \\[1ex]
    \hline
    \end{tabular}
    \label {table:Dataset}
    \end{table}

\subsection{Experimental Setup}
Now, we describe the experimental setup involved in our experiments. 
We consider the influence probability value to be fixed for all our experiments. In the billboard dataset provided by LAMAR, we have information about the panel size of the billboards. From all the billboards, we calculate the influence probability of a billboard as a ratio of that billboard's panel size and the largest billboard panel size. The influence probability of all billboard slots corresponding with that billboard is the same for all. However, there are some standard methods for calculating the value \cite{ali2022influential,zhang2020towards,zhang2021minimizing}. Next, we start to define the cost of the billboard slots.

\subsubsection{Billboard Slot Cost.}
Unfortunately, most billboard advertising companies do not provide the exact cost of billboard slots; instead, they provide a range of costs. So, as reported in the recent studies \cite{zhang2018trajectory,zhang2021minimizing,zahradka2021price}, we calculate the cost of billboard slots by a function proportional to the number of trajectories influenced by the billboard slots denoted as $C(bs_{i}) = \lfloor \delta * I(bs_{i})/10 \rfloor)$, where $\delta$ denotes a factor randomly chosen from 0.9 to 1.1 to simulate various costs and benefit ratio. $I(bs_{i})$ denotes the number of trajectories influenced by a billboard slot $bs_{i}$, where $bs_{i} \in \mathbb{BS}$.

\subsubsection{Performance measurement.}
We evaluate the performance of the existing and proposed method by the run time and the total regret. Total regret consists of two types of regret: unsatisfied regret and excessive regret. Each experiment for all existing and proposed methods is repeated thrice, and the average results are reported in this paper.

\subsubsection{Environment Setup.}
All codes are implemented in Python using the Jupyter Notebook environment, and experiments are conducted on a system with a $3.50$ GHz Intel $24$ Core CPU and $64$ GB memory running the Ubuntu operating system.

\subsection{Key Parameters}
This section will discuss all the parameters used in our experiment. The list of parameters are Demand-Supply ratio $\alpha$, Average-Individual Demand ratio $\mathcal{I}^{ID}$, Advertiser's Demand $\mathcal{I}$, Advertiser's Payment $\mathcal{W}$, Unsatisfied Penalty ratio $\gamma$ and distance $\lambda$. Now, we start describing the Demand-Supply ratio $\alpha$.

\subsubsection{Demand-Supply ratio $\alpha$. } It is the ratio of global demand of the advertisers over the influence provider supply, i.e., $\alpha = \mathcal{I}^{A} / \mathcal{I}^{H}$, where $\mathcal{I}^{A} = \sum_{a\in \mathcal{A} }\mathcal{I}_{a}$, denotes the total demand of all available advertisers and $ \mathcal{I}^{H} = \sum_{bs_{i}\in \mathbb{BS}} \mathcal{I}(bs_{i})$ denotes the supply of influence corresponding to the advertiser's demand. We report our experiment results on different Demand-Supply ratio ($\alpha$) values, i.e., low (40\%), average (60\%), high (80\%), full (100\%), and excessive (120\%).

\subsubsection{Average-Individual Demand ratio $\mathcal{I}^{ID}$. } It refers to the percentage of the average individual demand ($ \mathcal{I}^{ID} $) over the influence provider supply ($\mathcal{I}^{H}$), i.e., p$(\mathcal{I}^{ID}) = \mathcal{I}^{ID} / \mathcal{I}^{H}$, where $\mathcal{I}^{ID} =$ $ \mathcal{I}^{A} / \mathcal{\mathcal{|A|}} $ is the average individual demand. Here, we consider four different cases of p$(\mathcal{I}^{ID})$, i.e., 1\%, 2\%, 5\%, and 10\% to adjust the demand of the advertisers.

\subsubsection{Advertiser's Demand $\mathcal{I}$. } We generate the demand of all the advertisers individually based on $\mathcal{I}_{a} = \lfloor \beta * \mathcal{I}^{H} *  \mathcal{I}^{ID}  \rfloor$, where $\beta$ is the factor randomly chosen from 0.8 to 1.2 to generate payment for individual advertisers. When the Demand-Supply ratio $\alpha$ and Average-Individual Demand ratio p$(\mathcal{I}^{ID})$ are known, then we can calculate the average demand of advertisers. 


\subsubsection{Advertiser's Payment $\mathcal{W}$. } By following recent studies \cite{zhang2021minimizing}, we consider an advertiser's payment to be proportional to the influence demand of that particular advertiser. We calculate payment as $\mathcal{W}_{a} = \lfloor \tau * \mathcal{I}_{a} \rfloor$, where $\tau$ is the factor chosen randomly between 0.9 to 1.1 to generate different payments.

\subsubsection{Unsatisfied Penalty ratio $\gamma$. } 
The unsatisfied penalty ratio ($\gamma$) came into existence when the influence provider could not satisfy the desired influence demand of the advertiser. We assume $\gamma \in [0,1]$, and it controls the fraction of the penalty. Here, two extreme cases can occur, in case one when $\gamma = 0$ influence provider cannot receive any payment due to not fulfilling the advertiser's required demand. In the other case, when $\gamma = 1$, the influence provider can receive an equal fraction of payment as the amount of influence is satisfied to the advertisers. In our experiment, we have considered the default unsatisfied penalty ratio $\gamma = 0.5$.

\subsubsection{Distance $\lambda$. }
With the increase of $\lambda$, the influence of a billboard slot increases as a single billboard slot can influence more trajectories. But a threshold value, $\lambda$, determines the maximum distance a billboard can influence a trajectory. In our experiment, we have considered the distance as $\lambda$ = $100$ meters for calculating the influence of individual billboard slots.

\subsection{Goals of the Experiments}\label{Sec:Goals_of_Exp}
The goals of the experiments are four folded, and they are mentioned below:
\begin{itemize}
\item \textbf{The Effectiveness of Preprocessing}: Finding out the effectiveness of preprocessing and understanding its benefits from computational aspects is the first experimental goal of our study. As the influence function, $I()$ holds the property of submodularity; therefore, removing the billboard slots having zero influence, i.e., that particular slots do not influence any trajectory, will not make any difference in calculating the influence of billboard slots.

\item \textbf{Effectiveness of Splitting}: One of our goals is to study the percentage of the billboard slots pruned out after splitting the remaining unallocated billboard slots into three parts, i.e., most influential, medium influential, and less influential, and it is good if the remaining unallocated number of billboard slots are sufficiently large. Hence, without much computational burden, we can easily apply the Effective Billboard Driven One-by-One and One-by-Two Exchange Policy.
  
\item \textbf{Minimization of Regret}: Another goal of our study is to compare the proposed and existing methods based on regret minimization.
\item \textbf{Computational Time Requirement}: Finally, we aim to compare the computational time among the proposed and existing methods. 
\end{itemize}

\subsection{Algorithms Compared}
We compare the performance of the proposed solution approaches with the following methods.
\begin{itemize}
\item \textbf{Budget Effective Greedy (BG)}:
This is a greedy heuristic algorithm whose working principle is as follows. First, all the advertisers are sorted based on their budget over influence demand value. Then, initialize as many empty sets as the number of advertisers and keep allocating billboard slots that can minimize the regret of that particular advertiser. After satisfying the first advertiser, do the same for all the remaining advertisers until all are satisfied or the influence provider runs out of billboard slots. The main drawback of this greedy algorithm is that most influential billboard slots are exhausted first as these are allocated to the first few advertisers serially, and the rest of the advertisers may not get the ideal billboard slots.

\item \textbf{Synchronous Greedy (SG)}: 
This is the extension of the G-order algorithm introduced by Zhang et al. \cite{zhang2021minimizing}. Here, ideal billboard slots are assigned to each advertiser synchronously. If assigning new billboard slots minimizes the regret, it allocates them to the advertisers whose demand still needs to be satisfied. In the worst case, if there are no billboard slots to assign and more than two advertisers are still unsatisfied, remove less budget-effective advertisers one by one and allocate those released slots to the remaining unsatisfied advertisers until no fewer than two are unsatisfied.

\item \textbf{Advertiser Driven local Search (ALS)}:
The advertiser-driven local search method follows a baseline plan consisting of two steps. First, randomly assign billboard slots to the advertisers. Second, allocate the remaining billboard slots to the advertisers using the Synchronous Greedy method \cite{zhang2021minimizing}. After that, a local search technique is used to explore a neighborhood search space. Later, exchange a set of billboard slots from one advertiser with the set of billboard slots assigned to another if total regret minimizes. 
\end{itemize}

\section{Observations with Explanation}\label{Sec:Experimental_Evaluations}
In this section, we describe the effectiveness of the proposed approach and experimental observations, along with their explanations.

\subsection{Effectiveness Study}

\paragraph{\textbf{The Effectiveness of Preprocessing}}
During our experiments, we observed that in most cases, a significantly large number of billboard slots are removed from $\mathbb{BS}$ after the preprocessing step. The preprocessing step is significant in terms of the computational time requirement. When we consider the location type `Mall' among $24768$ billboard slots, only $938$ billboard slots remain left, and in percentage, more than $96\%$ of billboard slots are removed. These observations are consistent with other location types, e.g., `Beach', `Bank', `Park', and `Airport'. After the preprocessing step for different location types, the number and percentage of removed billboard slots are shown in Table \ref{table:Preprocessing}. Here, $\mathbb{BS}$ denotes the set of billboard slots before preprocessing, and $\mathbb{BS}^{'}$ denotes slots after preprocessing.

\begin{table} [h!]
    \centering
    \caption{Experimental results of the Preprocessing Step}
    \begin{tabular}{ || c c c c || }
    \hline
    Location Type & $|\mathbb{BS}|$ & $|\mathbb{BS}^{'}|$  & Percentage\\ [0.5 ex]
    \hline \hline
    Beach & 21888 & 455 & 97.92 \% \\
    Mall & 24768 & 938 & 96.21 \% \\
    Bank & 193248 & 2949 &  98.47 \% \\
    Park & 90144 & 4446 &  95.06 \% \\
    Airport & 154368 & 8749 & 94.33 \% \\ [1ex]
    \hline
    \end{tabular}
    \label {table:Preprocessing}
    \end{table}

\paragraph{\textbf{The Effectiveness of Splitting}}
As discussed previously, splitting is an important step for the proposed Algorithm \ref{Algo: Algorithm_Heuristic_3} and Algorithm \ref{Algo: Algorithm_Heuristic_4}. Now, for location type `Mall' number of billboard slots, $\mathbb{BS}^{'} $ after the preprocessing step is 938. After that, using $\mathbb{BS}^{'} $ in Algorithm \ref{Algo: Algorithm_Heuristic_3} allocate billboard slots effectively, and the remaining number of billboard slots in $\mathbb{BS}^{''}$ is 674. Now, if we apply `Splitting' on the remaining billboard slots, then the number of billboard slots in low influential billboard slot set $\mathbb{BS}^{''}_{3}$ will be 300. This $\mathbb{BS}^{''}_{3}$ further used in the Algorithm \ref{Algo: Algorithm_Heuristic_3} and Algorithm \ref{Algo: Algorithm_Heuristic_4} to reduce computational burden. These observations are also consistent with other location types, as shown in table \ref{table:Splitting}.

\begin{table} [h!]
    \centering
    \caption{Experimental results related to the Splitting Step}
    \begin{tabular}{ || c c c c c c|| }
    \hline
    Location Type & $|\mathbb{BS}^{'}|$ & $|\mathbb{BS}^{''}|$  & $|\mathbb{BS}^{''}_{3}|$ & Splitting (\%) & Preprocessing+Splitting (\%) \\ [0.5 ex]
    \hline \hline
    Beach & 455 & 254 & 144 & 55.11 \% & 99.47 \% \\
    Mall & 938 & 674 & 300 & 55.48 \%  & 98.78 \% \\
    Bank & 2949 & 1908 & 848 & 55.55 \% & 99.56 \% \\
    Park & 4446 & 3210 & 1427 &  55.54 \% & 98.41 \% \\
    Airport & 8749 & 6559 & 2916 & 55.54 \% & 98.11 \% \\ [1ex]
    \hline
    \end{tabular}
    \label {table:Splitting}
    \end{table}

\paragraph{\textbf{Minimization of Regret}}
Our primary goal in this experimental study is to minimize the total regret from the influence provider's perspective. Hence, billboard slots are allocated so that excessive and unsatisfied regret, i.e., total regret, should be minimized. Now, when we consider the location type `Mall' in the given figure \ref{Fig:Mall}, it is clear that our proposed algorithms outperform existing methods. These observations are consistent with other location types also, as shown in Figure \ref{Fig:Beach},\ref{Fig:Bank},\ref{Fig:Park},\ref{Fig:Airport}.

\subsection{Parameter Study}
In our experiments, we first fix two parameters, unsatisfied penalty ratio $(\gamma)$ and distance $(\lambda)$, as 0.5 and 100 meters, respectively. Then, we evaluate how the varying Demand-Supply ratio $(\alpha)$ and Average-Individual demand ratio $(\mathcal{I}^{ID})$ impact in total regret. In our experiment, we report unsatisfied regret from unsatisfied advertisers and excessive regret over satisfied advertisers as total regret according to Definition No. \ref{Def:5}. Now, one question arises in our mind: if the global demand for advertisers is deficient, close to, or over the influence provider influence supply, then what will happen? We vary the Demand-Supply ratio, $\alpha$, to address this question from $40\%$ to $120\%$. During our experiment, we observed that when the value of $\alpha$ is low, global demand is also low, and all advertisers are satisfied by the influence provider. However, when the value of $\alpha$ is very large, the global demand exceeds the supplied influence by the influence provider. So, some advertisers may not be satisfied due to billboard slots running out. We have also varied Average-Individual Demand ratio, $\mathcal{I}^{ID}$, from $1\%$ to $10\%$ to check which type of advertisers are most profitable for influence provider, i.e., a minimal number of advertisers with high individual influence demand or a vast number of advertisers with small individual influence demands.

\par In the following, we will present the results of the different location types with four cases. For better understanding, we used stacked bars to represent total regret, including excessive and unsatisfied regret.

\begin{figure}[h!]
\centering
    \begin{tabular}{lclc}
       Unsatisfied Regret & \includegraphics[width=0.11\linewidth]{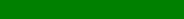} \  & \ Excessive Regret & \includegraphics[width=0.11\linewidth]{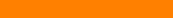} \\
    \end{tabular}
\graphicspath{{/home/user/Dropbox/Second Work/Regreat Analysis in Billboard Advertisement/manuscript/Basic Draft/Beach/}}

\begin{tabular}{cccc}
\includegraphics[scale=0.11]{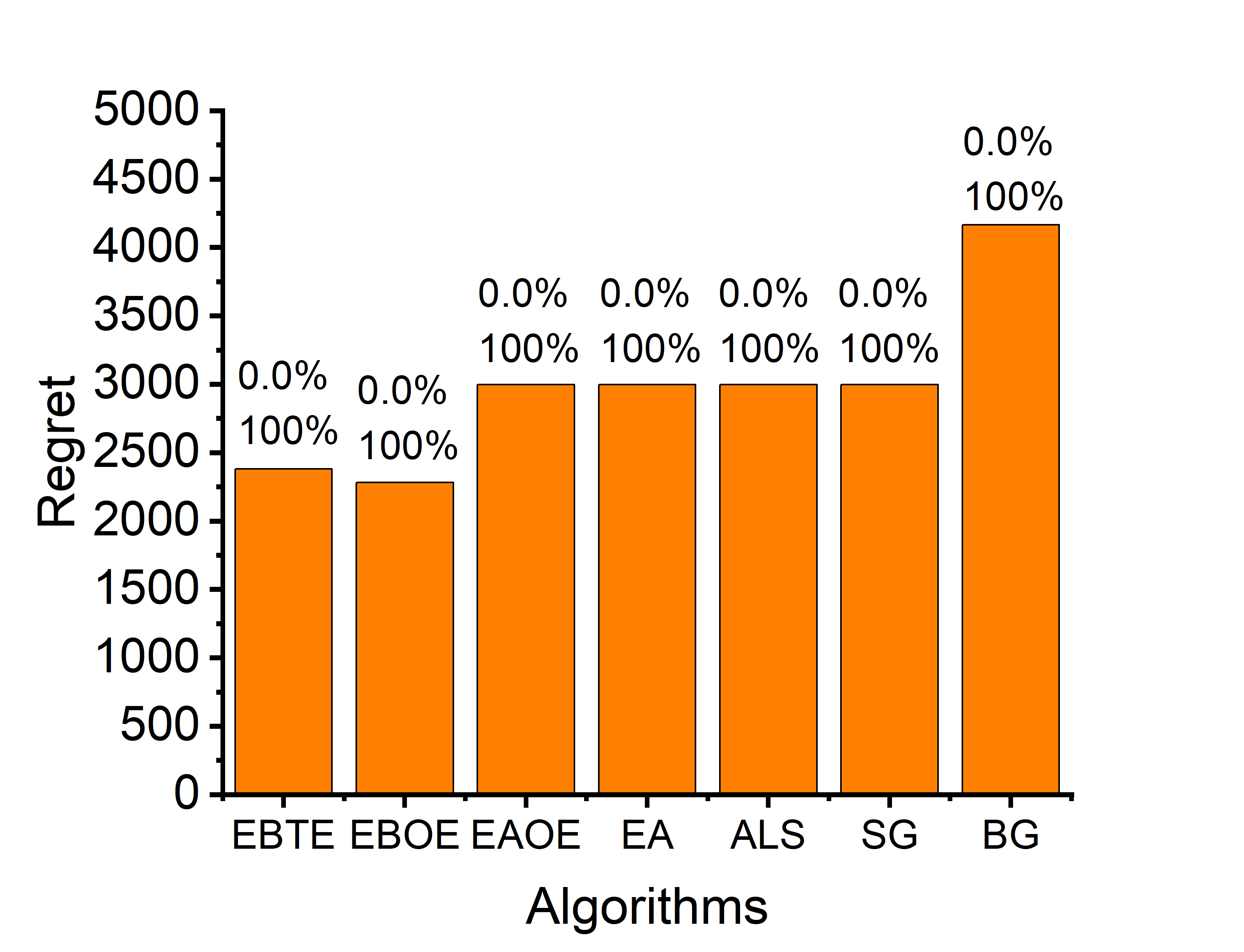} & \includegraphics[scale=0.11]{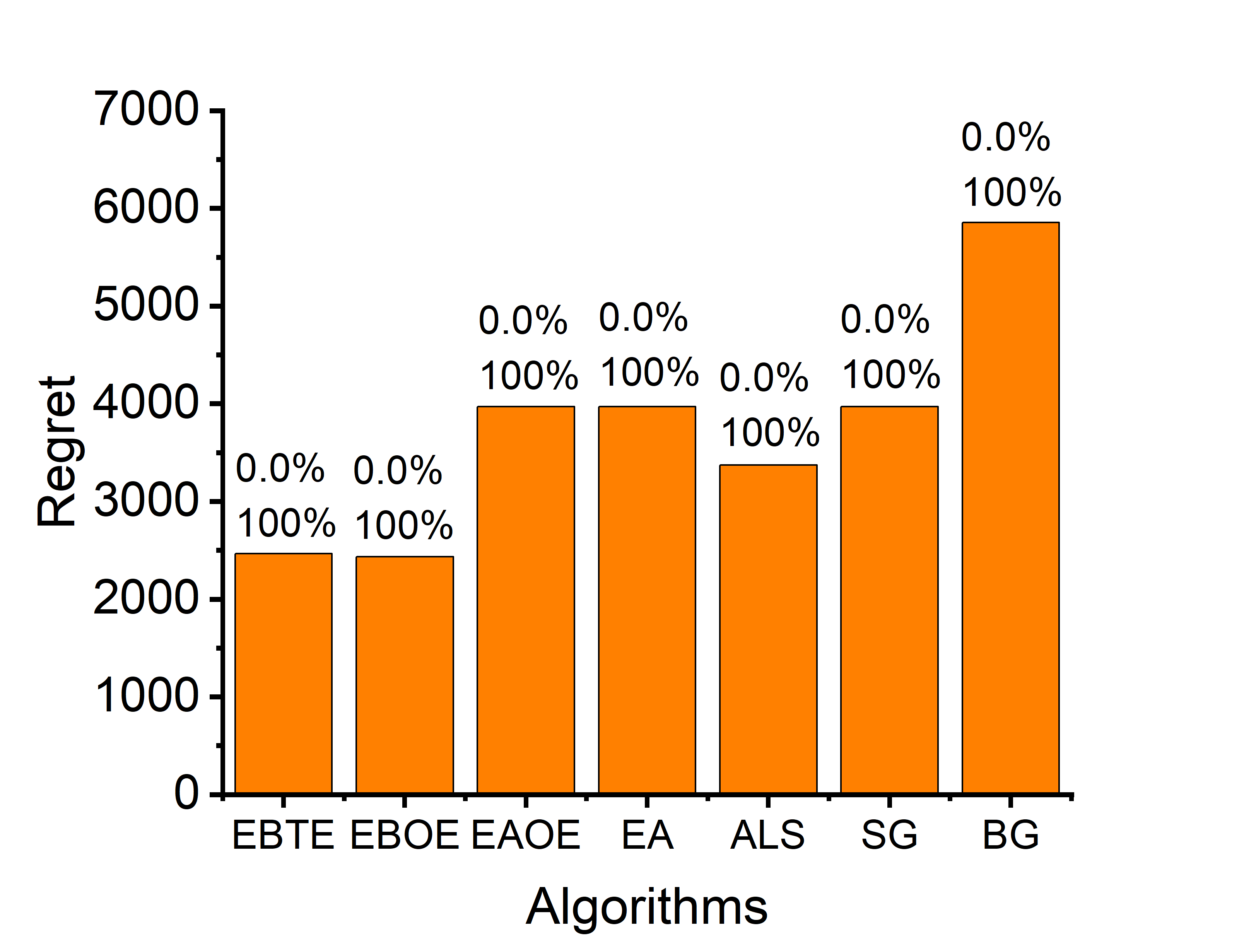}  &\includegraphics[scale=0.11]{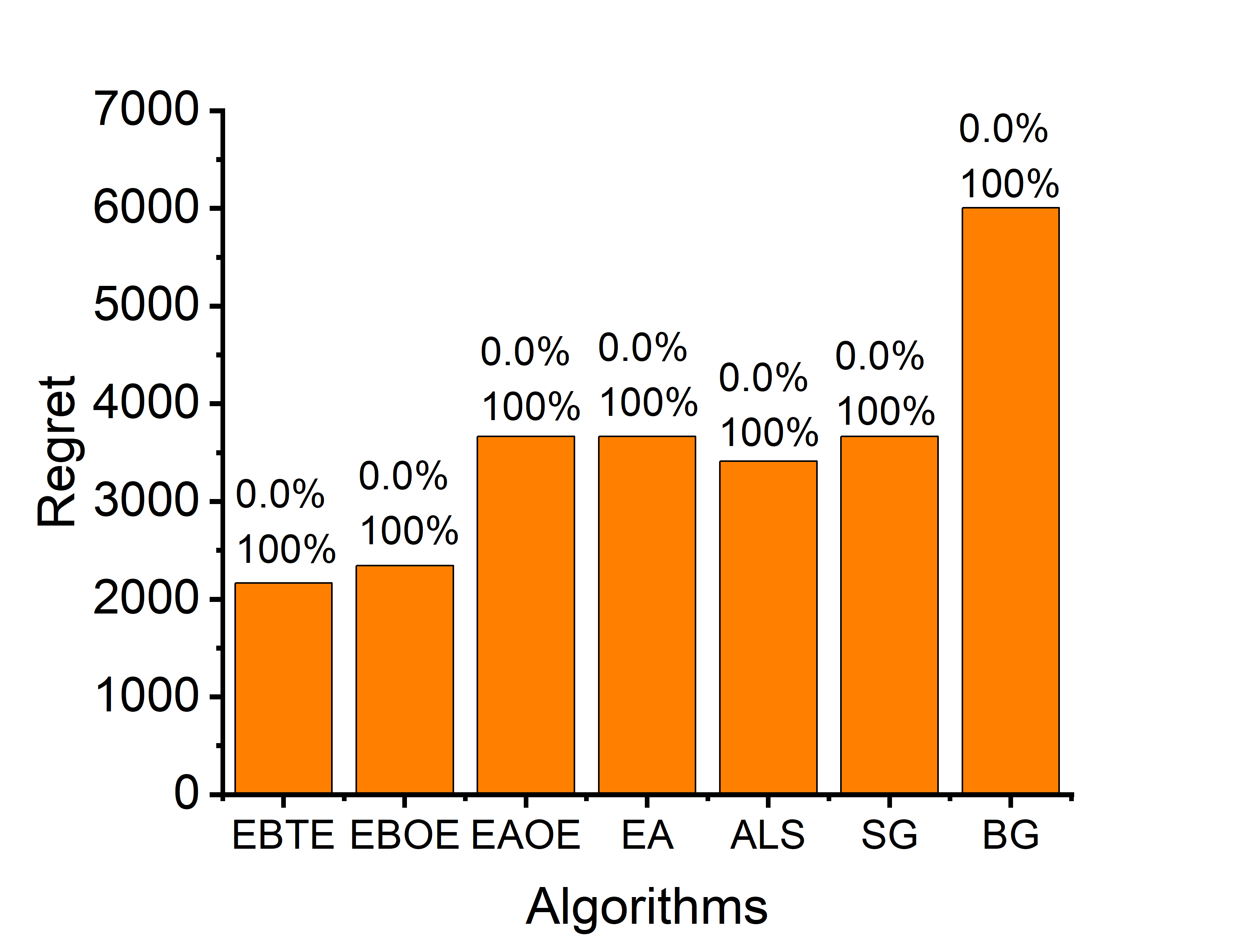} & \includegraphics[scale=0.11]{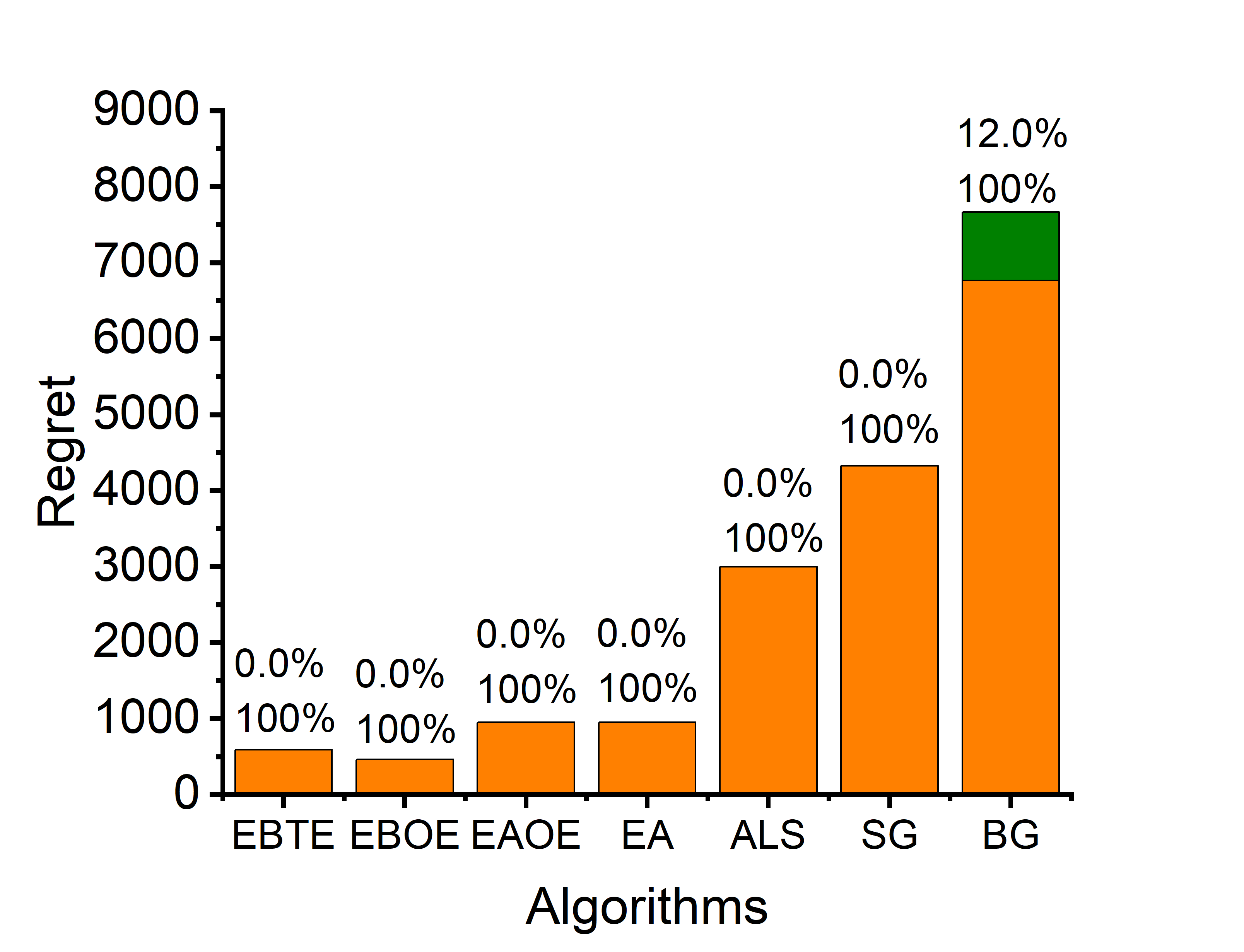} \\
{\tiny (a) $\alpha = 40 \%$} & {\tiny (b) $\alpha = 60 \%$} & {\tiny (c) $\alpha = 80 \%$} & {\tiny (d) $\alpha = 100 \%$} \\


 \includegraphics[scale=0.11]{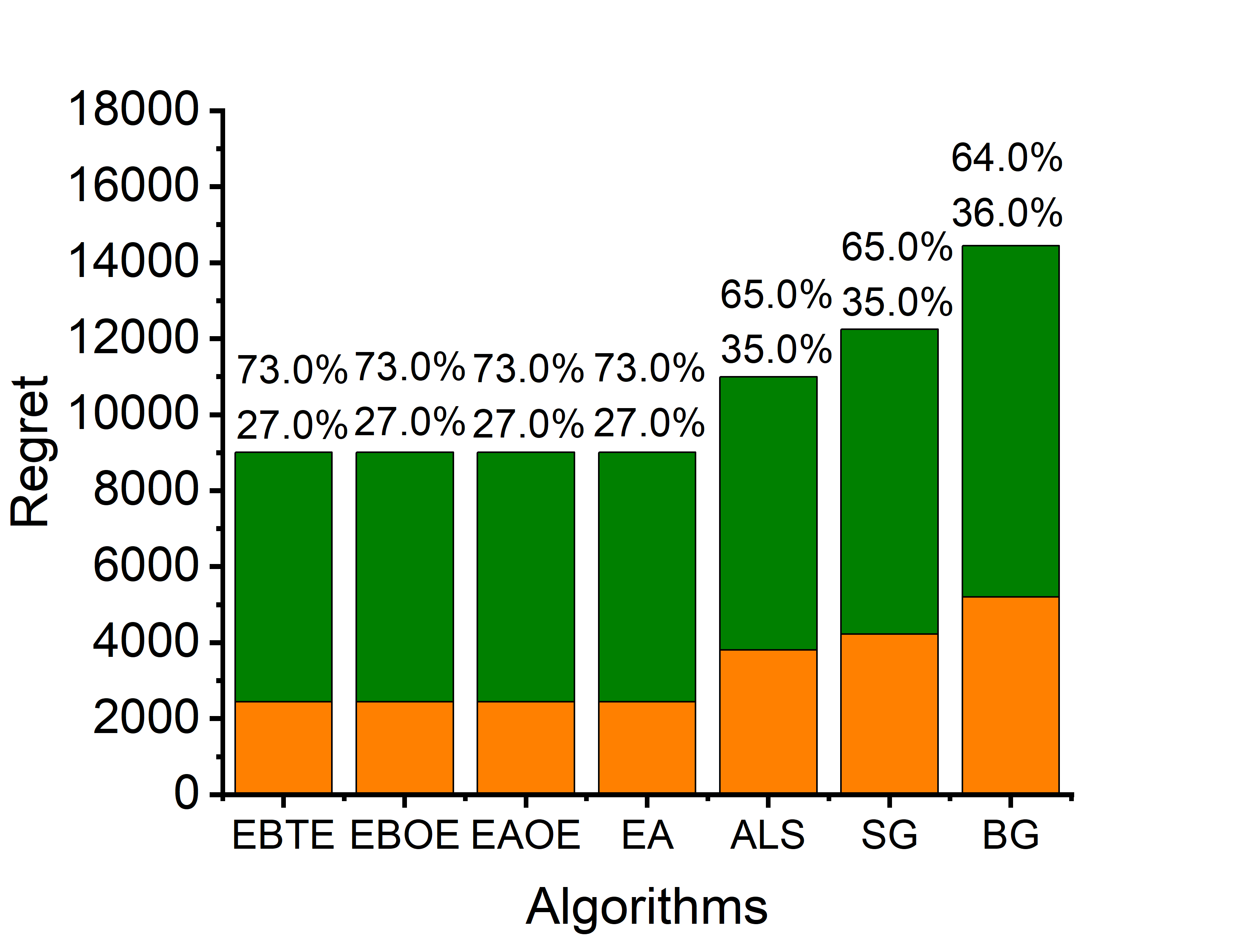} & \includegraphics[scale=0.11]{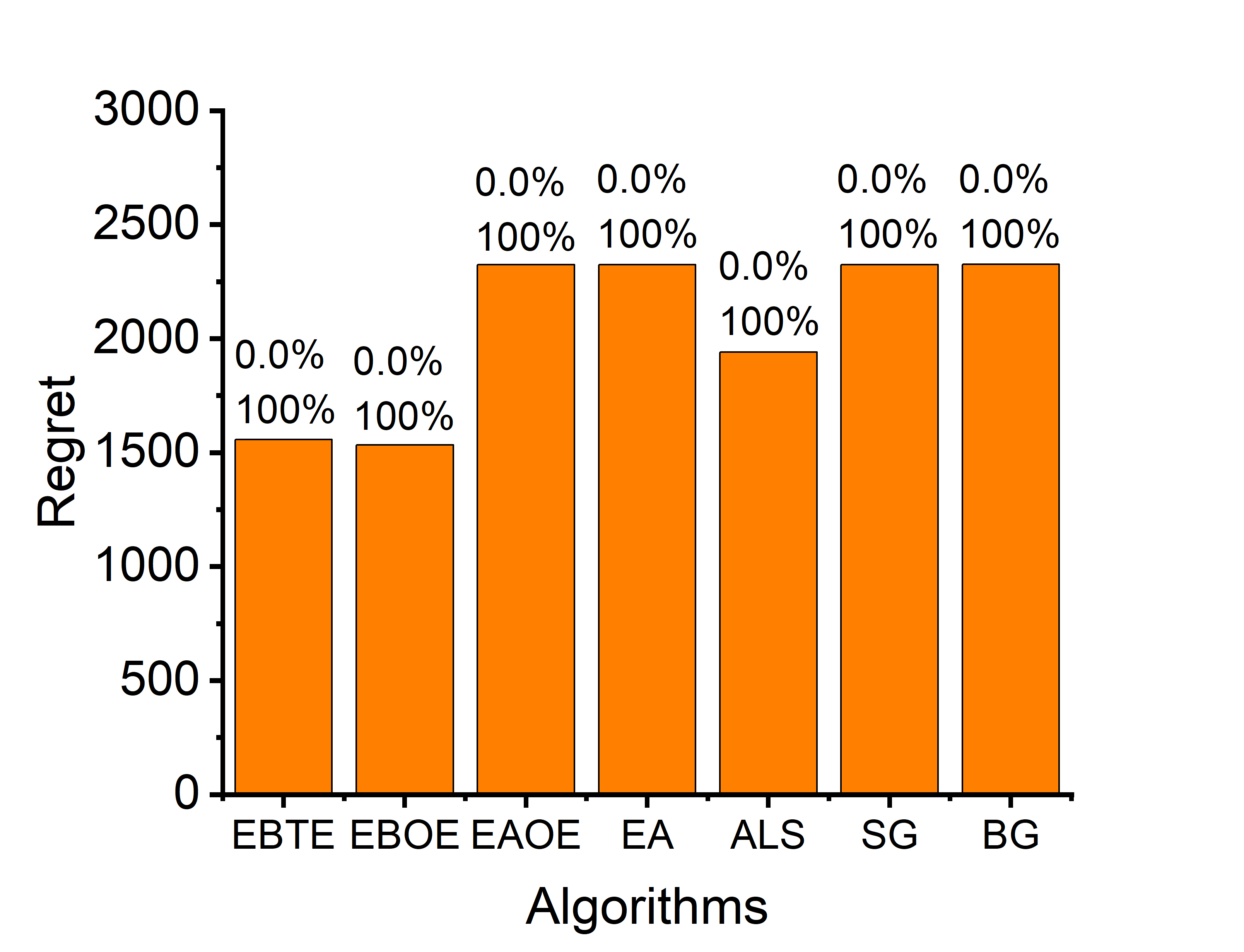} & \includegraphics[scale=0.11]{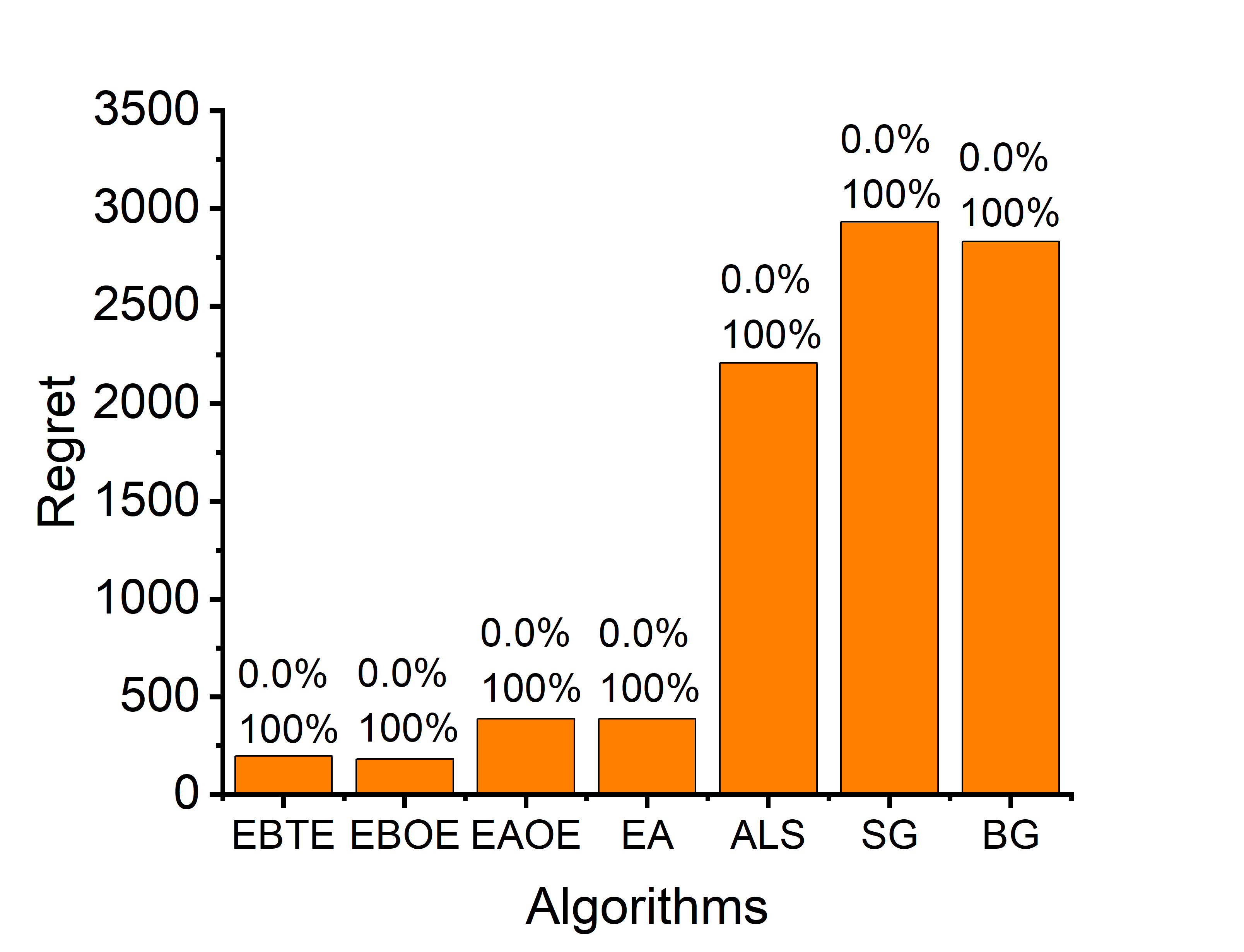}  &\includegraphics[scale=0.11]{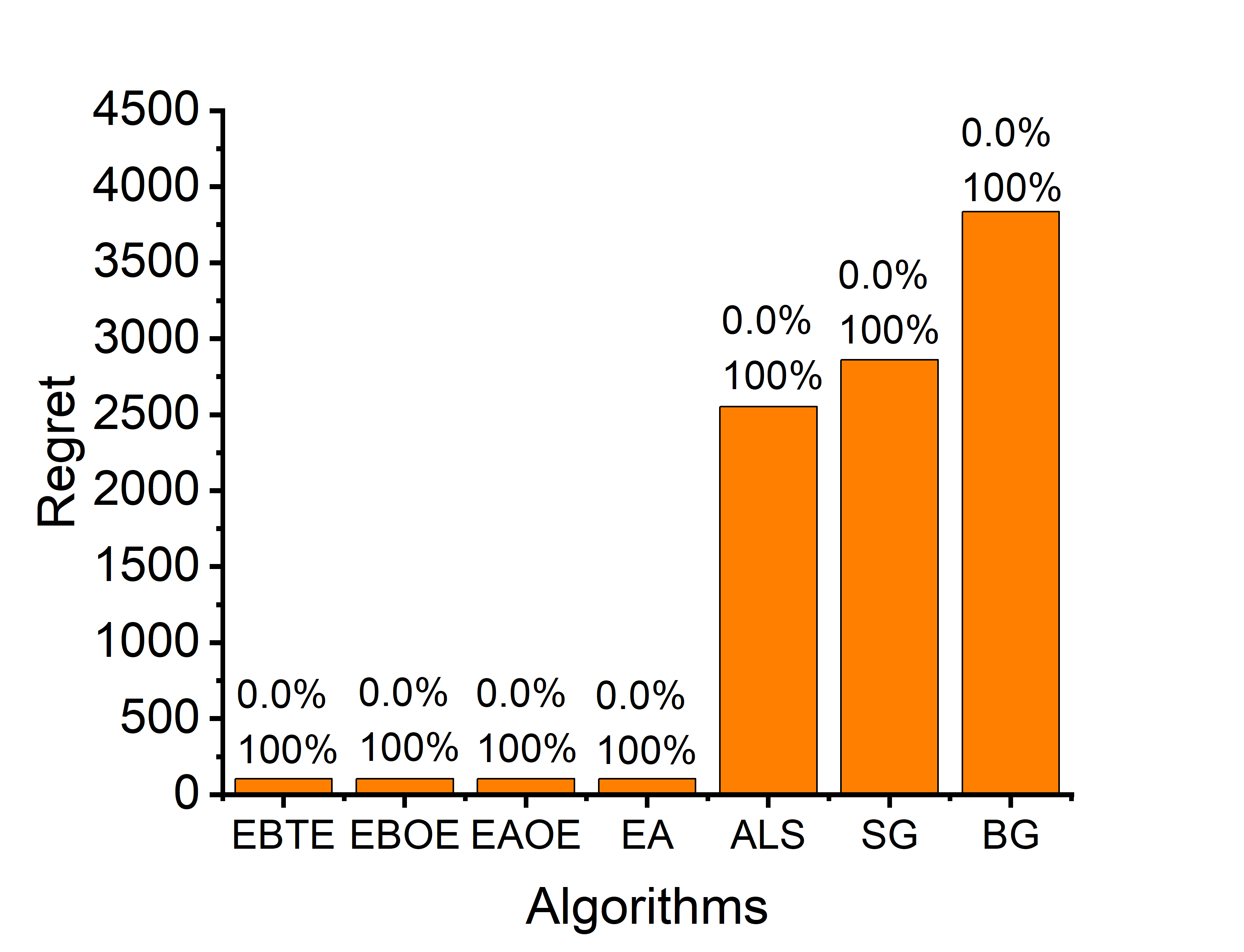} \\
{\tiny (e) $\alpha = 120 \%$} & {\tiny (f) $\alpha = 40 \%$} & {\tiny (g) $\alpha = 60 \%$} & {\tiny (h) $\alpha = 80 \%$} \\

 \includegraphics[scale=0.11]{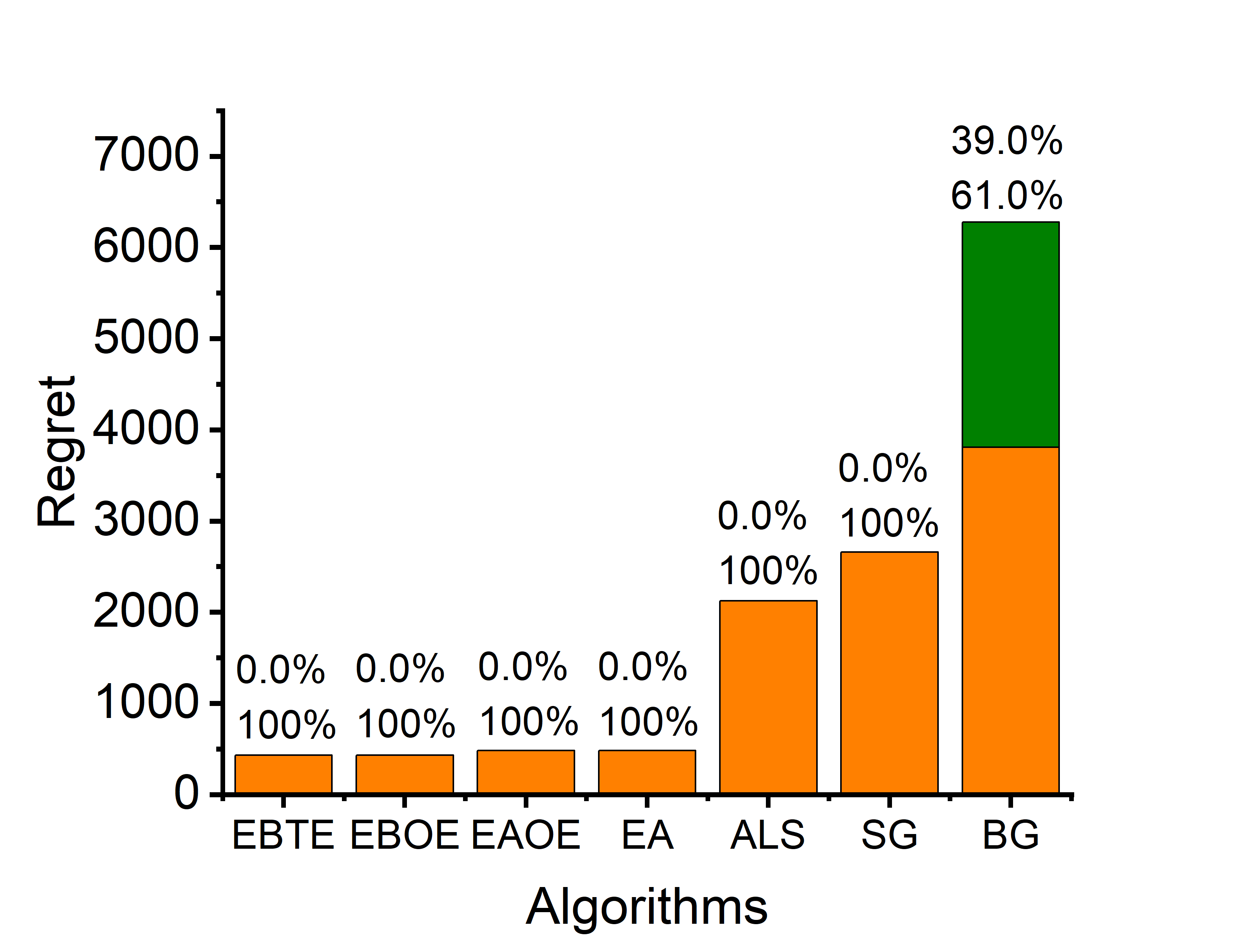} & \includegraphics[scale=0.11]{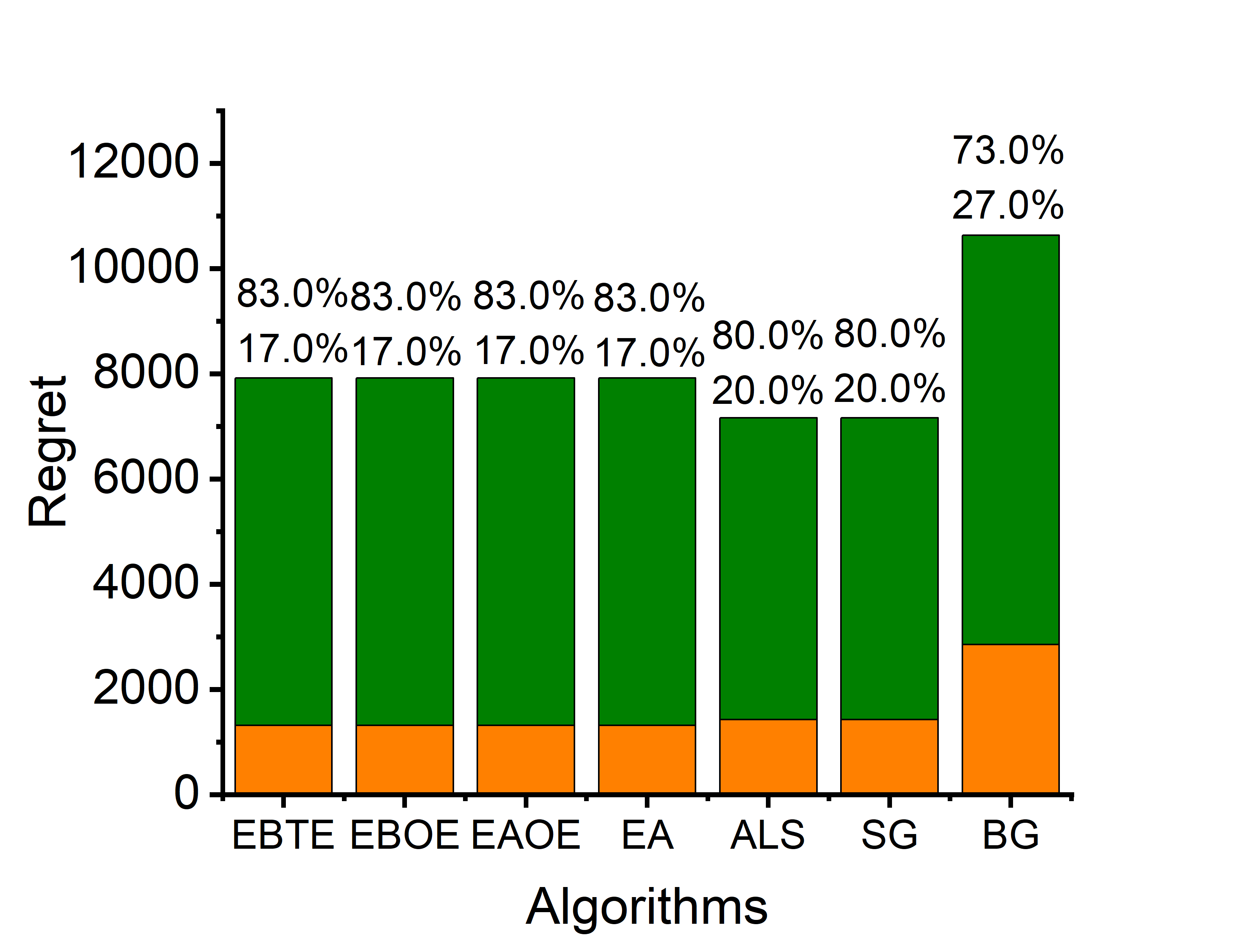} &
\includegraphics[scale=0.11]{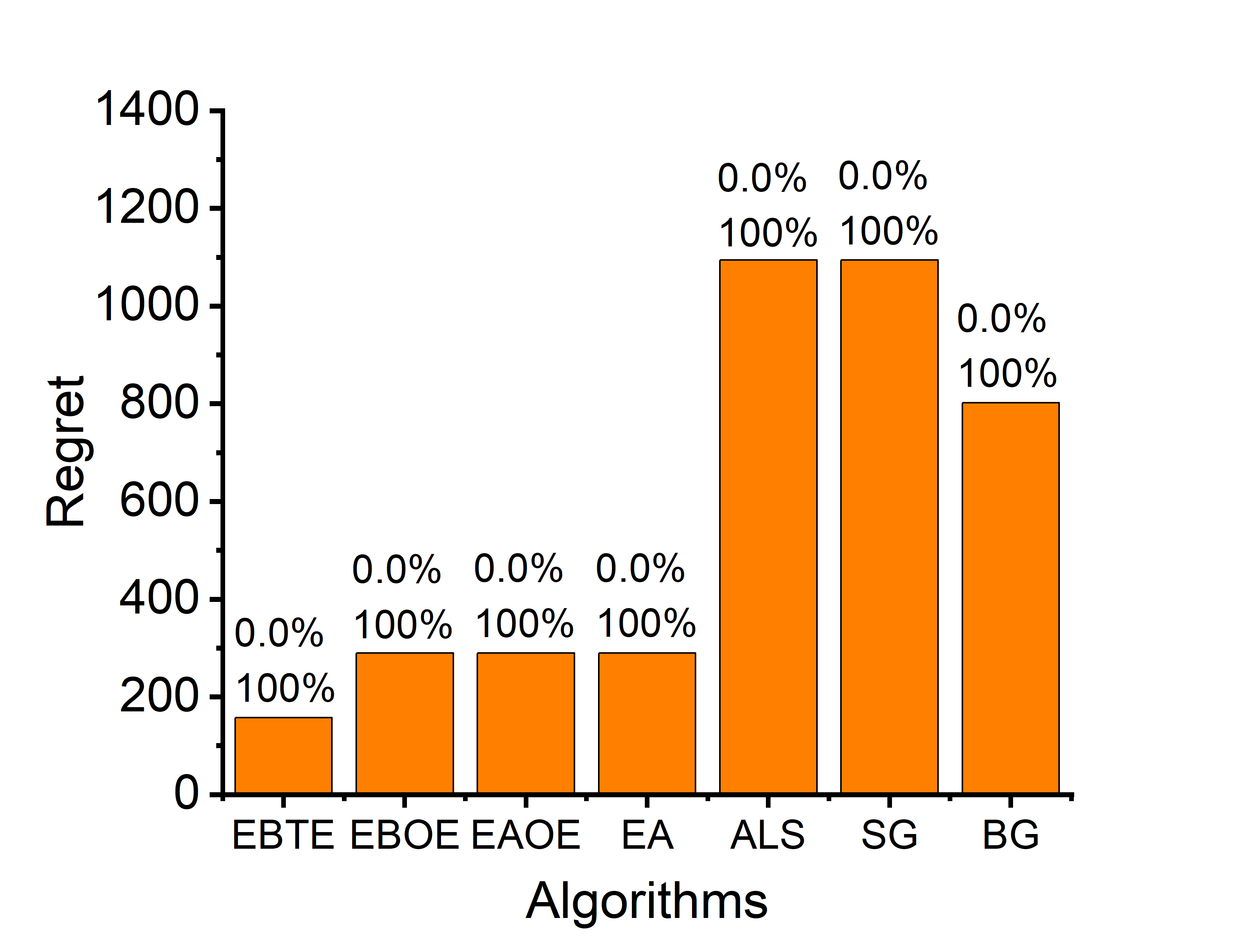} & \includegraphics[scale=0.11]{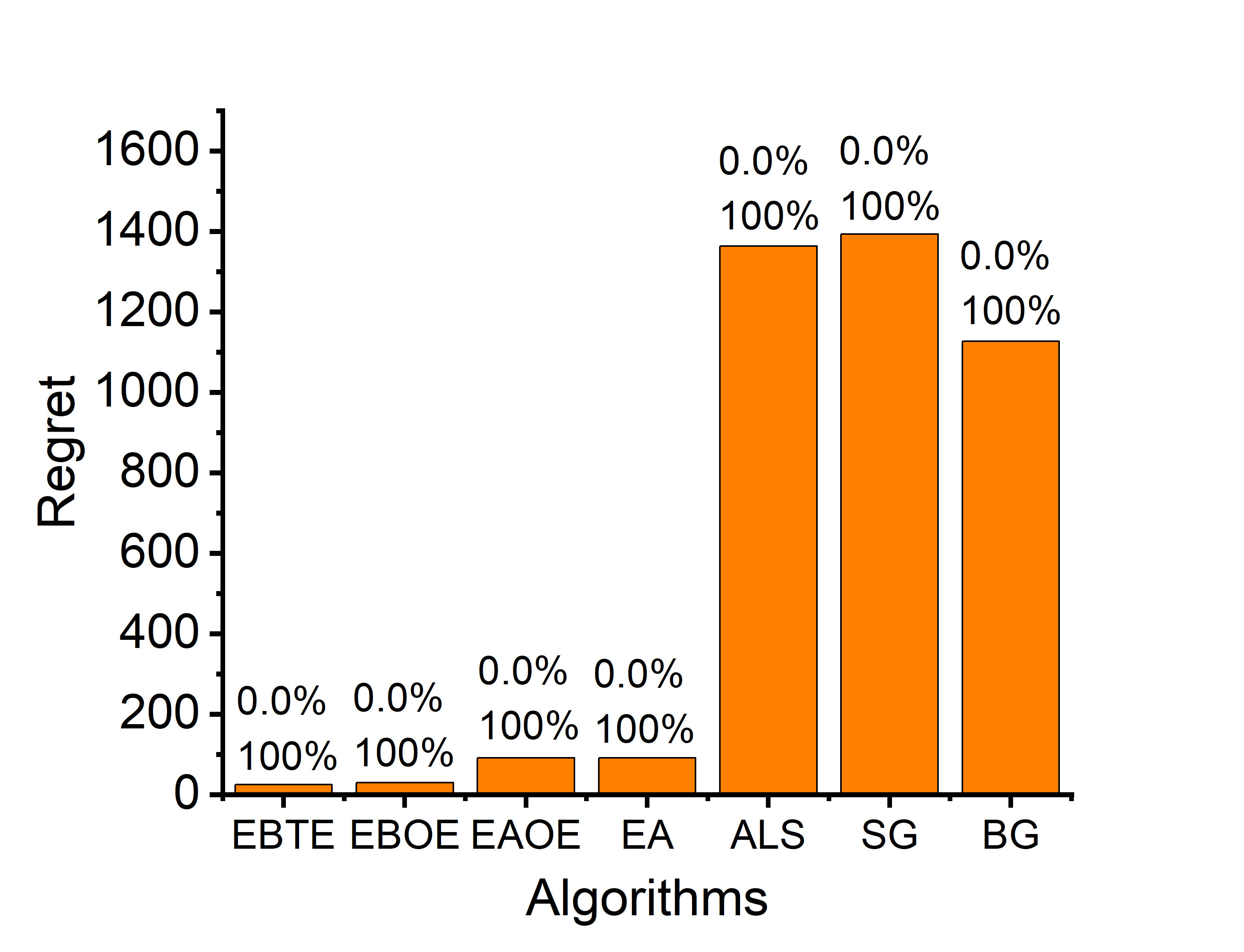} \\
{\tiny (i) $\alpha = 100 \%$} &{\tiny (j) $\alpha = 120 \%$} & {\tiny (k) $\alpha = 40 \%$} & {\tiny (l) $\alpha = 60 \%$} \\

\includegraphics[scale=0.11]{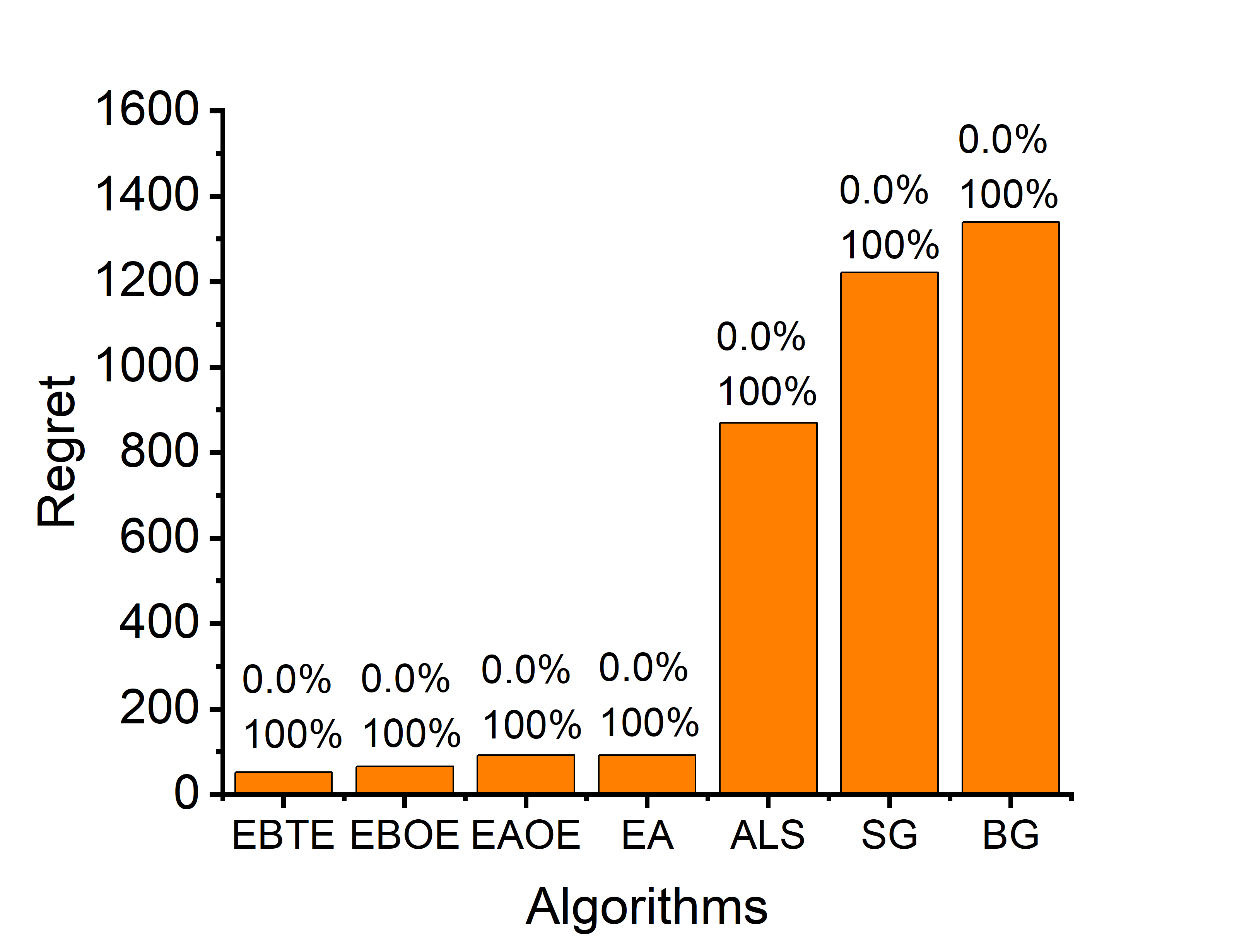} & \includegraphics[scale=0.11]{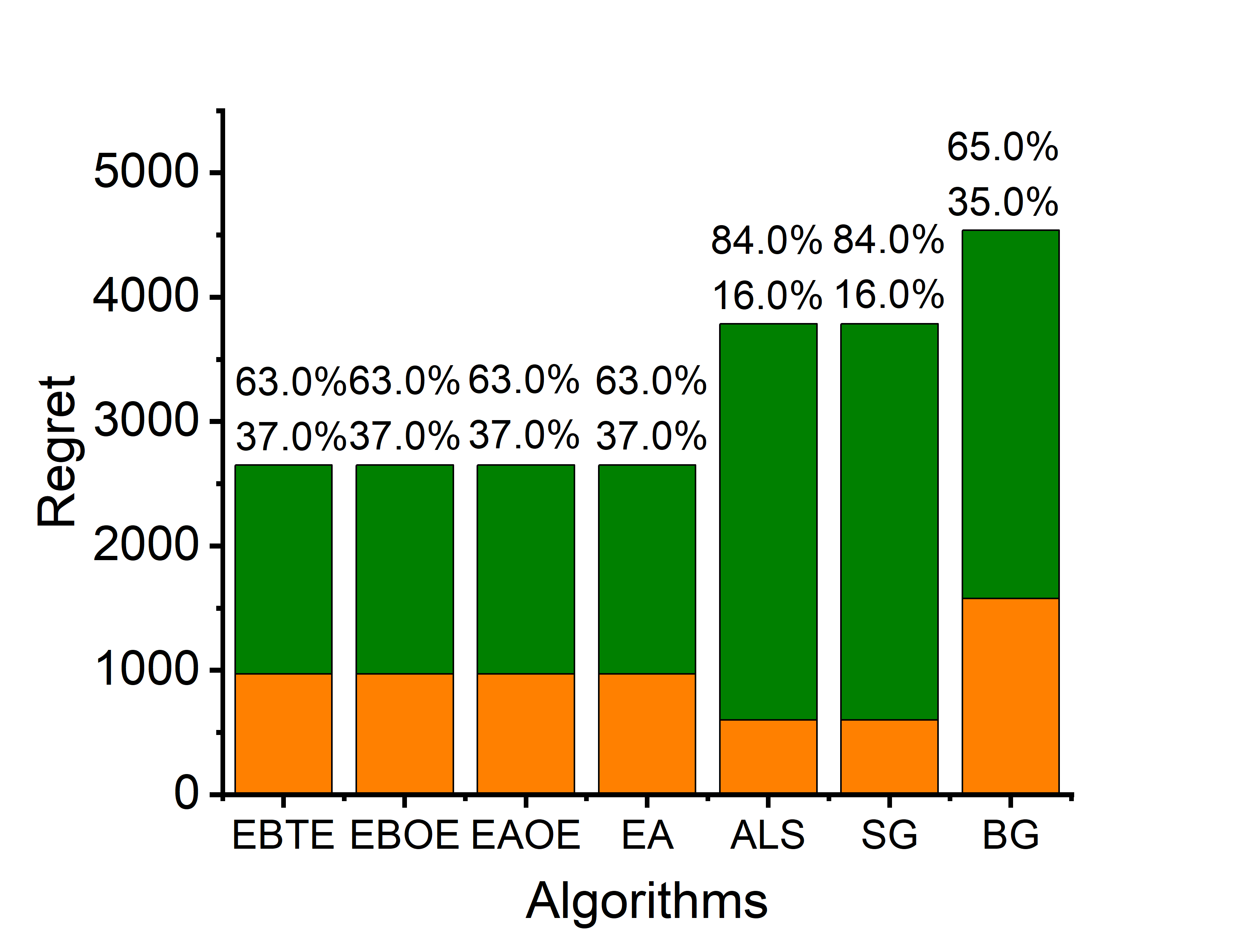} & \includegraphics[scale=0.11]{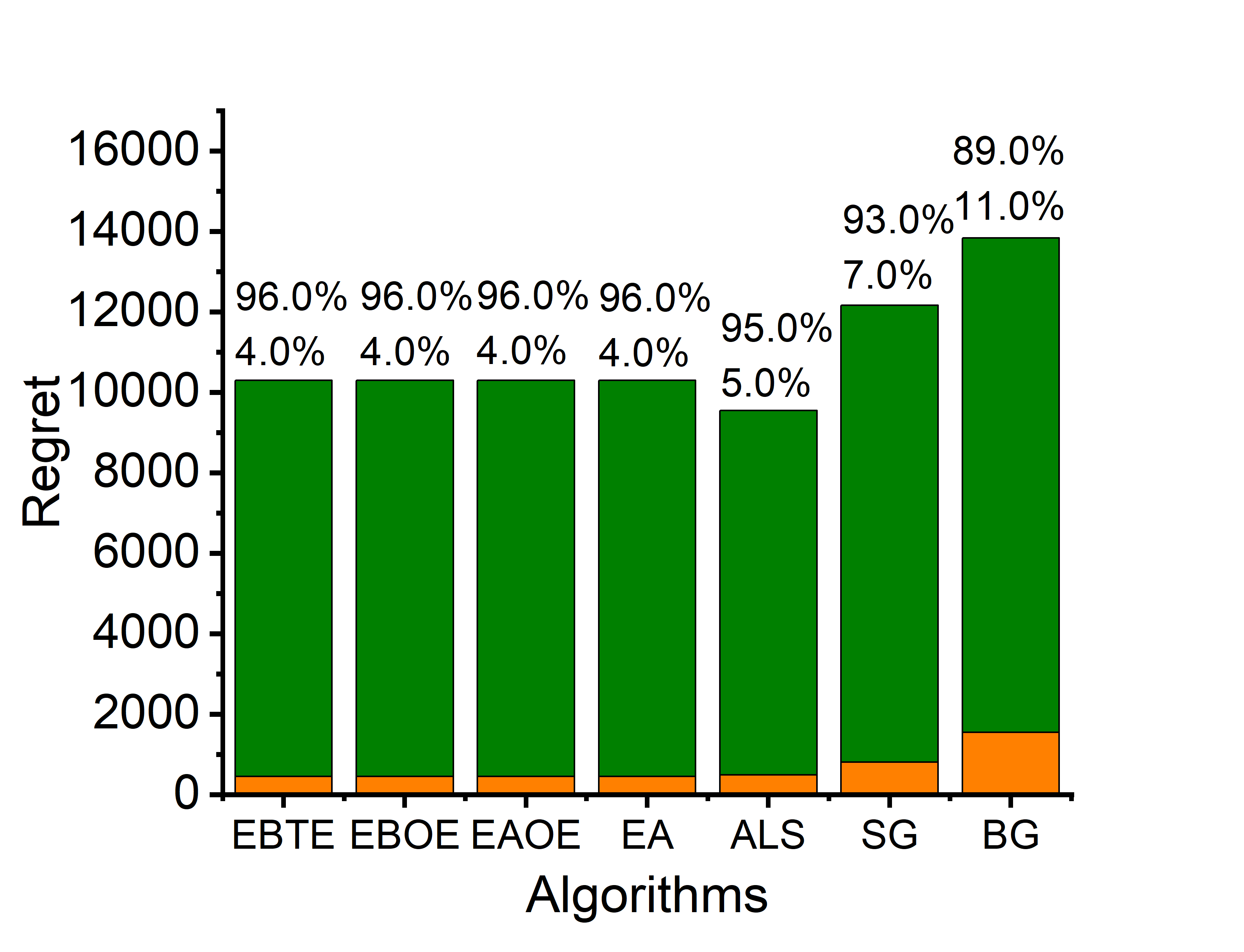} &
\includegraphics[scale=0.11]{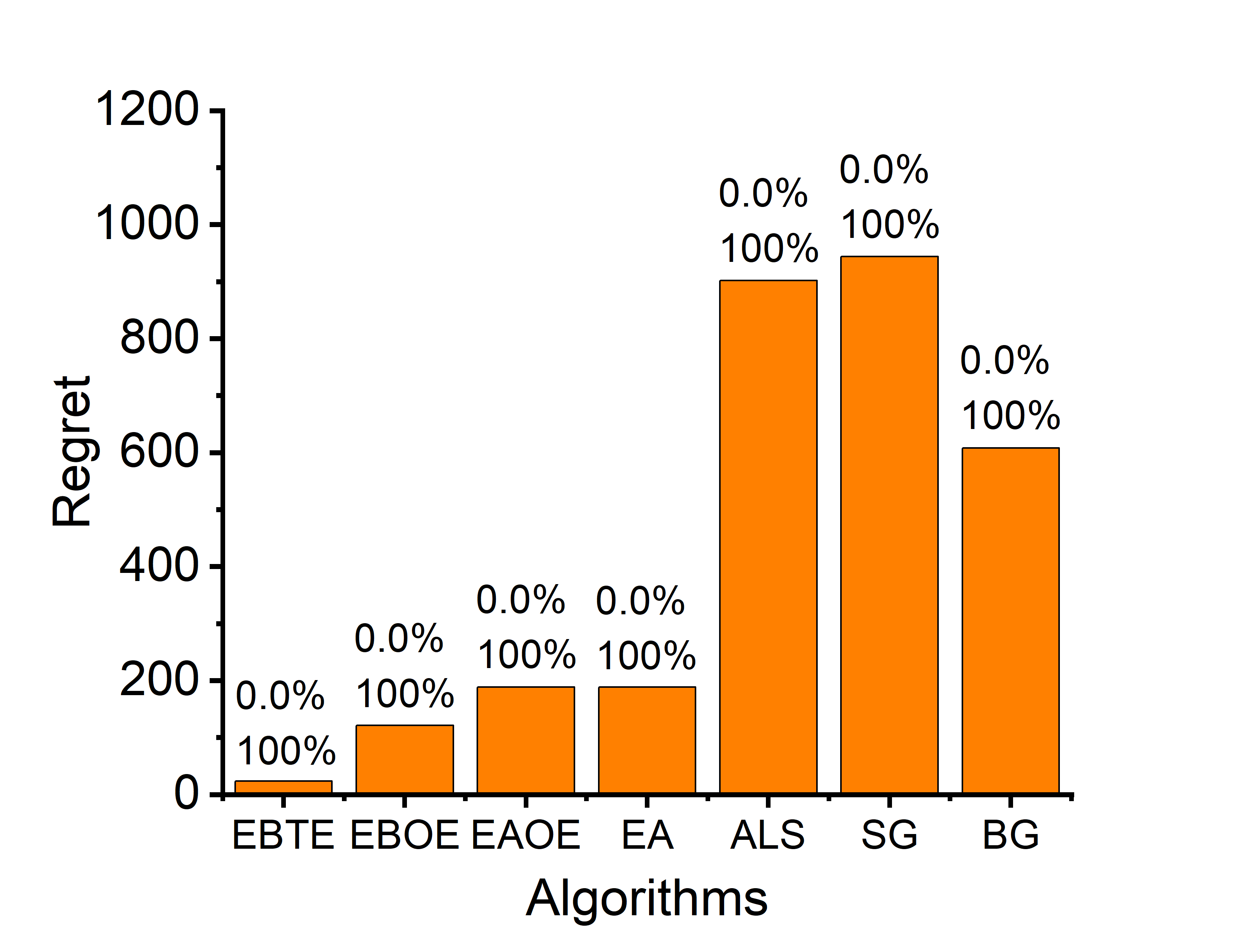} \\
{\tiny (m) $\alpha = 80 \%$} & {\tiny (n) $\alpha = 100 \%$} &{\tiny (o) $\alpha = 120 \%$} & {\tiny (p) $\alpha = 40 \%$} \\

\includegraphics[scale=0.11]{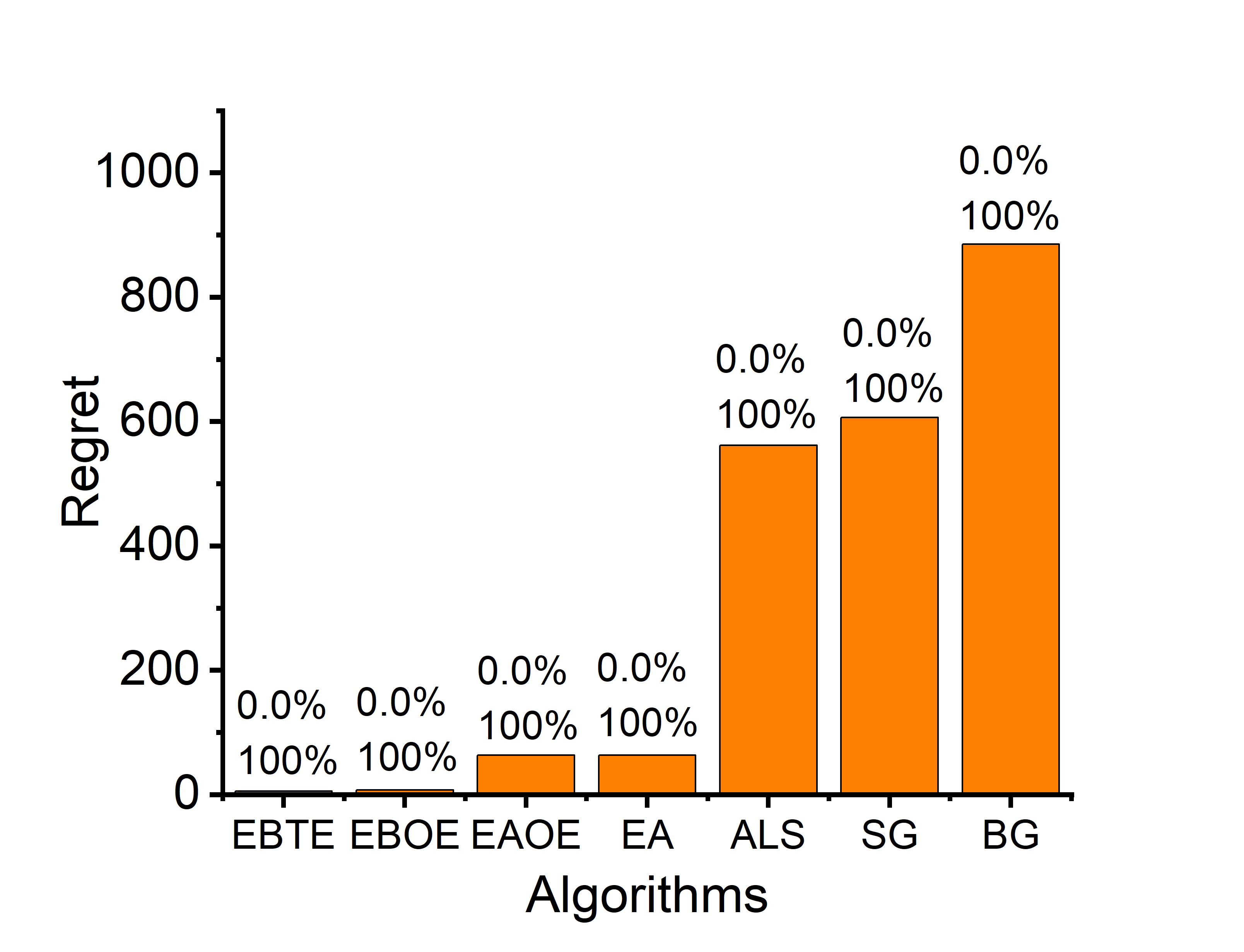}  &\includegraphics[scale=0.11]{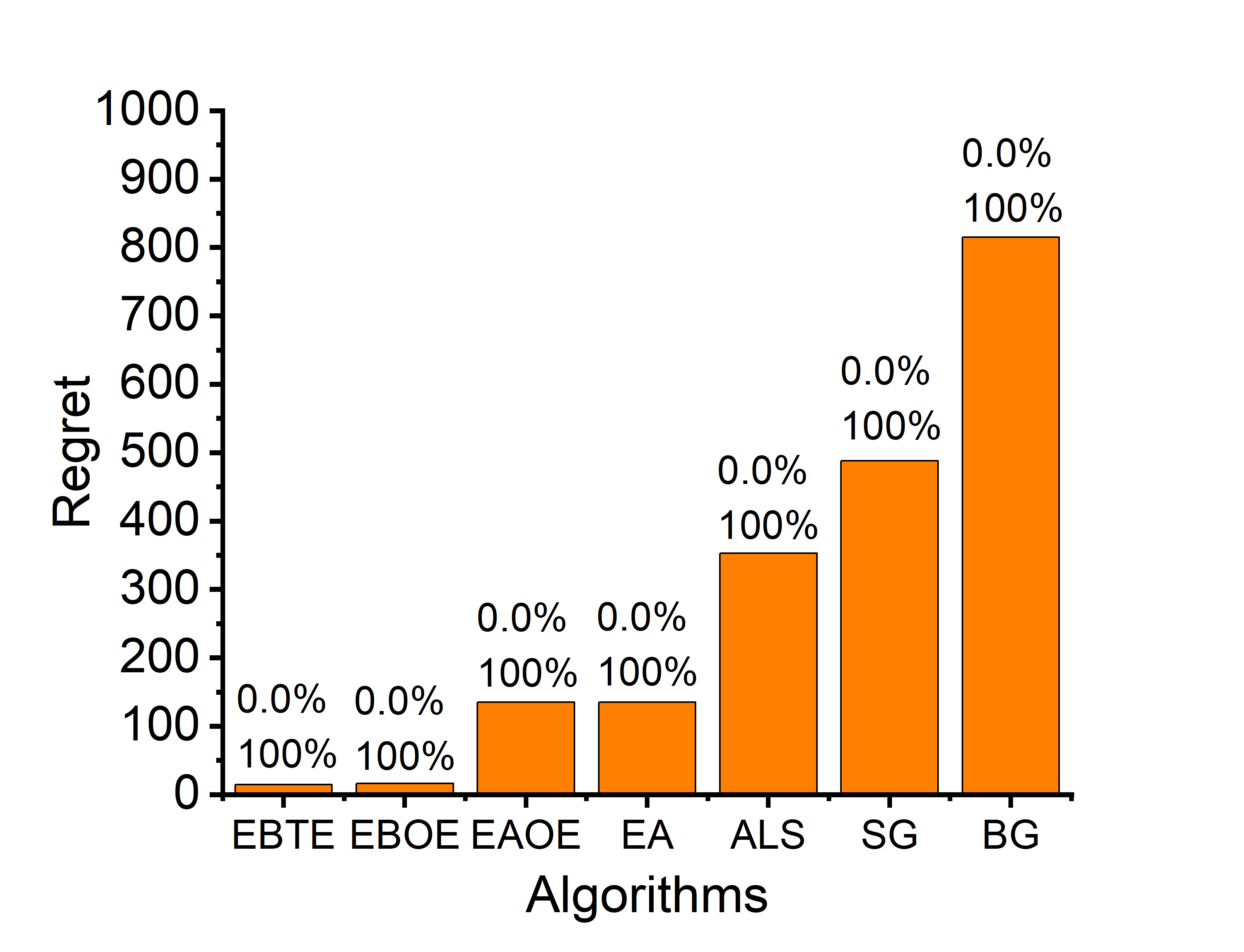} & \includegraphics[scale=0.11]{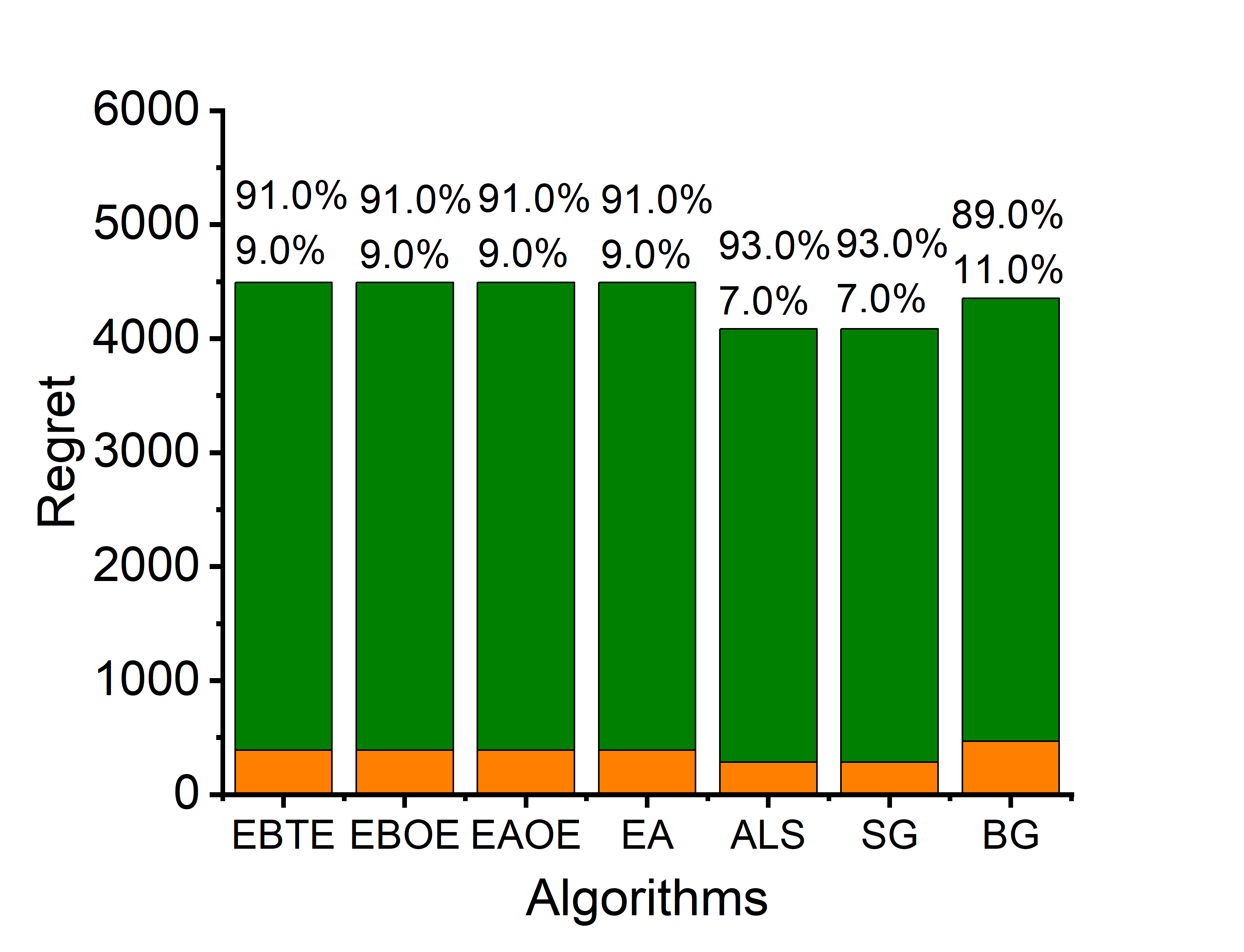} & \includegraphics[scale=0.11]{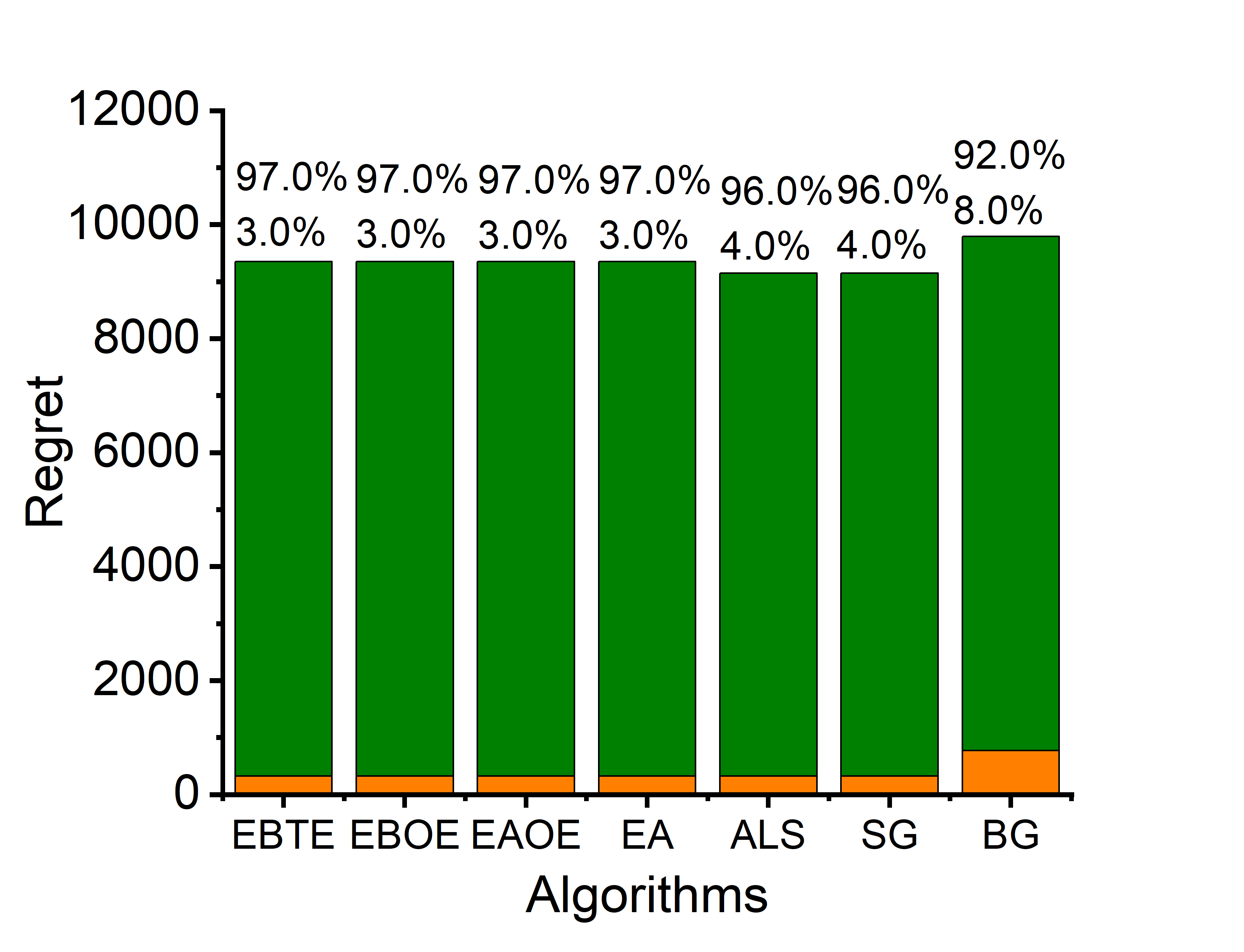}\\
 {\tiny (q) $\alpha = 60 \%$} & {\tiny (r) $\alpha = 80 \%$} & {\tiny (s) $\alpha = 100 \%$} &{\tiny (t) $\alpha = 120 \%$}\\

\end{tabular}
\caption{Regret varying $\alpha$ when $\mathcal{I}^{ID} = 1\%, \mathcal{|A|} = 100$ (a, b, c, d, e), when $\mathcal{I}^{ID} = 2\%, \mathcal{|A|} = 50$ (f,g,h,i,j), when $\mathcal{I}^{ID} = 5\%, \mathcal{|A|} = 20$ (k,l,m,n,o) and when $\mathcal{I}^{ID} = 10\%, \mathcal{|A|} = 10$ (p, q, r, s, t)for Beach location type }
\label{Fig:Beach}
\end{figure}

\begin{figure}[h!]
\centering
    \begin{tabular}{lclc}
       Unsatisfied Regret & \includegraphics[width=0.11\linewidth]{Unsatisfied.png} \  & \ Excessive Regret & \includegraphics[width=0.11\linewidth]{Excessive.png} \\
    \end{tabular}

\begin{tabular}{cccc}
\includegraphics[scale=0.11]{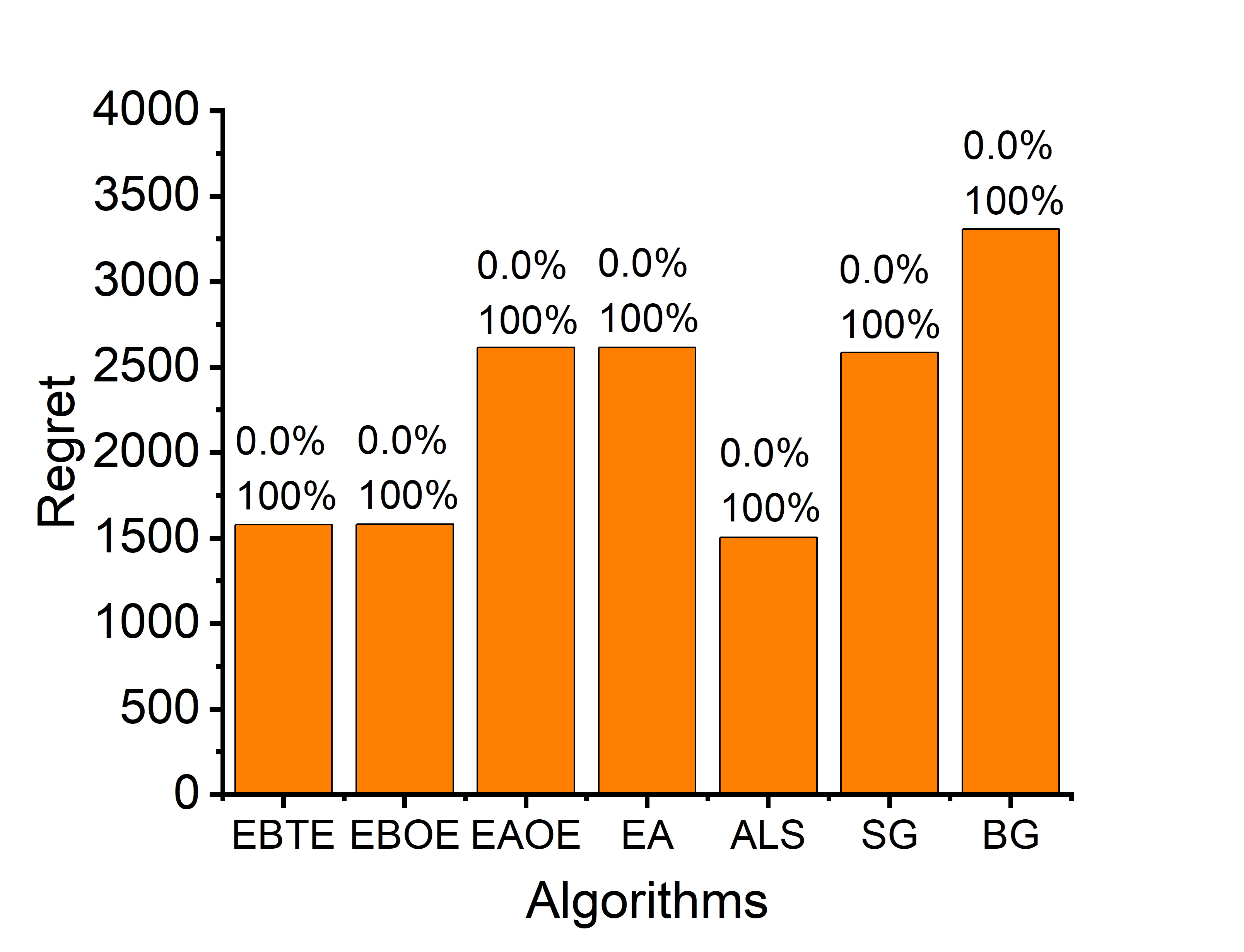} & \includegraphics[scale=0.11]{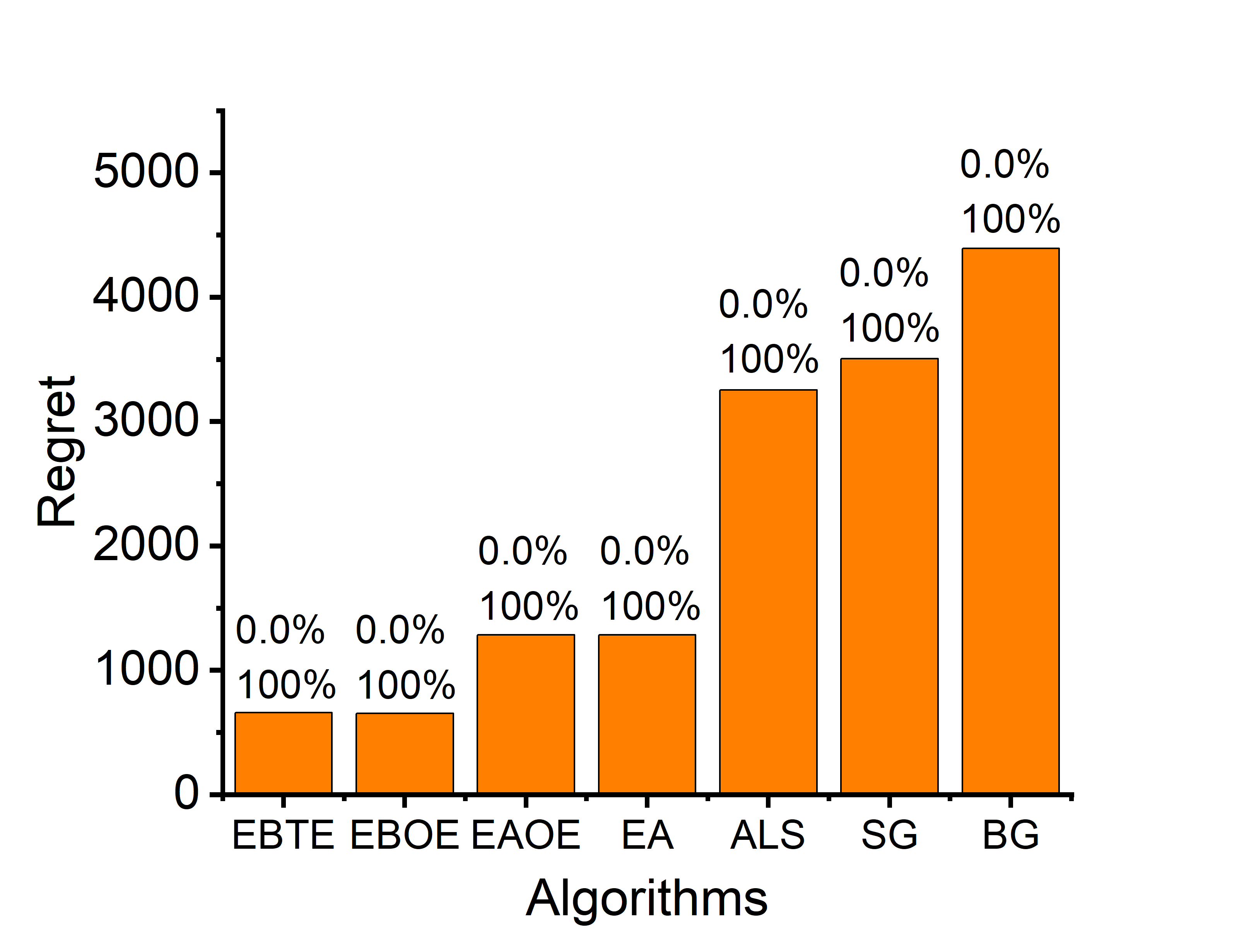}  &\includegraphics[scale=0.11]{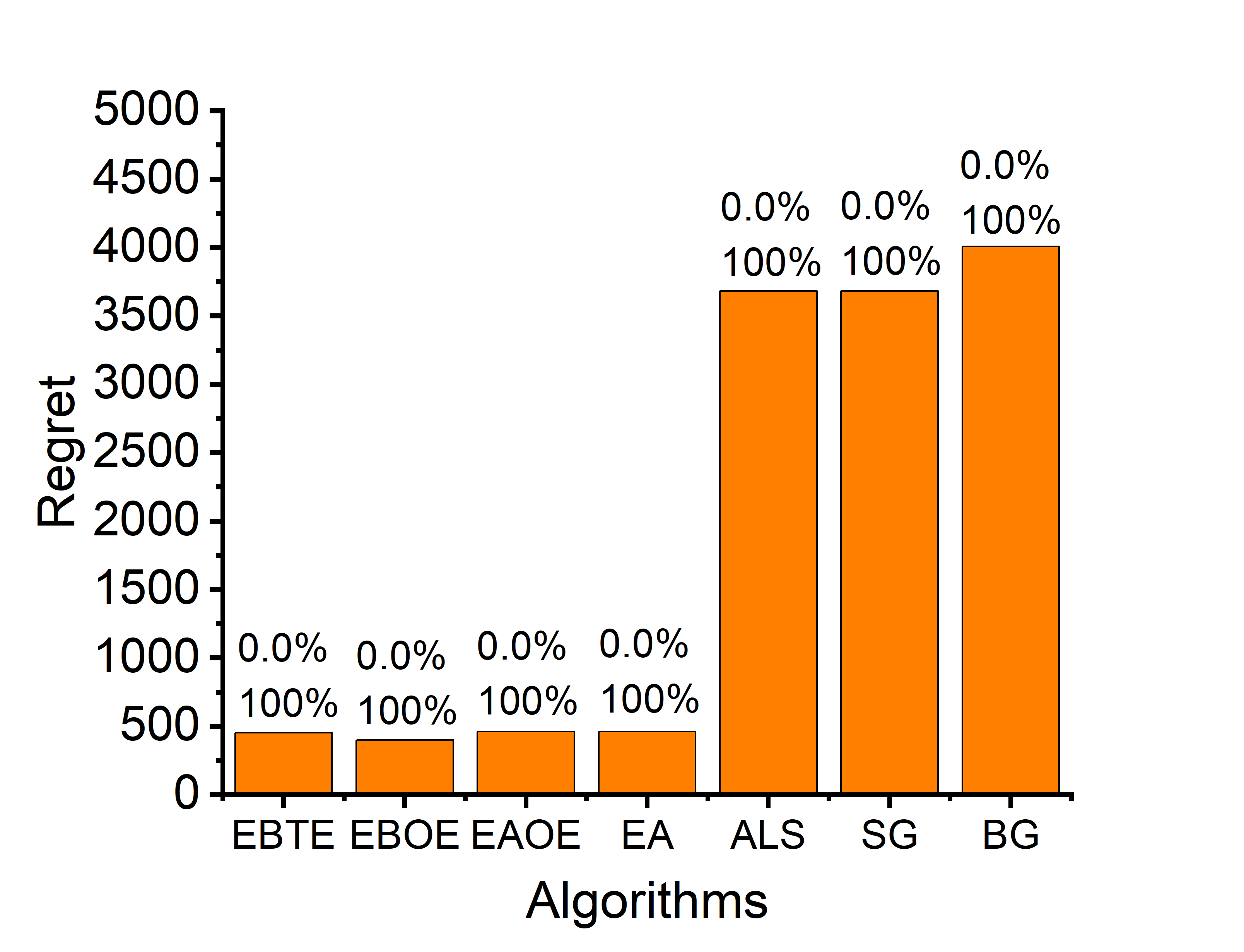} & \includegraphics[scale=0.11]{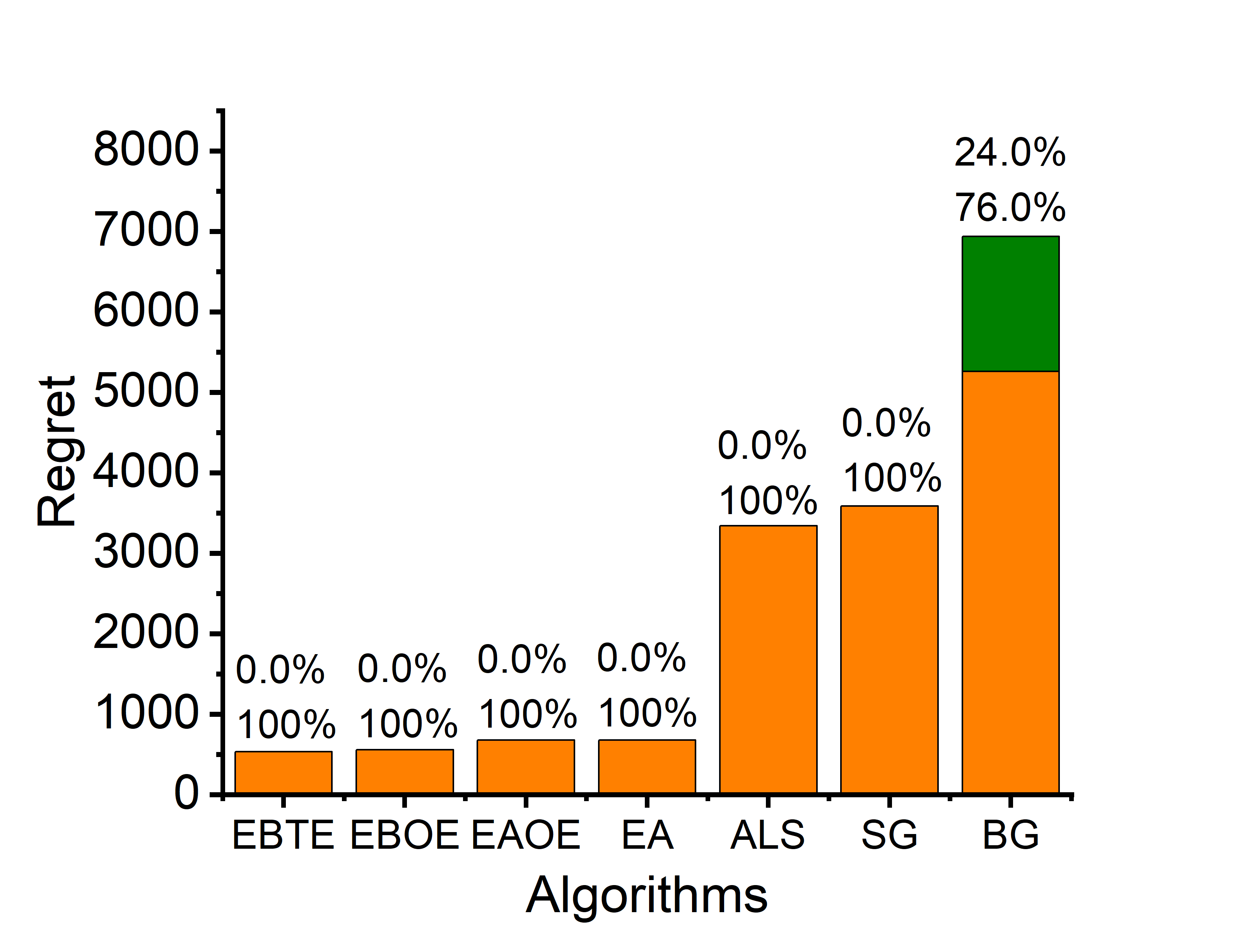} \\
{\tiny (a) $\alpha = 40 \%$} & {\tiny (b) $\alpha = 60 \%$} & {\tiny (c) $\alpha = 80 \%$} & {\tiny (d) $\alpha = 100 \%$} \\


 \includegraphics[scale=0.11]{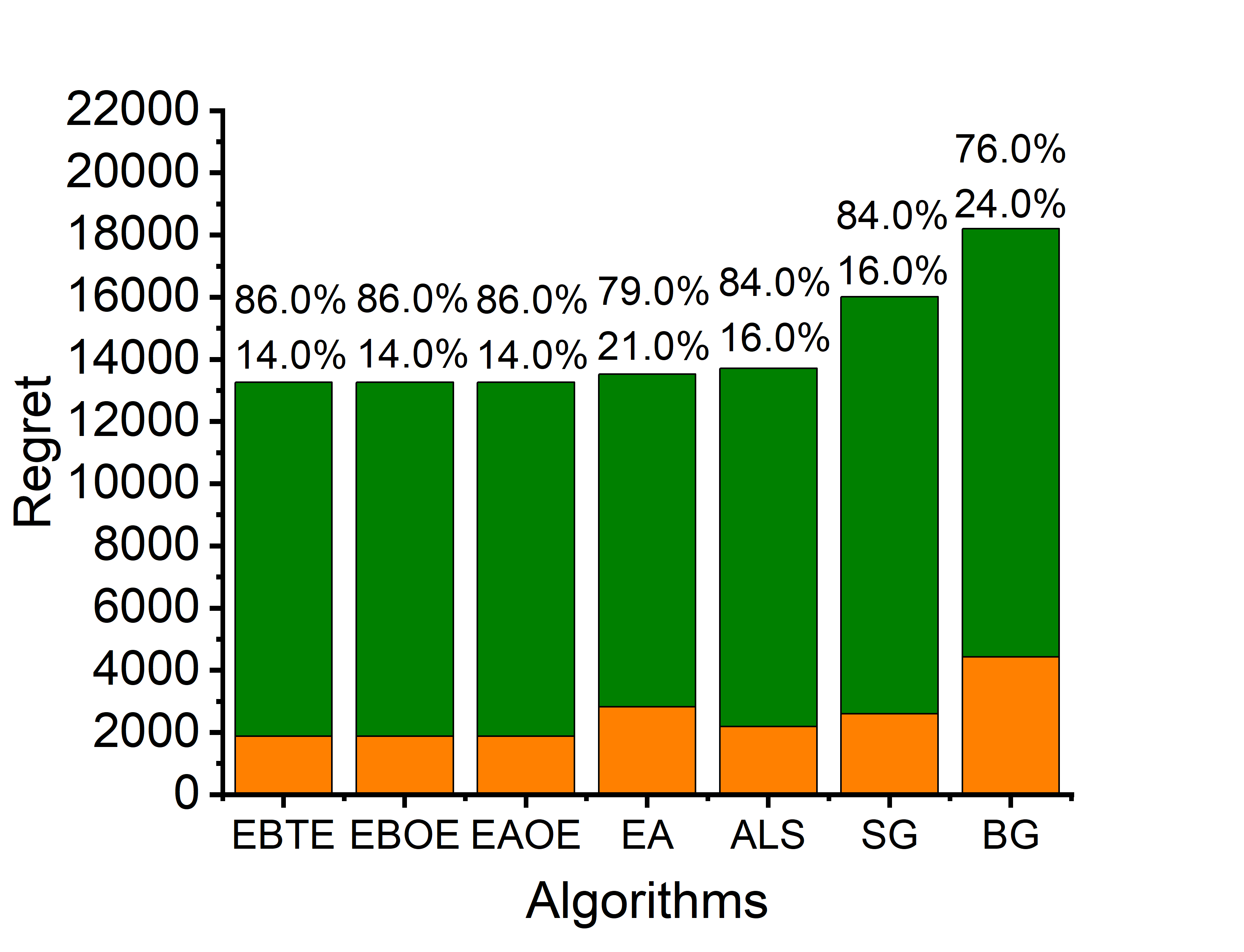} & \includegraphics[scale=0.11]{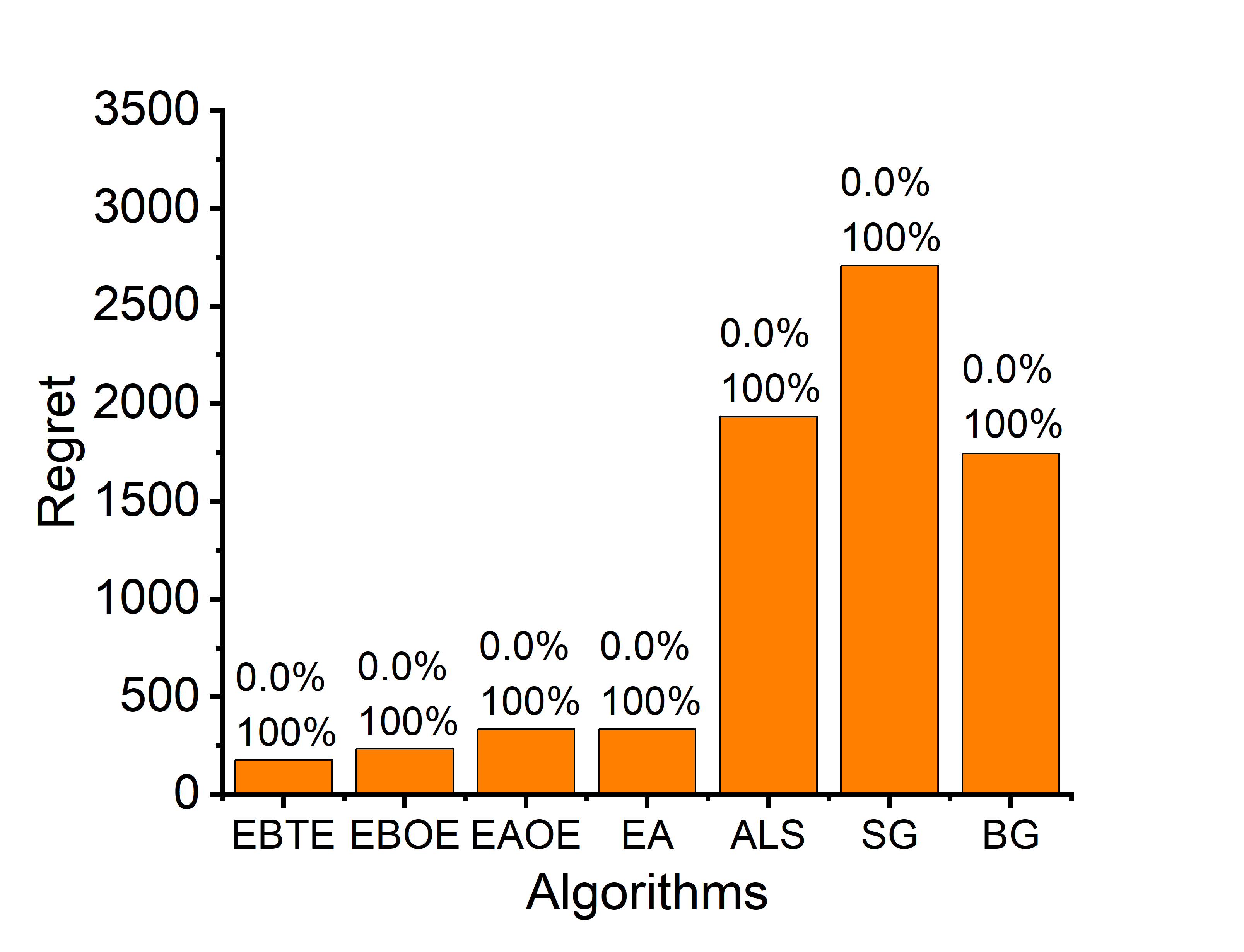} & \includegraphics[scale=0.11]{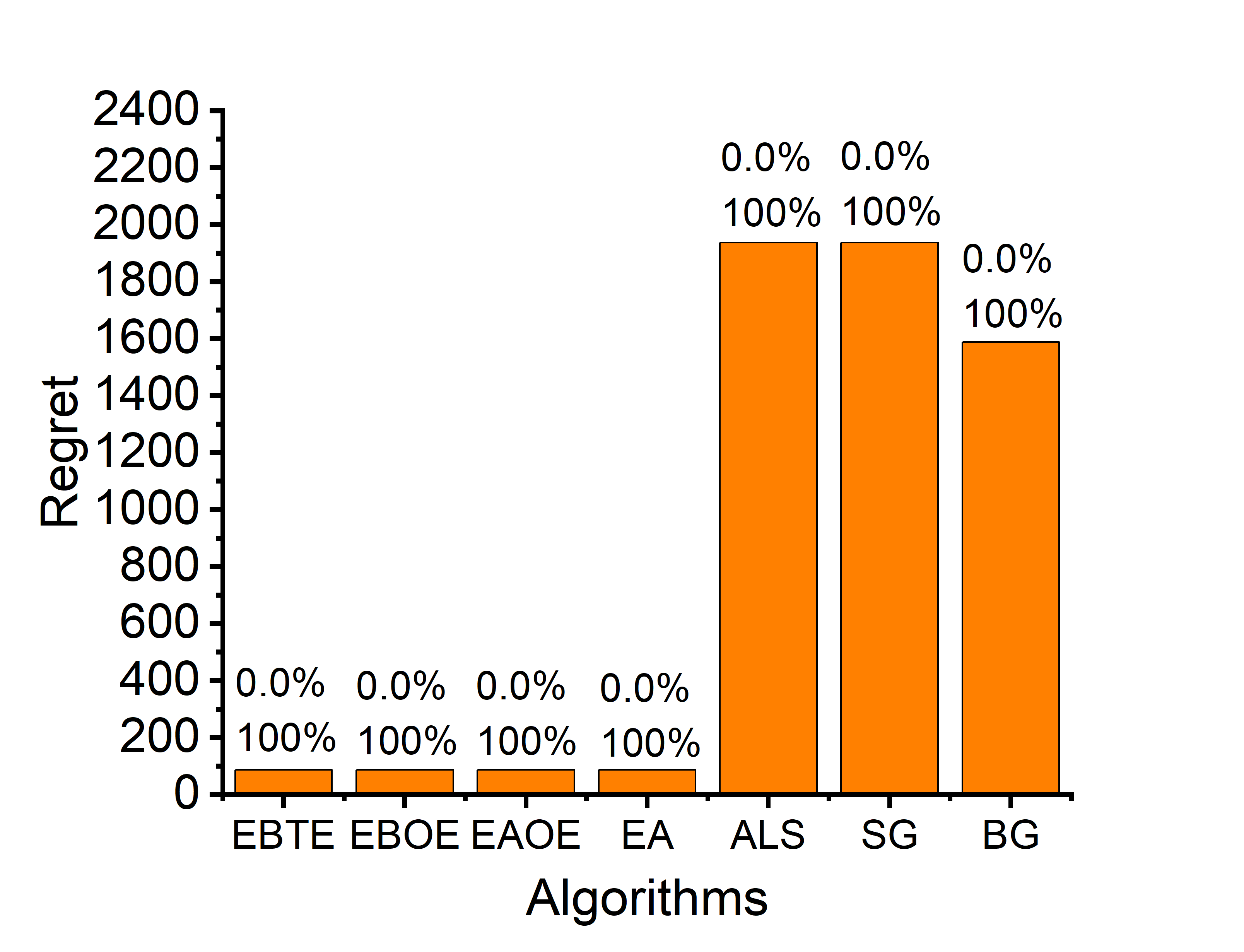}  &\includegraphics[scale=0.11]{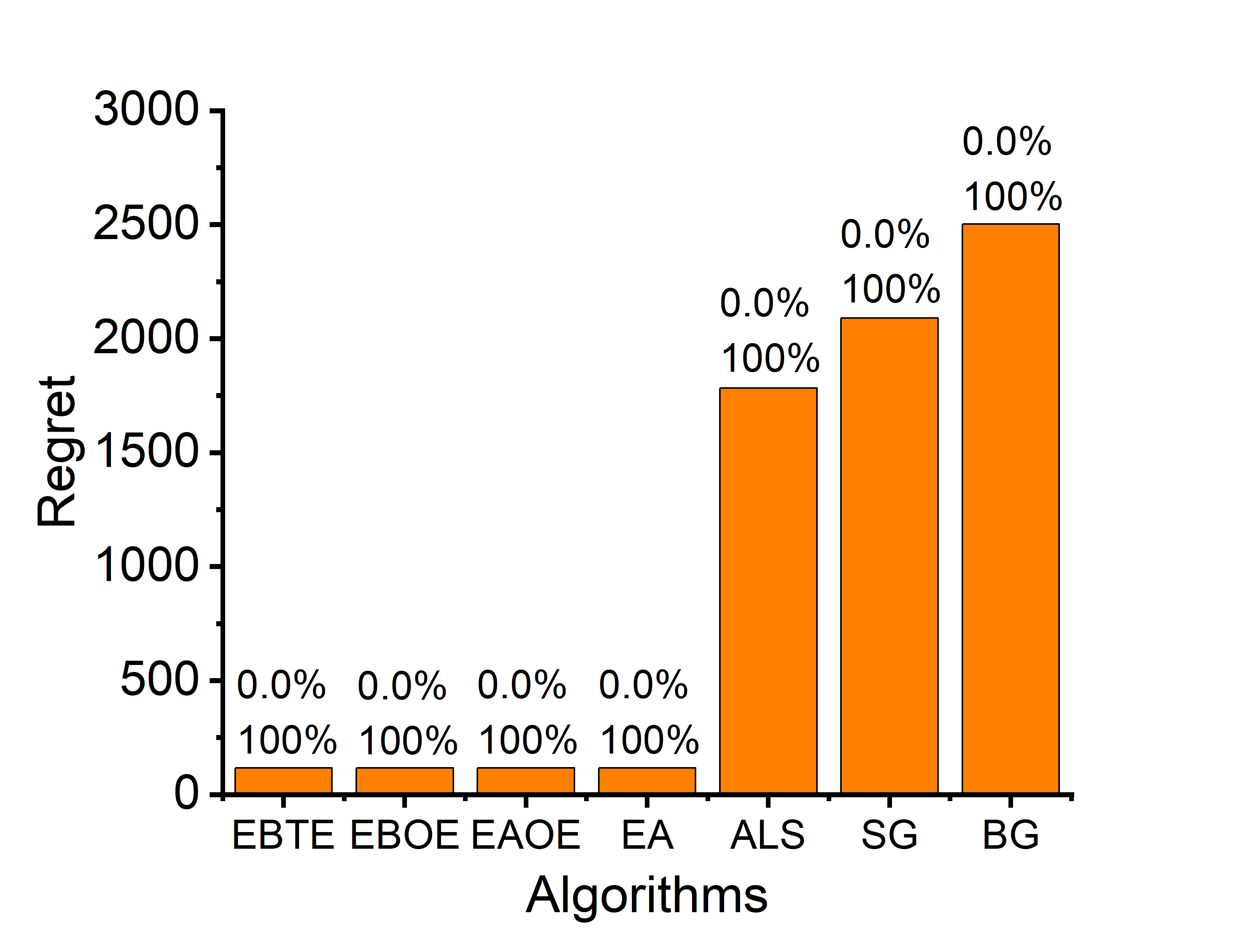} \\
{\tiny (e) $\alpha = 120 \%$} & {\tiny (f) $\alpha = 40 \%$} & {\tiny (g) $\alpha = 60 \%$} & {\tiny (h) $\alpha = 80 \%$} \\

 \includegraphics[scale=0.11]{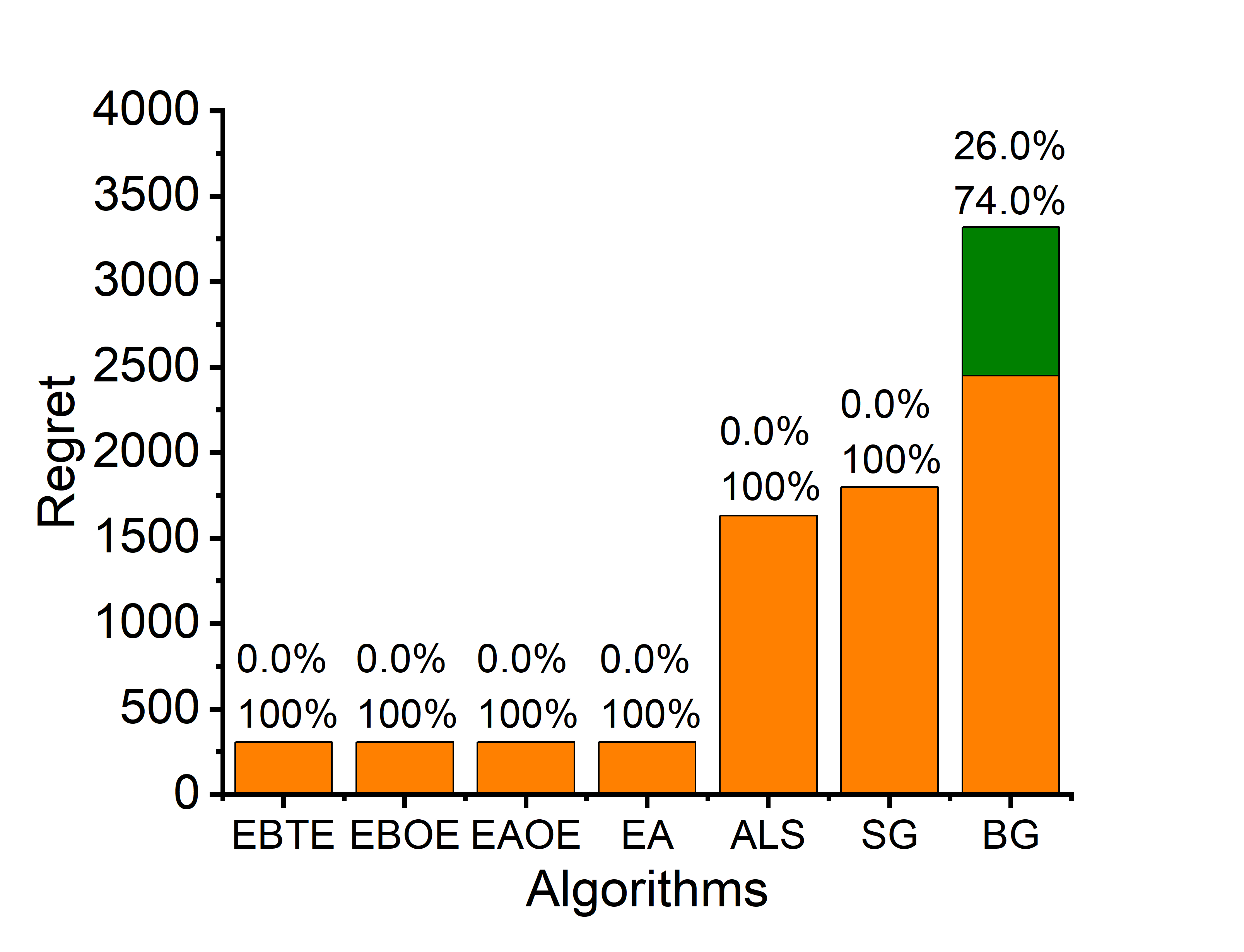} & \includegraphics[scale=0.11]{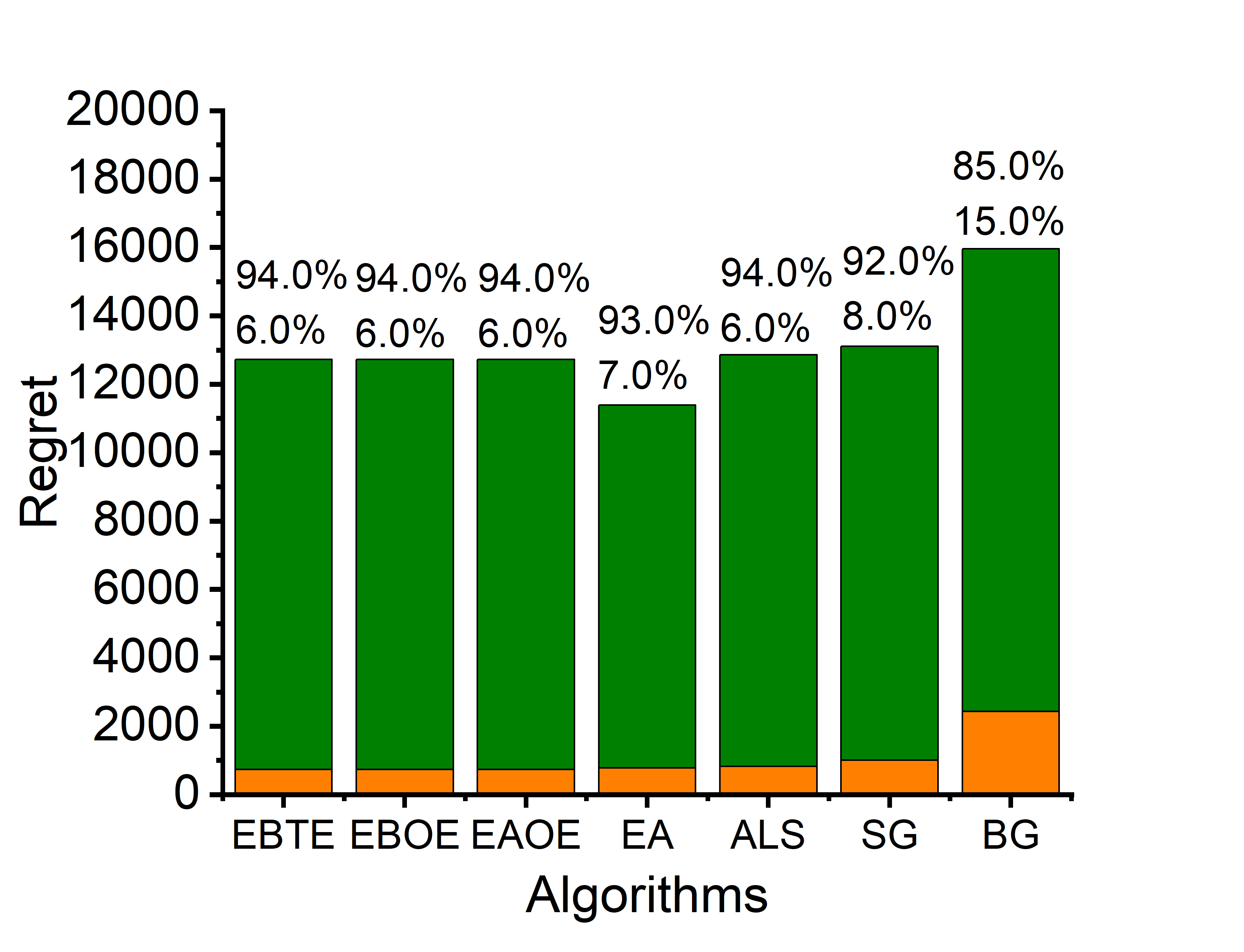} &
\includegraphics[scale=0.11]{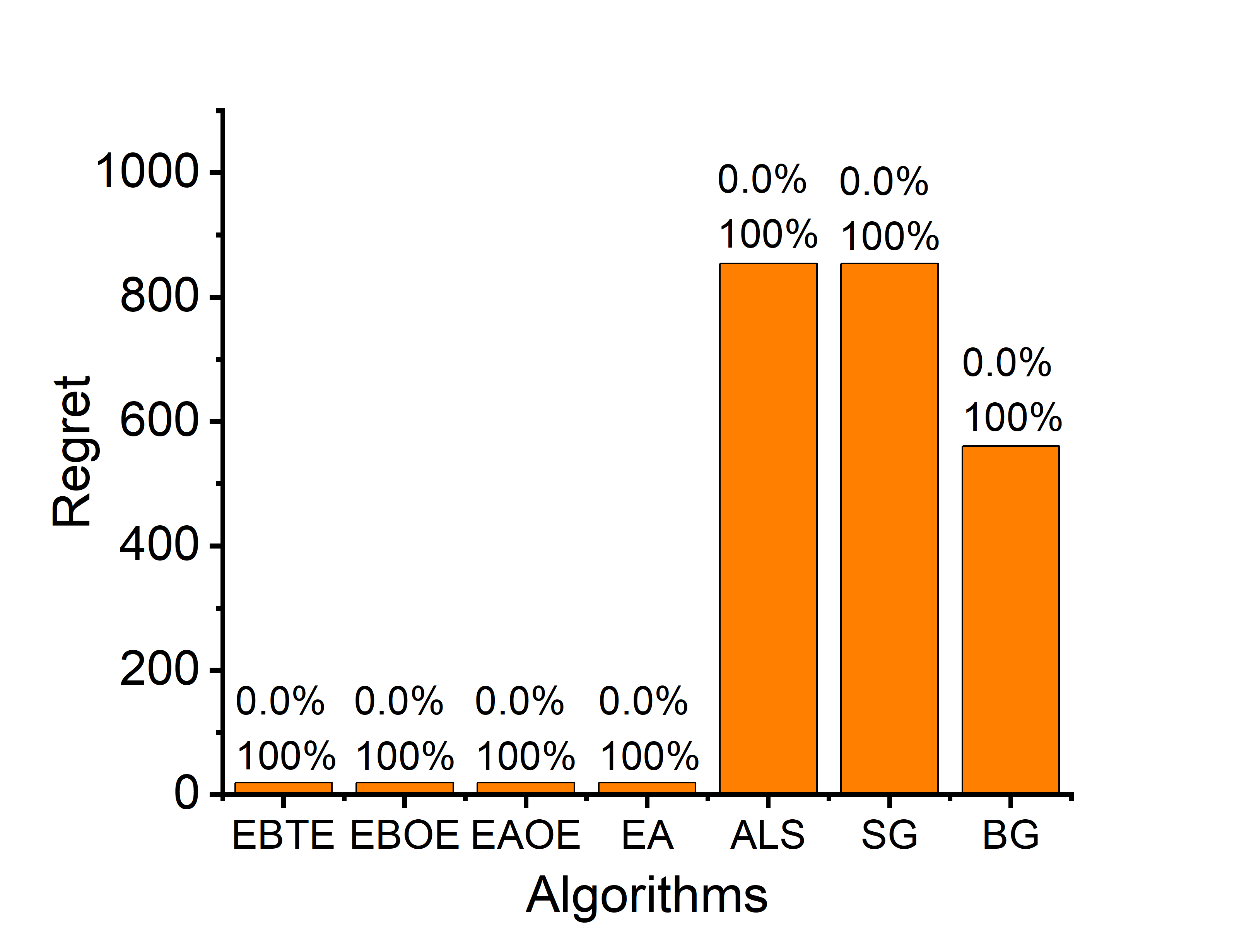} & \includegraphics[scale=0.11]{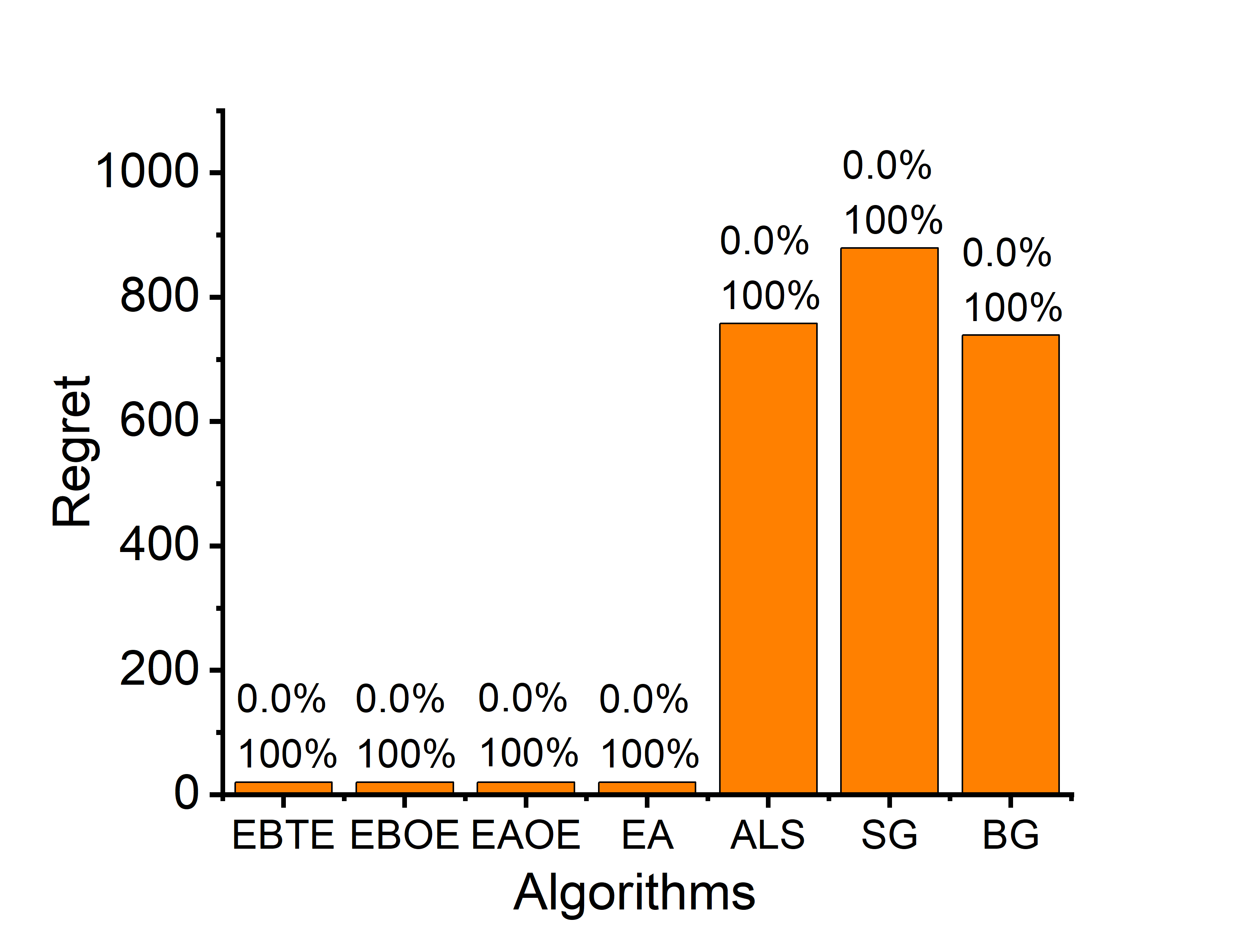} \\
{\tiny (i) $\alpha = 100 \%$} &{\tiny (j) $\alpha = 120 \%$} & {\tiny (k) $\alpha = 40 \%$} & {\tiny (l) $\alpha = 60 \%$} \\

\includegraphics[scale=0.11]{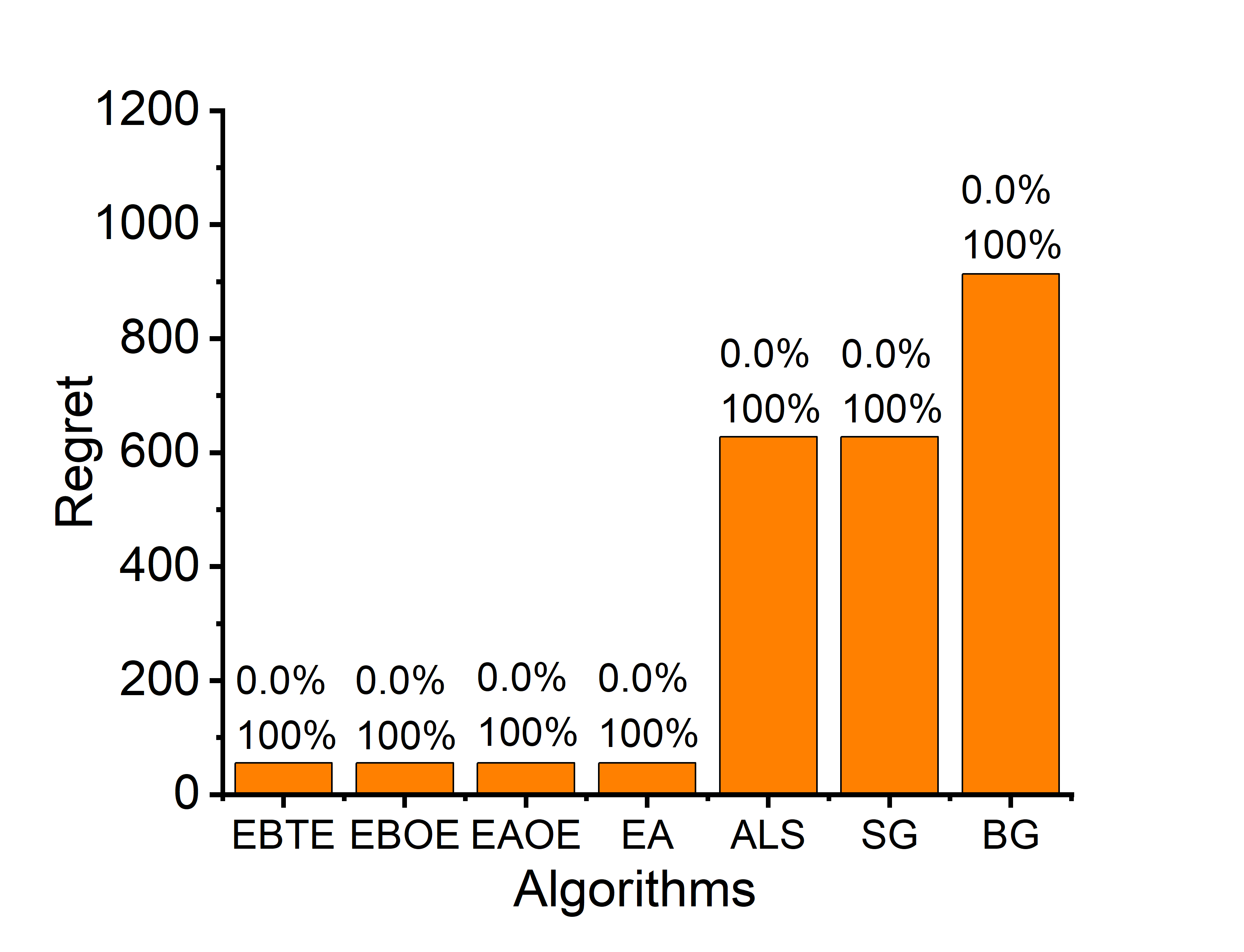} & \includegraphics[scale=0.11]{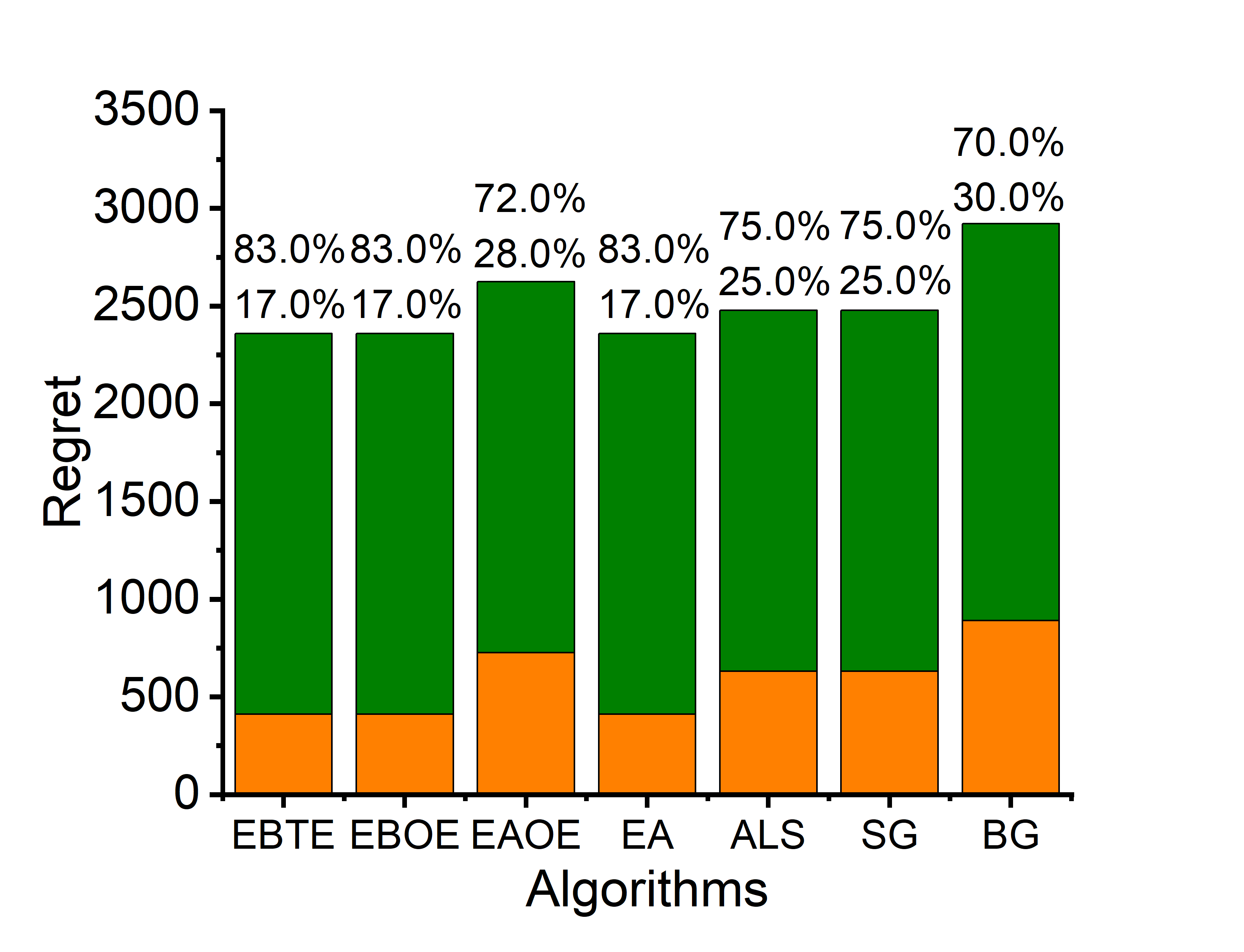} & \includegraphics[scale=0.11]{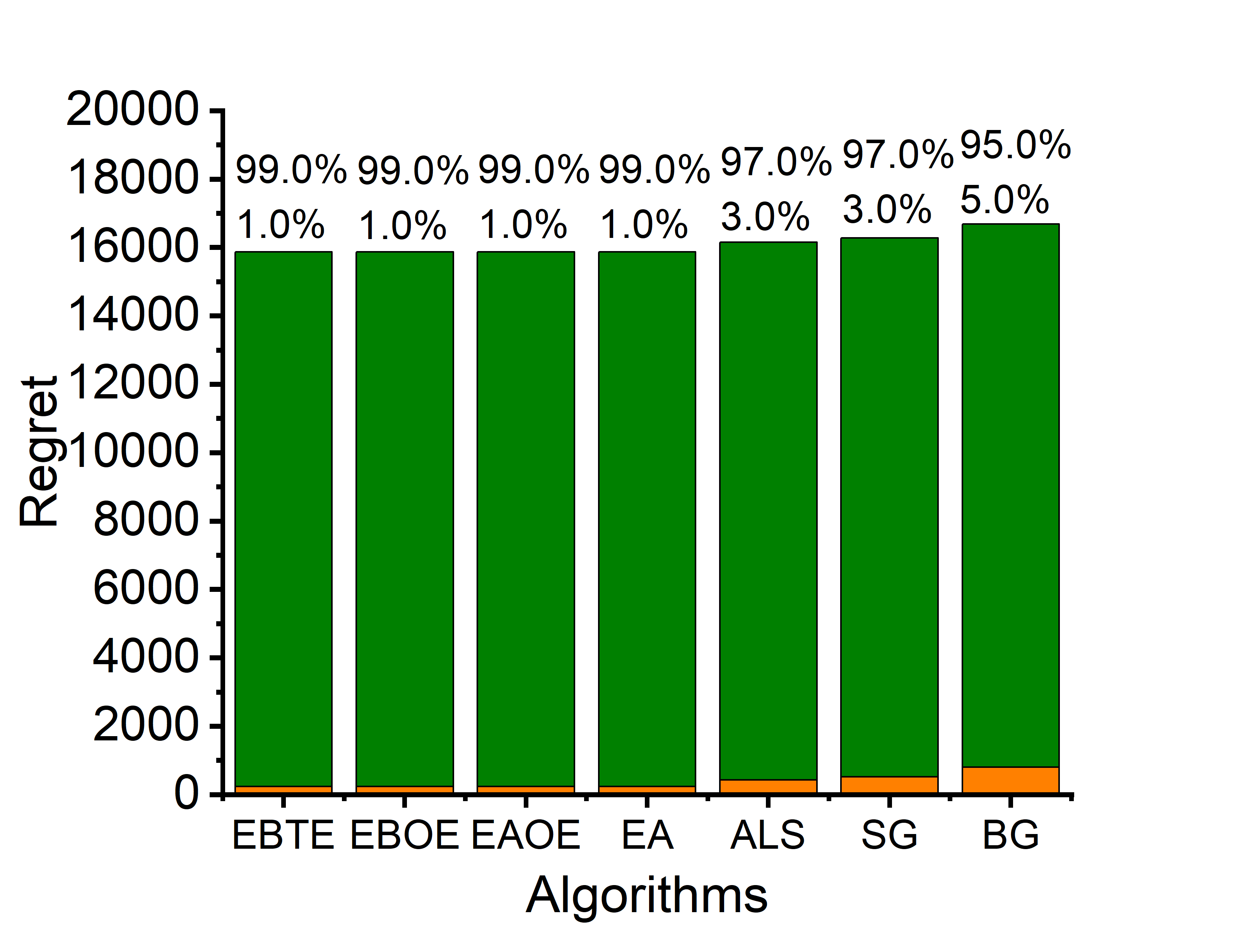} &
\includegraphics[scale=0.11]{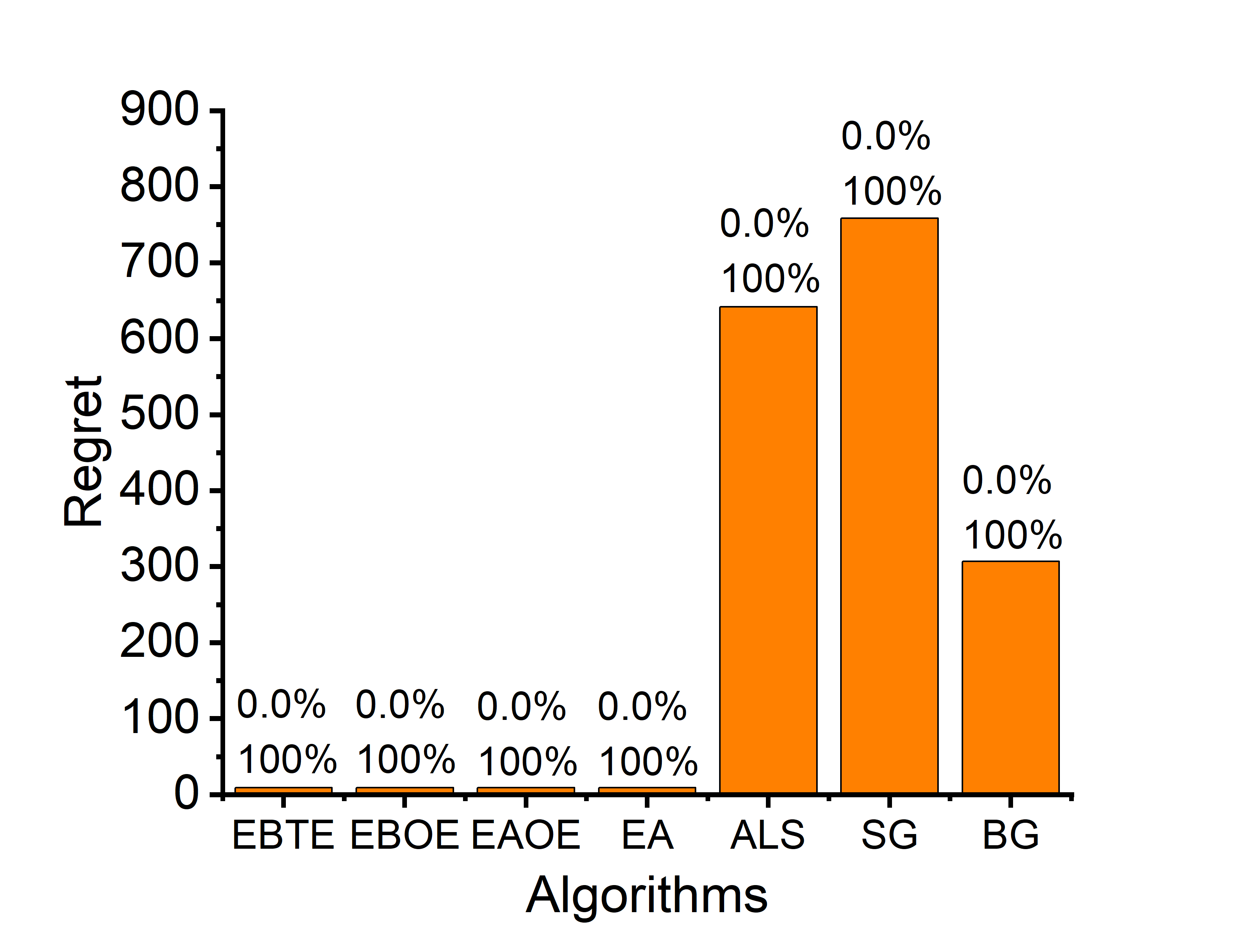} \\
{\tiny (m) $\alpha = 80 \%$} & {\tiny (n) $\alpha = 100 \%$} &{\tiny (o) $\alpha = 120 \%$} & {\tiny (p) $\alpha = 40 \%$} \\

\includegraphics[scale=0.11]{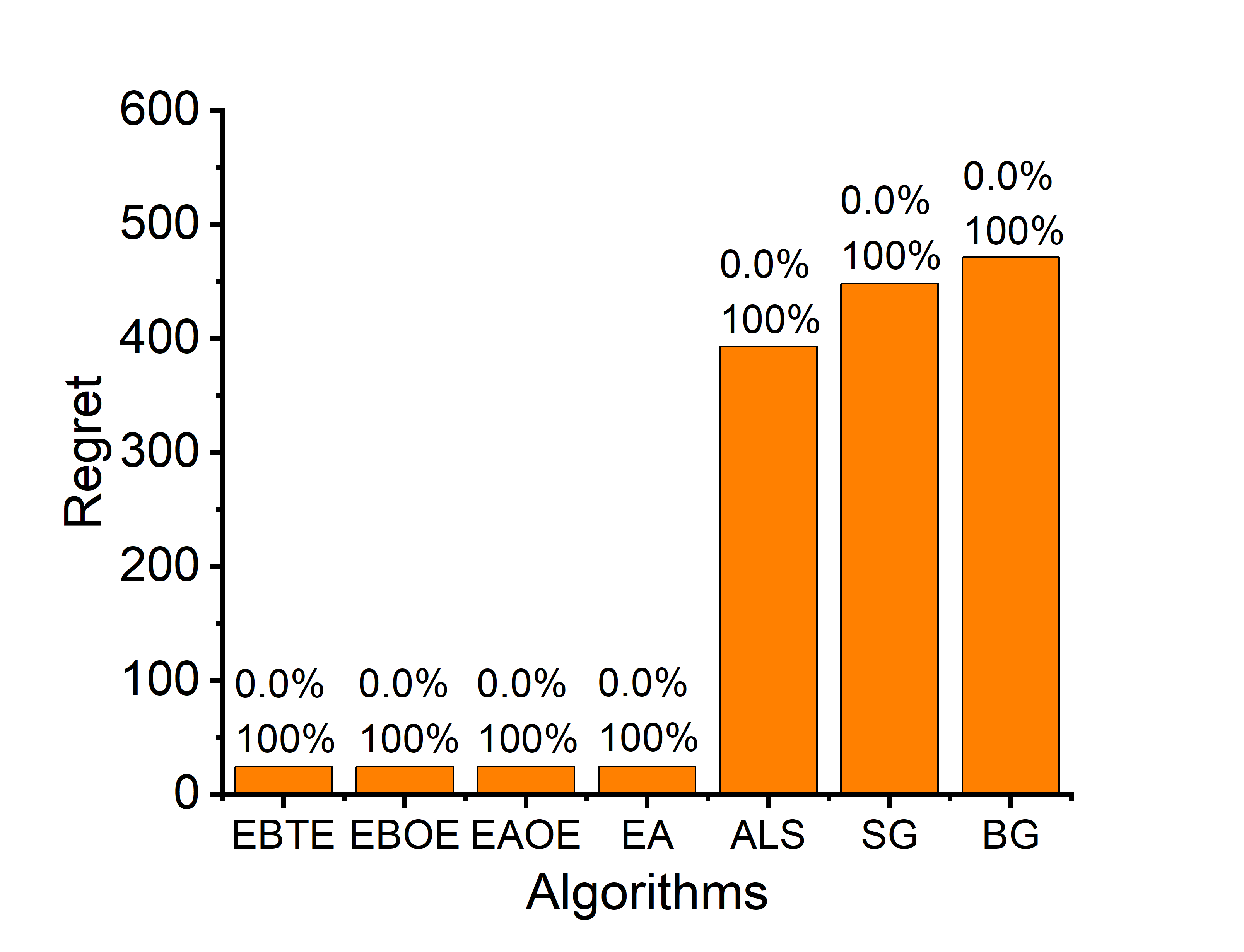}  &\includegraphics[scale=0.11]{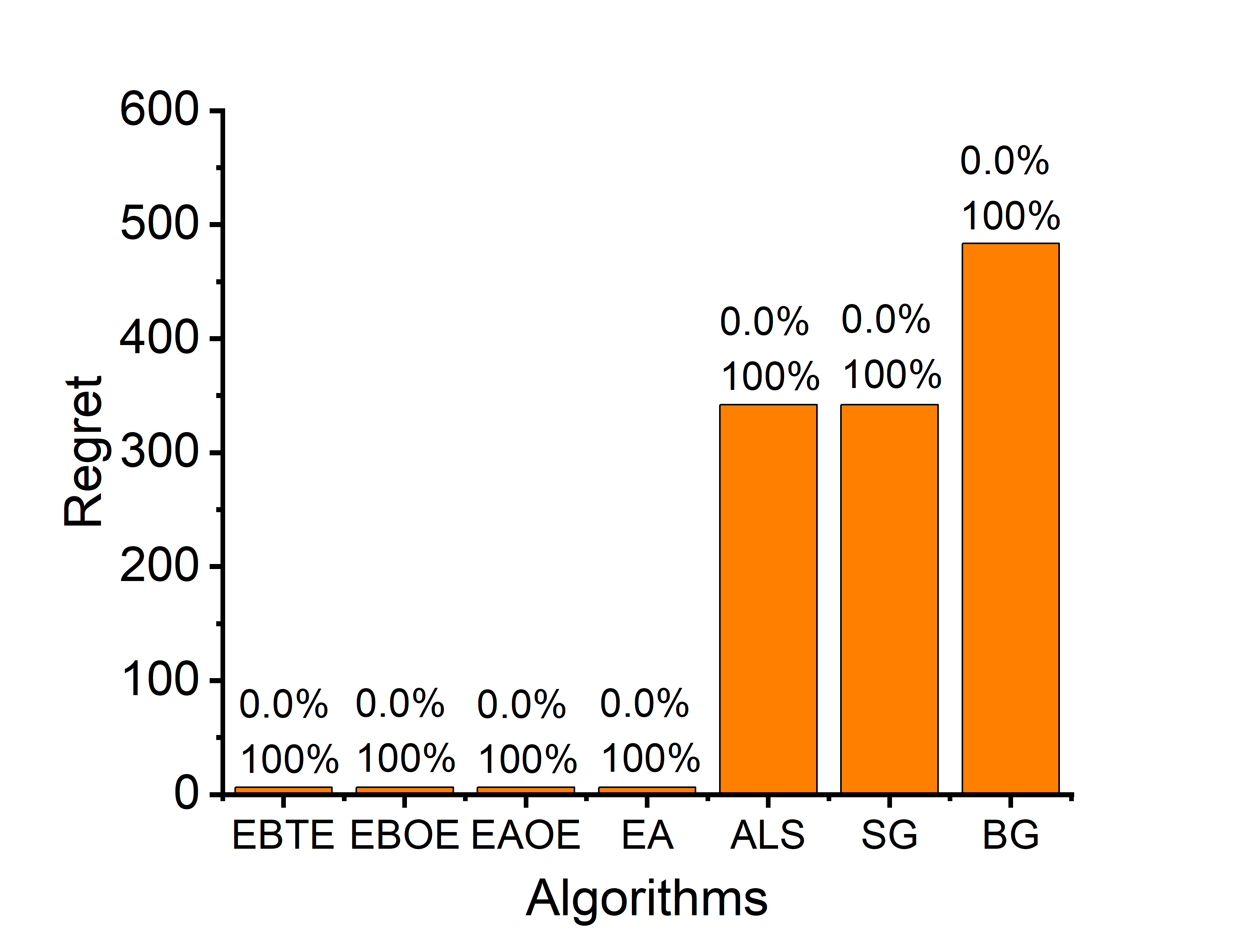} & \includegraphics[scale=0.11]{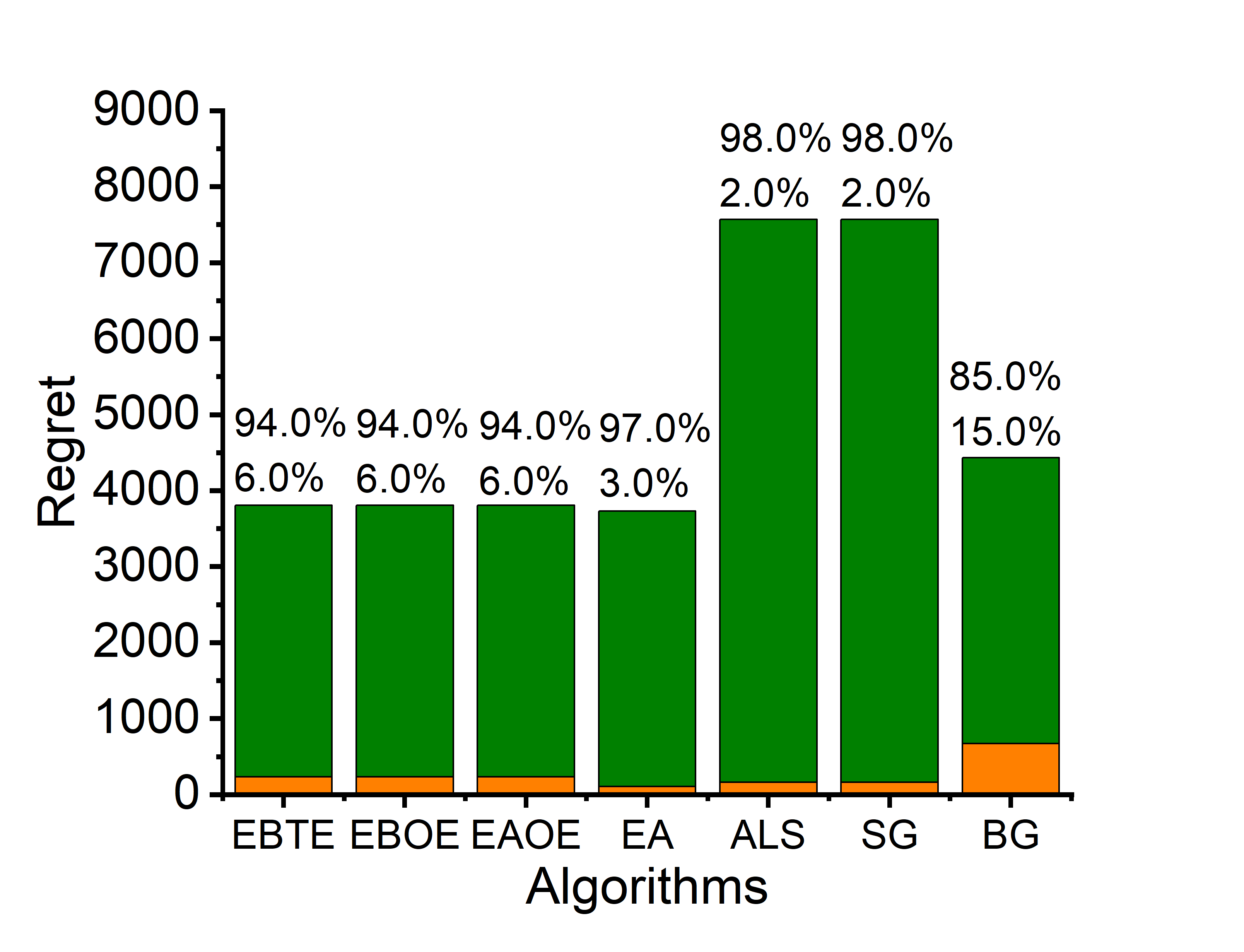} & \includegraphics[scale=0.11]{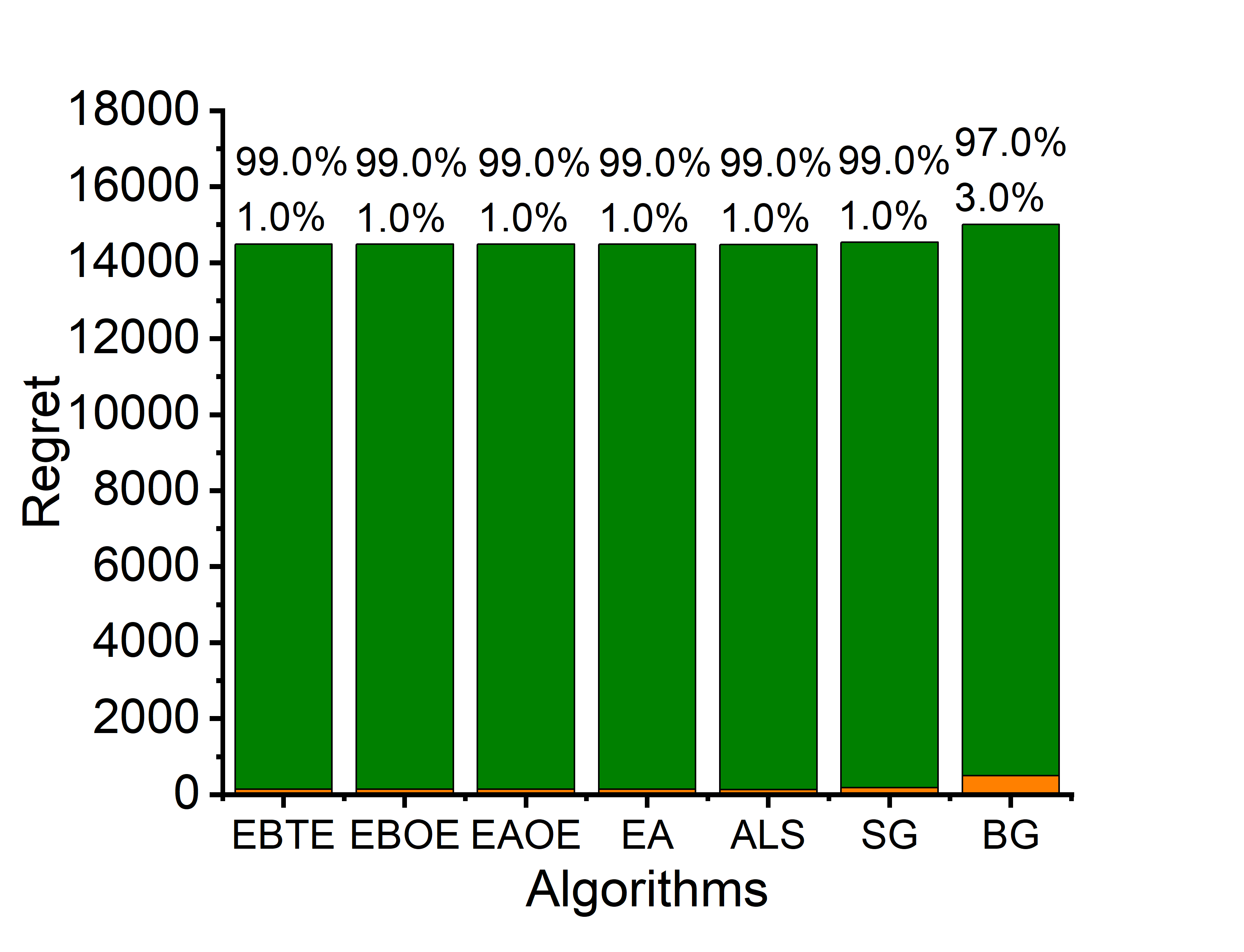}\\
 {\tiny (q) $\alpha = 60 \%$} & {\tiny (r) $\alpha = 80 \%$} & {\tiny (s) $\alpha = 100 \%$} &{\tiny (t) $\alpha = 120 \%$}\\

\end{tabular}
\caption{Regret varying $\alpha$ when $\mathcal{I}^{ID} = 1\%, \mathcal{|A|} = 100$ (a, b, c, d, e), when $\mathcal{I}^{ID} = 2\%, \mathcal{|A|} = 50$ (f,g,h,i,j), when $\mathcal{I}^{ID} = 5\%, \mathcal{|A|} = 20$ (k,l,m,n,o) and when $\mathcal{I}^{ID} = 10\%, \mathcal{|A|} = 10$ (p, q, r, s, t)for Mall location type }
\label{Fig:Mall}
\end{figure}

\begin{figure}[h!]
\centering
   \begin{tabular}{lclc}
       Unsatisfied Regret & \includegraphics[width=0.11\linewidth]{Unsatisfied.png} \  & \ Excessive Regret & \includegraphics[width=0.11\linewidth]{Excessive.png} \\
    \end{tabular}

\begin{tabular}{cccc}
\includegraphics[scale=0.11]{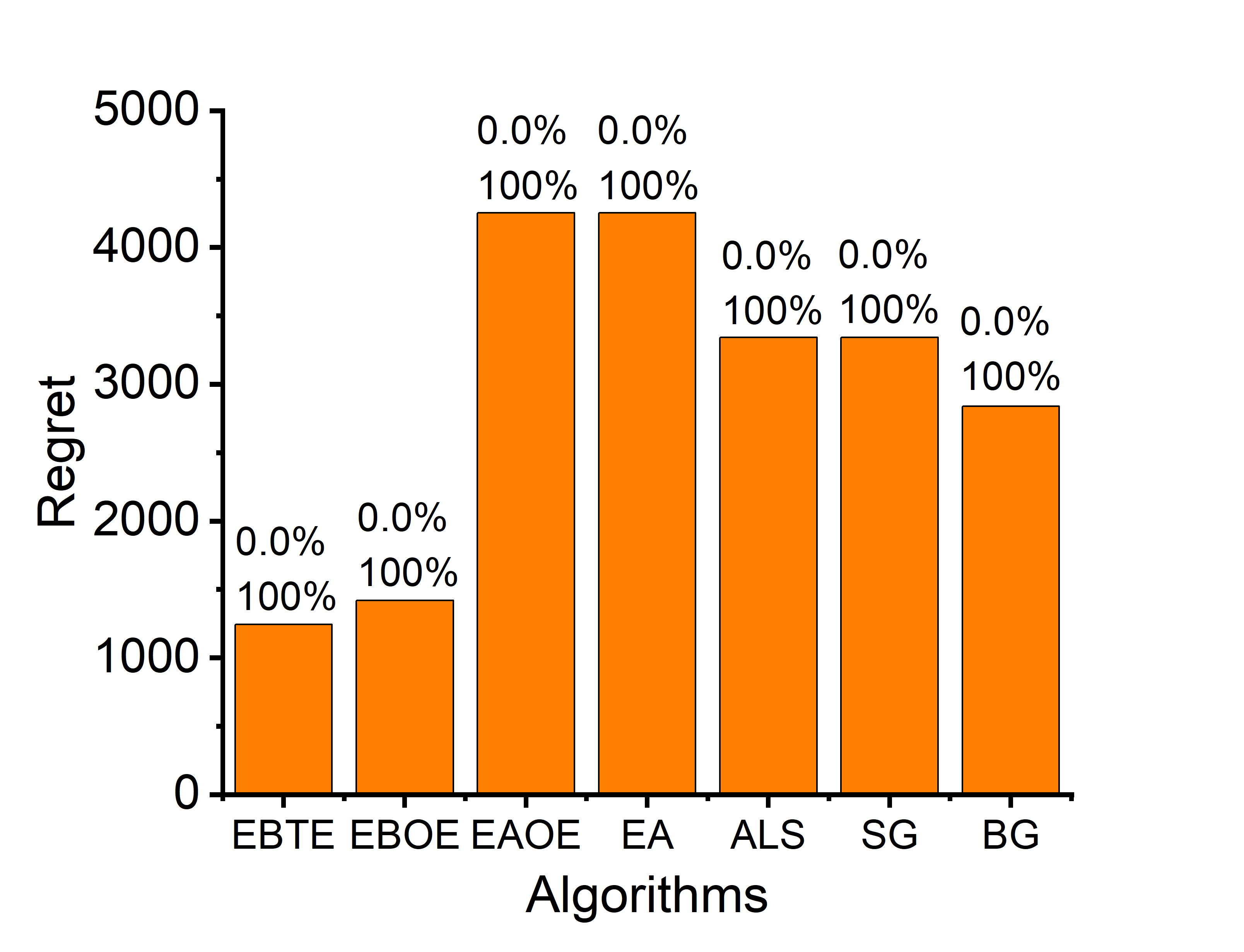} & \includegraphics[scale=0.11]{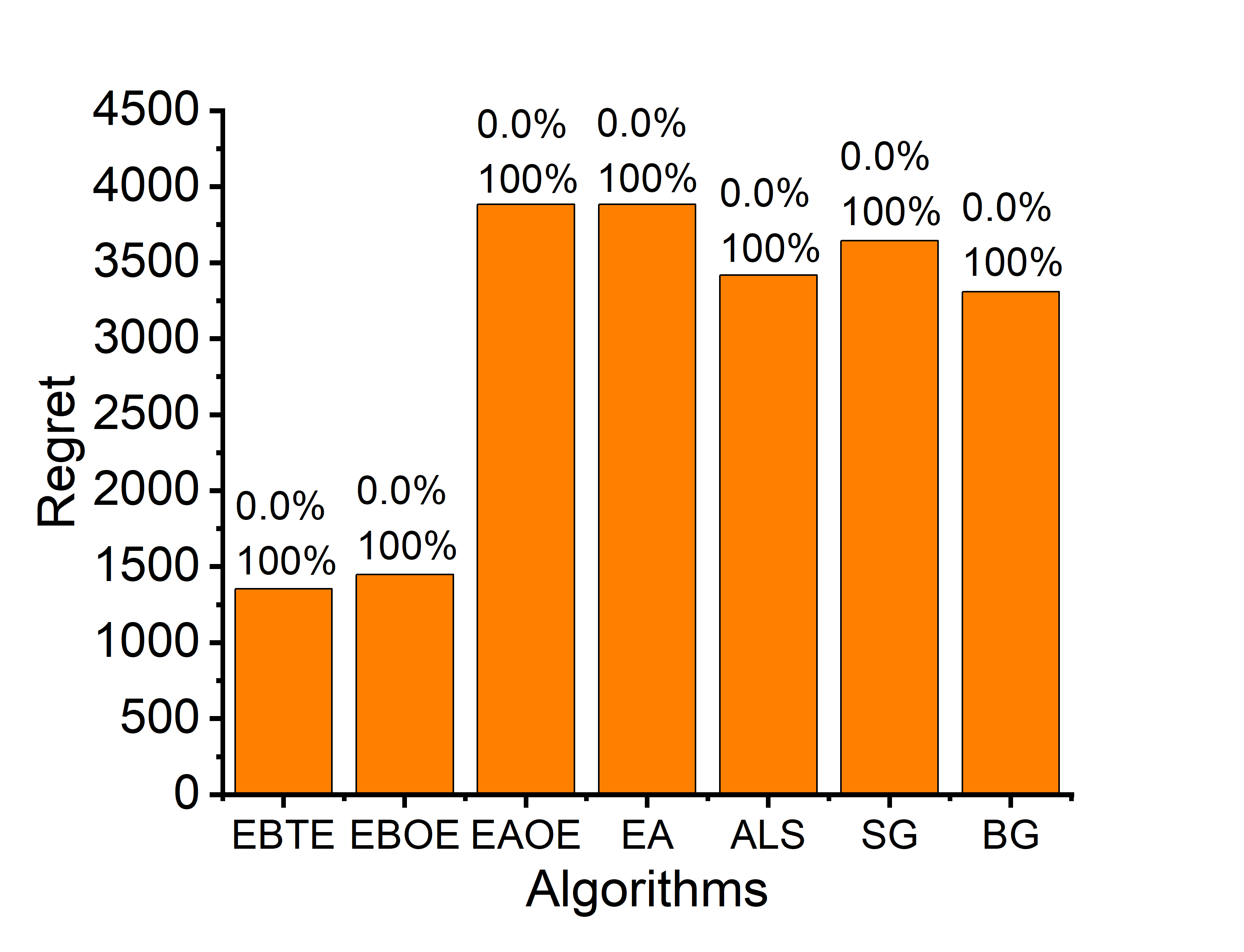}  &\includegraphics[scale=0.11]{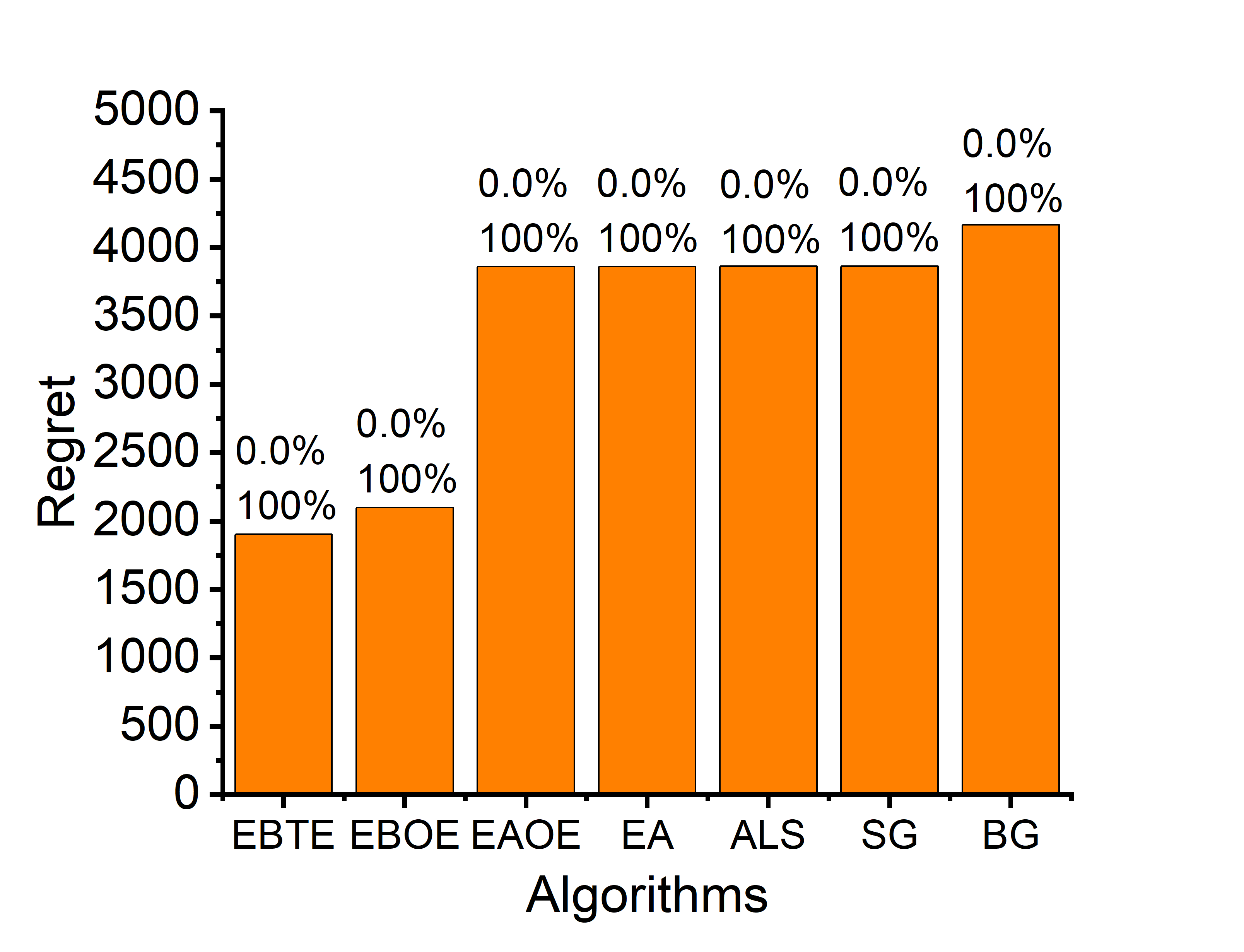} & \includegraphics[scale=0.11]{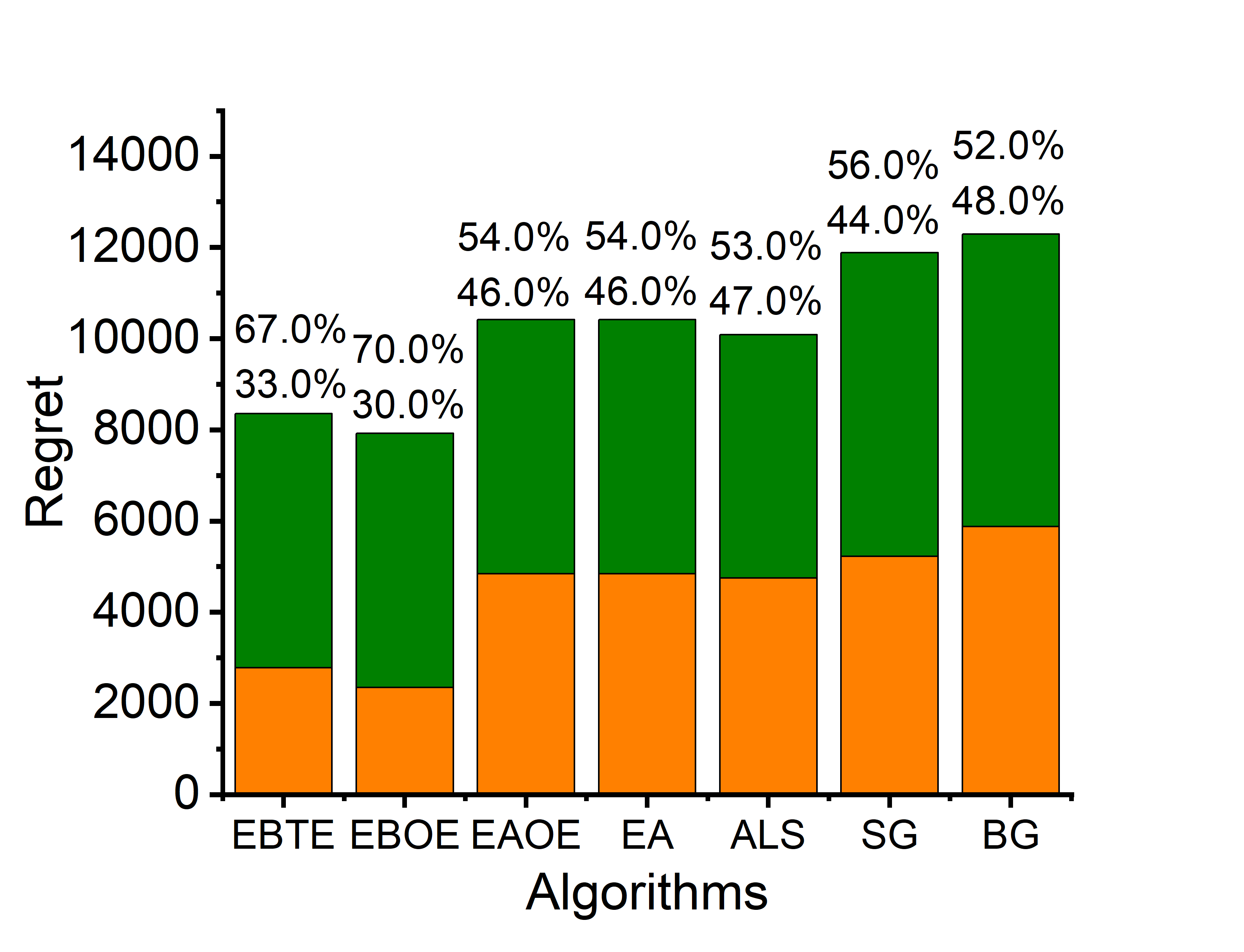} \\
{\tiny (a) $\alpha = 40 \%$} & {\tiny (b) $\alpha = 60 \%$} & {\tiny (c) $\alpha = 80 \%$} & {\tiny (d) $\alpha = 100 \%$} \\


 \includegraphics[scale=0.11]{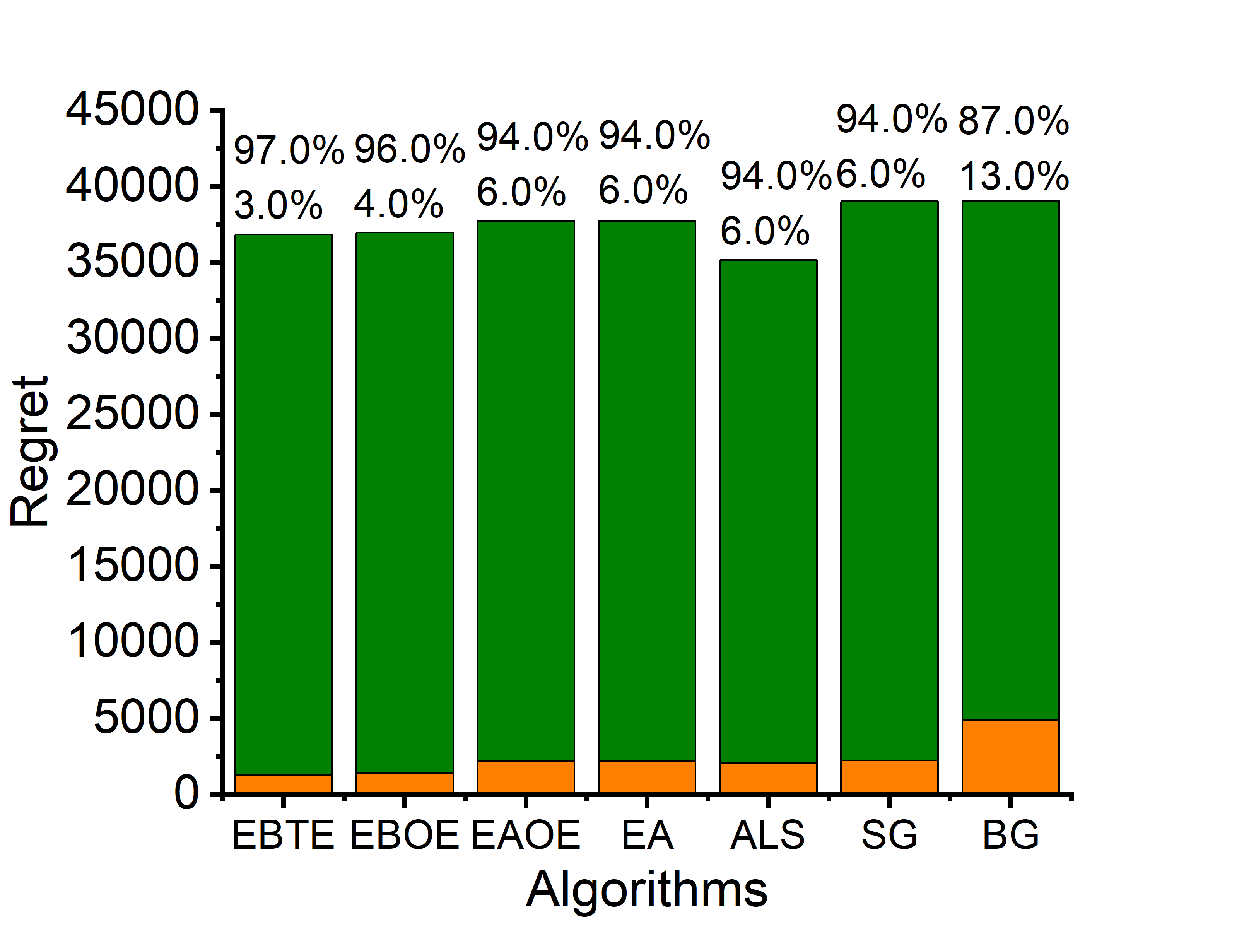} & \includegraphics[scale=0.11]{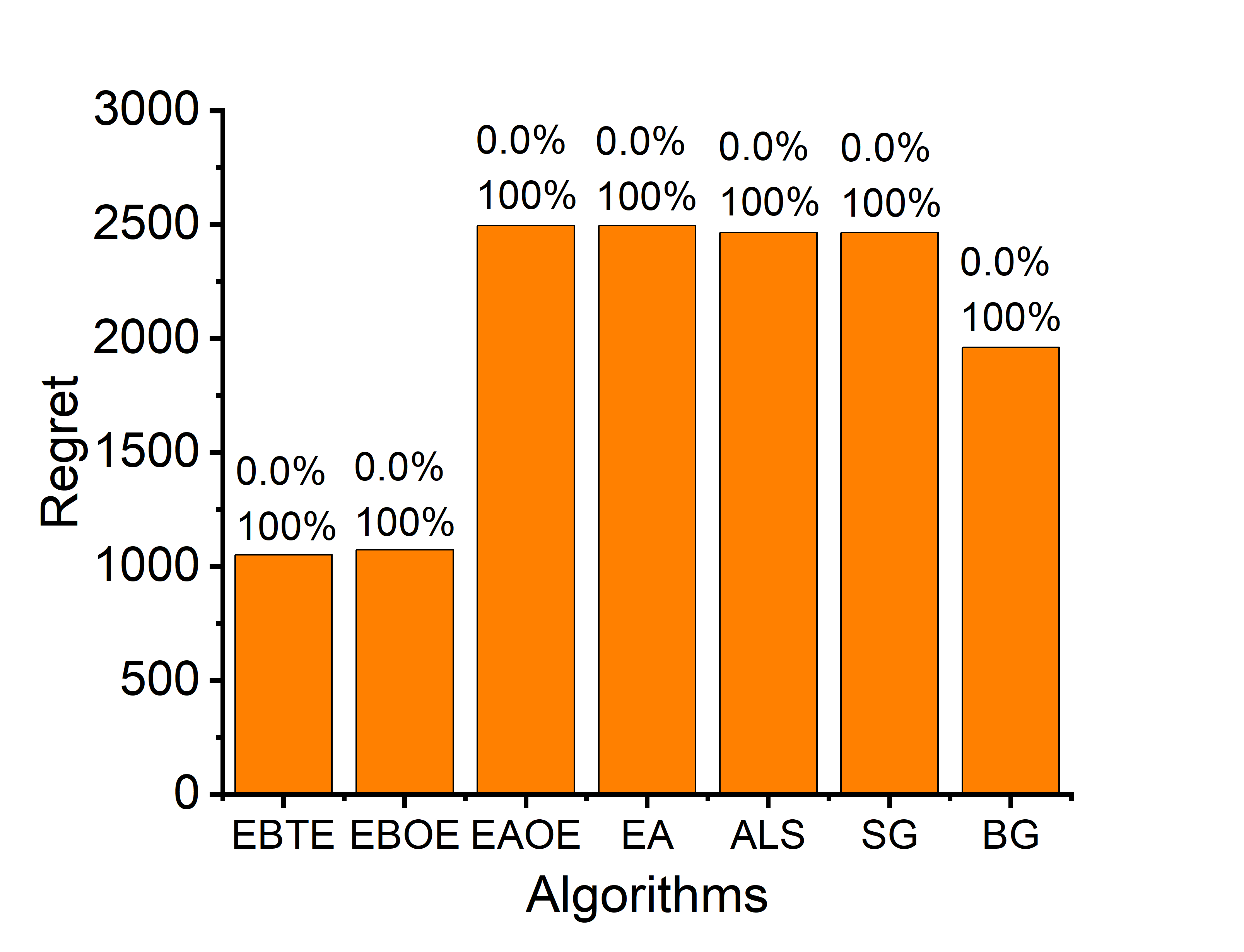} & \includegraphics[scale=0.11]{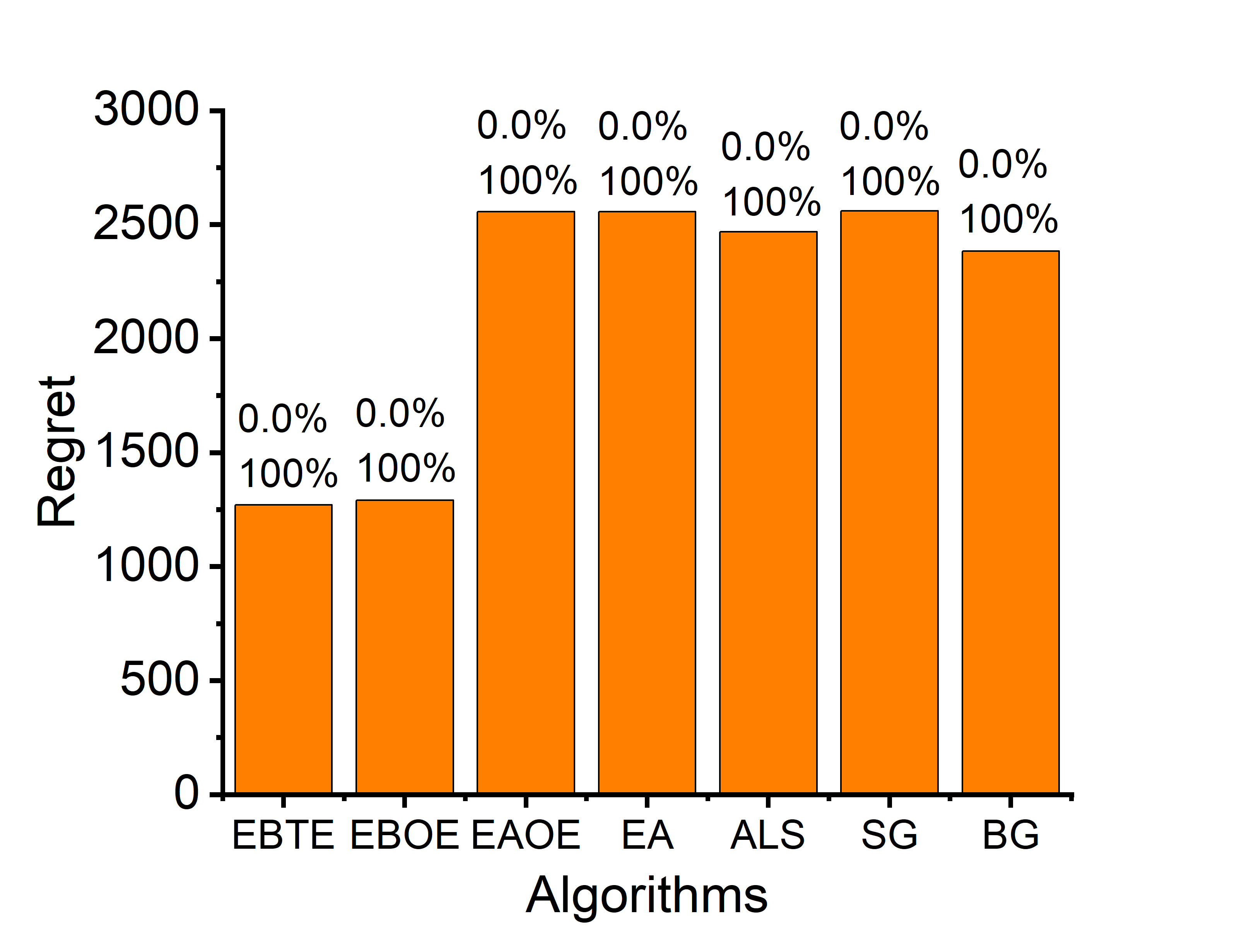}  &\includegraphics[scale=0.11]{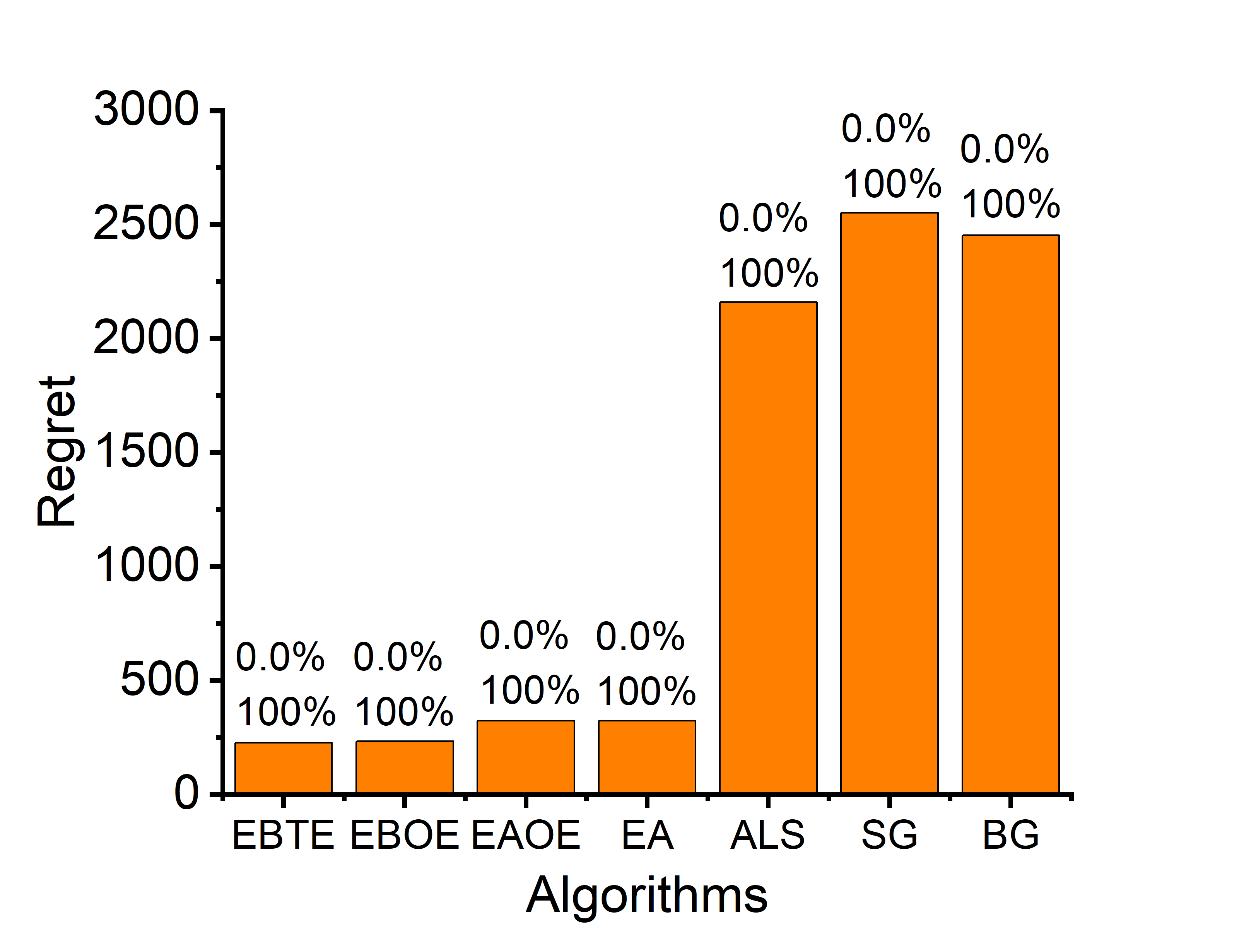} \\
{\tiny (e) $\alpha = 120 \%$} & {\tiny (f) $\alpha = 40 \%$} & {\tiny (g) $\alpha = 60 \%$} & {\tiny (h) $\alpha = 80 \%$} \\

 \includegraphics[scale=0.11]{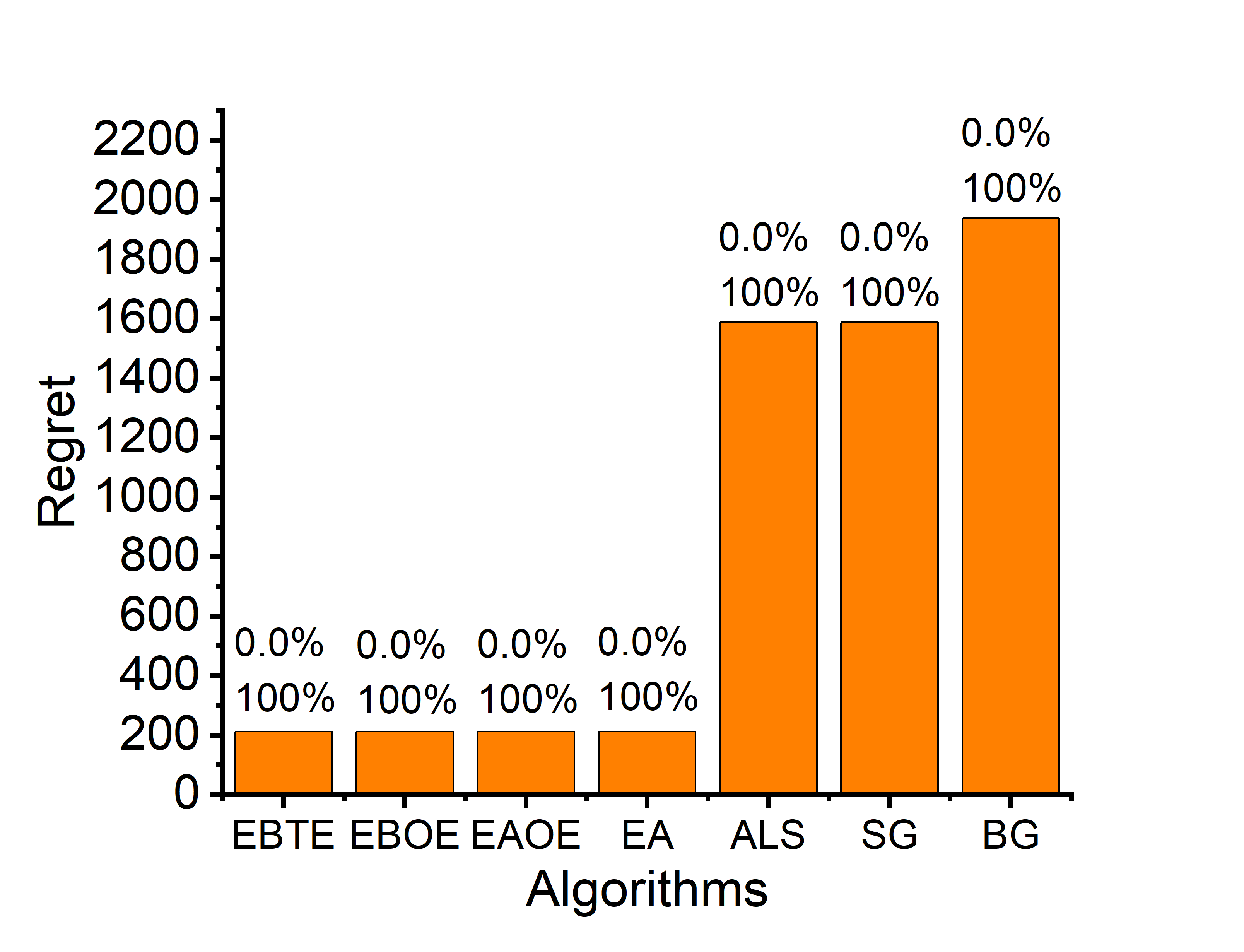} & \includegraphics[scale=0.11]{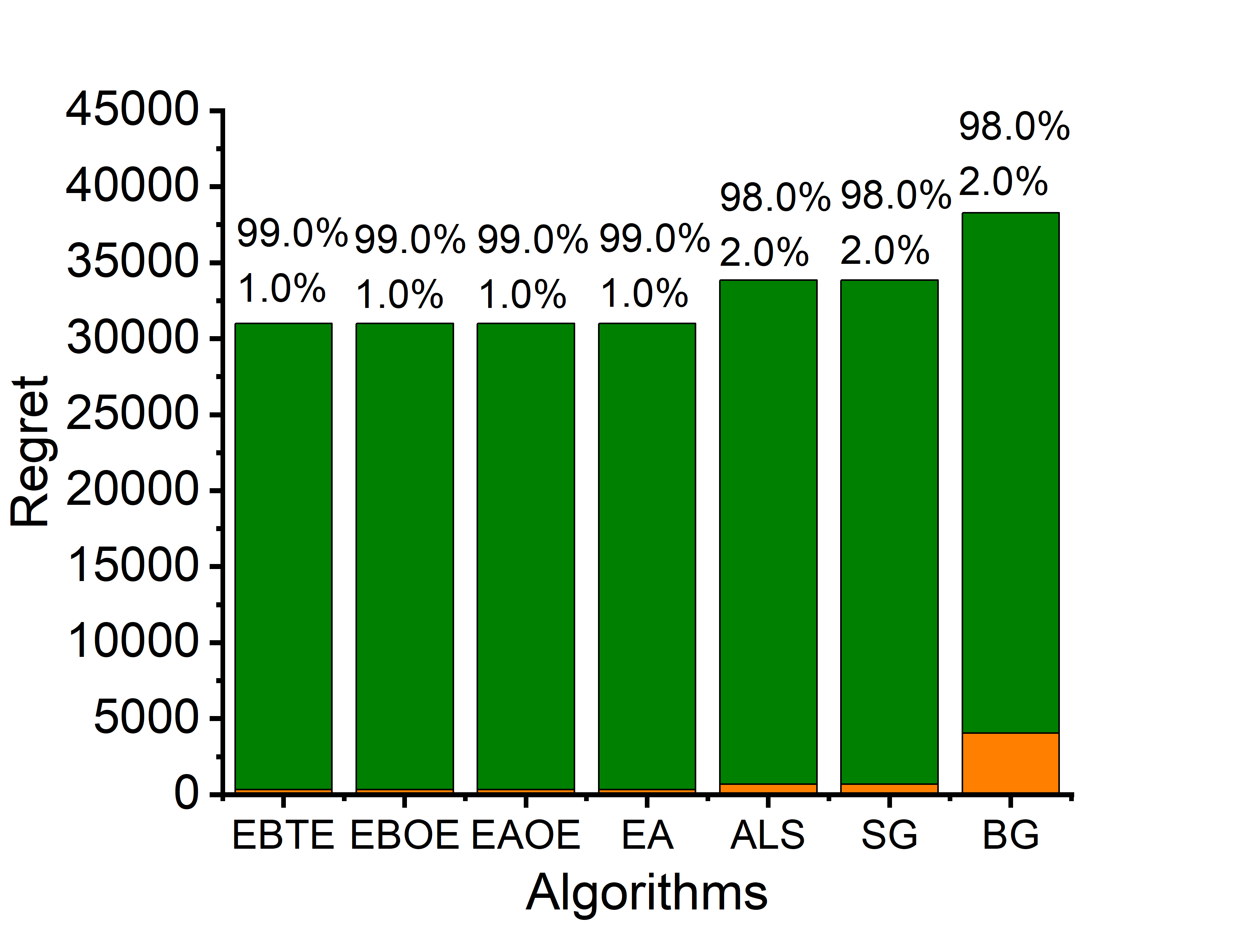} &
\includegraphics[scale=0.11]{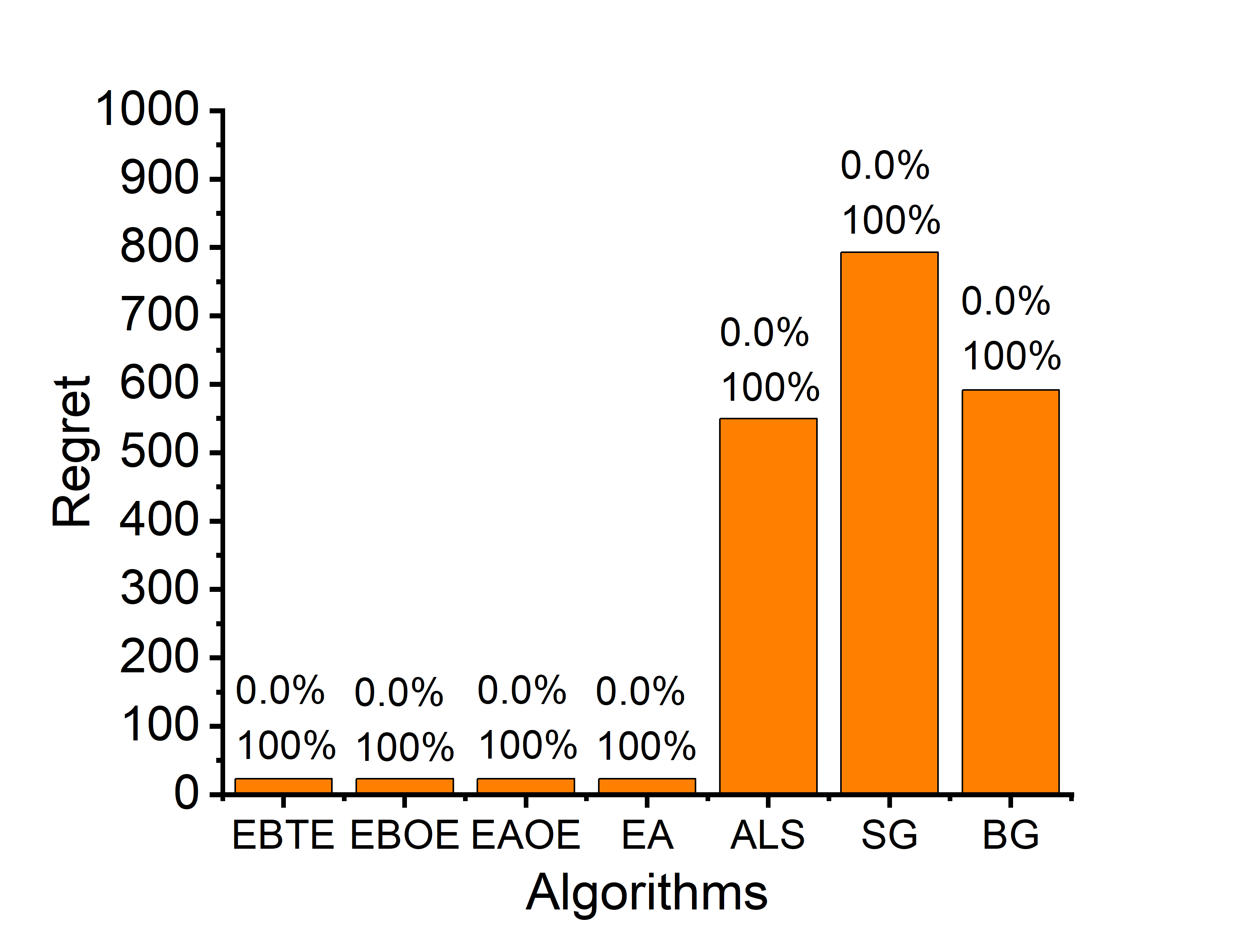} & \includegraphics[scale=0.11]{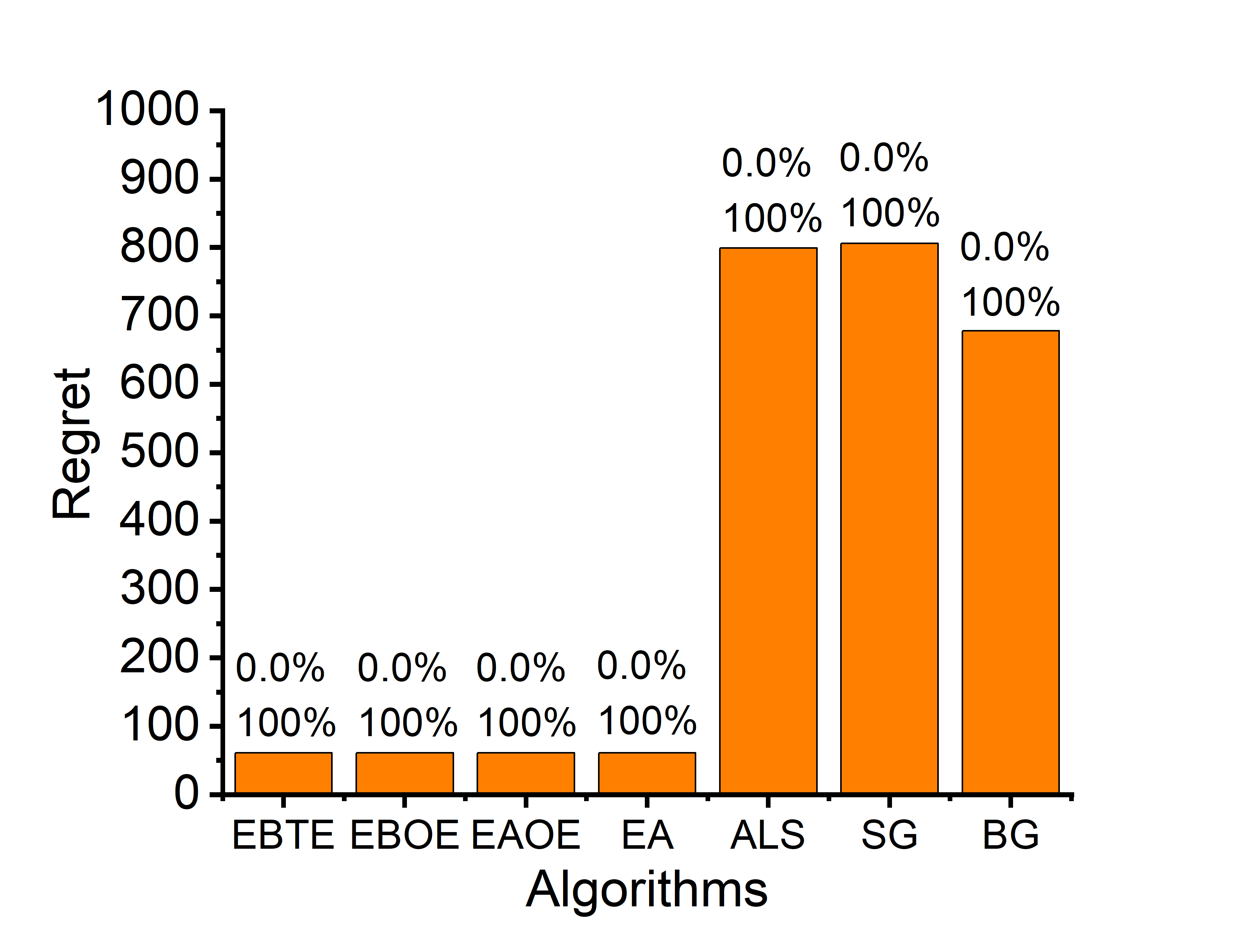} \\
{\tiny (i) $\alpha = 100 \%$} &{\tiny (j) $\alpha = 120 \%$} & {\tiny (k) $\alpha = 40 \%$} & {\tiny (l) $\alpha = 60 \%$} \\

\includegraphics[scale=0.11]{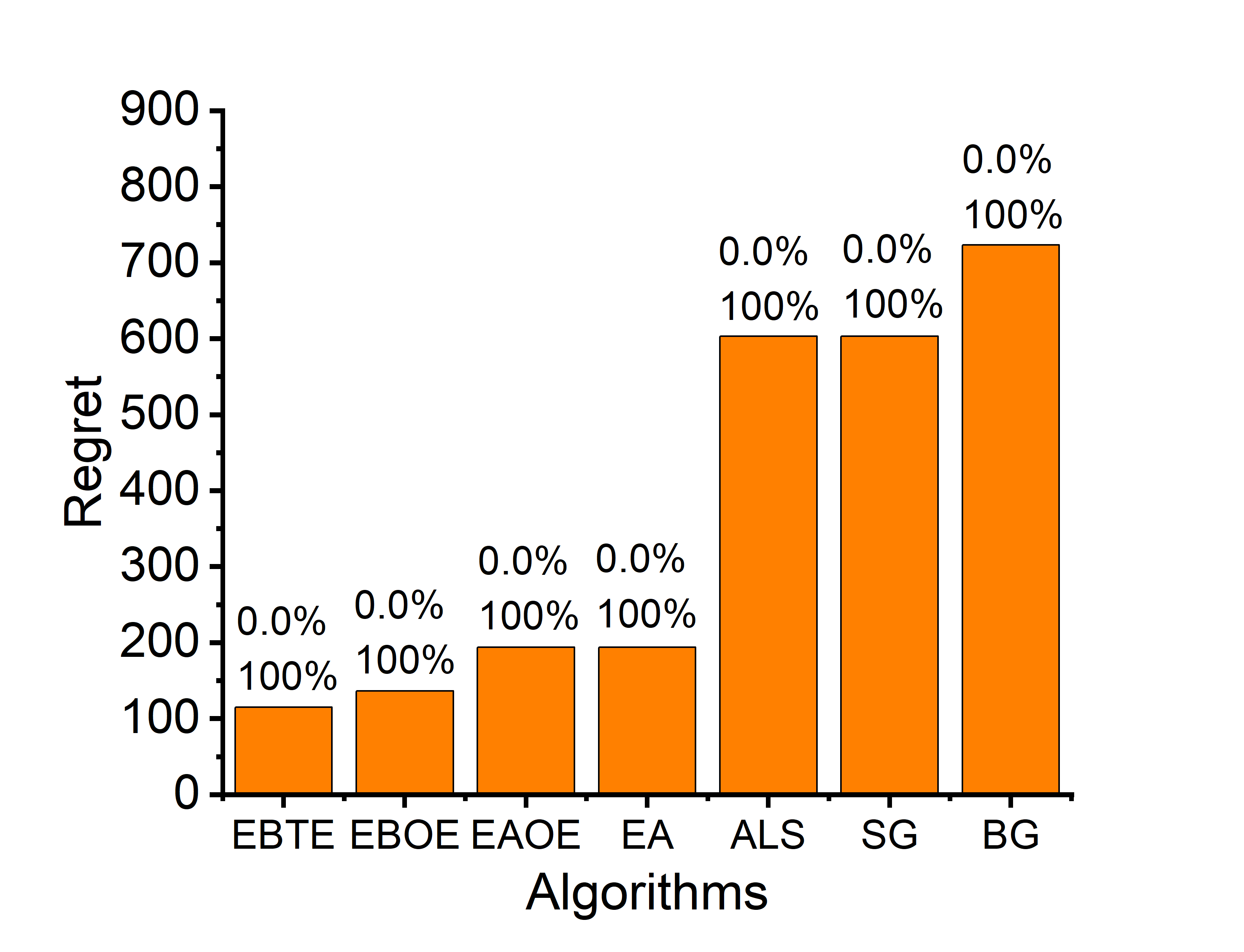} & \includegraphics[scale=0.11]{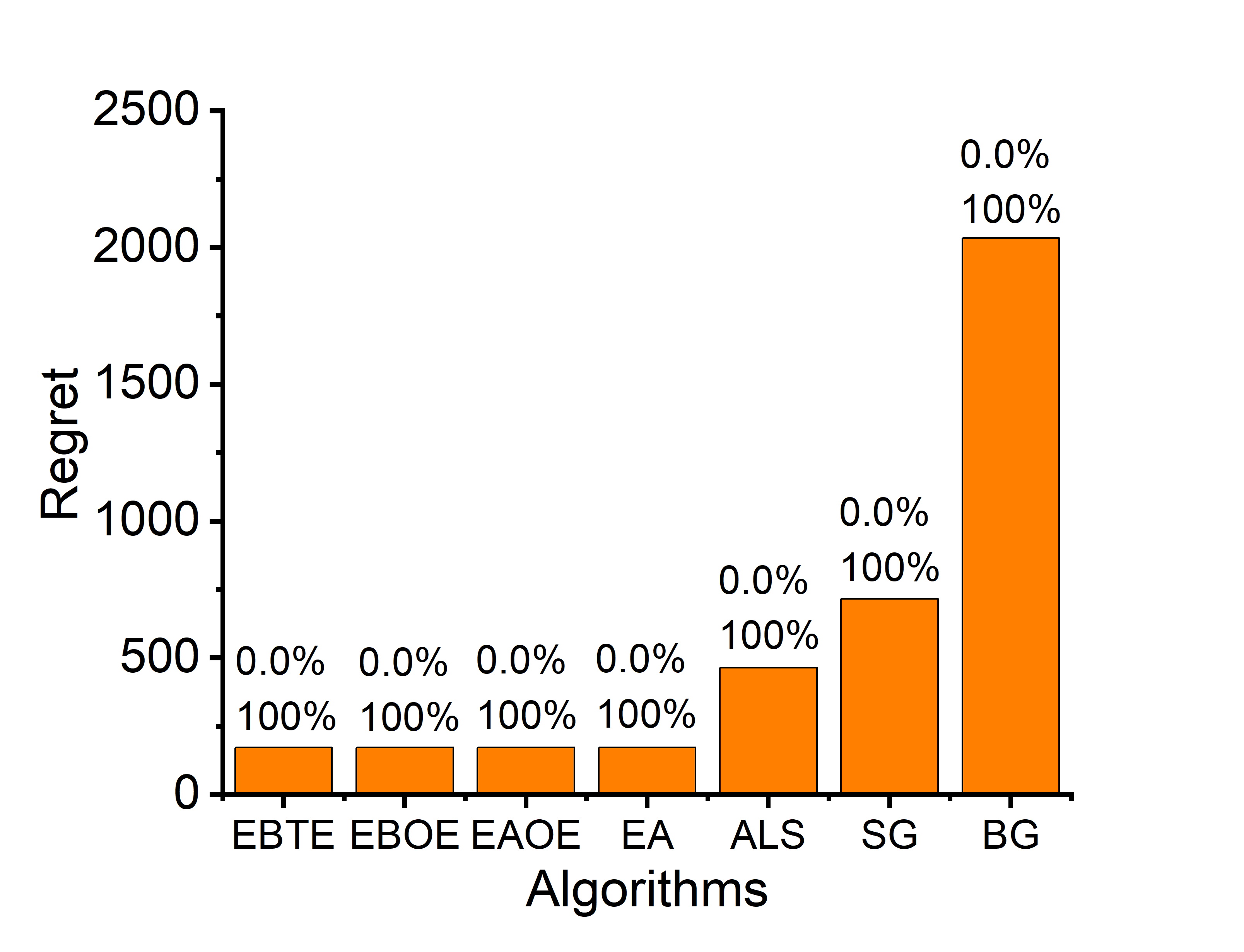} & \includegraphics[scale=0.11]{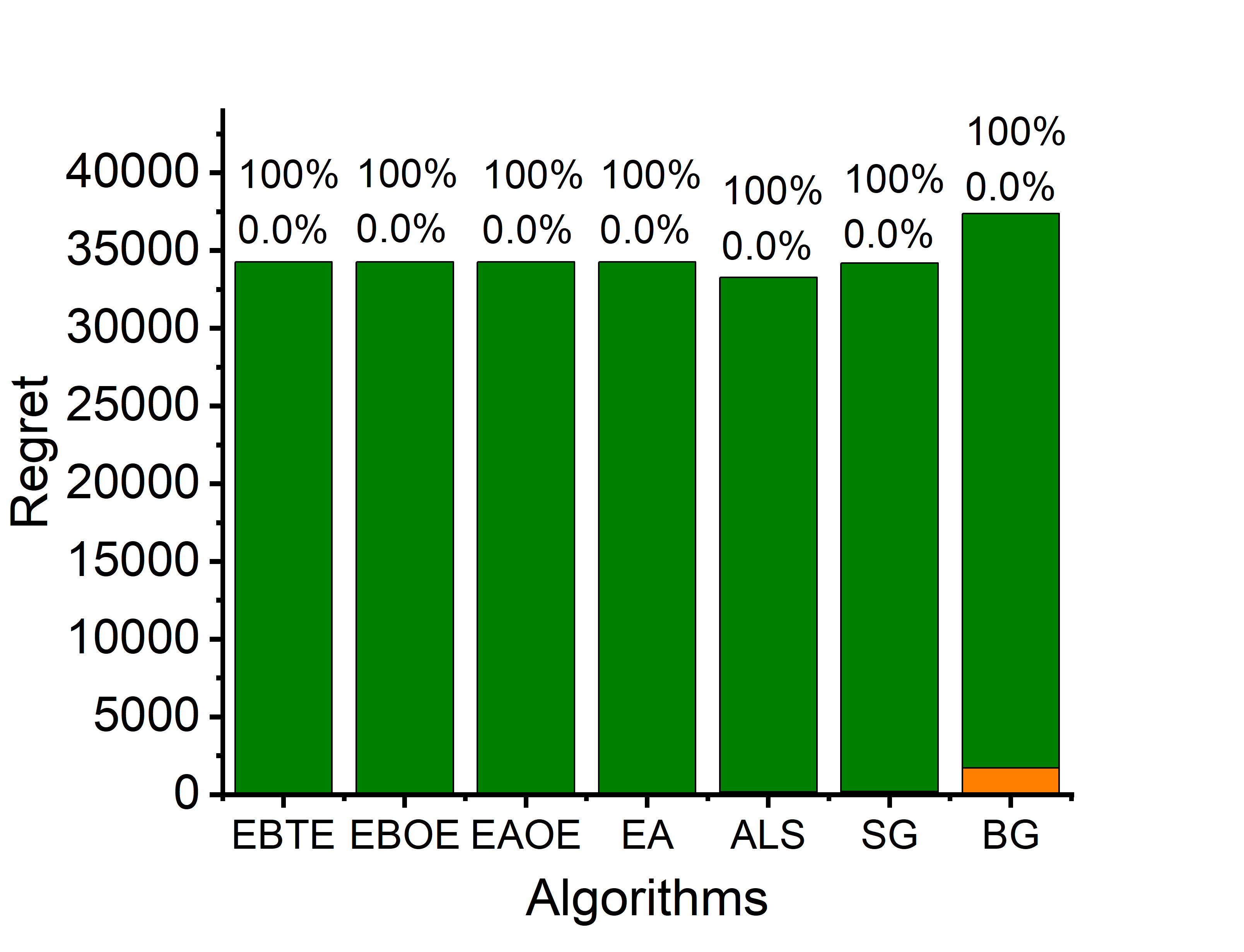} &
\includegraphics[scale=0.11]{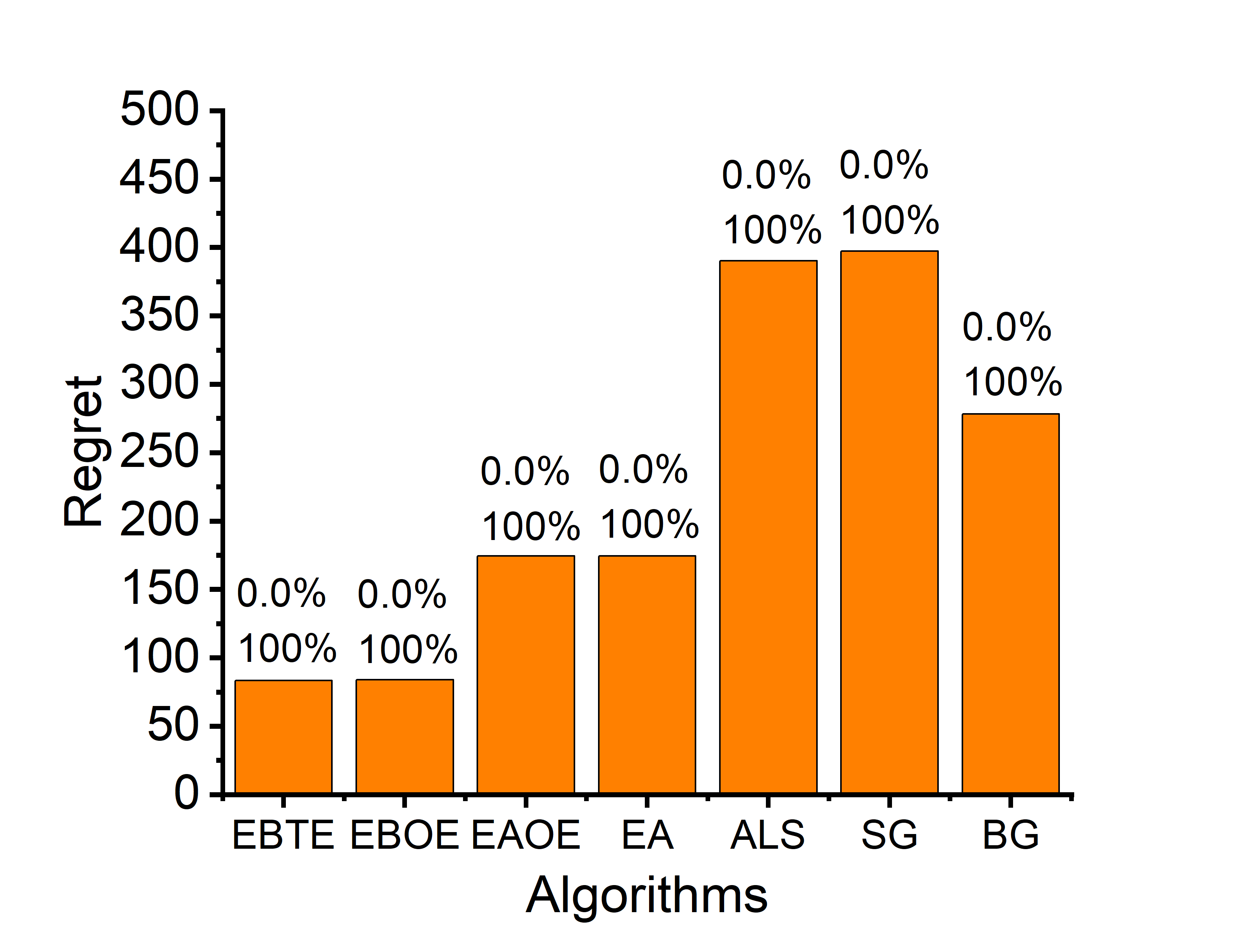} \\
{\tiny (m) $\alpha = 80 \%$} & {\tiny (n) $\alpha = 100 \%$} &{\tiny (o) $\alpha = 120 \%$} & {\tiny (p) $\alpha = 40 \%$} \\

\includegraphics[scale=0.11]{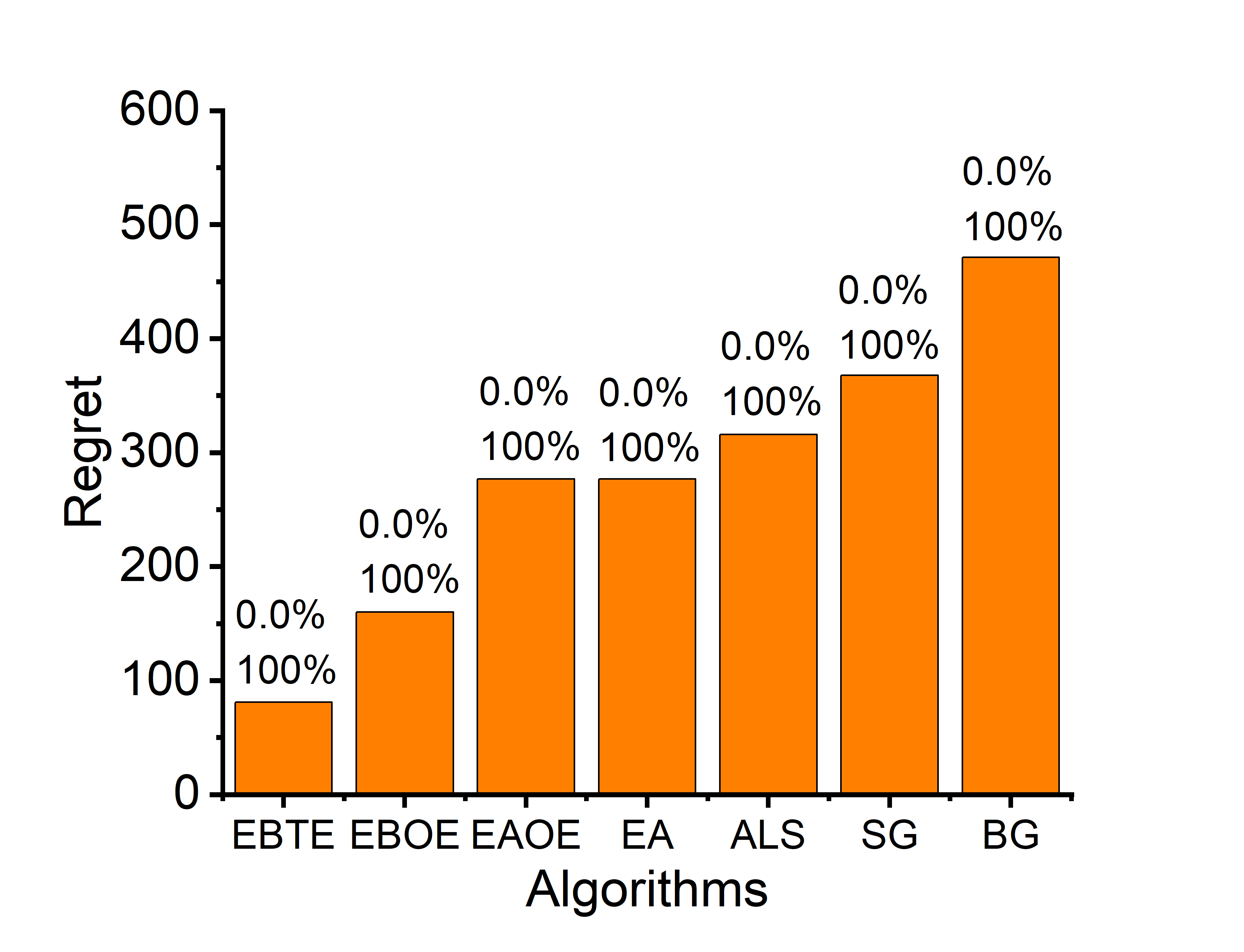}  &\includegraphics[scale=0.11]{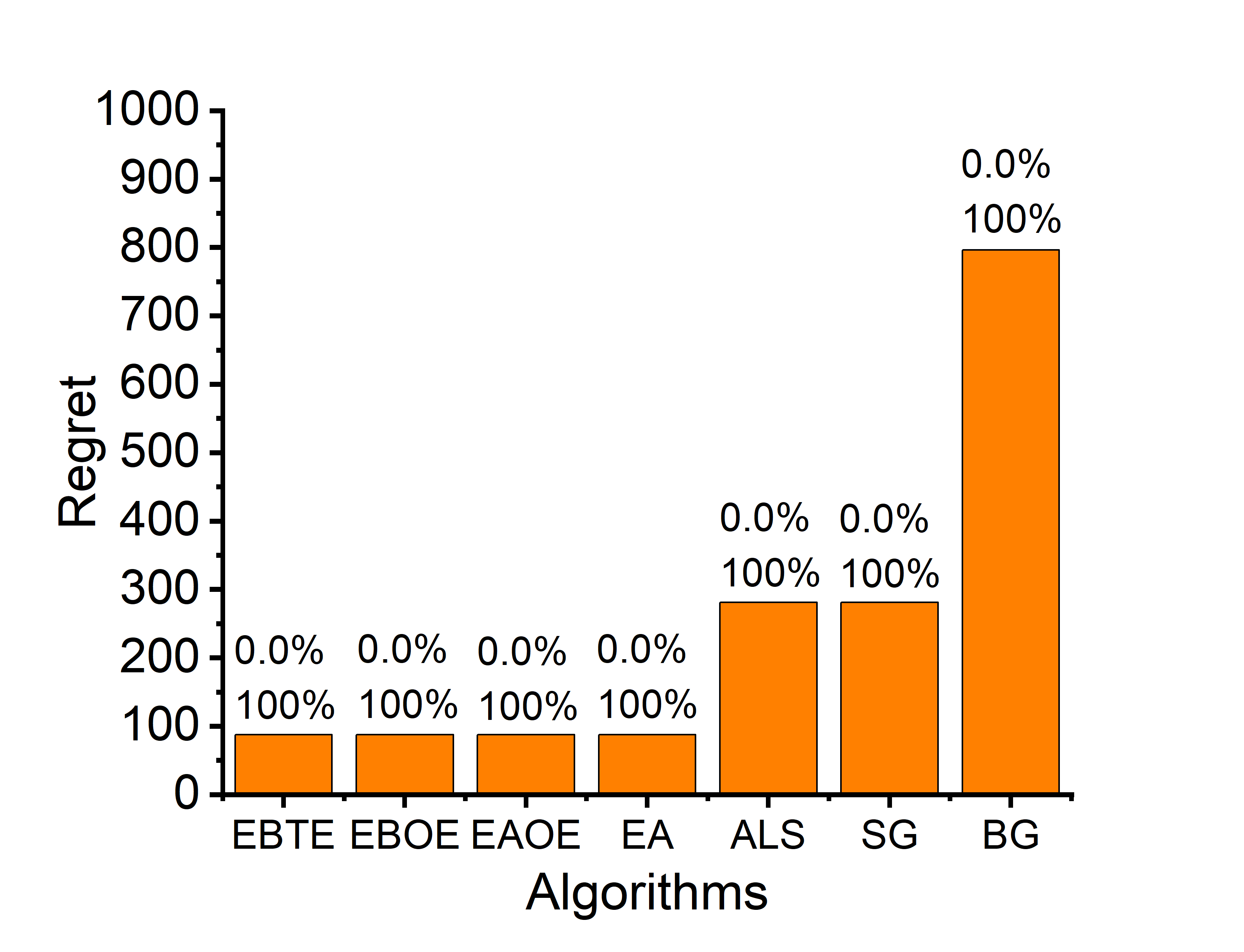} & \includegraphics[scale=0.11]{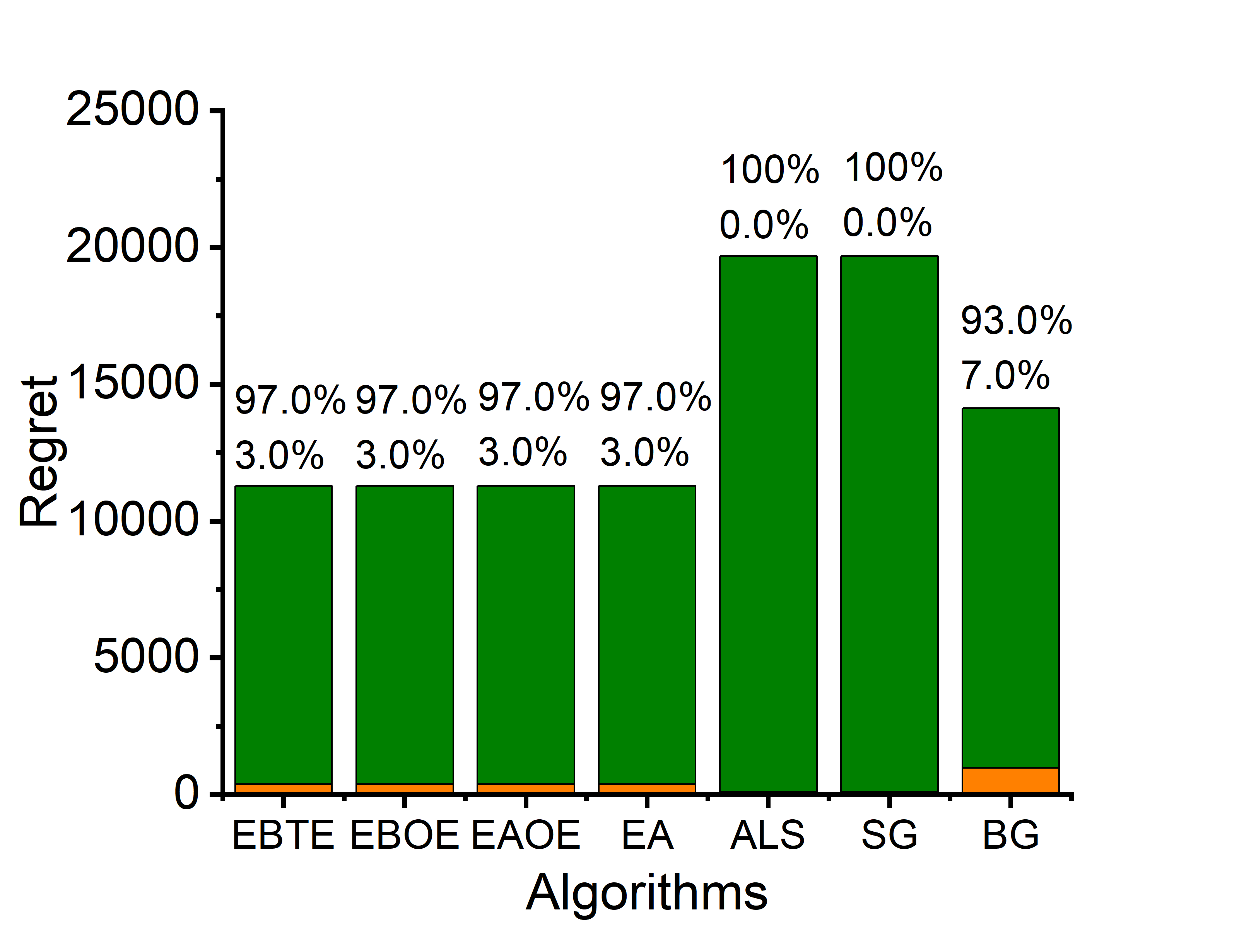} & \includegraphics[scale=0.11]{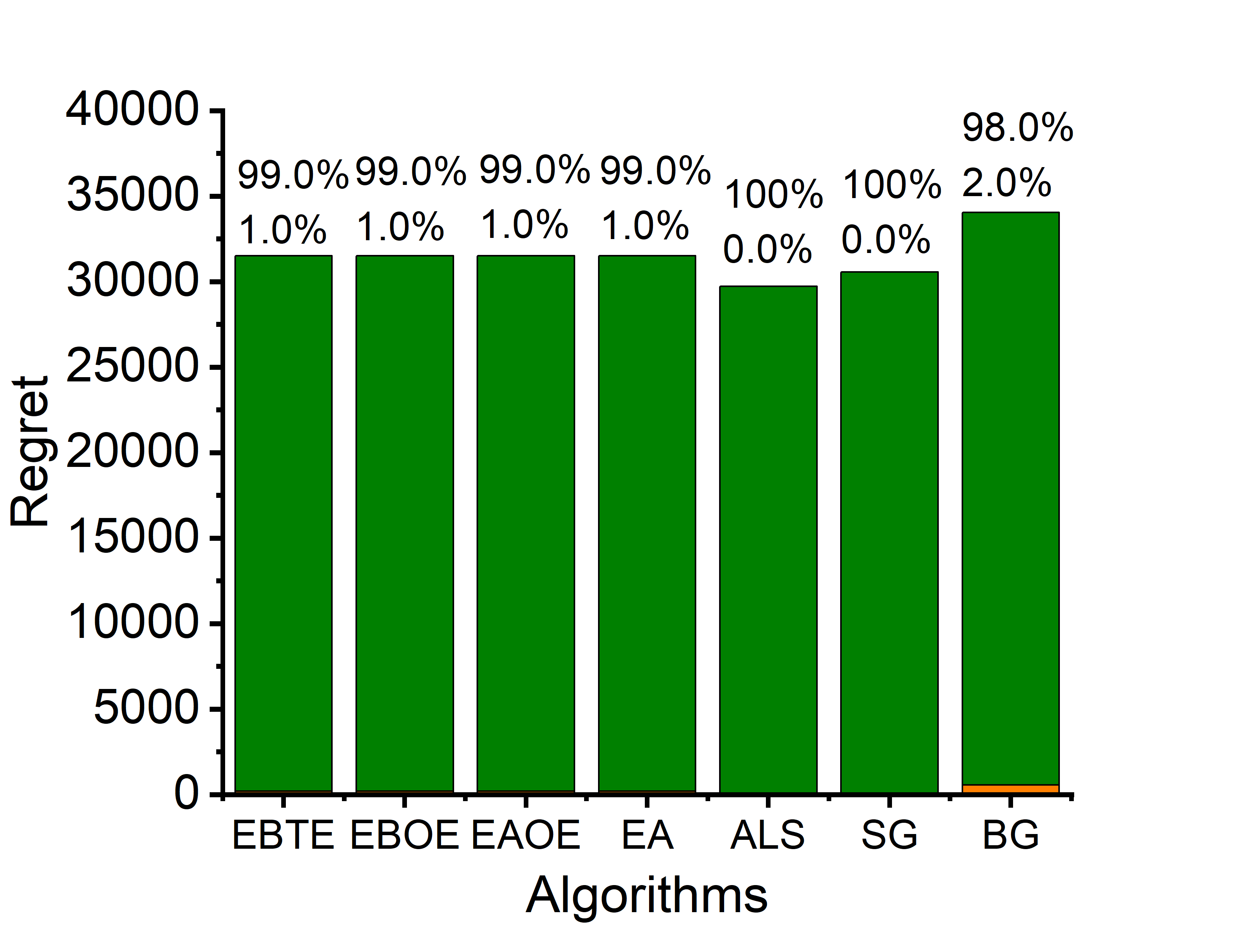}\\
 {\tiny (q) $\alpha = 60 \%$} & {\tiny (r) $\alpha = 80 \%$} & {\tiny (s) $\alpha = 100 \%$} &{\tiny (t) $\alpha = 120 \%$}\\

\end{tabular}
\caption{Regret varying $\alpha$ when $\mathcal{I}^{ID} = 1\%, \mathcal{|A|} = 100$ (a, b, c, d, e), when $\mathcal{I}^{ID} = 2\%, \mathcal{|A|} = 50$ (f,g,h,i,j), when $\mathcal{I}^{ID} = 5\%, \mathcal{|A|} = 20$ (k,l,m,n,o) and when $\mathcal{I}^{ID} = 10\%, \mathcal{|A|} = 10$ (p, q, r, s, t)for Bank location type }
\label{Fig:Bank}
\end{figure}

\begin{figure}[h!]
\centering
   \begin{tabular}{lclc}
       Unsatisfied Regret & \includegraphics[width=0.11\linewidth]{Unsatisfied.png} \  & \ Excessive Regret & \includegraphics[width=0.11\linewidth]{Excessive.png} \\
    \end{tabular}

\begin{tabular}{cccc}
\includegraphics[scale=0.11]{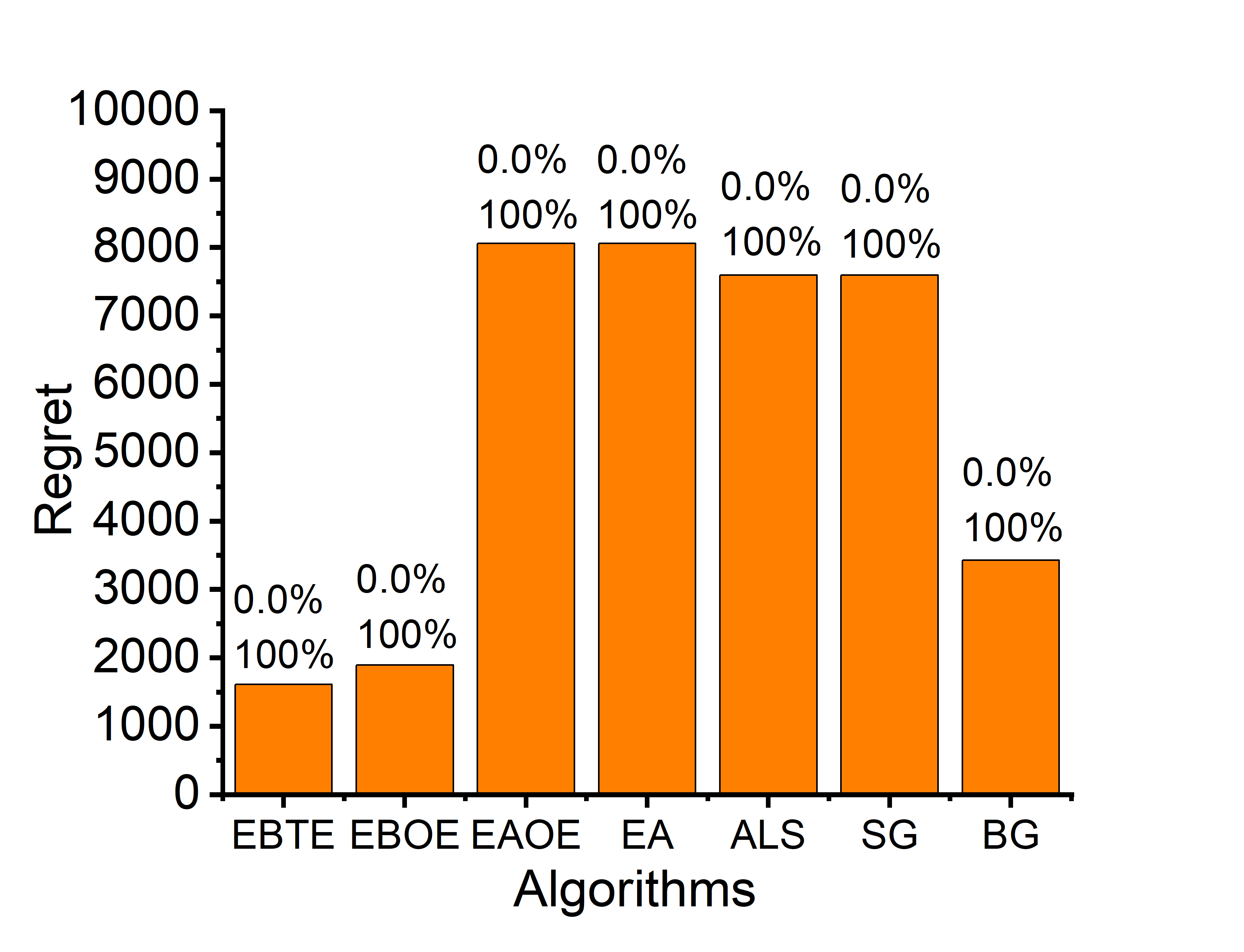} & \includegraphics[scale=0.11]{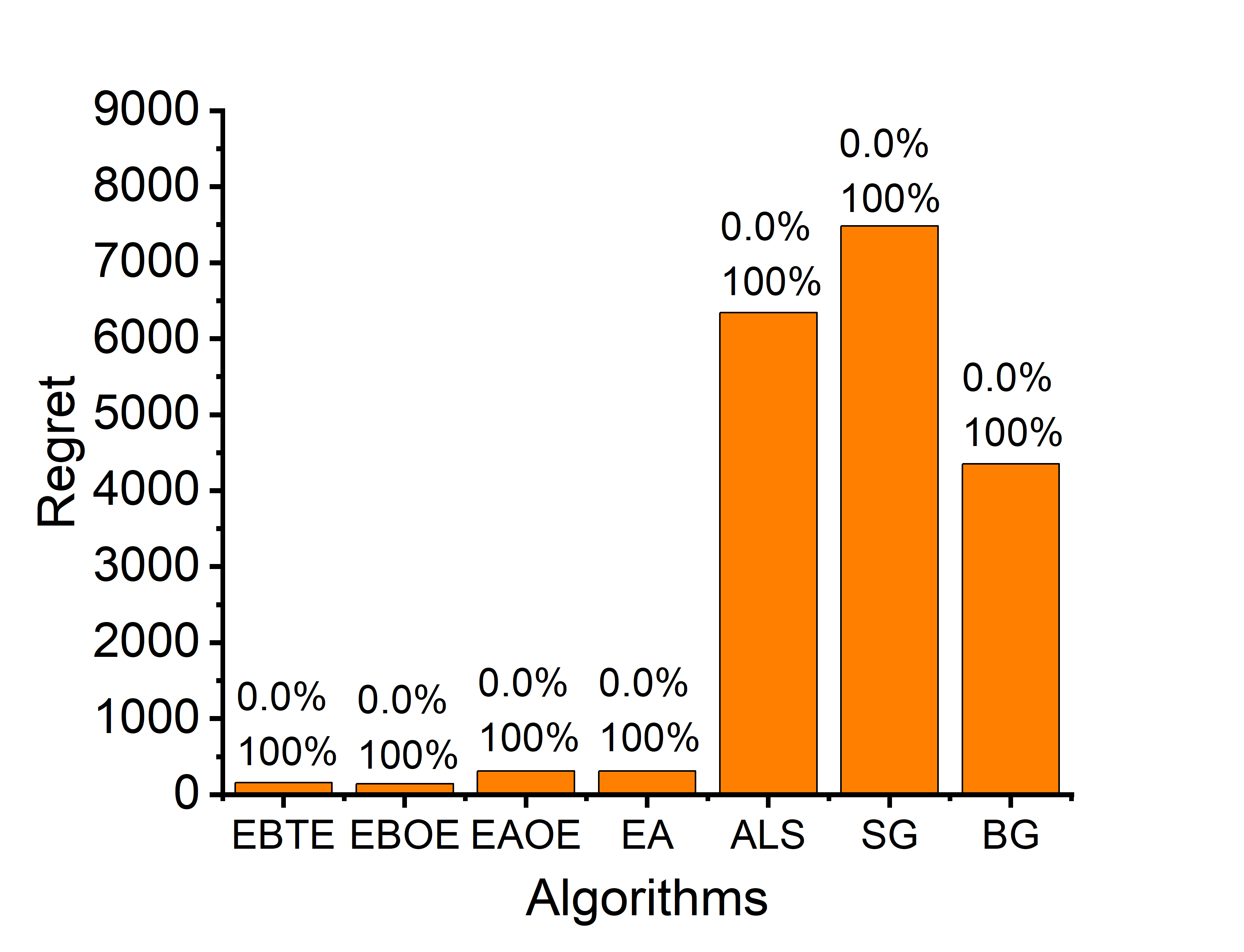}  &\includegraphics[scale=0.11]{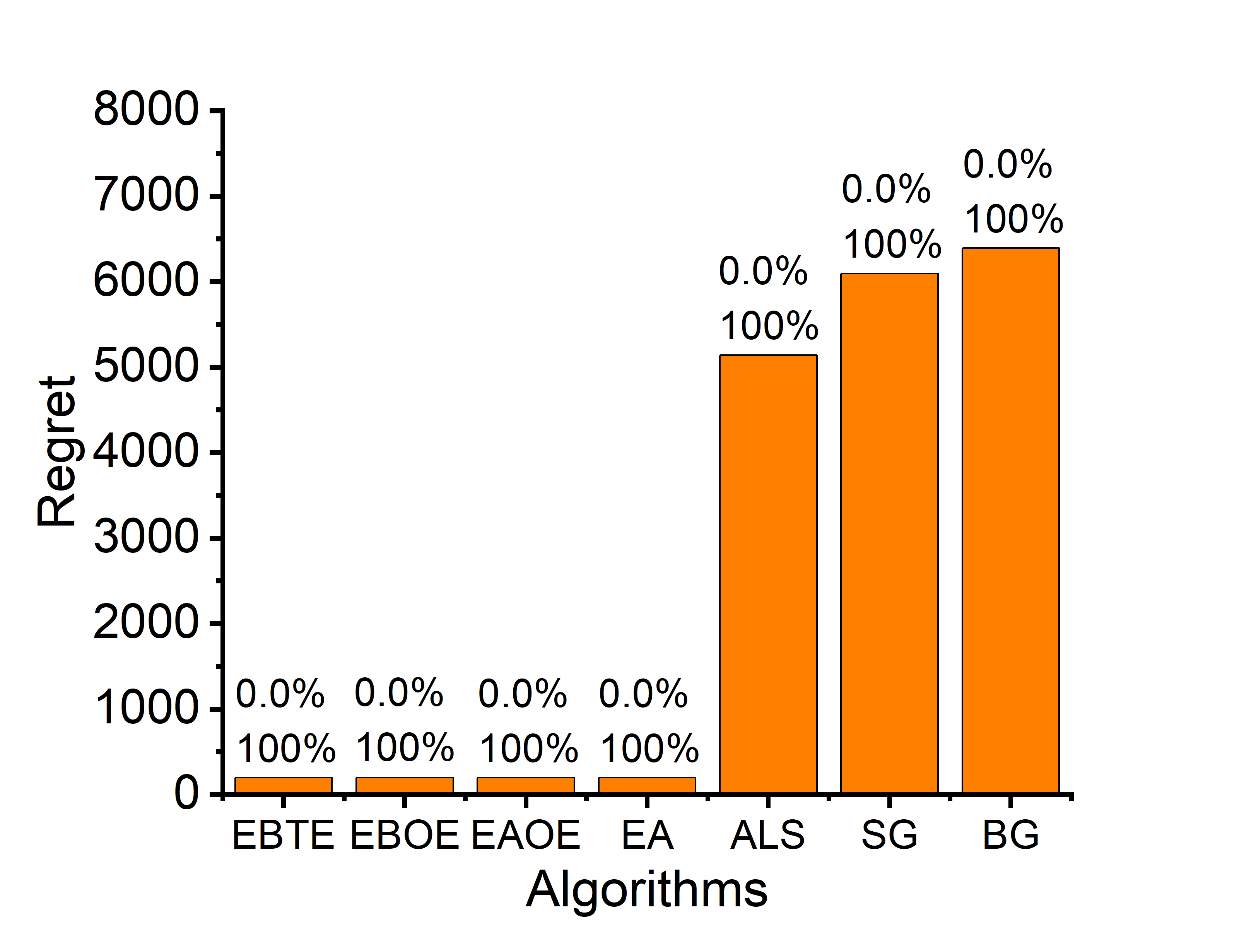} & \includegraphics[scale=0.11]{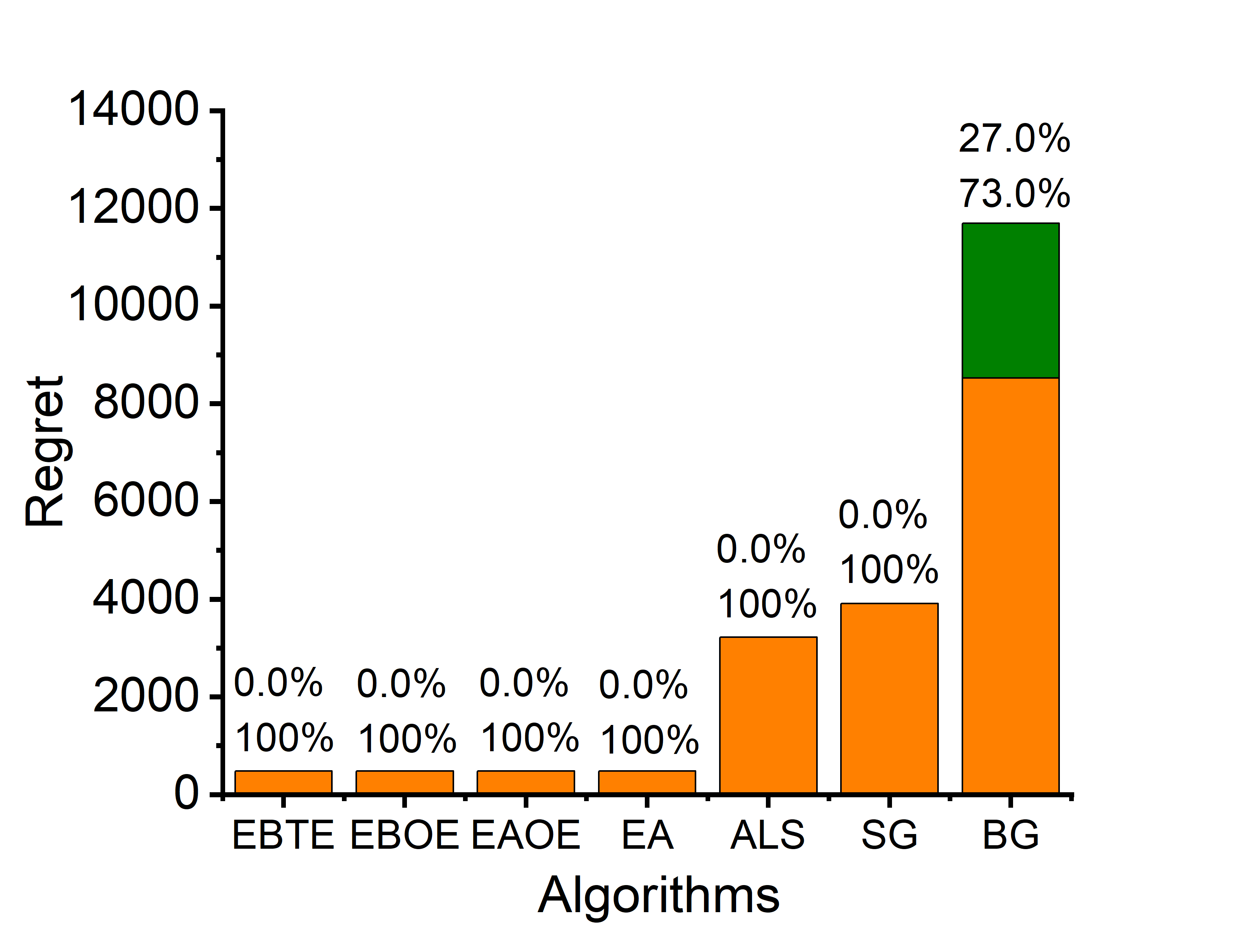} \\
{\tiny (a) $\alpha = 40 \%$} & {\tiny (b) $\alpha = 60 \%$} & {\tiny (c) $\alpha = 80 \%$} & {\tiny (d) $\alpha = 100 \%$} \\


 \includegraphics[scale=0.11]{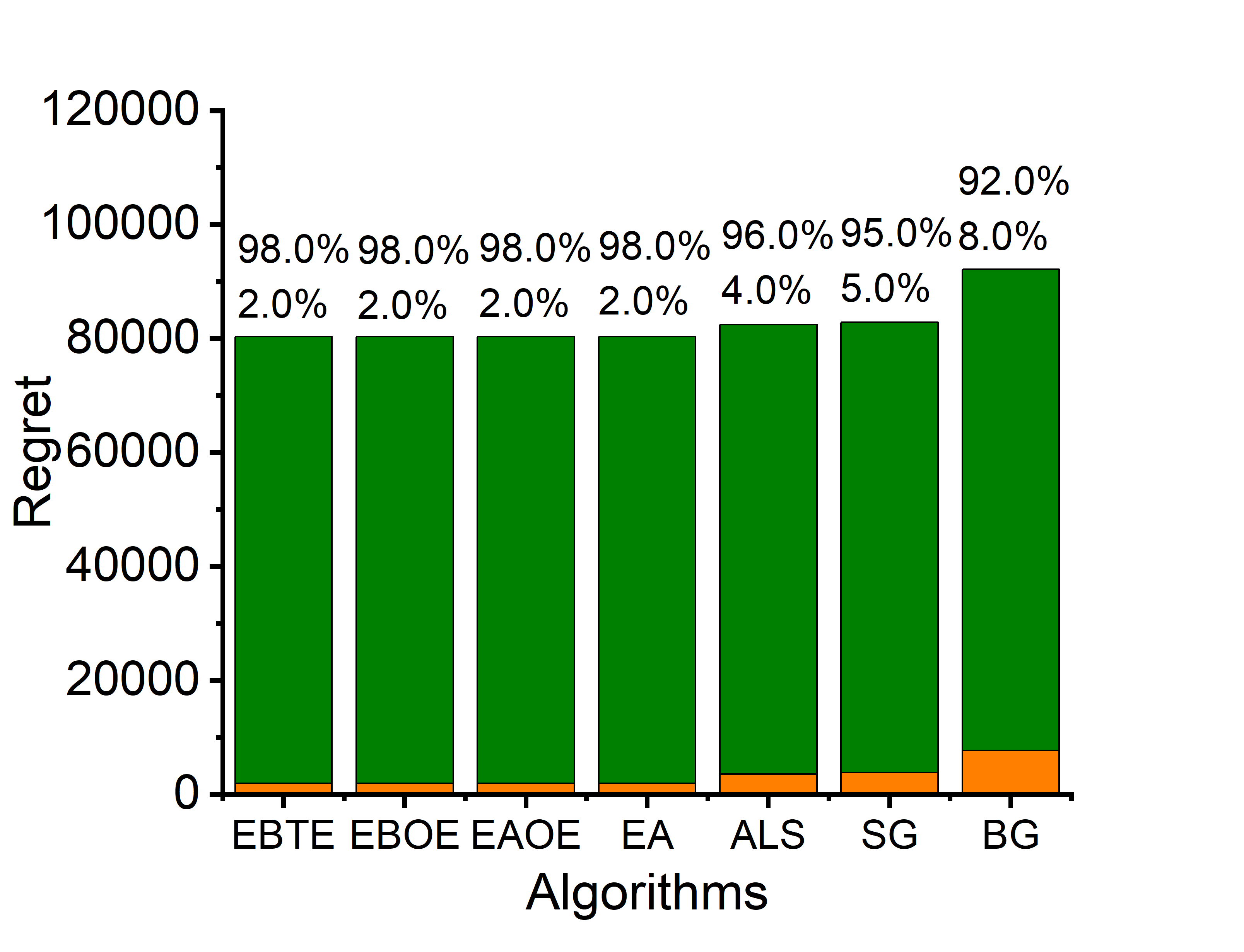} & \includegraphics[scale=0.11]{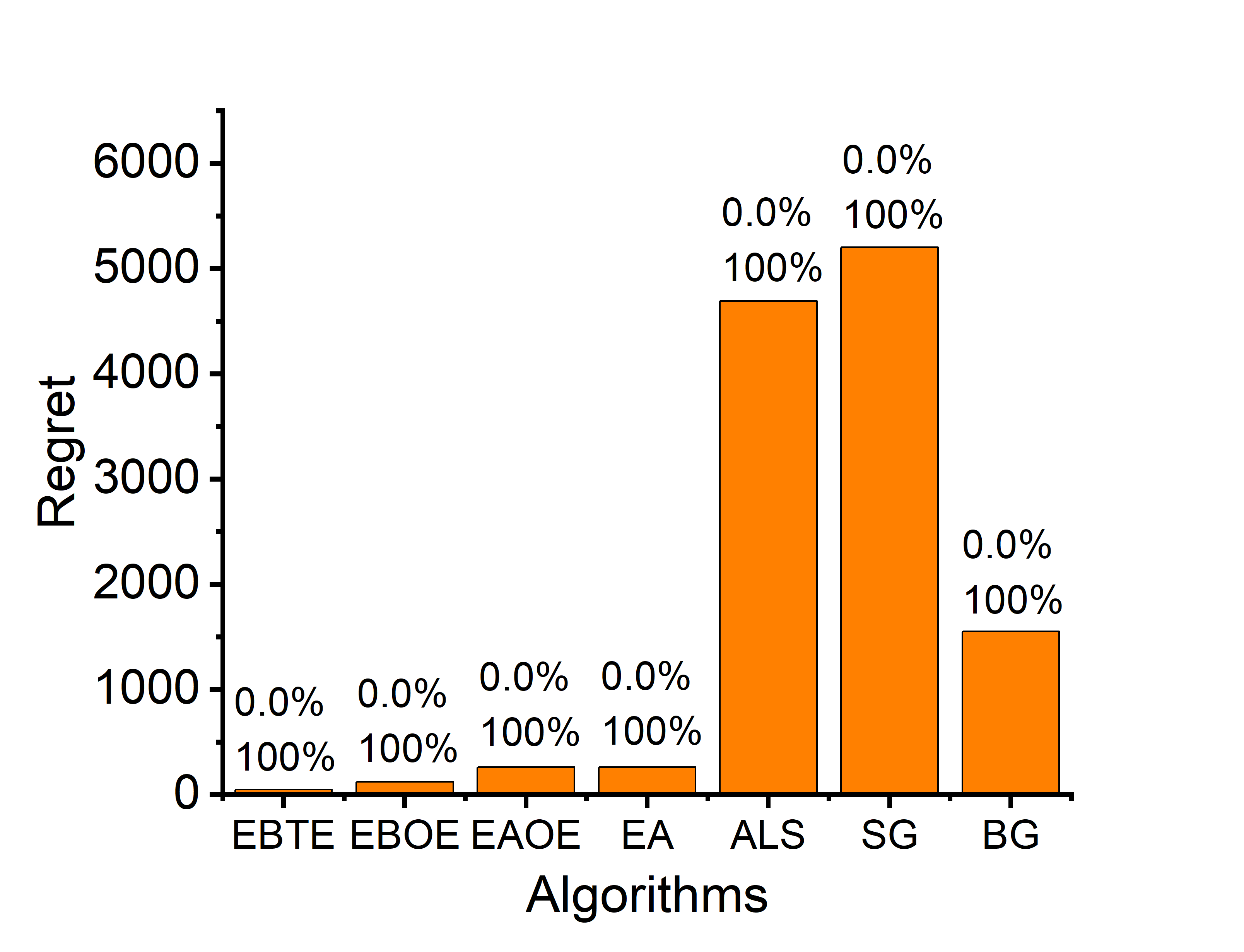} & \includegraphics[scale=0.11]{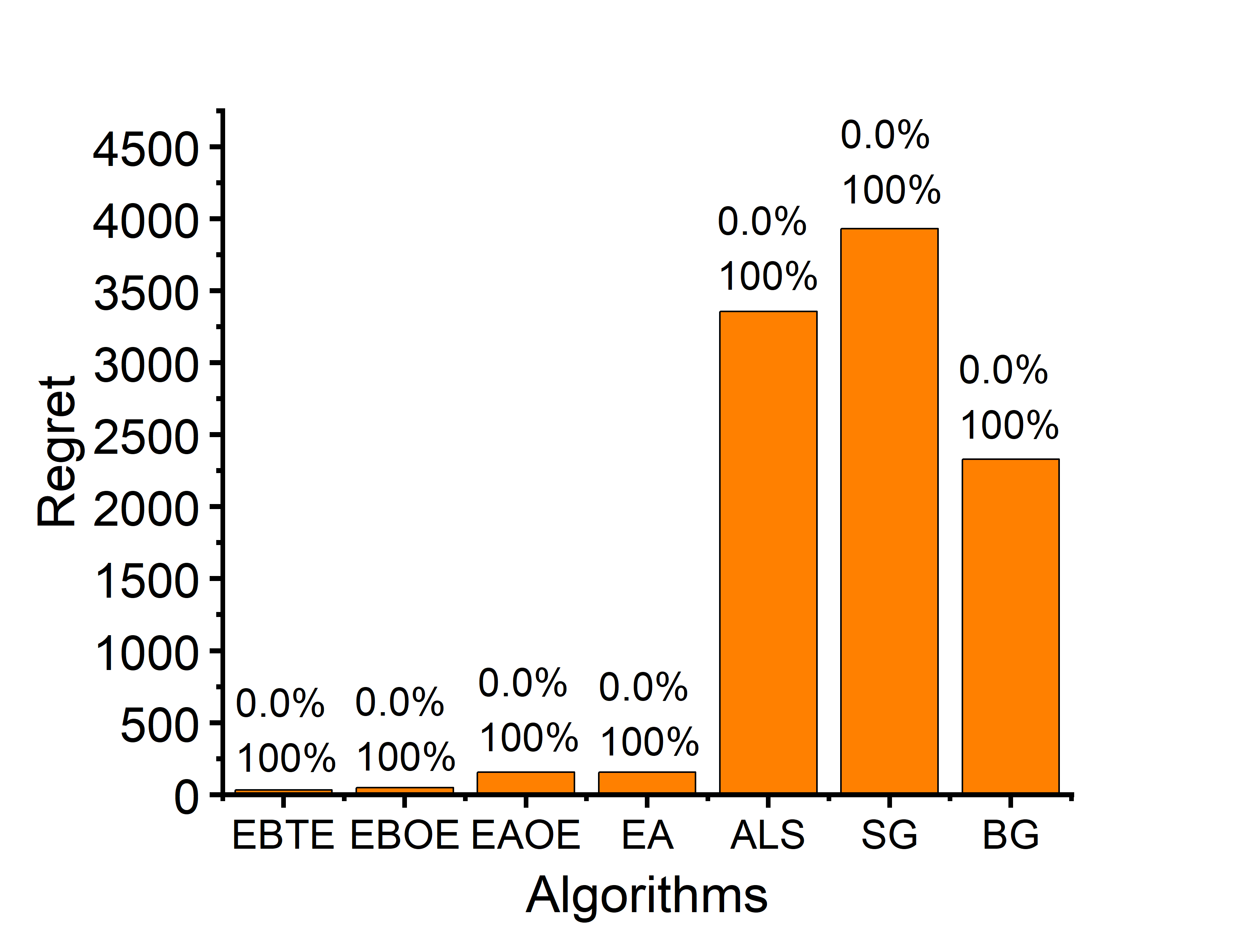}  &\includegraphics[scale=0.11]{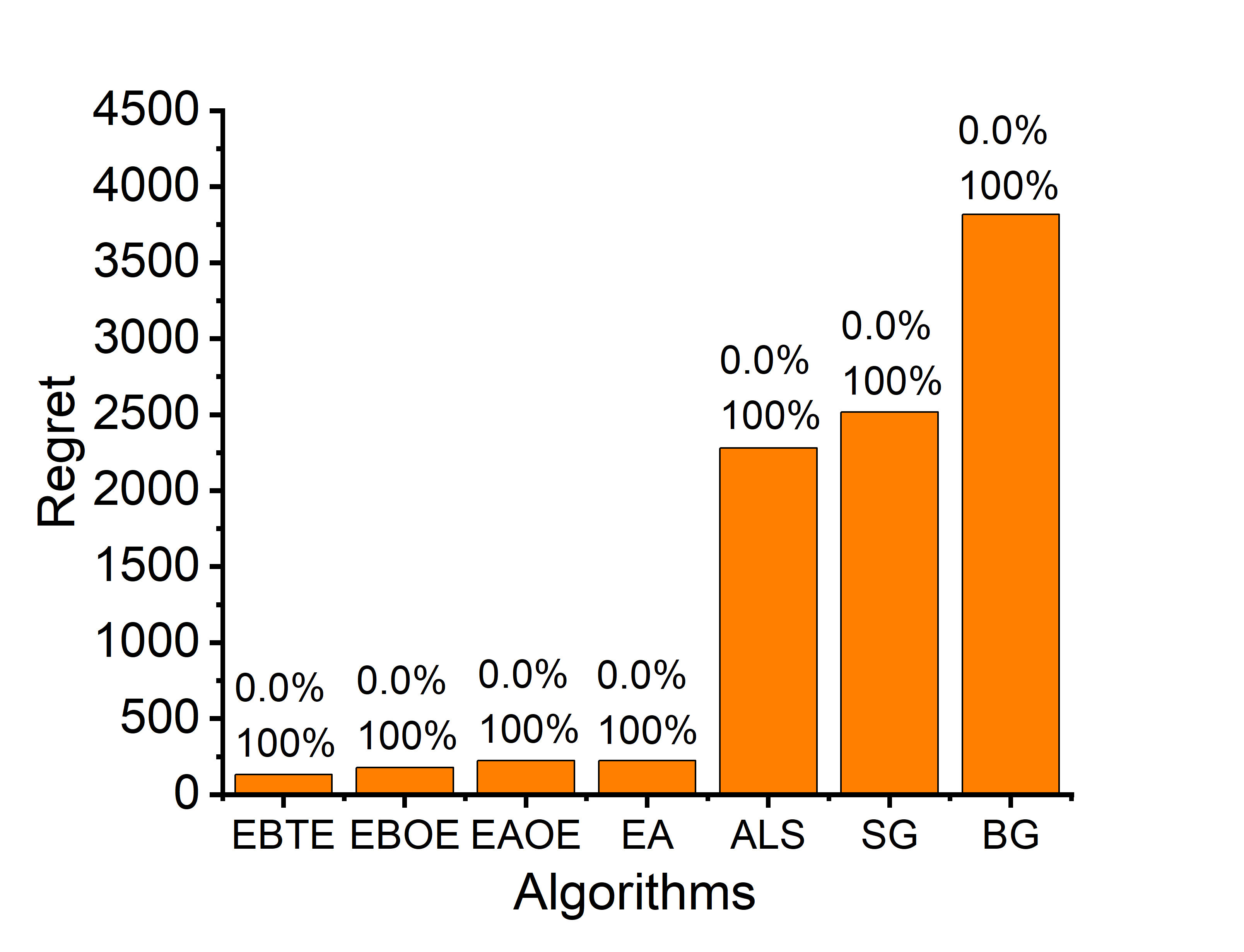} \\
{\tiny (e) $\alpha = 120 \%$} & {\tiny (f) $\alpha = 40 \%$} & {\tiny (g) $\alpha = 60 \%$} & {\tiny (h) $\alpha = 80 \%$} \\

 \includegraphics[scale=0.11]{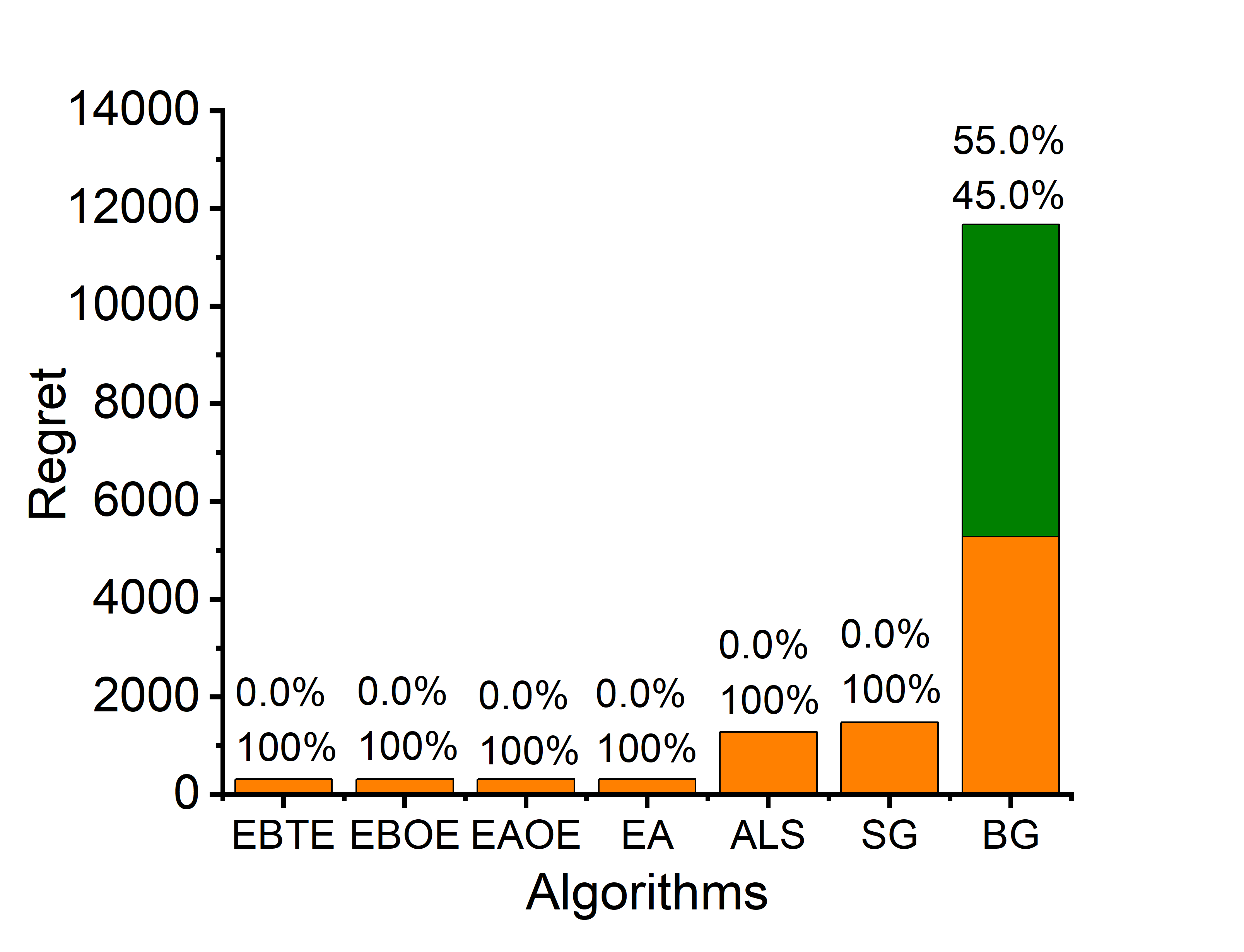} & \includegraphics[scale=0.11]{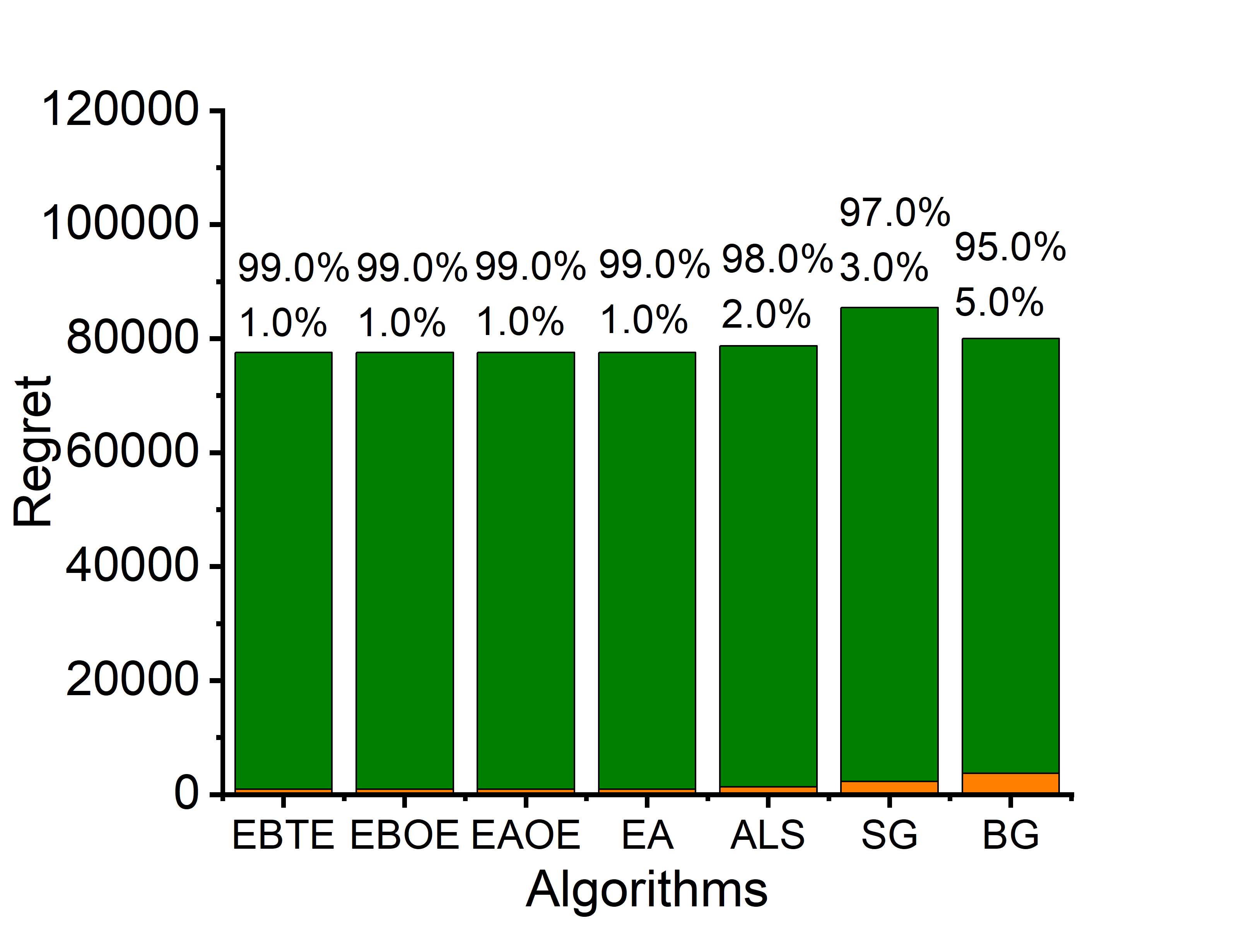} &
\includegraphics[scale=0.11]{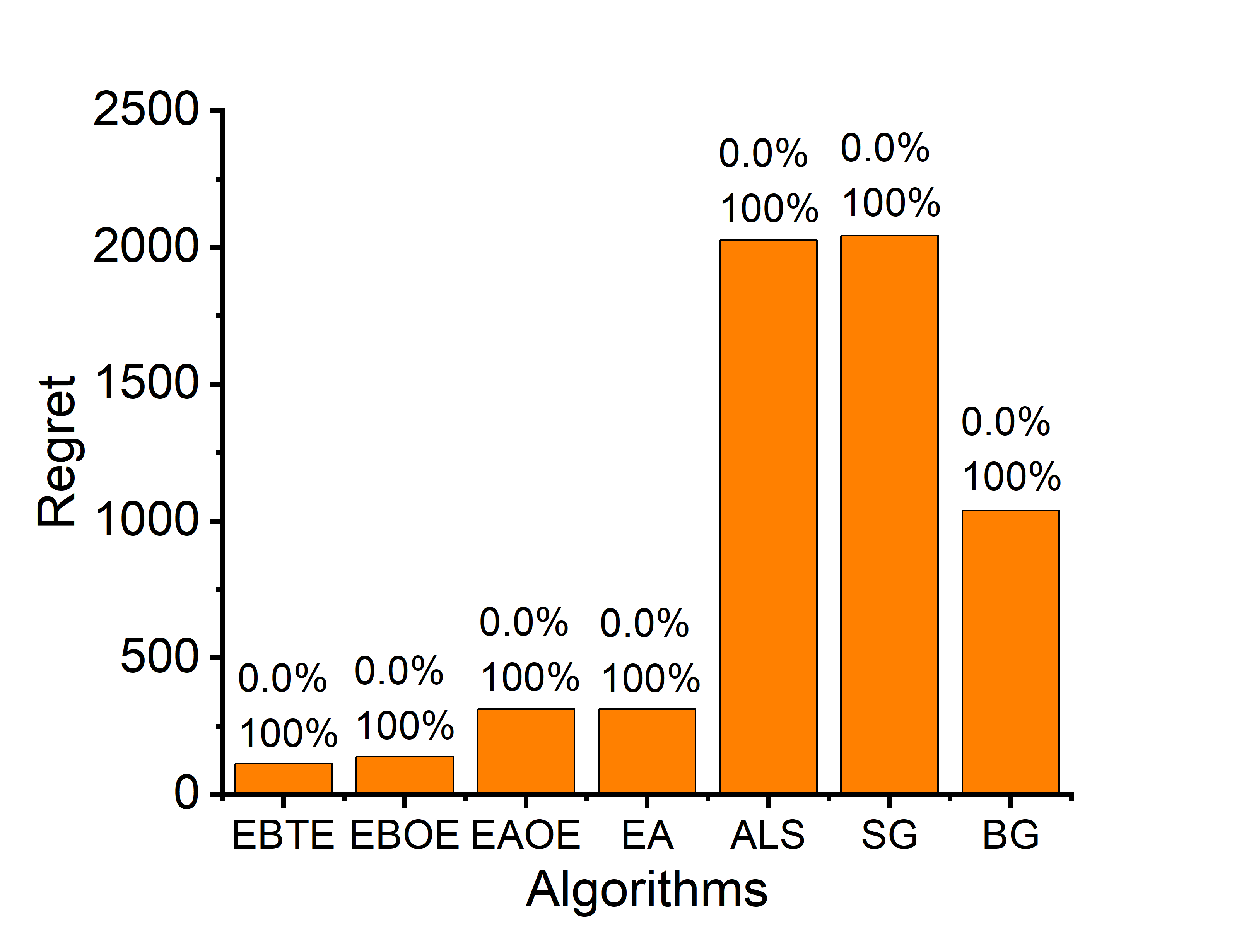} & \includegraphics[scale=0.11]{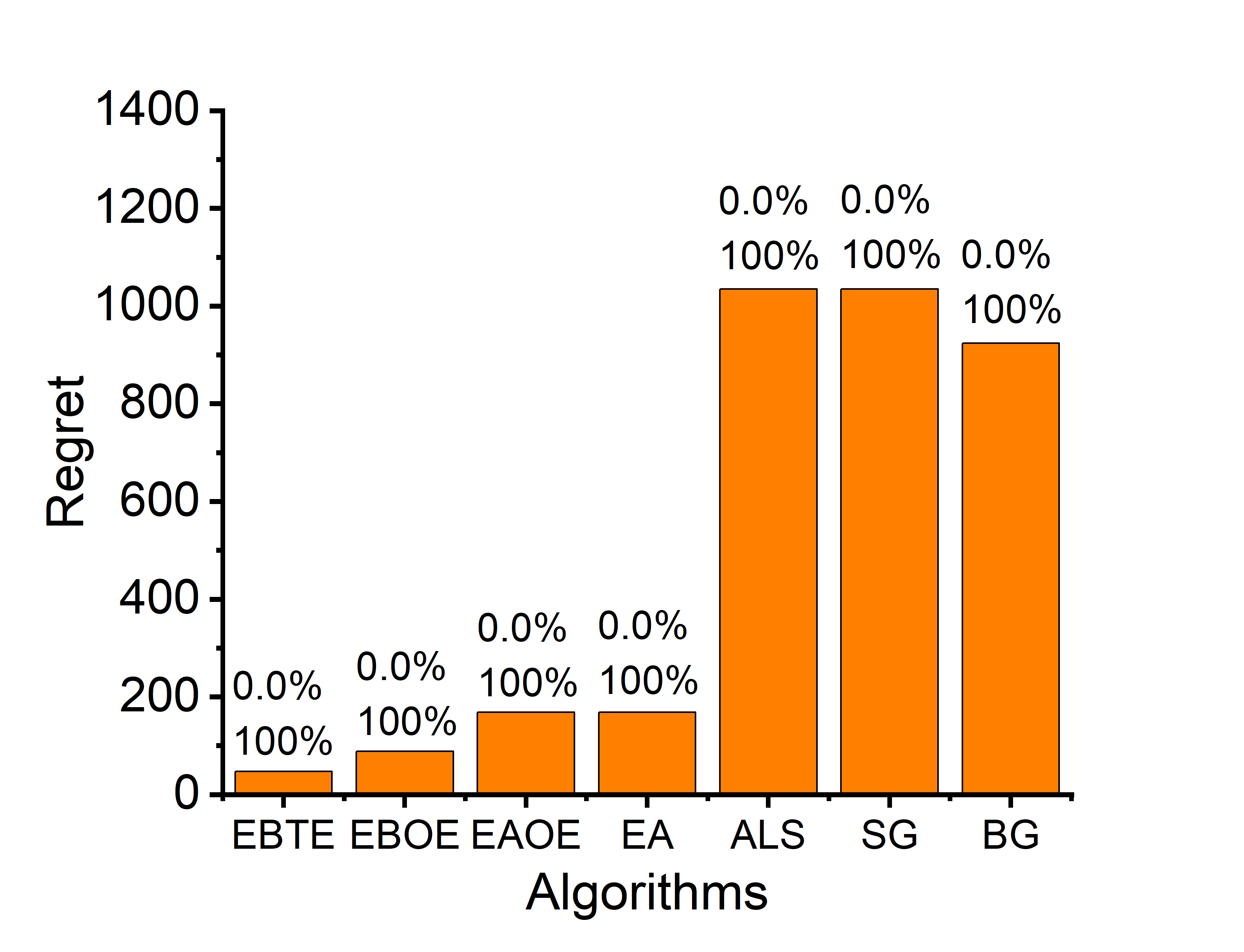} \\
{\tiny (i) $\alpha = 100 \%$} &{\tiny (j) $\alpha = 120 \%$} & {\tiny (k) $\alpha = 40 \%$} & {\tiny (l) $\alpha = 60 \%$} \\

\includegraphics[scale=0.11]{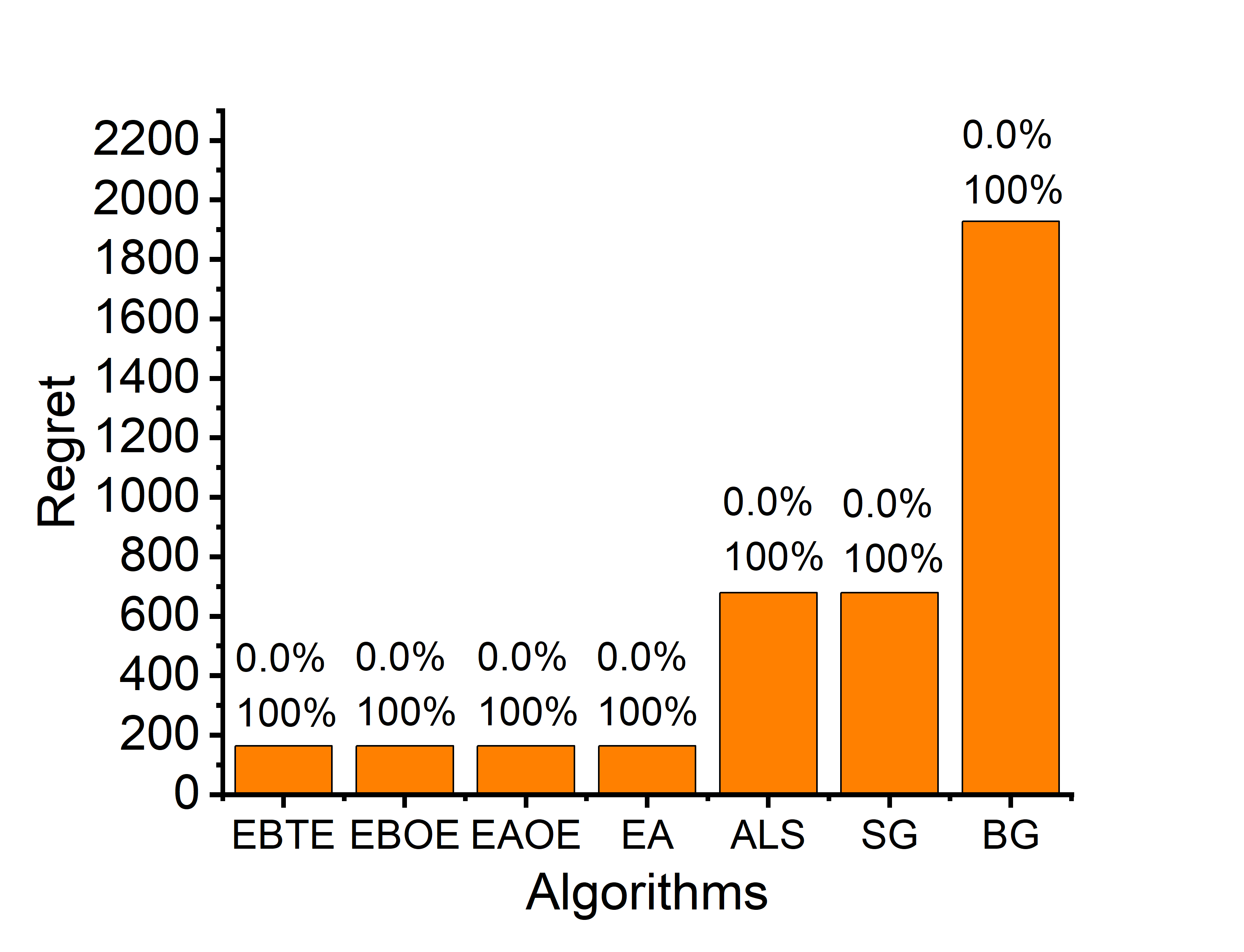} & \includegraphics[scale=0.11]{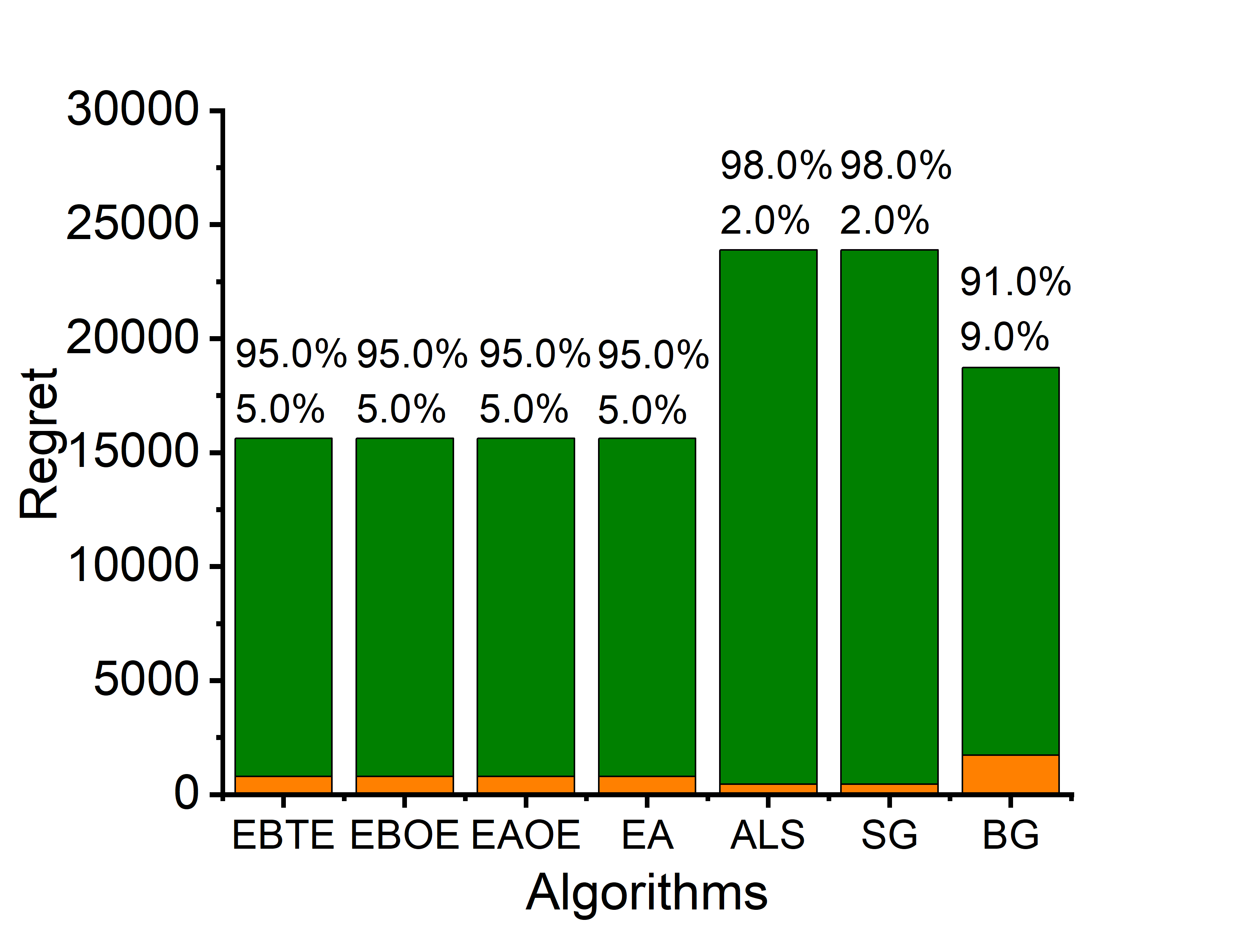} & \includegraphics[scale=0.11]{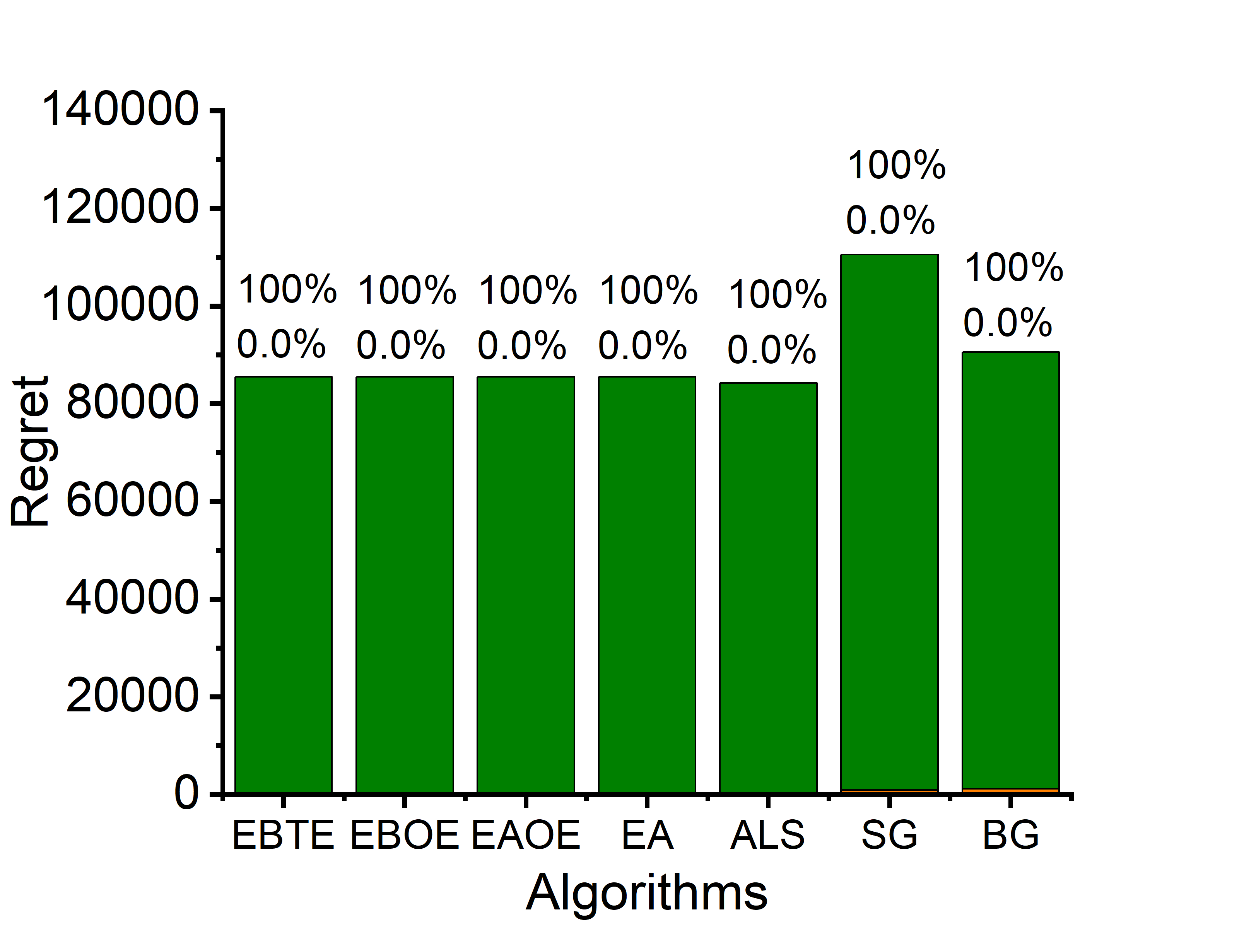} &
\includegraphics[scale=0.11]{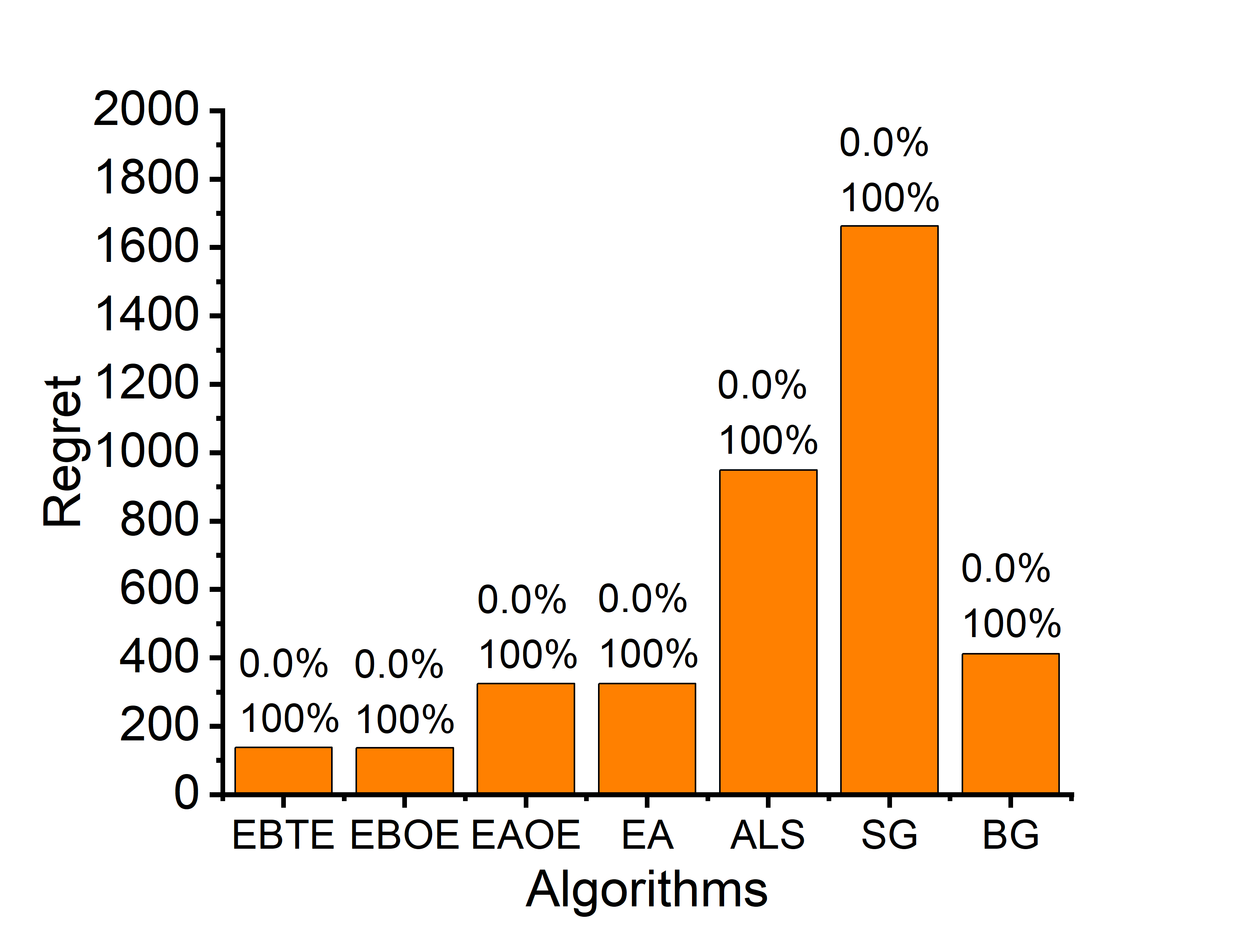} \\
{\tiny (m) $\alpha = 80 \%$} & {\tiny (n) $\alpha = 100 \%$} &{\tiny (o) $\alpha = 120 \%$} & {\tiny (p) $\alpha = 40 \%$} \\

\includegraphics[scale=0.11]{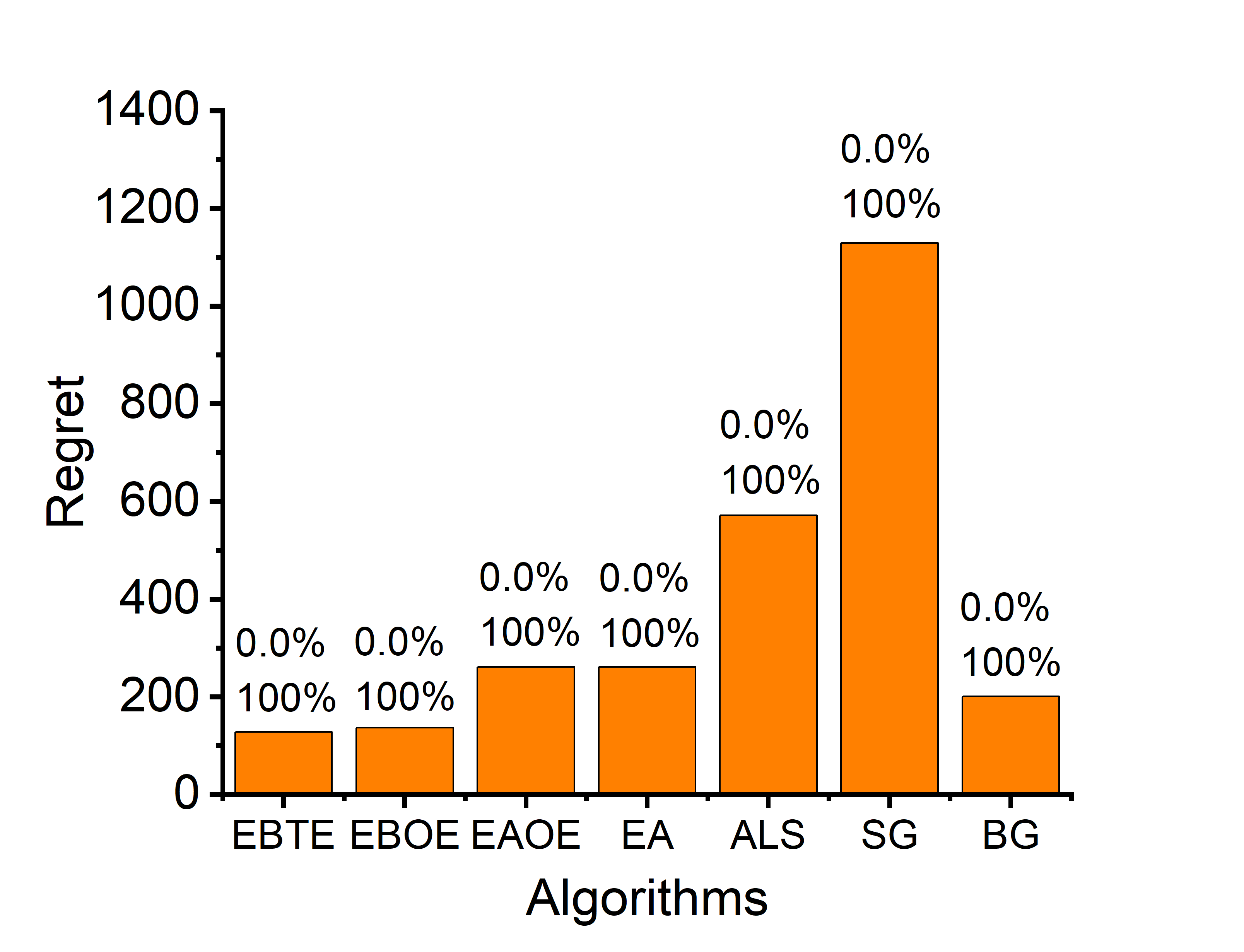}  &\includegraphics[scale=0.11]{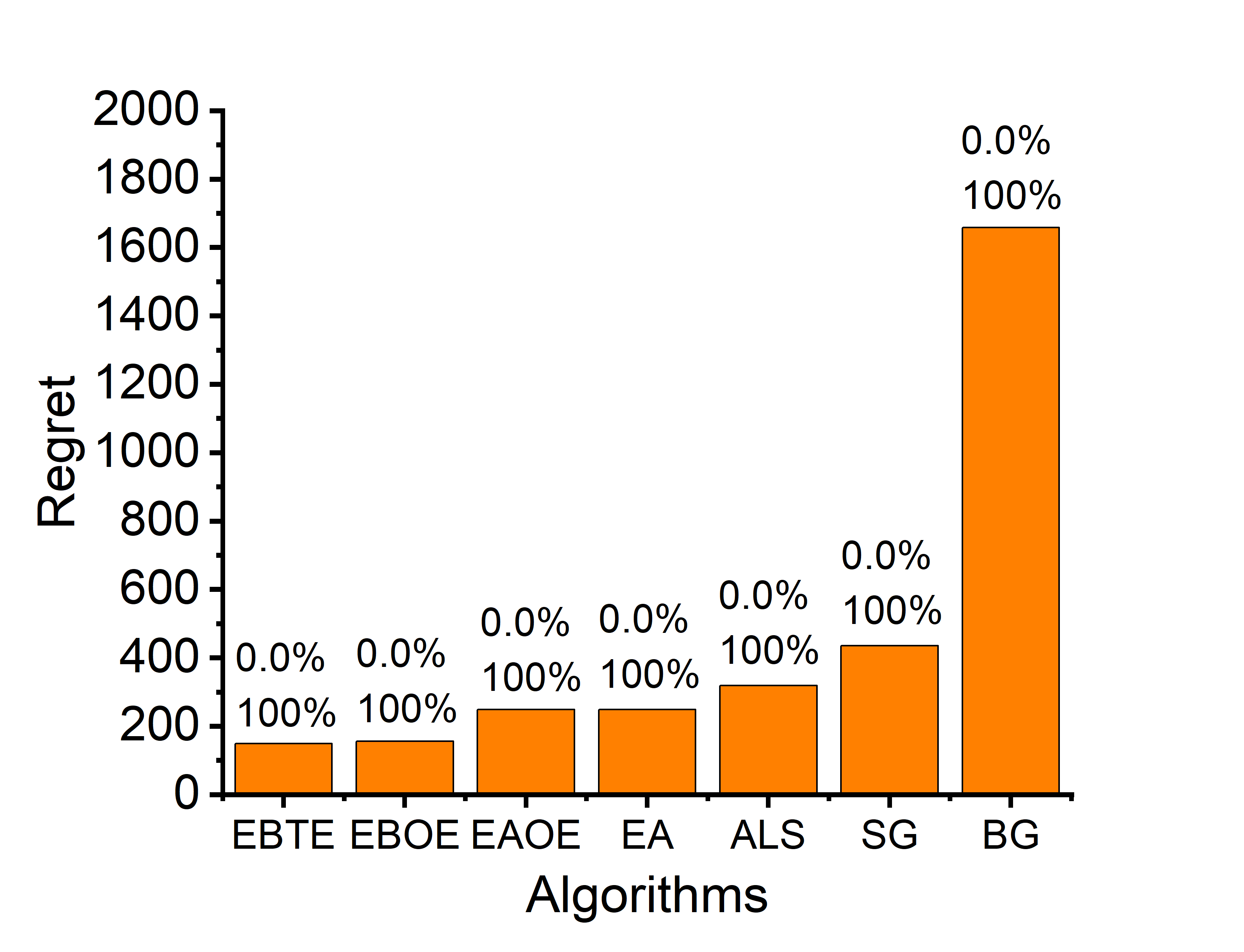} & \includegraphics[scale=0.11]{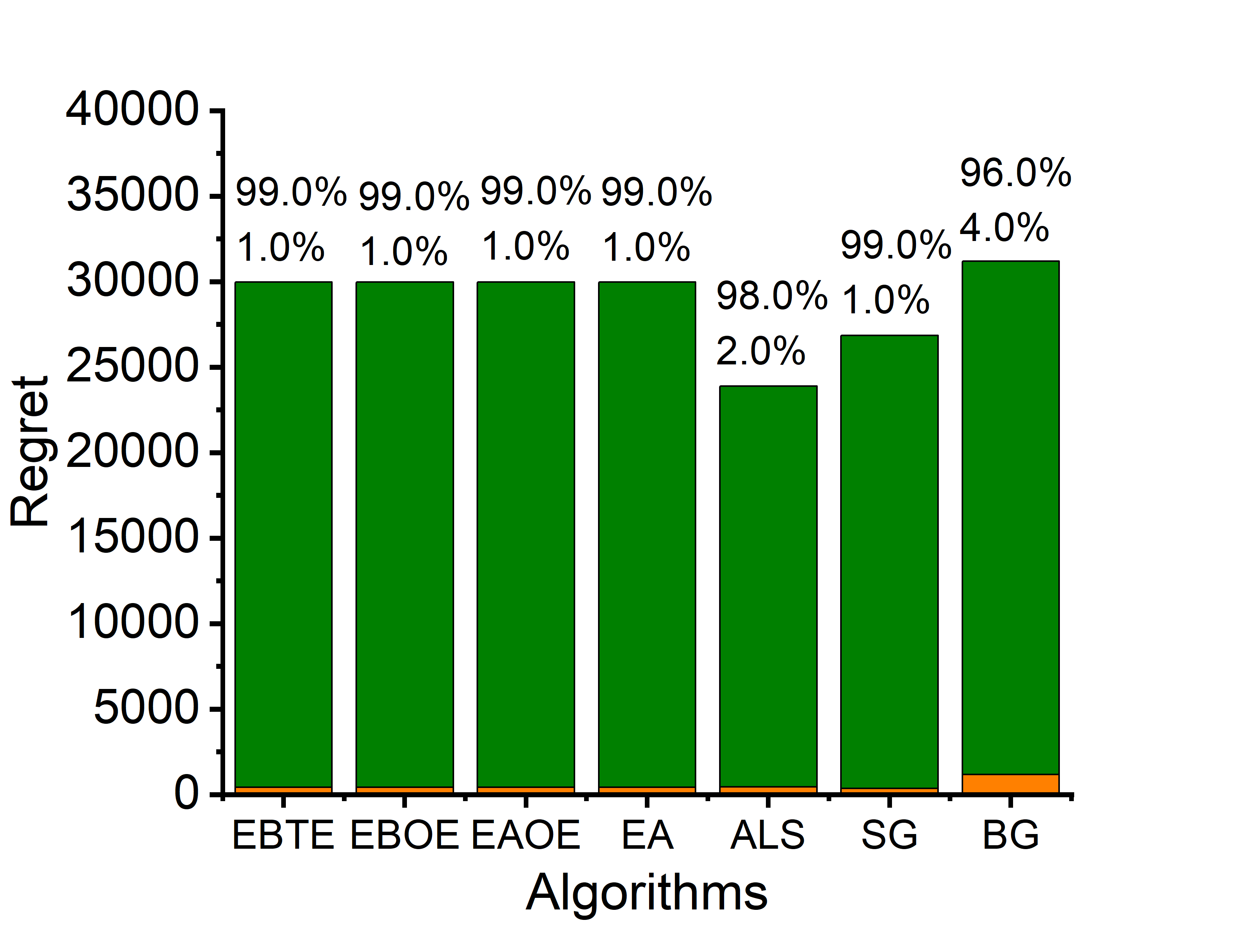} & \includegraphics[scale=0.11]{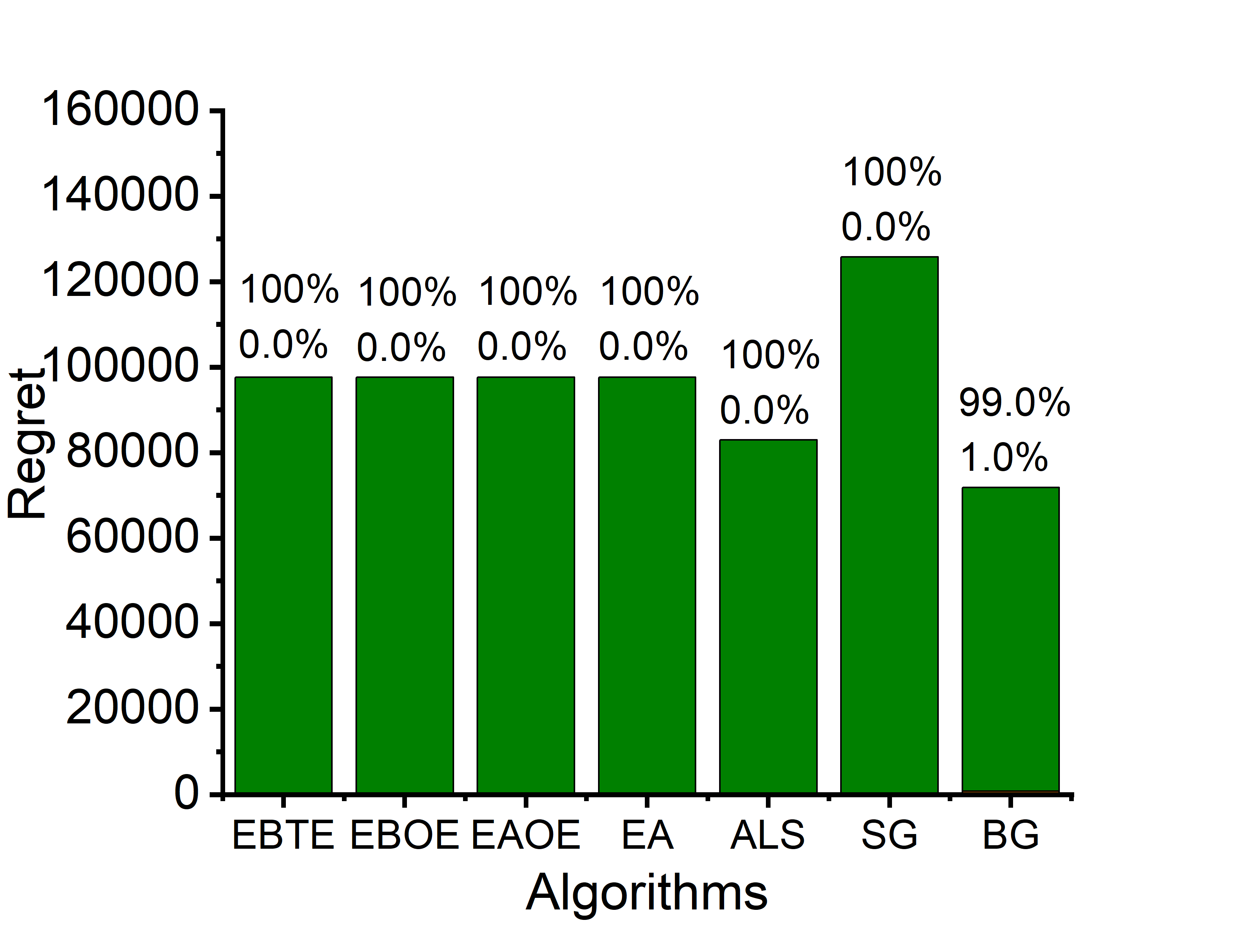}\\
 {\tiny (q) $\alpha = 60 \%$} & {\tiny (r) $\alpha = 80 \%$} & {\tiny (s) $\alpha = 100 \%$} &{\tiny (t) $\alpha = 120 \%$}\\

\end{tabular}
\caption{Regret varying $\alpha$ when $\mathcal{I}^{ID} = 1\%, \mathcal{|A|} = 100$ (a, b, c, d, e), when $\mathcal{I}^{ID} = 2\%, \mathcal{|A|} = 50$ (f,g,h,i,j), when $\mathcal{I}^{ID} = 5\%, \mathcal{|A|} = 20$ (k,l,m,n,o) and when $\mathcal{I}^{ID} = 10\%, \mathcal{|A|} = 10$ (p, q, r, s, t)for Park location type }
\label{Fig:Park}
\end{figure}

\begin{figure}[h!]
\centering
   \begin{tabular}{lclc}
       Unsatisfied Regret & \includegraphics[width=0.11\linewidth]{Unsatisfied.png} \  & \ Excessive Regret & \includegraphics[width=0.11\linewidth]{Excessive.png} \\
    \end{tabular}

\begin{tabular}{cccc}
\includegraphics[scale=0.11]{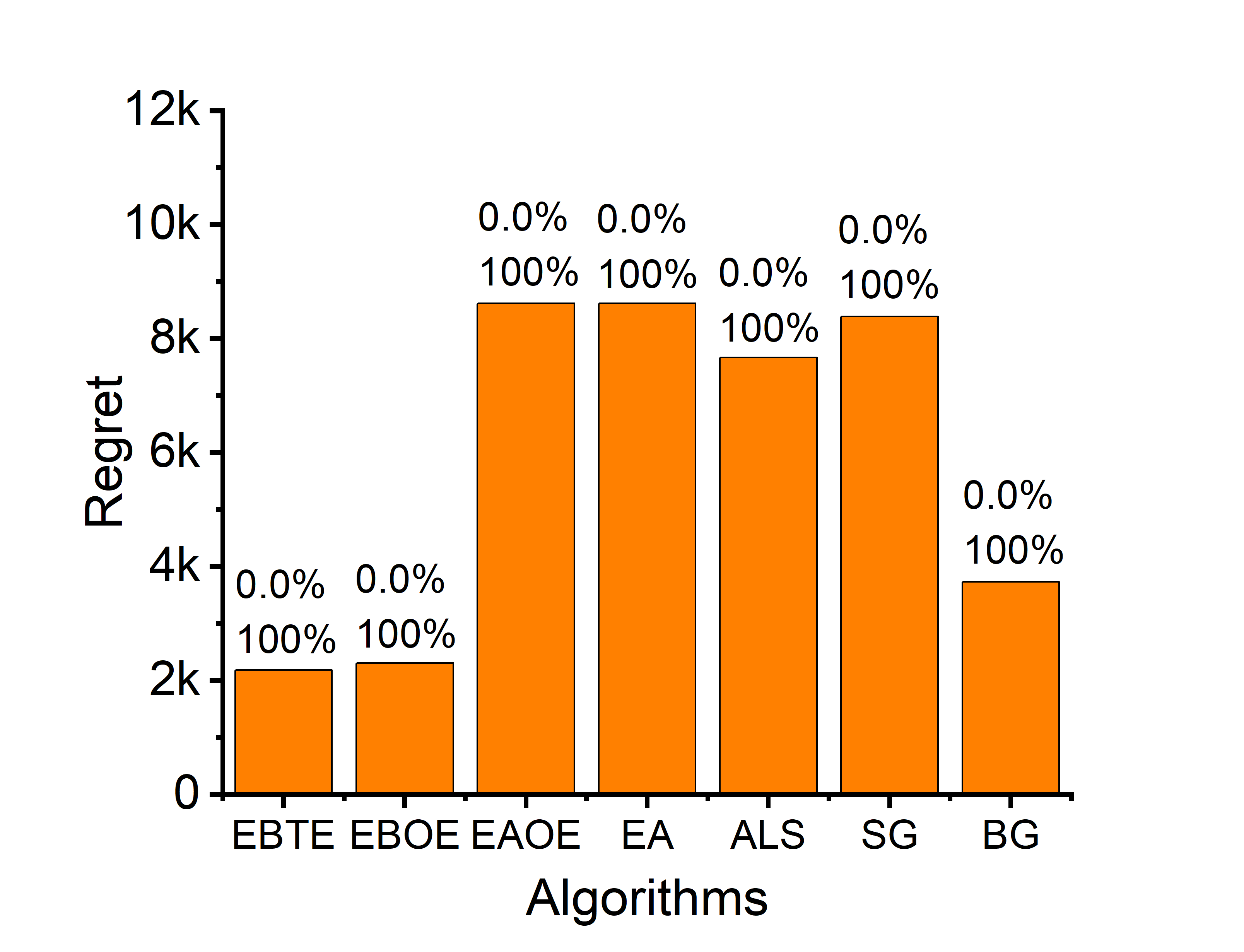} & \includegraphics[scale=0.11]{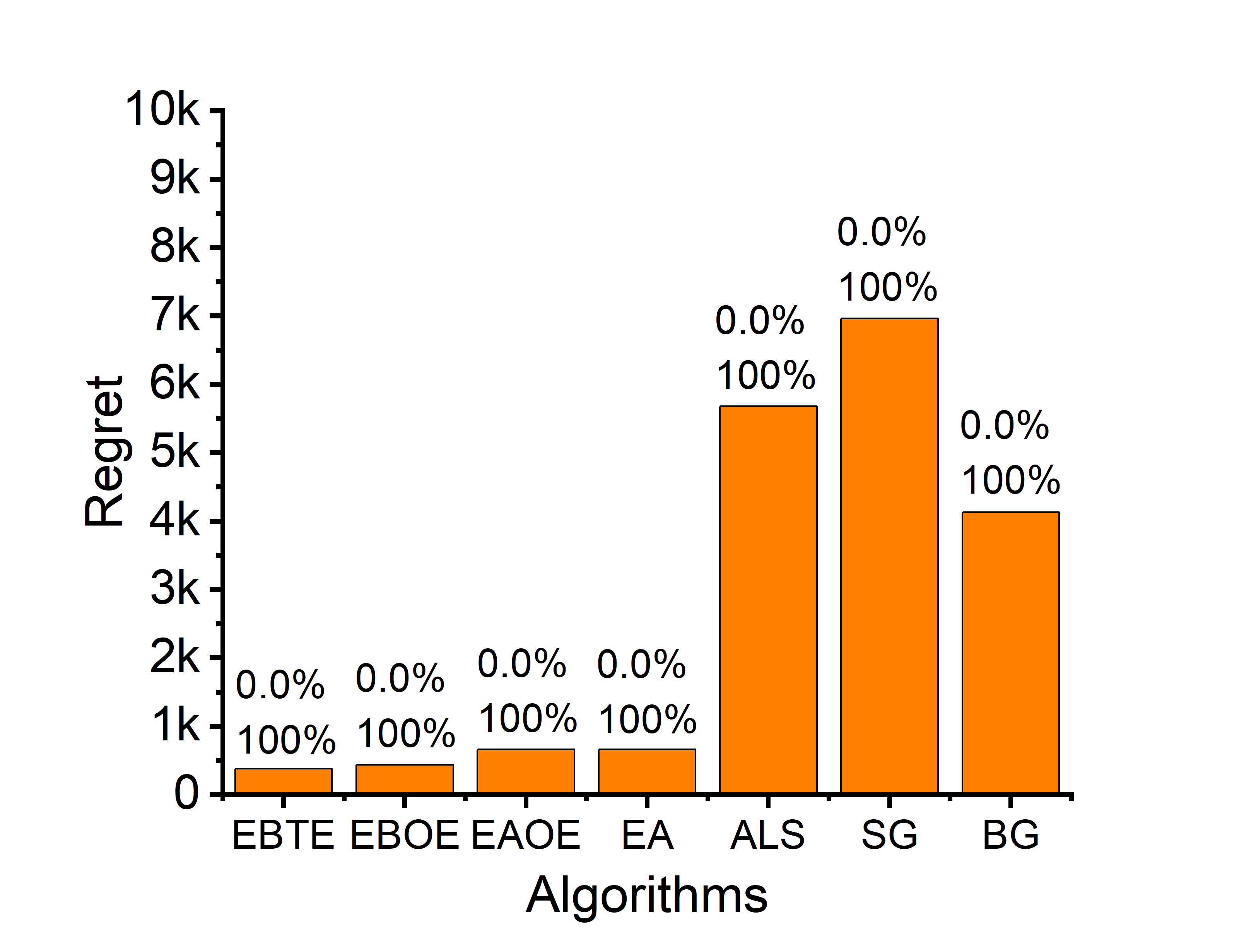}  &\includegraphics[scale=0.11]{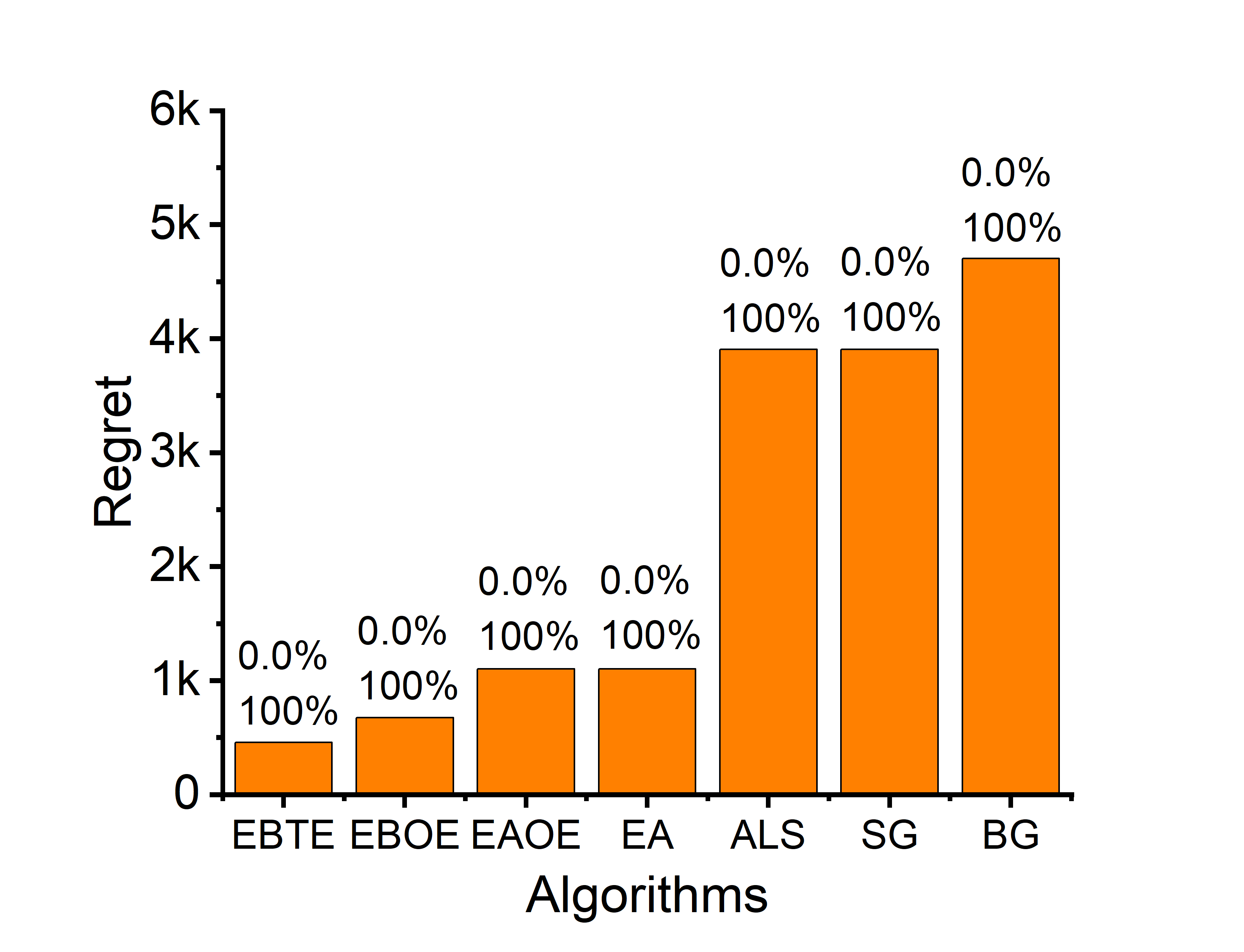} & \includegraphics[scale=0.11]{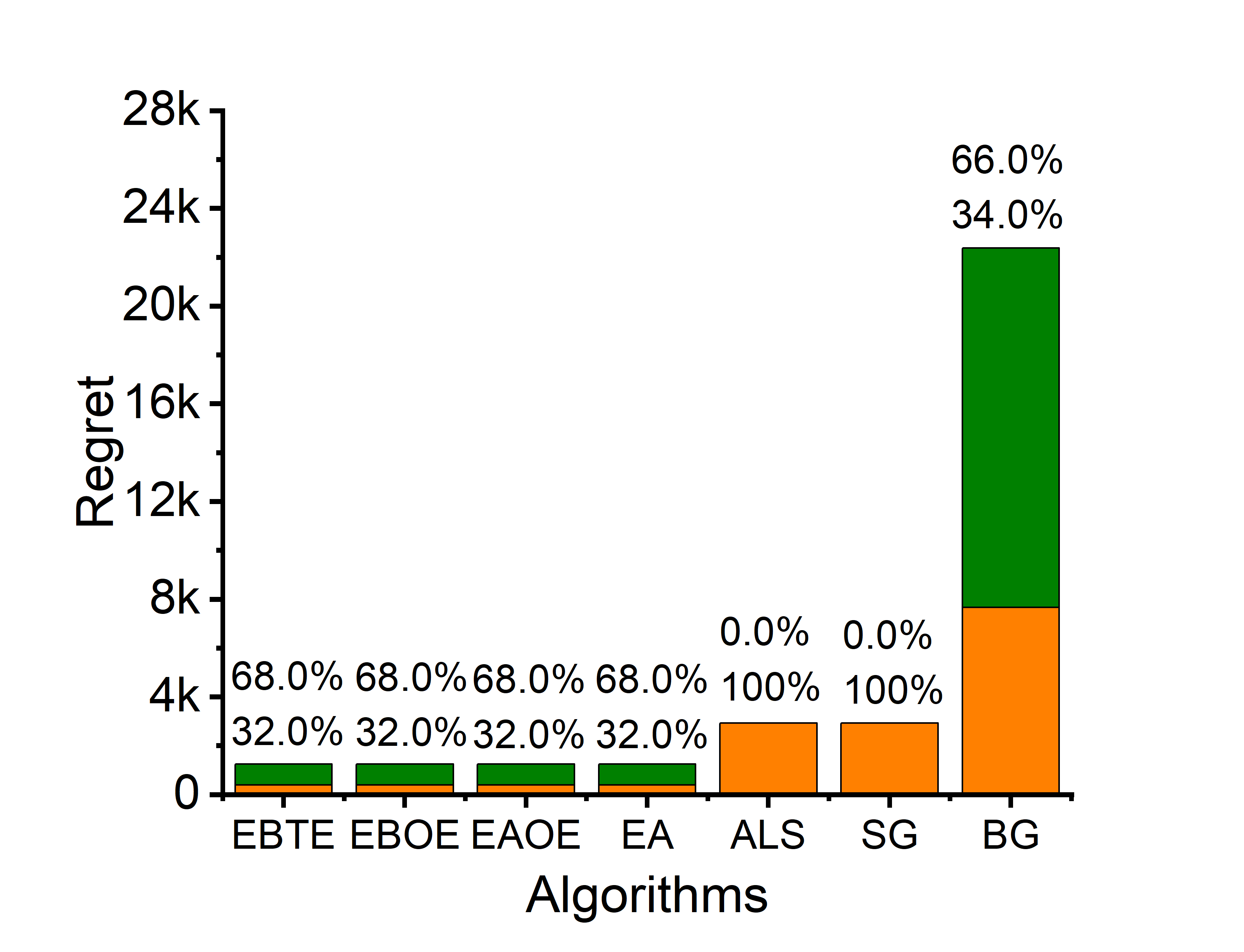} \\
{\tiny (a) $\alpha = 40 \%$} & {\tiny (b) $\alpha = 60 \%$} & {\tiny (c) $\alpha = 80 \%$} & {\tiny (d) $\alpha = 100 \%$} \\


 \includegraphics[scale=0.11]{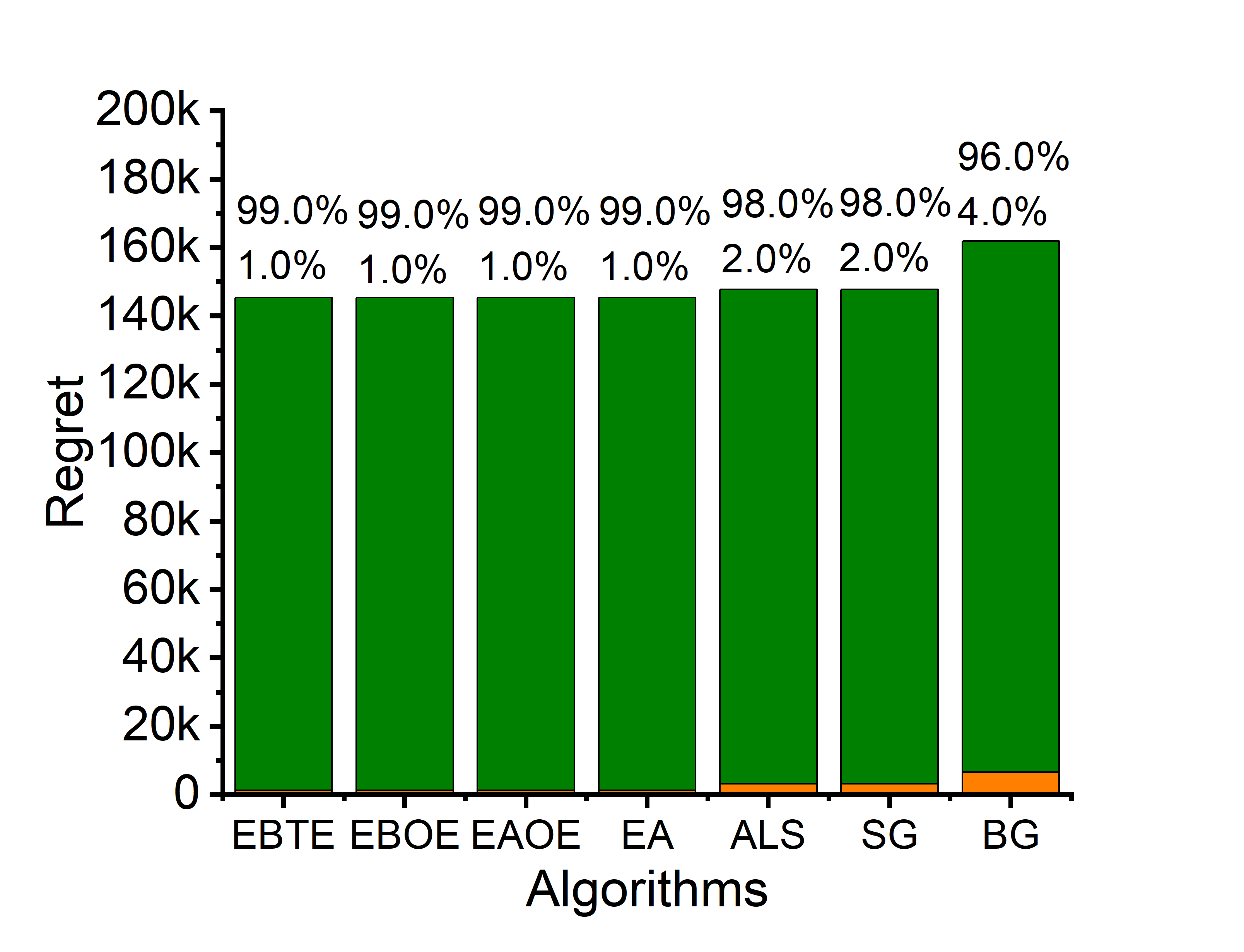} & \includegraphics[scale=0.11]{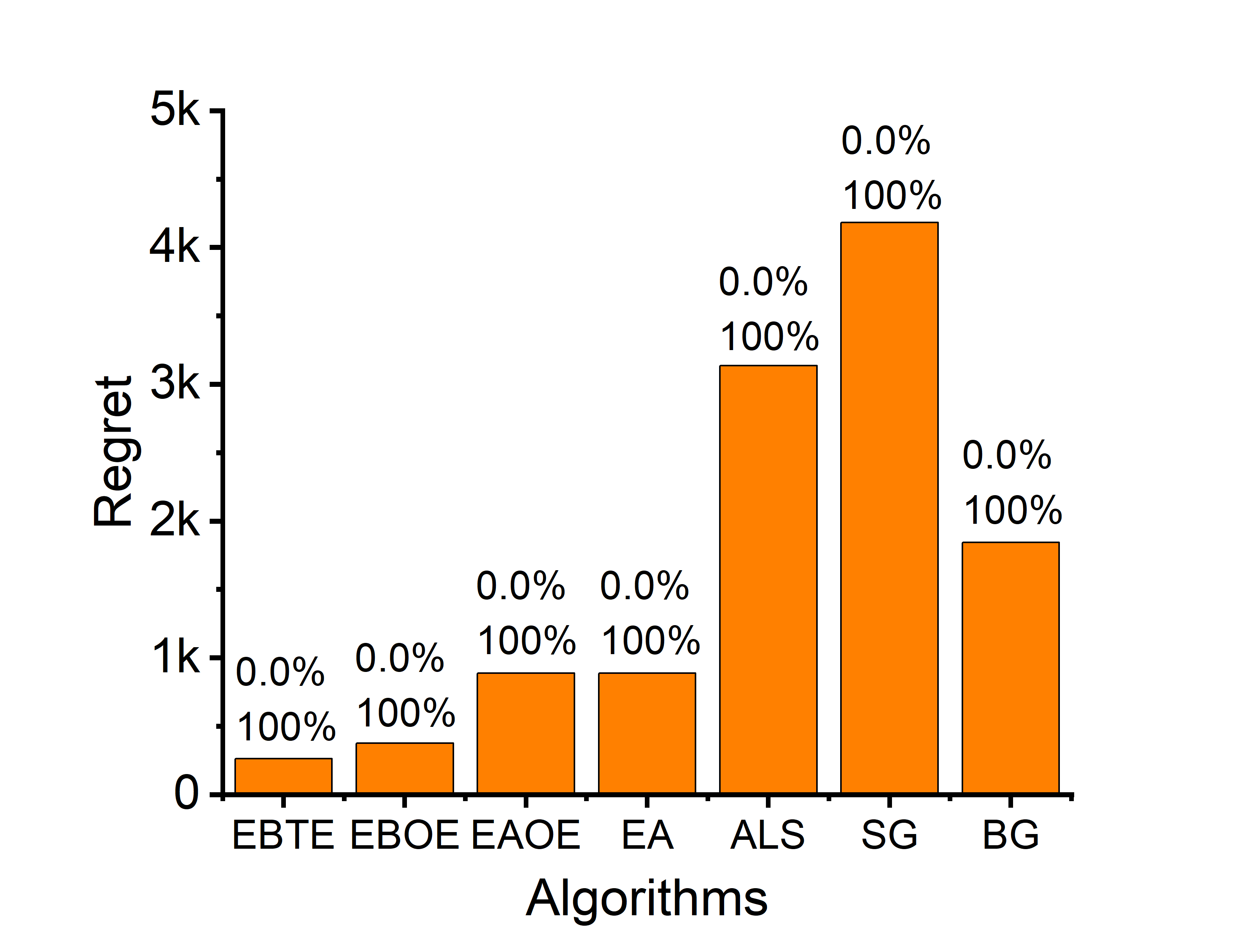} & \includegraphics[scale=0.11]{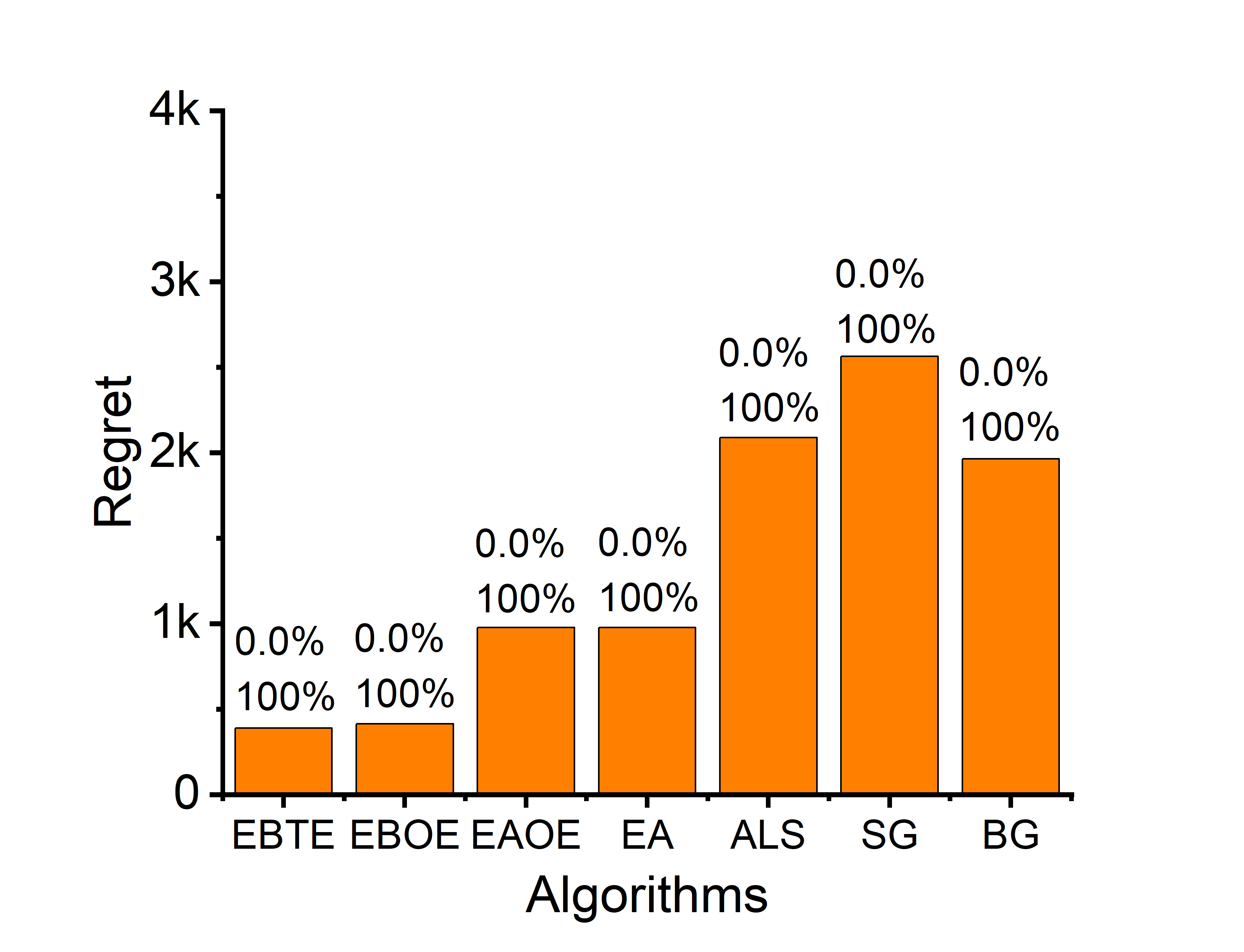}  &\includegraphics[scale=0.11]{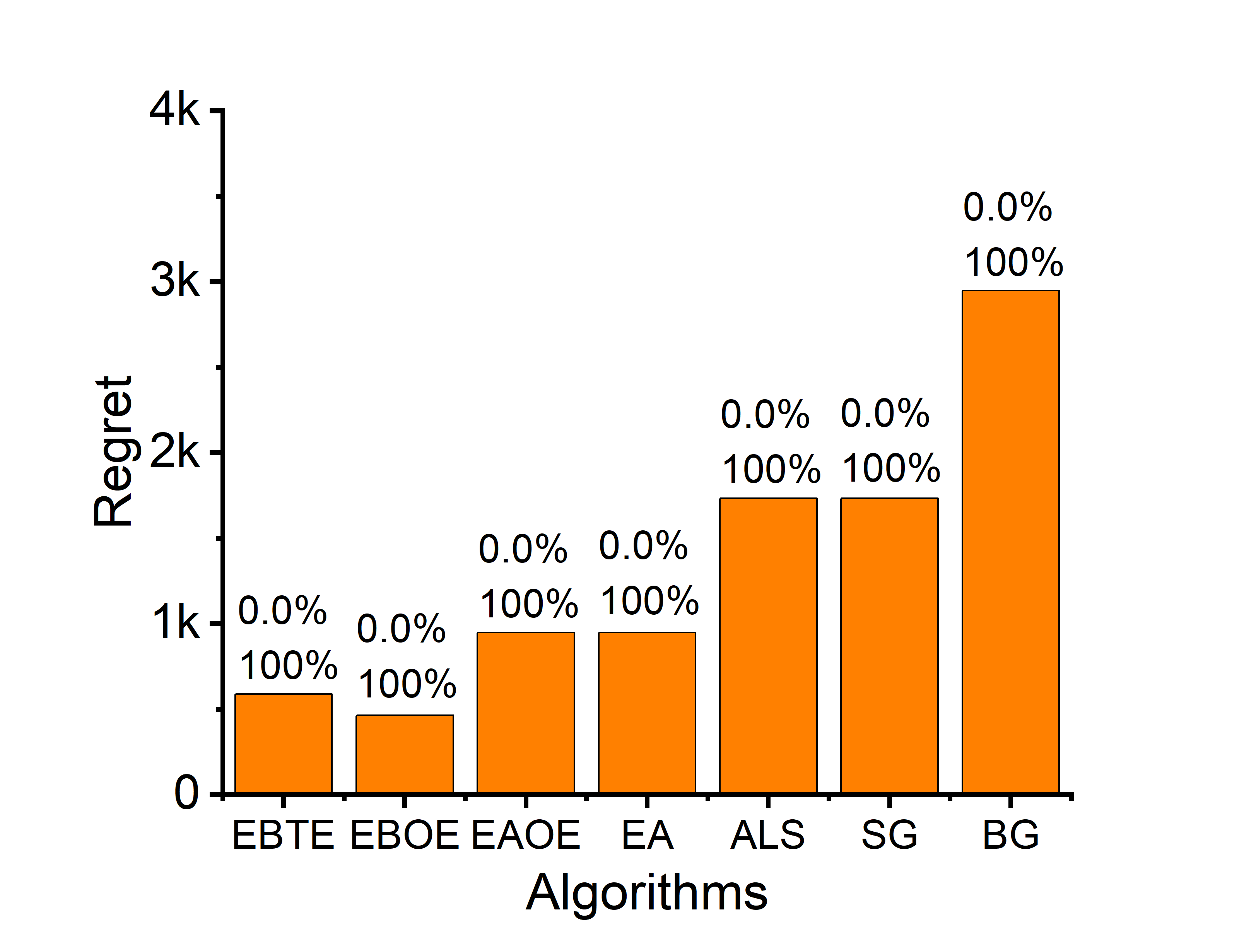} \\
{\tiny (e) $\alpha = 120 \%$} & {\tiny (f) $\alpha = 40 \%$} & {\tiny (g) $\alpha = 60 \%$} & {\tiny (h) $\alpha = 80 \%$} \\

 \includegraphics[scale=0.11]{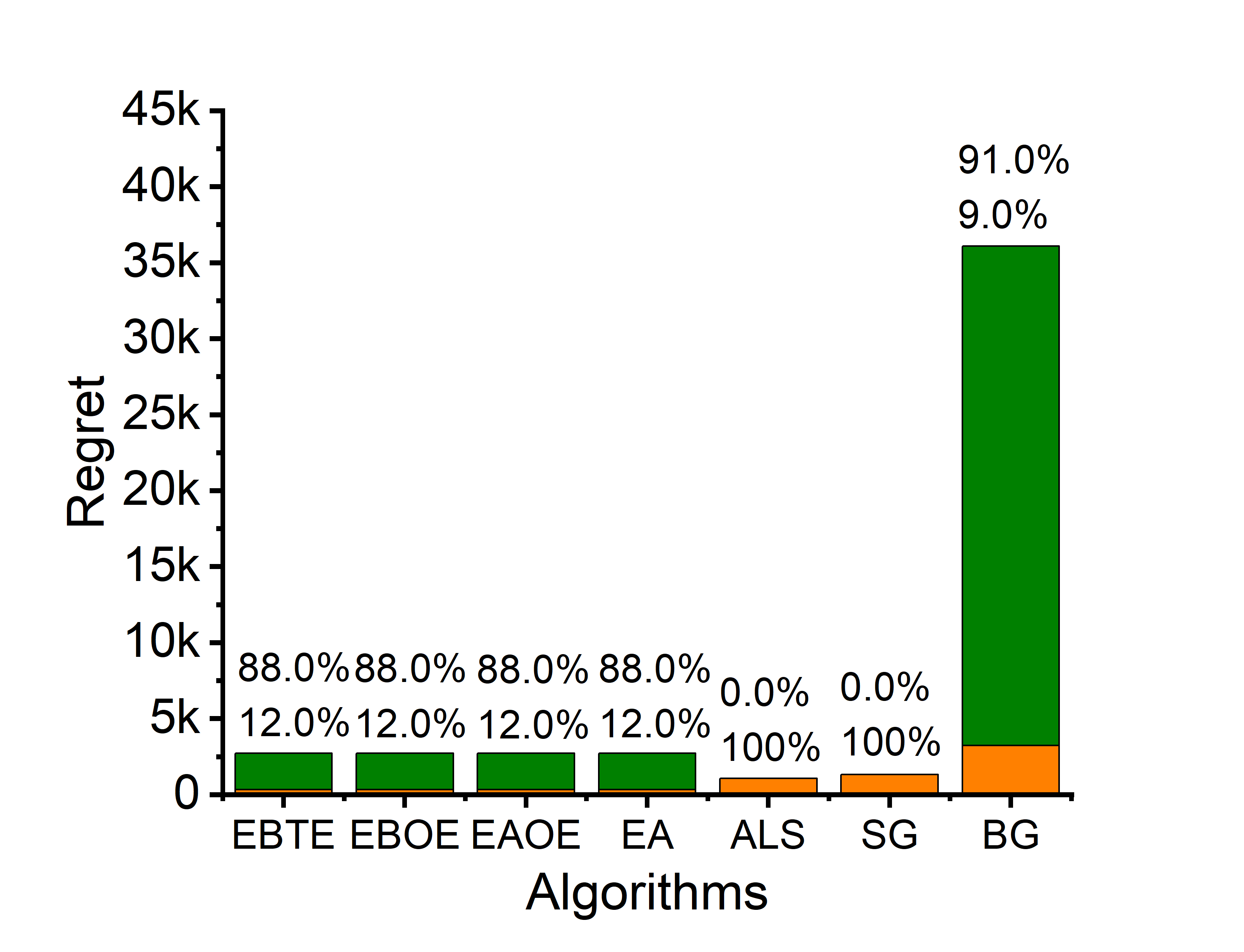} & \includegraphics[scale=0.11]{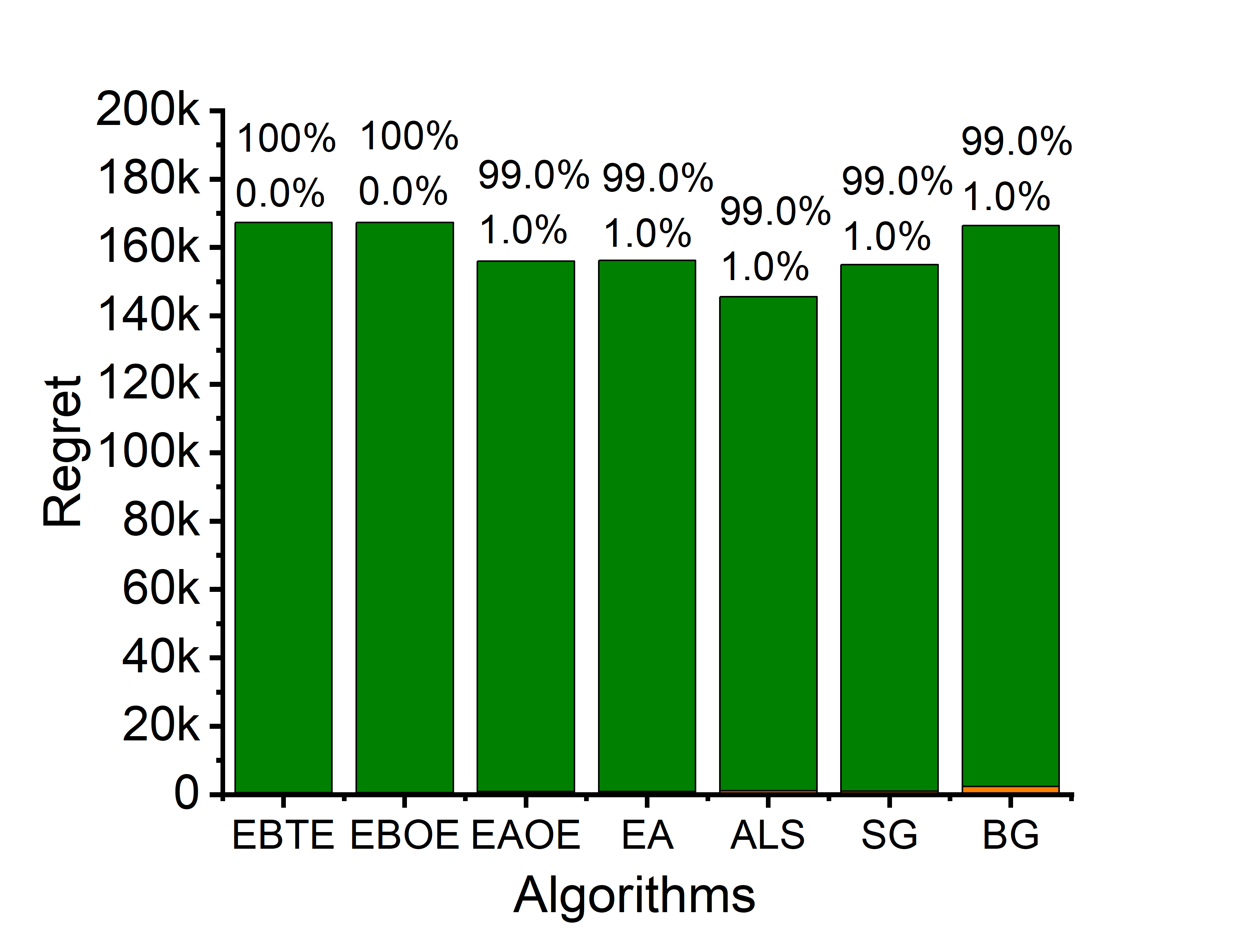} &
\includegraphics[scale=0.11]{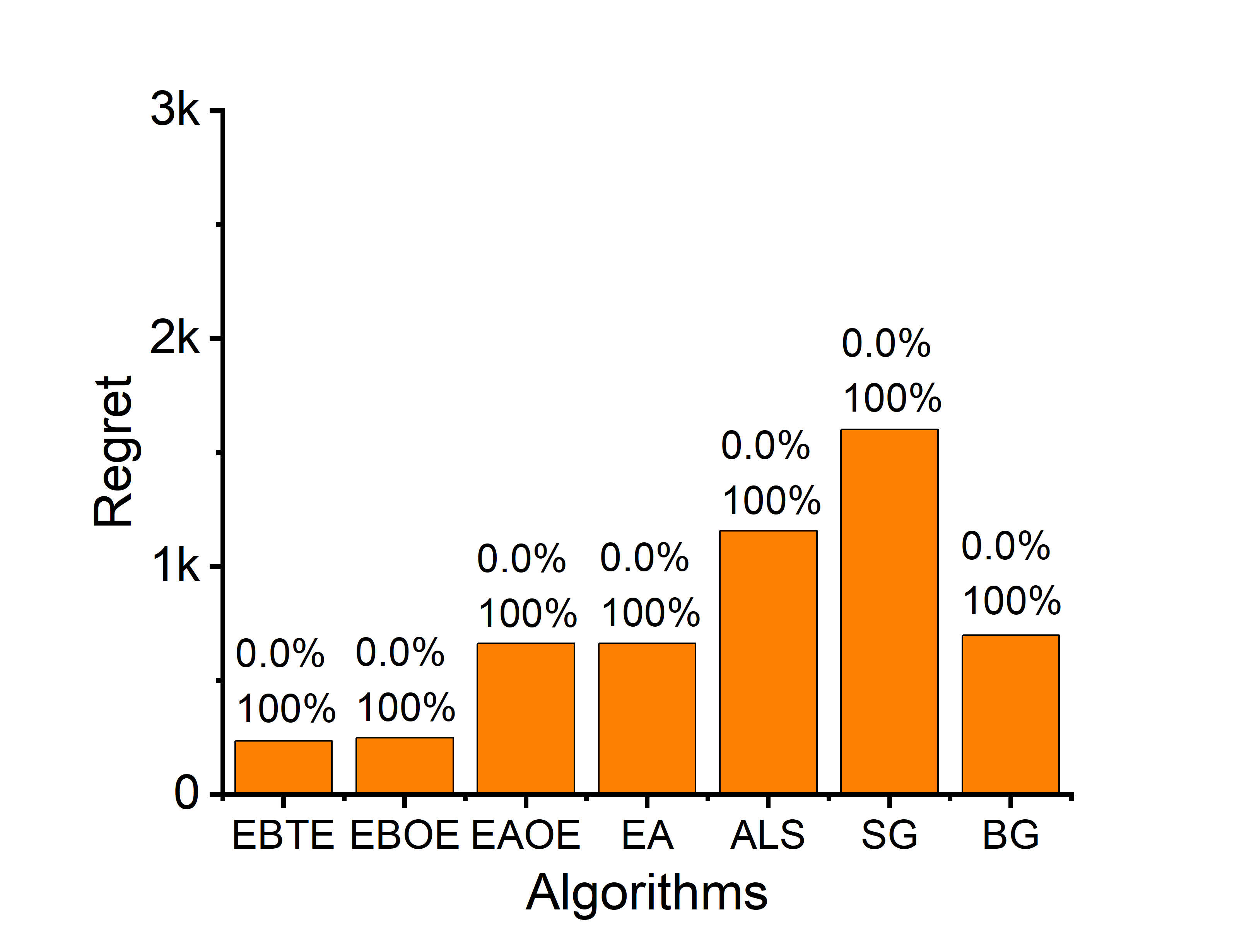} & \includegraphics[scale=0.11]{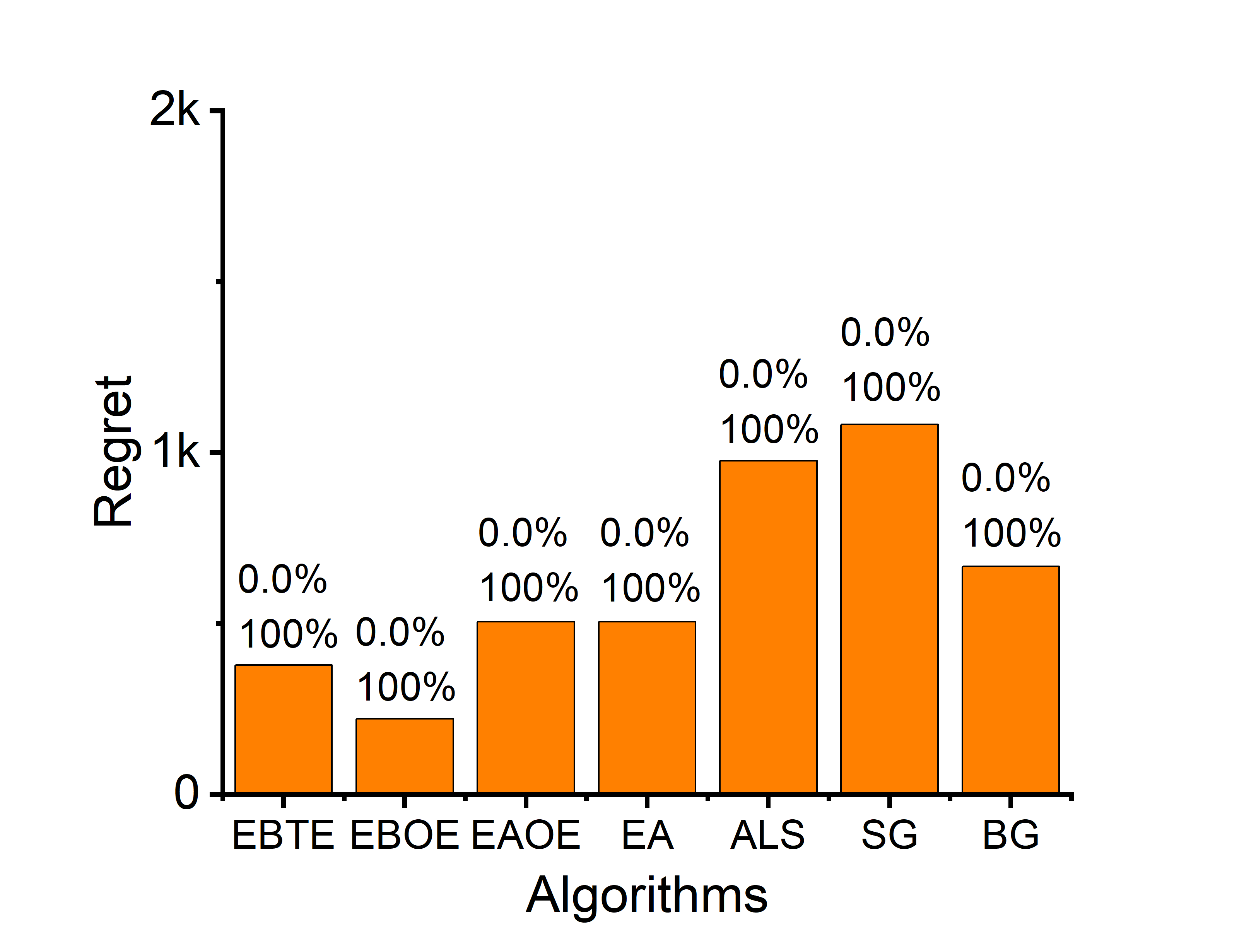} \\
{\tiny (i) $\alpha = 100 \%$} &{\tiny (j) $\alpha = 120 \%$} & {\tiny (k) $\alpha = 40 \%$} & {\tiny (l) $\alpha = 60 \%$} \\

\includegraphics[scale=0.11]{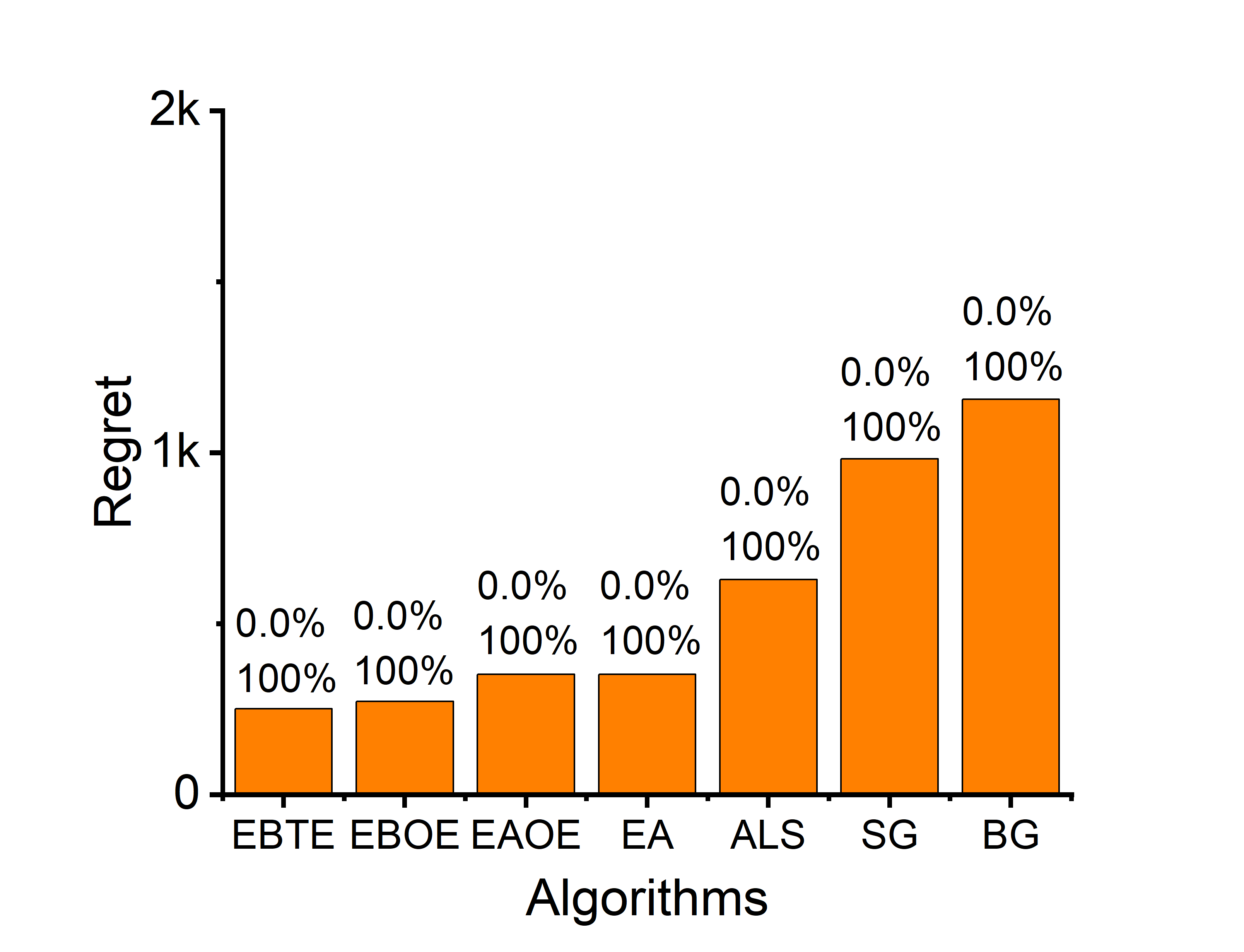} & \includegraphics[scale=0.11]{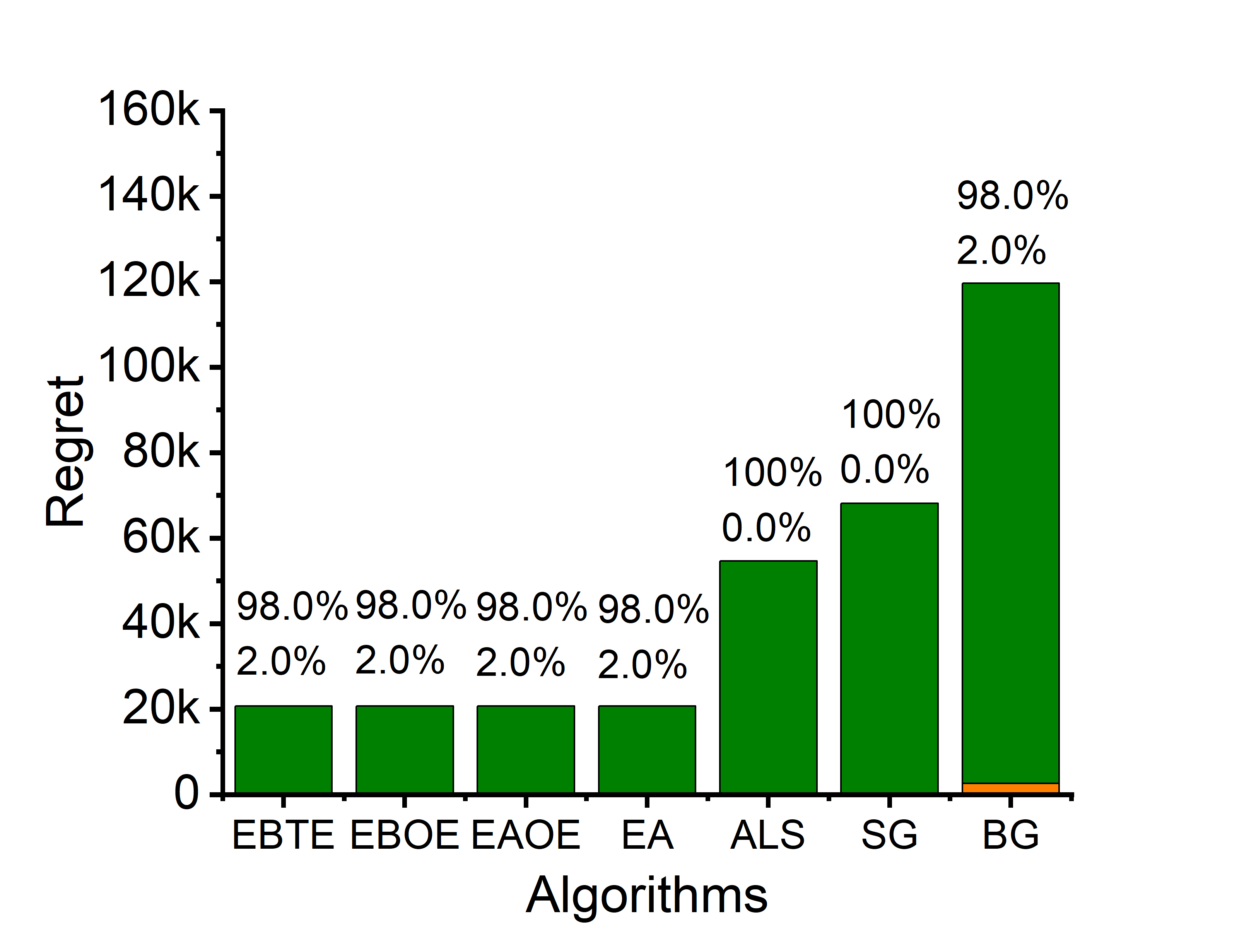} & \includegraphics[scale=0.11]{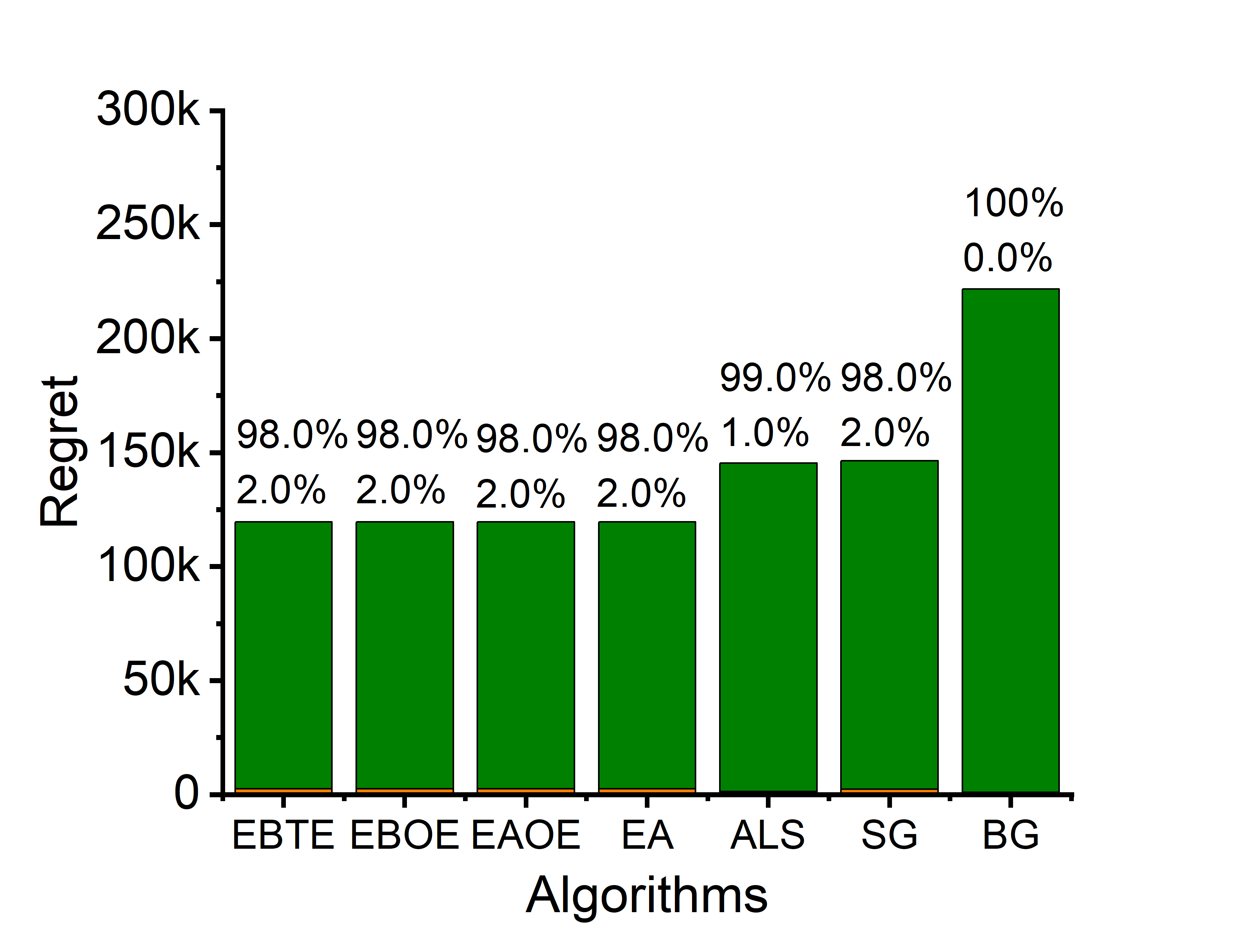} &
\includegraphics[scale=0.11]{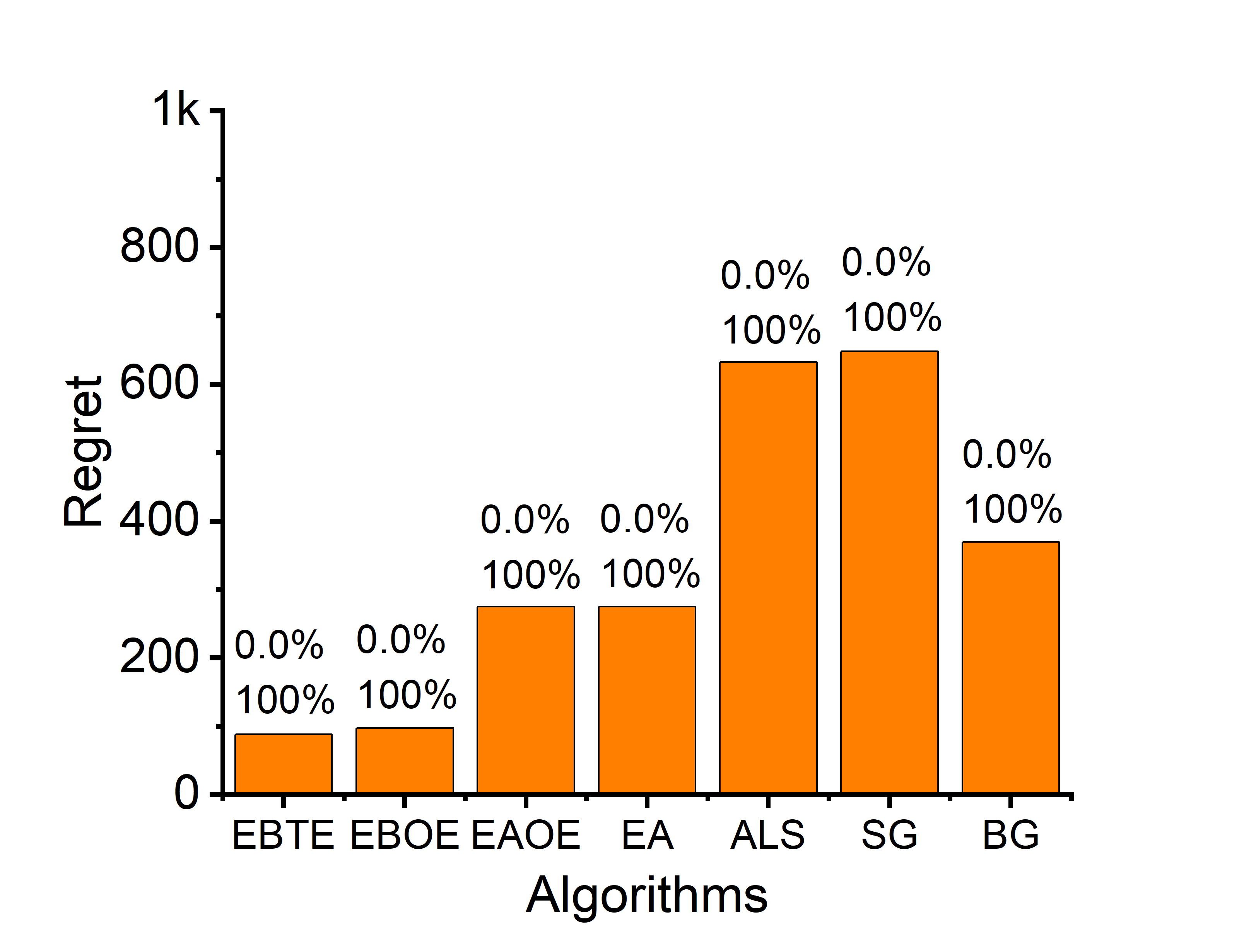} \\
{\tiny (m) $\alpha = 80 \%$} & {\tiny (n) $\alpha = 100 \%$} &{\tiny (o) $\alpha = 120 \%$} & {\tiny (p) $\alpha = 40 \%$} \\

\includegraphics[scale=0.11]{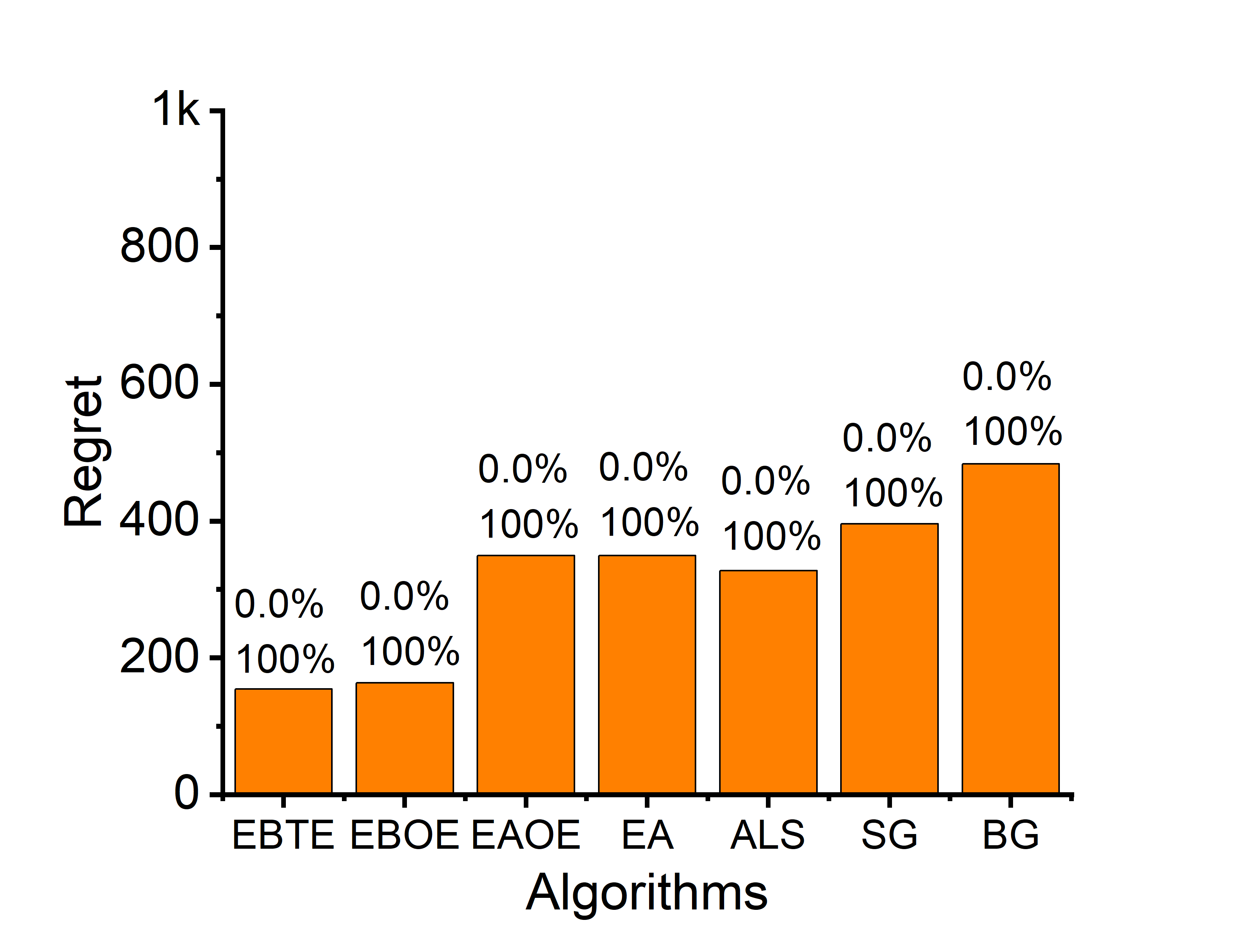}  &\includegraphics[scale=0.11]{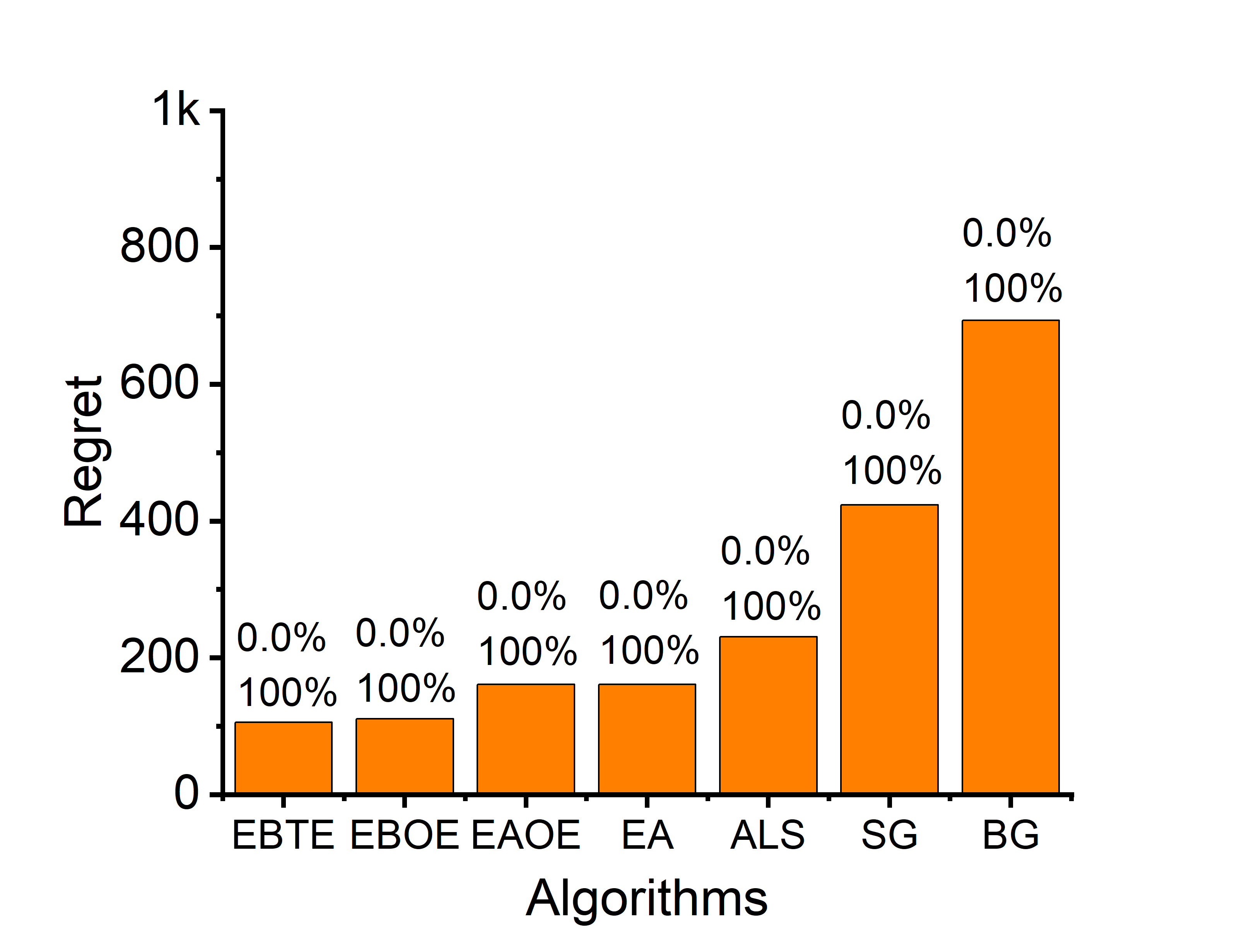} & \includegraphics[scale=0.11]{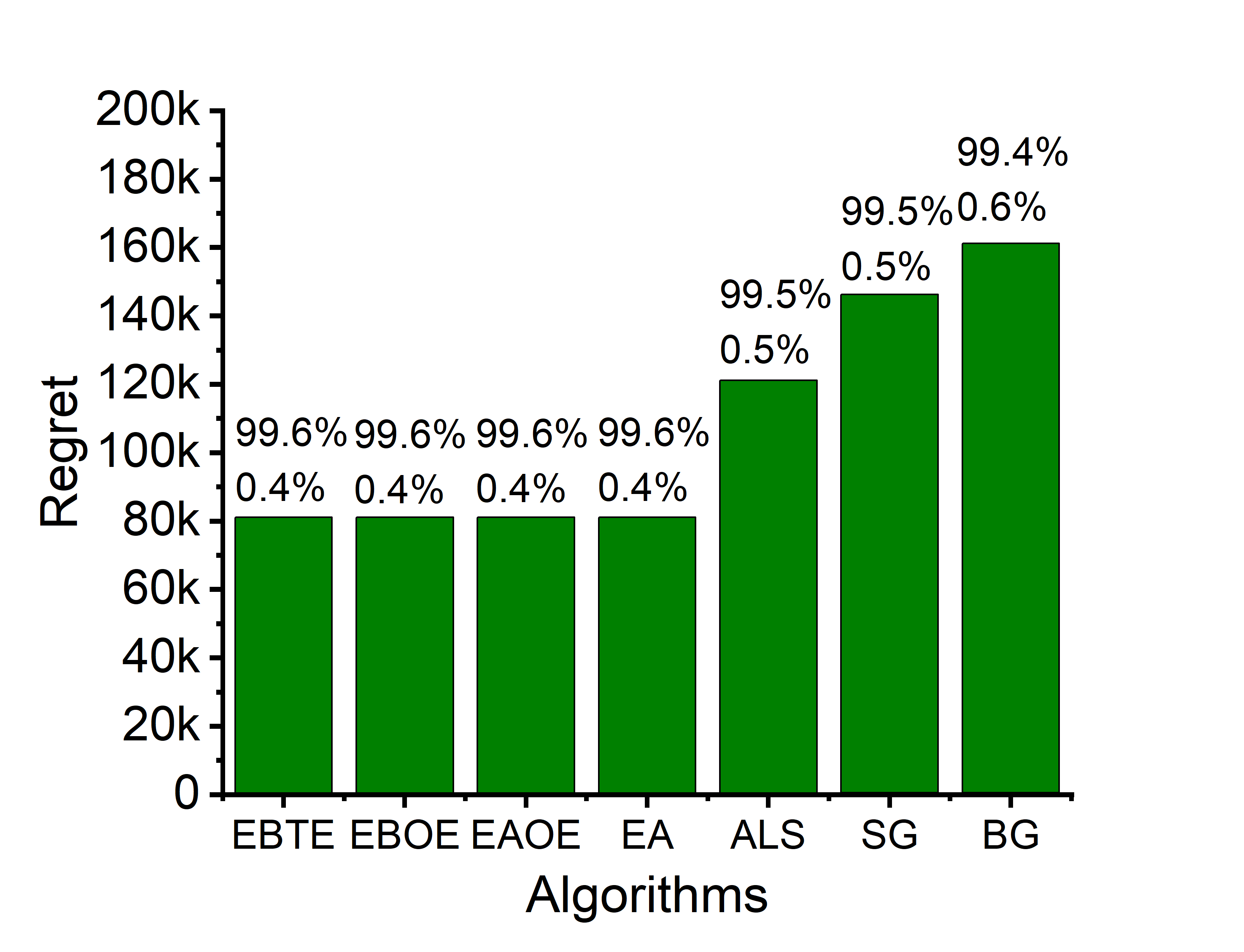} & \includegraphics[scale=0.11]{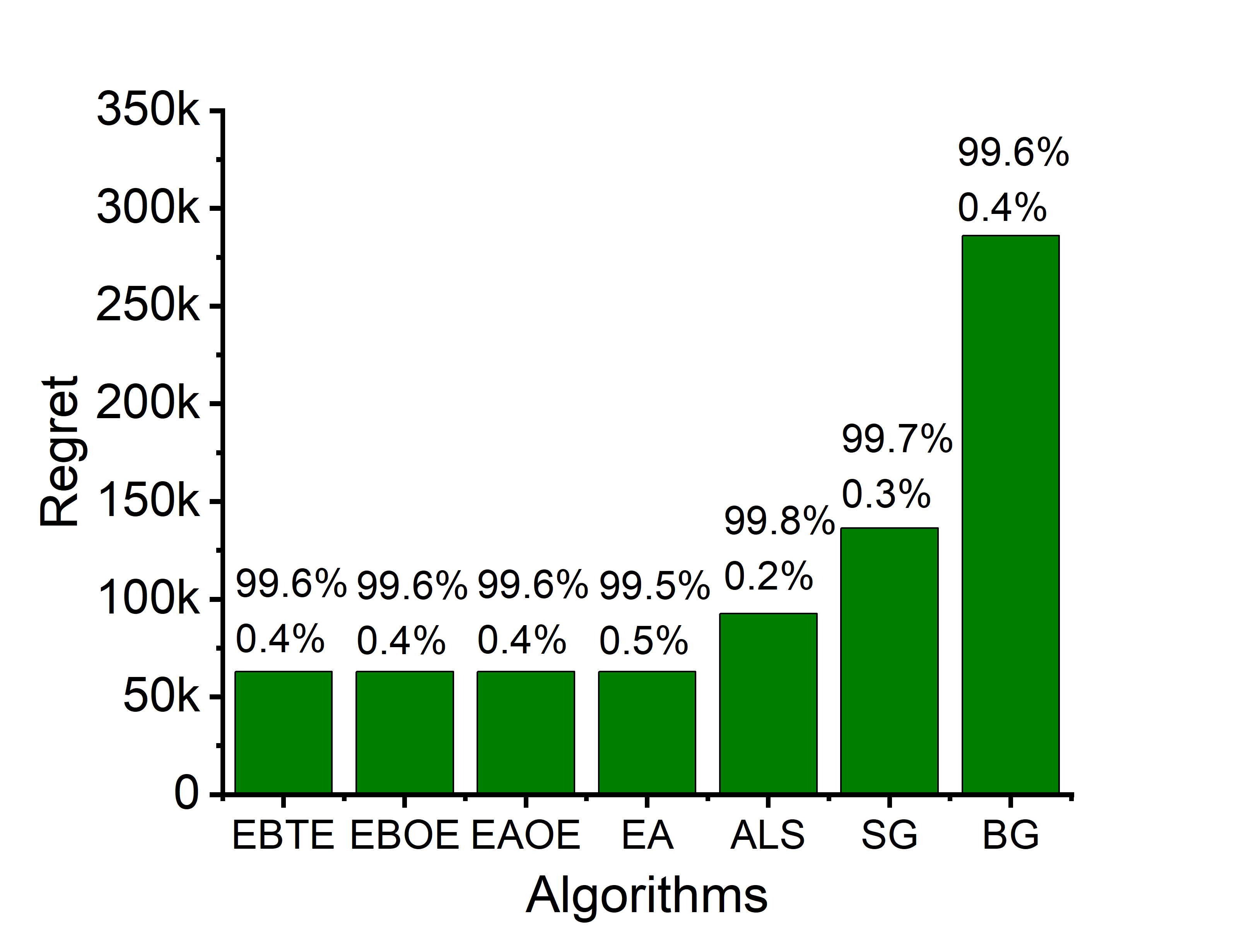}\\
 {\tiny (q) $\alpha = 60 \%$} & {\tiny (r) $\alpha = 80 \%$} & {\tiny (s) $\alpha = 100 \%$} &{\tiny (t) $\alpha = 120 \%$}\\

\end{tabular}
\caption{Regret varying $\alpha$ when $\mathcal{I}^{ID} = 1\%, \mathcal{|A|} = 100$ (a, b, c, d, e), when $\mathcal{I}^{ID} = 2\%, \mathcal{|A|} = 50$ (f,g,h,i,j), when $\mathcal{I}^{ID} = 5\%, \mathcal{|A|} = 20$ (k,l,m,n,o) and when $\mathcal{I}^{ID} = 10\%, \mathcal{|A|} = 10$ (p, q, r, s, t)for Airport location type }
\label{Fig:Airport}
\end{figure}

\paragraph{\textbf{Case 1: Low $\alpha$, Low $\mathcal{I}^{ID}$ (parts a, b, c, f, g, h of Figure \ref{Fig:Beach},\ref{Fig:Mall},\ref{Fig:Bank},\ref{Fig:Park},\ref{Fig:Airport})}}
In case $1$, we have a Demand-Supply ratio, $\alpha \leq 80\%$, and Average-Individual Demand ratio, $\mathcal{I}^{ID} \leq 2\%$. This means that both $\alpha$ and $\mathcal{I}^{ID}$ are low, i.e., global and individual demand both are low. As $\alpha$ is very small, influence providers have many small advertisers, and all advertisers are satisfied. Therefore, the total regret consists only of excessive influence regret except for the budget-effective greedy approach. We have three main observations.

First, as the $\alpha$ value increases, the excessive regret of all the algorithms also decreases because when both $\alpha$ and $\mathcal{I}^{ID}$ are small, the influence demand of all the advertisers is also small. Hence, it is easy for the influence provider to satisfy the influence demand of all the advertisers by allocating billboard slots. However, the influence provider suffers from higher excessive regret when the influence supply crosses the influence demand of the advertisers.

Second, in our experiment, the `EA', `EBOE', and `EBTE' controls the excessive regret better because `EBOE' and `EBTE' satisfy the advertisers with fewer billboard slots. The `EBOE' and `EBTE' achieve equal or less regret than `EA' because `EBOE' and `EBTE' try to exchange billboard slots between all the advertisers. The `EBOE' and `EBTE' explore finer-grained exchange compared to `EAOE', reducing the gap between the advertiser's influence demand and its corresponding influence supply from the influence provider.

Third, we observe the behavior of different algorithms on several datasets. In the `Beach' dataset, only excessive regret is there. However, only the `BG' has excessive and unsatisfied regret in the case of $\alpha = 100\%$. Among baseline methods, the `ALS' performs better than the `SG' and `BG'. Next, in the `Bank' dataset, all the algorithms suffer from both excessive and unsatisfied regret when $\alpha = 100\%$. The same observations were observed in the case of the `Airport' dataset as in the `Bank' dataset. At last, the `Mall' and `Park' datasets behave similarly to the `Beach' dataset.

\paragraph{\textbf{Case 2: Low $\alpha$, High $\mathcal{I}^{ID}$ (parts k, l, m, p, q, r of Figure \ref{Fig:Beach},\ref{Fig:Mall},\ref{Fig:Bank},\ref{Fig:Park},\ref{Fig:Airport})}}
Corresponding to case $2$, we have the Demand-Supply ratio, $\alpha \leq 80\%$, and Average-Individual Demand ratio, $\mathcal{I}^{ID} \geq 5\%$. This refers to individual demand being high but global demand being low, i.e., influence provider have a small number of big advertisers. Here, we have three main observations.

First, the excessive regret for all the algorithms decreased compared to case $1$ because the individual demand of each advertiser increased. Also, with the increase of $\mathcal{I}^{ID}$, the number of advertisers decreased, but the individual demand of the advertisers was still higher. The influence supply to the advertisers is closer to the advertiser's individual influence demand, which is the reason behind the decrease in advertisers.

Second, as individual influence demand is high, the influence provider must deploy many billboard slots to satisfy the advertiser's demand. Hence, `EAOE', `EBOE', and `EBTE' need to explore more neighborhood search space to exchange billboard slots, and these methods hugely outperform `BG', `SG', and `ALS'.

Third, when $\alpha$ increases from $40\%$ to $80\%$ among baseline methods, the performance of `BG' degrades rapidly. However, EBTE' and `EBOE' outperform the other proposed methods. Now, consider a case when $\alpha$ value is $40\%$, $60\%$ and $\mathcal{I}^{ID} = 5\%$ `BG' performs better compared to `SG' and `ALS' methods. Among the proposed methods, `EAOE' performs equal to or better than `EA' as `EA' is used as the initial allocation for the remaining proposed methods. When $\mathcal{I}^{ID}$ increases from $5\%$ to $10\%$, the `ALS' performs better than `SG' and `BG'. 

\paragraph{\textbf{Case 3: High $\alpha$, Low $\mathcal{I}^{ID}$ (parts d, e, i, j of Figure \ref{Fig:Beach},\ref{Fig:Mall},\ref{Fig:Bank},\ref{Fig:Park},\ref{Fig:Airport})}}
In case $3$, we have the Demand-Supply ratio, $\alpha \geq 100\%$, and the Average-Individual Demand ratio, $\mathcal{I}^{ID} \leq 2\%$. This refers to a situation where individual demand is low, but the global demand is very high, i.e., the influence provider has many advertisers whose influence demand is less. We have made some observations. 

First, when $\alpha \geq 100\%$, i.e., very high global demand, none of our proposed and existing algorithms can satisfy all the advertisers. Even when $\alpha = 100 \%$, i.e., the global demand is equal to the influence provider supply, all advertisers are unsatisfied because the excessive regrets cannot be fully minimized. Most of the portion of total regret comes from unsatisfied regret. When $\alpha = 120\%$, the advertiser's influence demand exceeds the supply, and the unsatisfied regret becomes very high. 

Second, as the influence provider needs a sufficient billboard slot to fulfill the influence demand of the advertisers, there is a risk of losing some advertisers. Hence, all allocations are critical, and the `EA', `EBTE', and `EBOE' carefully handled these risks of losing advertisers. The advantage of the `EA', `EBTE', and `EBOE' in minimizing total regret is more efficient than existing baseline approaches. However, these approaches can satisfy more advertisers, and the influence provider suffers from minor unsatisfied regret, i.e., the overall regret of the influence provider decreases. 

Third, in most cases when $\alpha = 100\%$ and $\mathcal{I}^{ID} = 1\%$ or $2\%$, only the `BG' suffers from both excessive and unsatisfied regret. However, when $\alpha = 120\%$, all the proposed and baseline methods suffer from excessive and unsatisfied regret. These observations are shown in the `Beach', `Mall', and `Park' datasets. In the case of `Bank' and `Airport' datasets, when $\mathcal{I}^{ID}$ increases from $1\%$ to $2\%$, unsatisfied regret increases because global influence demand is already high. In addition, when individual influence increases, unsatisfied regret also increases. If we evaluate the proposed and baseline methods, obviously, `EBOE' and `EBTE' reduce the overall regret best. Among baseline methods, `ALS' and `SG' outperform `BG' and reduce total regret by almost $40\%$ to $45\%$ and $50\%$ to $60\%$, respectively. So, among baseline, `ALS' performs better. Next, among proposed approaches, `EA', `EBOE', and `EBTE' reduces overall regret compared to `ALS' are $60\%$ to $65\%$, $79\%$ to $84\%$ and $75\%$ to $80\%$, respectively.

\paragraph{\textbf{Case 4: High $\alpha$, High $\mathcal{I}^{ID}$ (parts n, o, s, t of Figure \ref{Fig:Beach},\ref{Fig:Mall},\ref{Fig:Bank},\ref{Fig:Park},\ref{Fig:Airport})}}
In case $4$, we have the Demand-Supply ratio, $\alpha \geq 100\%$, and the Average-Individual Demand ratio, $\mathcal{I}^{ID} \geq 5\%$. This refers to individual and global demand being very high, i.e., influence provider is over-demanded by a small number of big advertisers. We have three main observations.

First, every unsatisfied penalty from the advertisers leads to higher regret for the influence provider due to large $\alpha$, and $\mathcal{I}^{ID}$ value. Hence, all the proposed and baseline methods suffer from higher overall regret, and the advantage of the `EBOE' and `EBTE' is not very effective, especially compared to the `EA' approach.

Second, when $\mathcal{I}^{ID}$ increases from $5\%$ to $10\%$ total regret also increases as shown in Figure \ref{Fig:Beach},\ref{Fig:Mall},\ref{Fig:Bank},\ref{Fig:Park},\ref{Fig:Airport}. All the algorithms suffer from excessive and unsatisfied regret because of higher individual and global influence demand. The `BG' approach suffers from more excessive regret than other approaches because the ideal billboard slots are allocated first to the advertisers individually until all the advertisers are satisfied or billboard slots run out.

Third, the advantage of `EBOE' and `EBTE' becomes less significant than `ALS' because of higher global and individual demand. Hence, if the influence provider has sufficient billboard slots to satisfy all the advertisers, then a large number of advertisers with small individual influence demand is desirable for the influence provider. Because in this setting, the unsatisfied regret will be much less. However, excessive regret will be there as it can not be fully diminished.




\subsection{Efficiency Study}
\begin{figure}[h!]
    \centering
    \begin{tabular}{lclclclclclclc}
        EBTE & \includegraphics[width=0.11\linewidth]{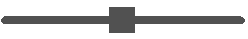} & EBOE & \includegraphics[width=0.068\linewidth]{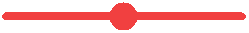} & EAOE & \includegraphics[width=0.068\linewidth]{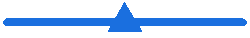} & EA & \includegraphics[width=0.068\linewidth]{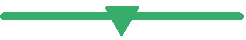}& ALS & \includegraphics[width=0.068\linewidth]{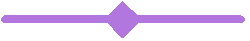}& SG & \includegraphics[width=0.068\linewidth]{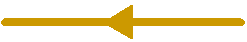}& BG & \includegraphics[width=0.068\linewidth]{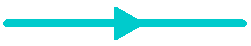}\\
    \end{tabular}
    \begin{tabular}{cccc}     
        \includegraphics[width=0.25\linewidth]{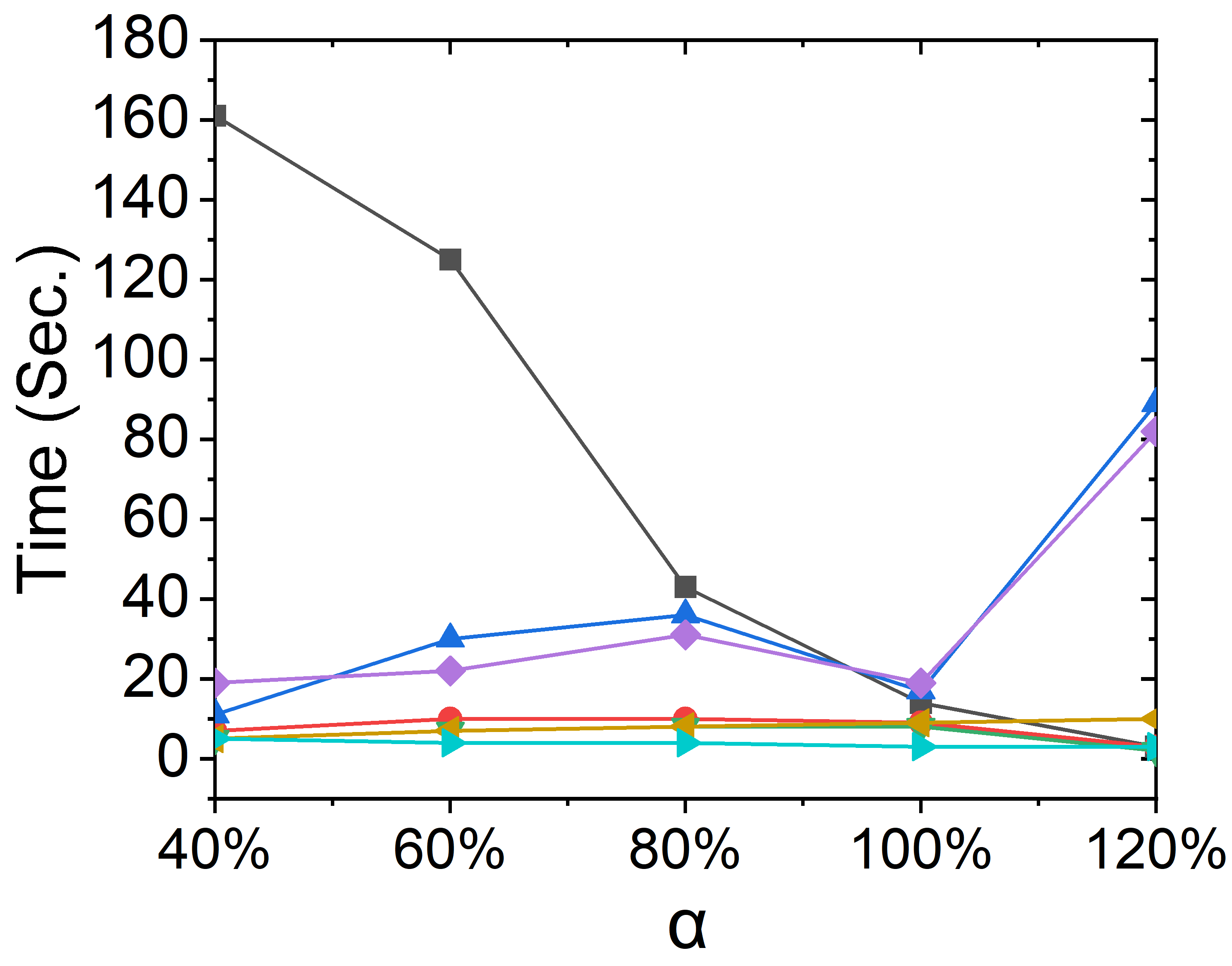} &
        \includegraphics[width=0.25\linewidth]{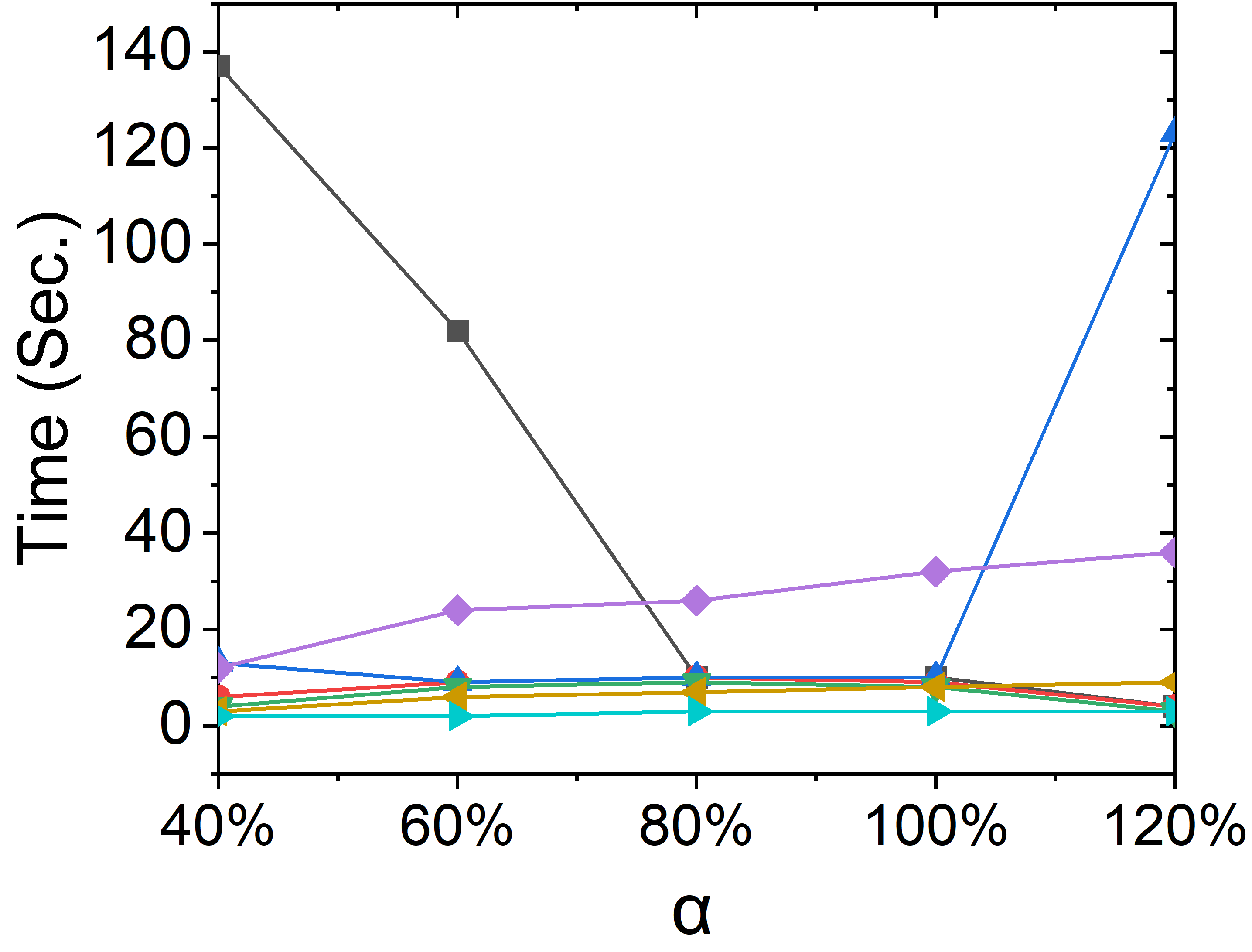} &
        \includegraphics[width=0.25\linewidth]{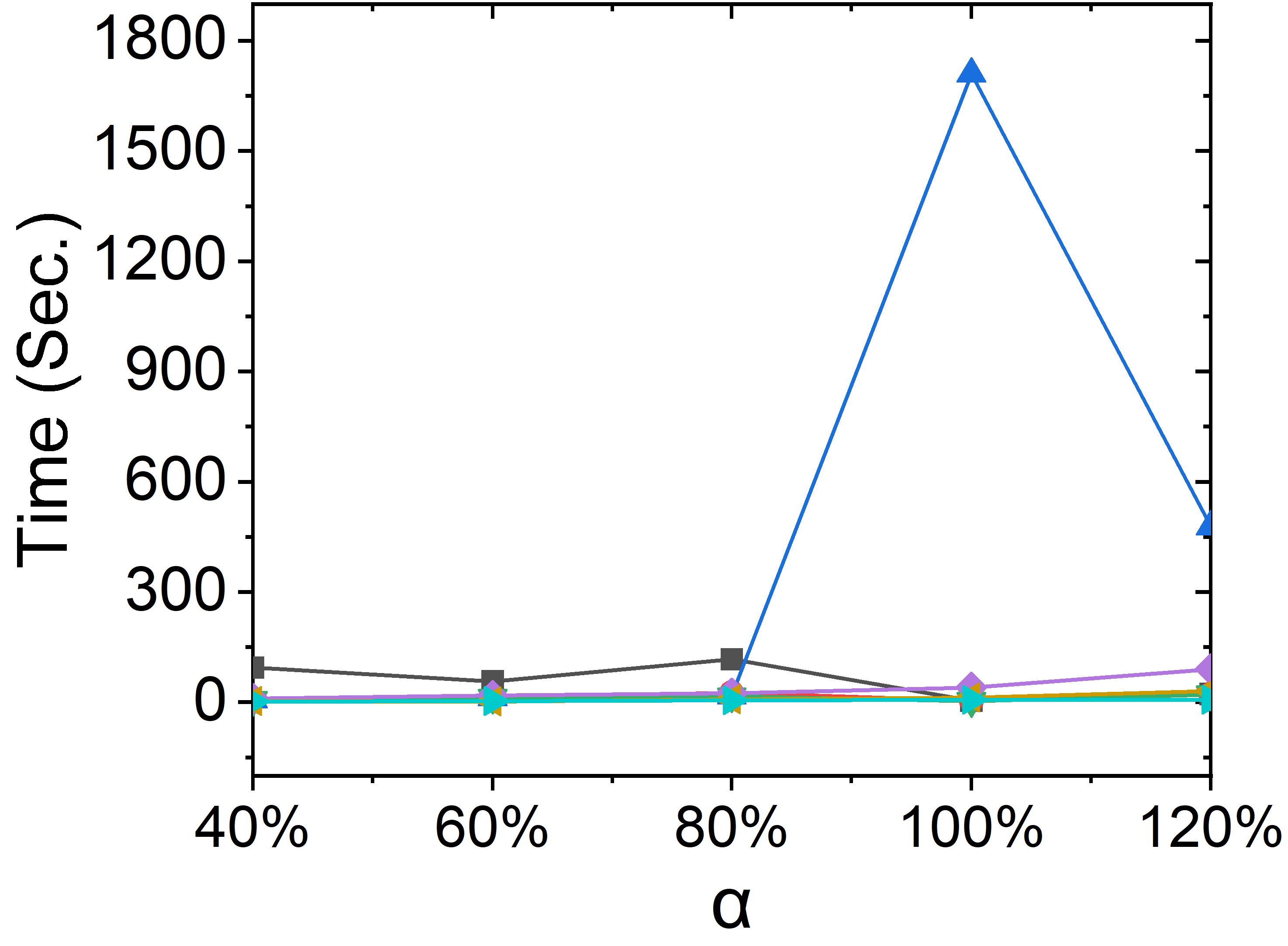} &
        \includegraphics[width=0.25\linewidth]{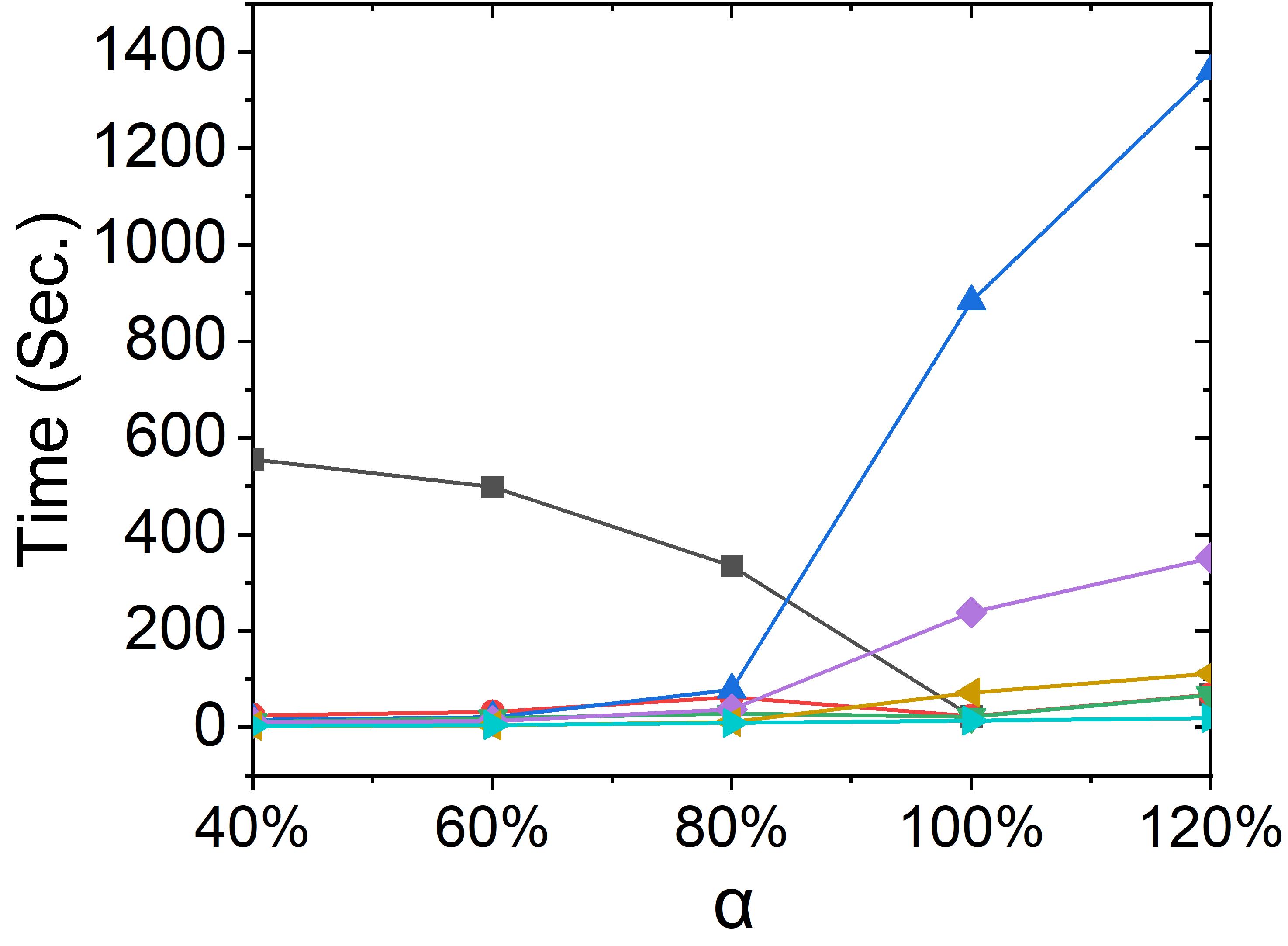}\\
        {\tiny (a) $\mathcal{I}^{ID} = 1\%$, $|\mathcal{A}| = 100$} &
        {\tiny (b) $\mathcal{I}^{ID} = 2\%$, $|\mathcal{A}| = 50$} &
        {\tiny (c) $\mathcal{I}^{ID} = 5\%$, $|\mathcal{A}| = 20$} &
        {\tiny (d) $\mathcal{I}^{ID} = 10\%$, $|\mathcal{A}| = 10$}\\
    \end{tabular}
    \caption{Efficiency Study on Beach location type}
    \label{Fig:Beach_Time}
\end{figure}

\begin{figure}[h!]
    \centering
    \begin{tabular}{lclclclclclclc}
        EBTE & \includegraphics[width=0.11\linewidth]{EBTE.png} & EBOE & \includegraphics[width=0.068\linewidth]{EBOE.png} & EAOE & \includegraphics[width=0.068\linewidth]{EAOE.png} & EA & \includegraphics[width=0.068\linewidth]{EA.png}& ALS & \includegraphics[width=0.068\linewidth]{ALS.png}& SG & \includegraphics[width=0.068\linewidth]{SG.png}& BG & \includegraphics[width=0.068\linewidth]{BG.png}\\
    \end{tabular}
    \begin{tabular}{cccc}     
        \includegraphics[width=0.25\linewidth]{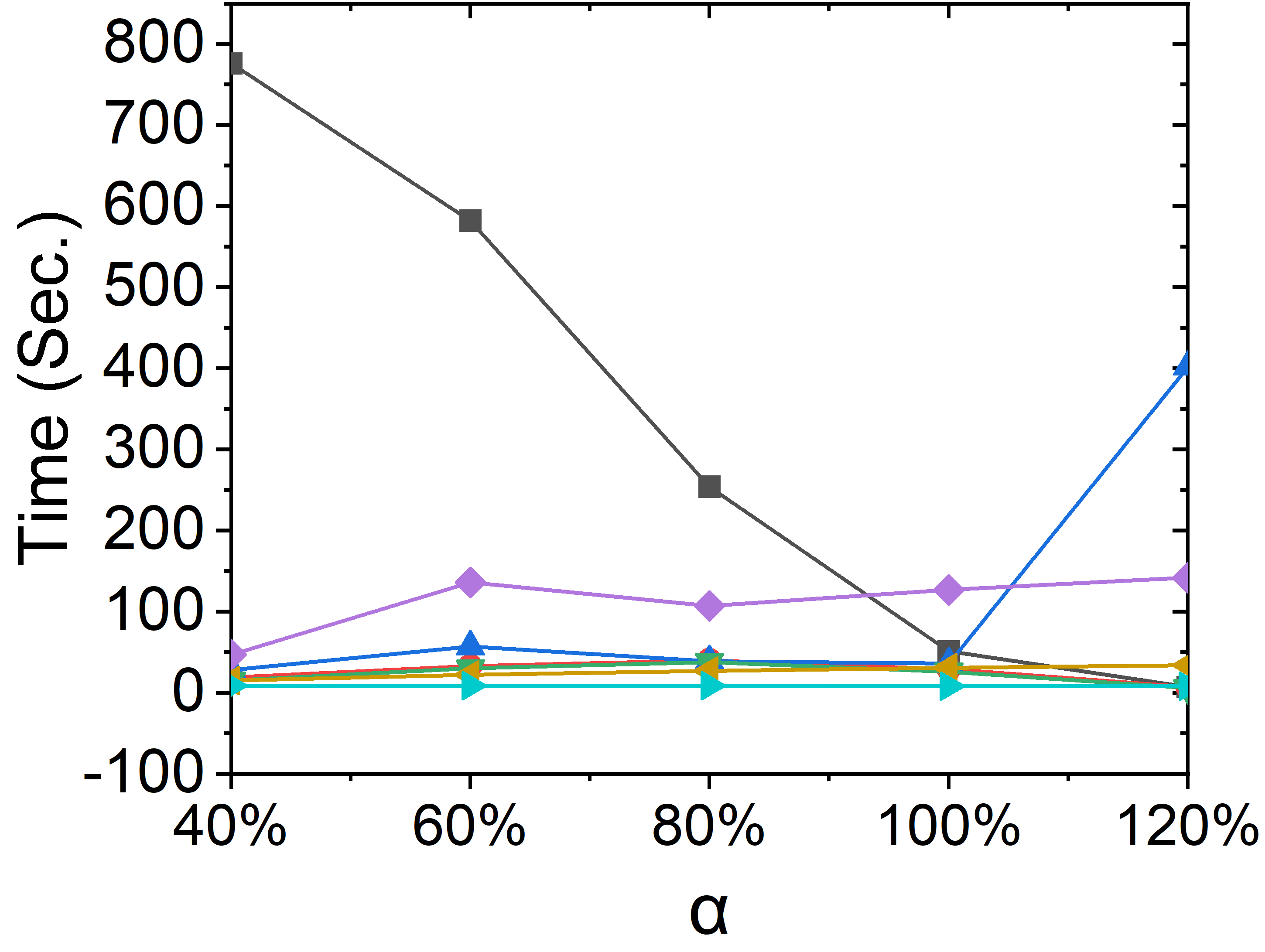} &
        \includegraphics[width=0.25\linewidth]{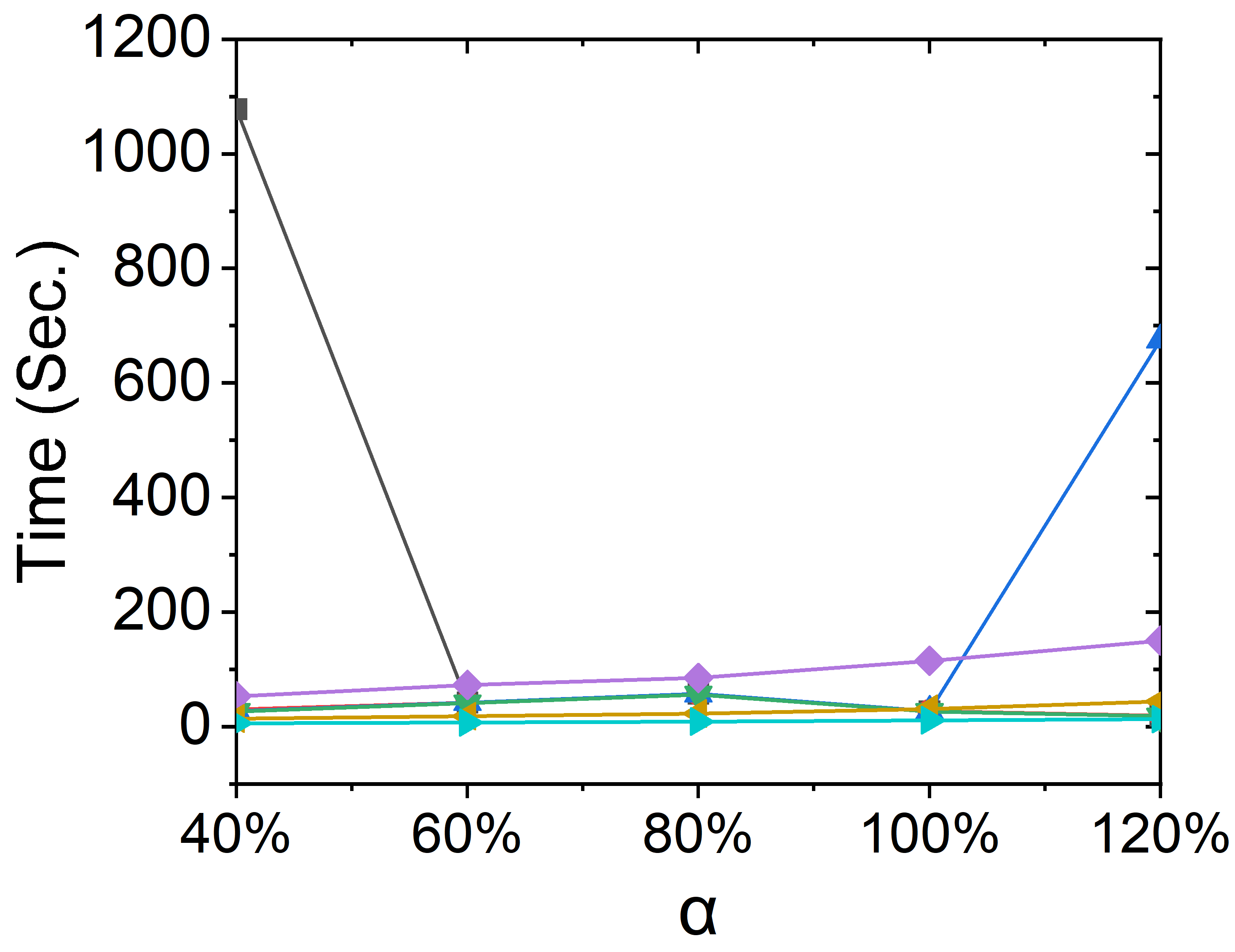} &
        \includegraphics[width=0.25\linewidth]{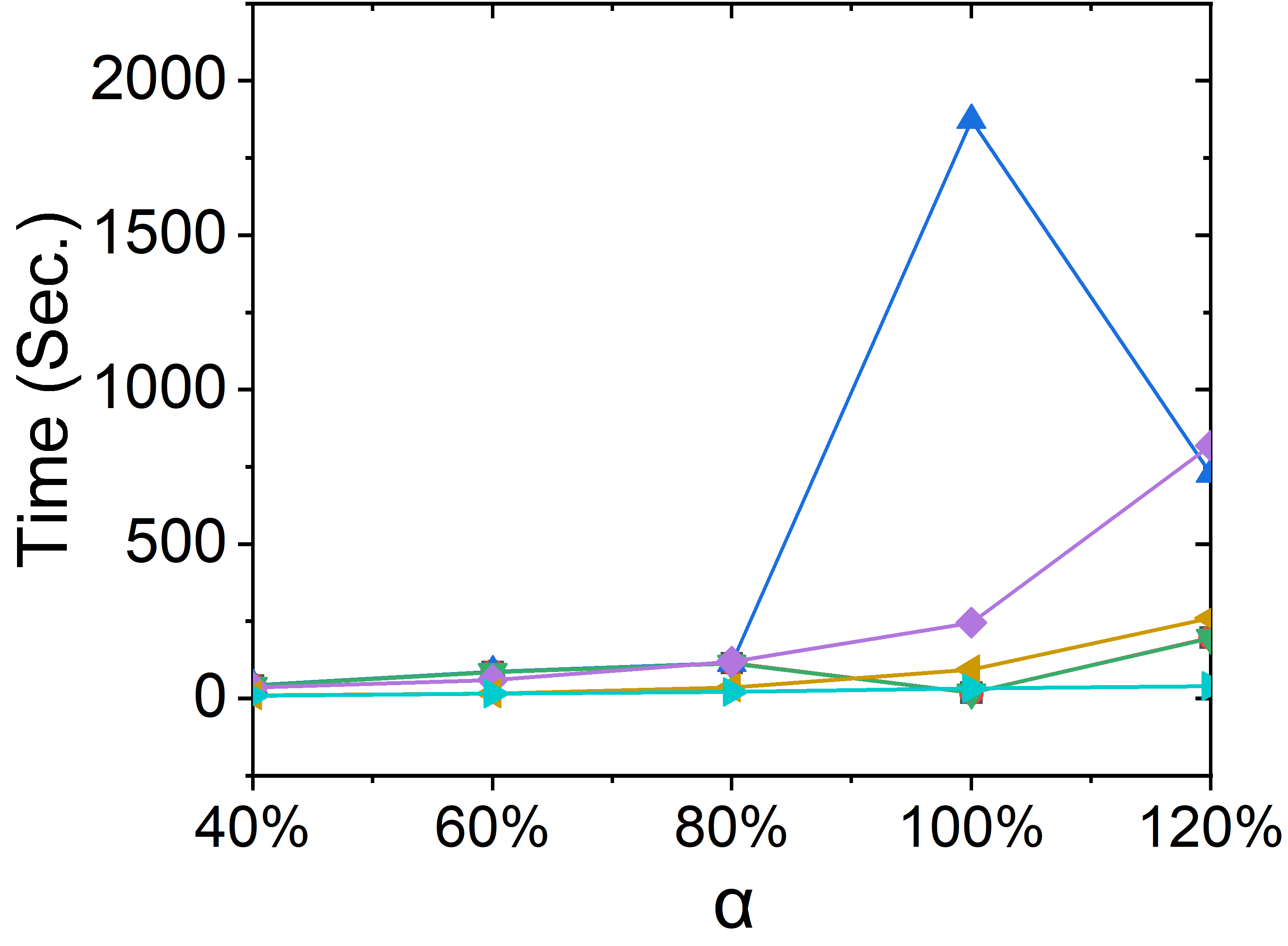} &
        \includegraphics[width=0.25\linewidth]{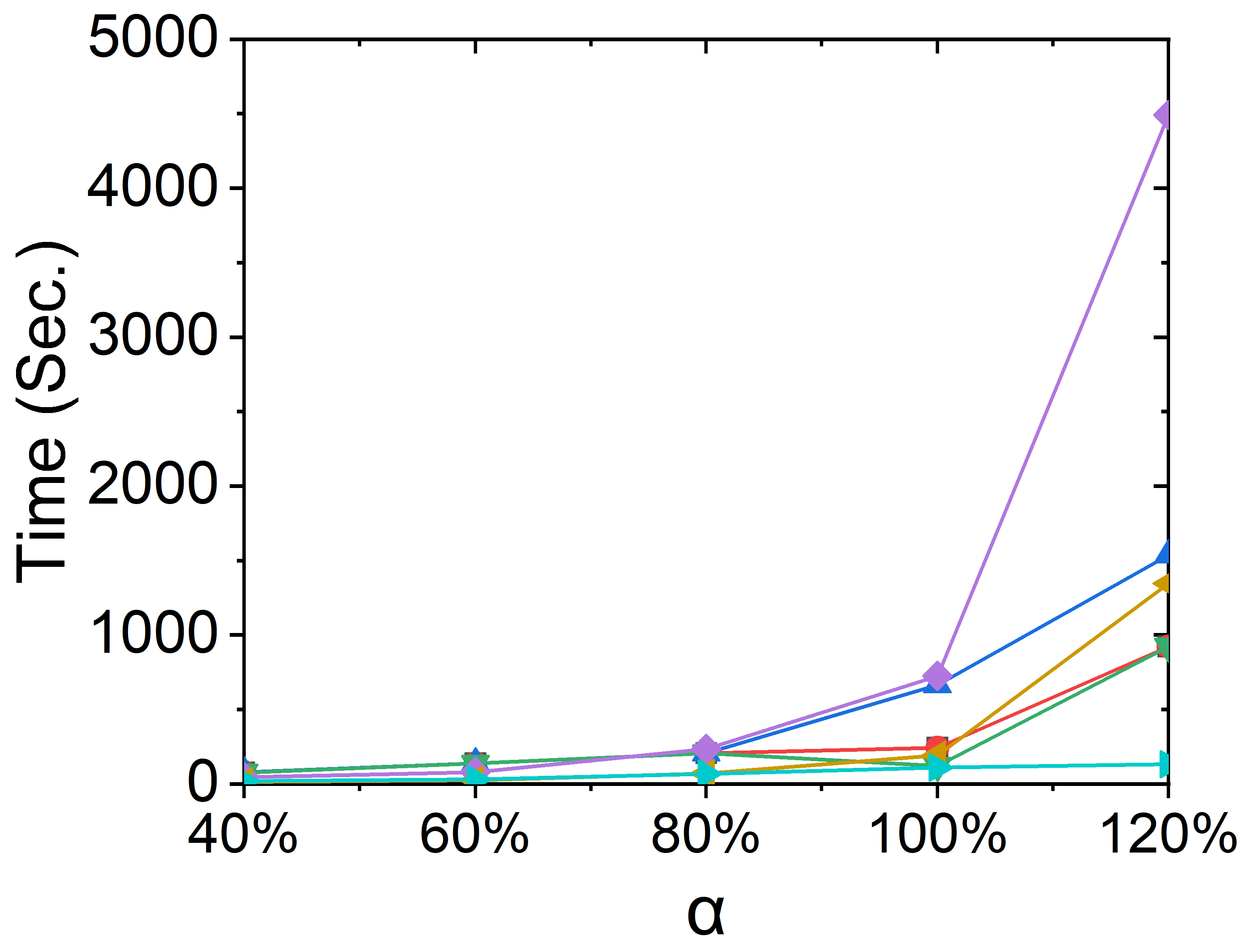}\\
        {\tiny (a) $\mathcal{I}^{ID} = 1\%$, $|\mathcal{A}| = 100$} &
        {\tiny (b) $\mathcal{I}^{ID} = 2\%$, $|\mathcal{A}| = 50$} &
        {\tiny (c) $\mathcal{I}^{ID} = 5\%$, $|\mathcal{A}| = 20$} &
        {\tiny (d) $\mathcal{I}^{ID} = 10\%$, $|\mathcal{A}| = 10$}\\
    \end{tabular}
    \caption{Efficiency Study on Mall location type}
    \label{Fig:Mall_Time}
\end{figure}

\begin{figure}[h!]
    \centering
    \begin{tabular}{lclclclclclclc}
        EBTE & \includegraphics[width=0.11\linewidth]{EBTE.png} & EBOE & \includegraphics[width=0.068\linewidth]{EBOE.png} & EAOE & \includegraphics[width=0.068\linewidth]{EAOE.png} & EA & \includegraphics[width=0.068\linewidth]{EA.png}& ALS & \includegraphics[width=0.068\linewidth]{ALS.png}& SG & \includegraphics[width=0.068\linewidth]{SG.png}& BG & \includegraphics[width=0.068\linewidth]{BG.png}\\
    \end{tabular}
    \begin{tabular}{cccc}     
        \includegraphics[width=0.25\linewidth]{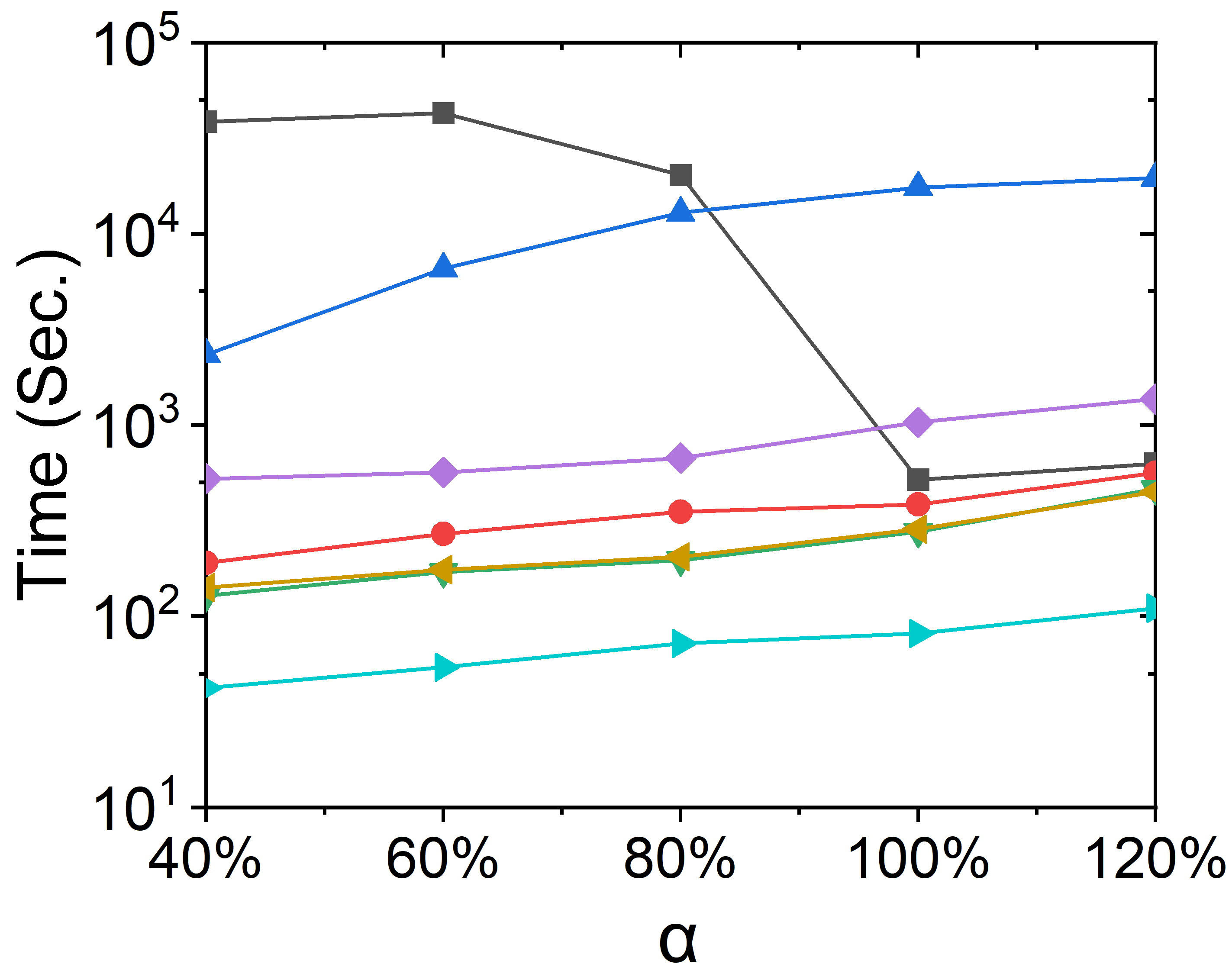} &
        \includegraphics[width=0.25\linewidth]{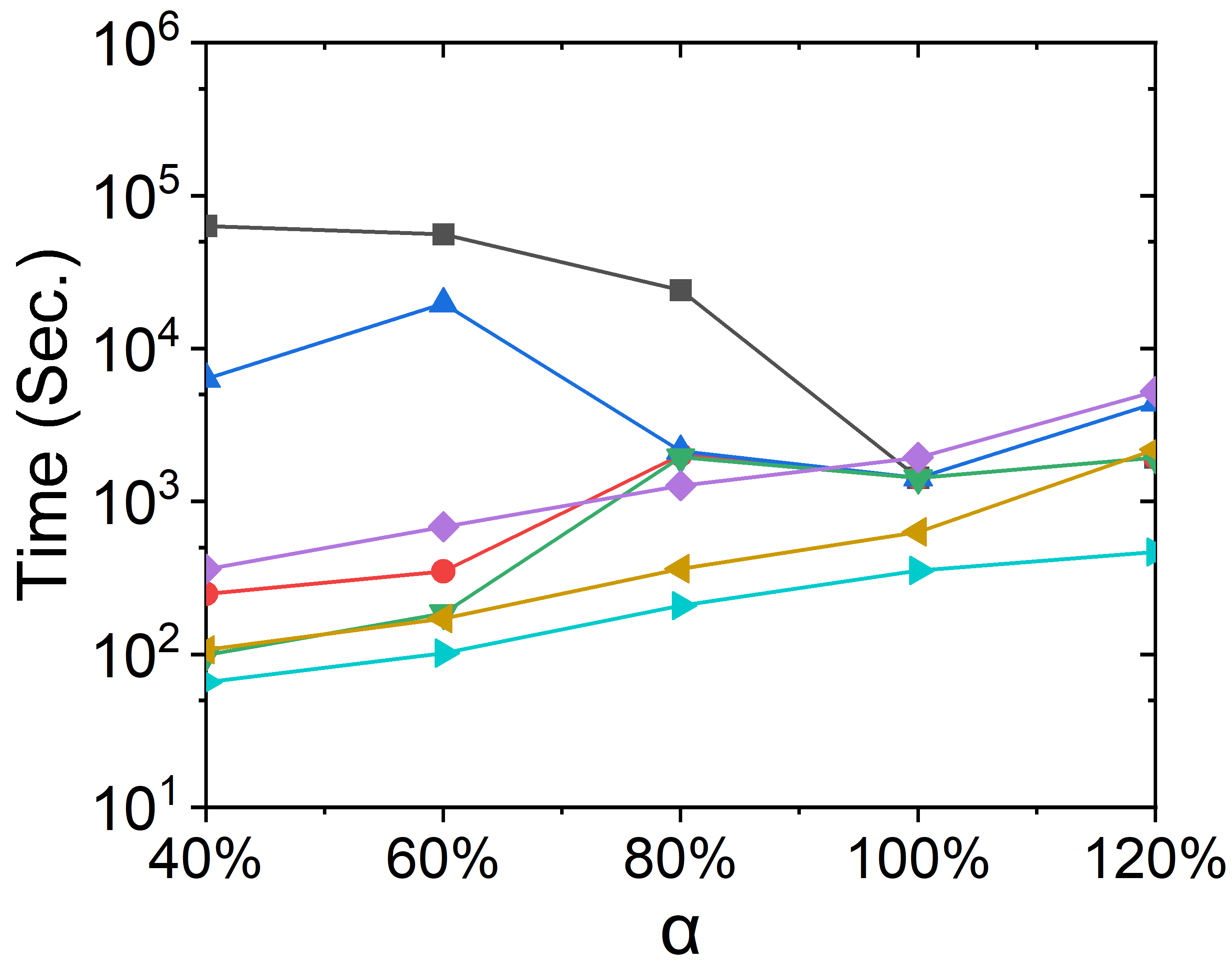} &
        \includegraphics[width=0.25\linewidth]{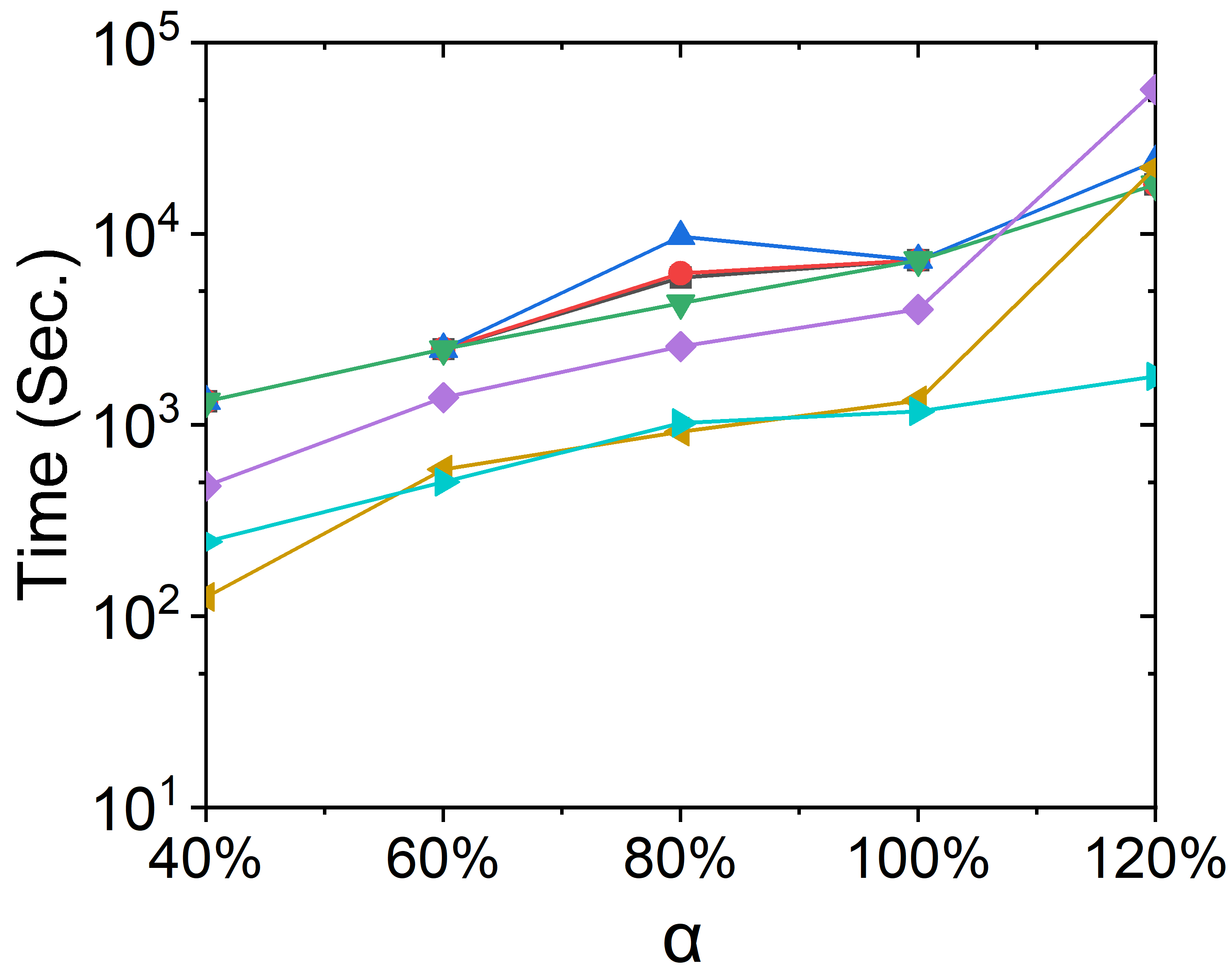} &
        \includegraphics[width=0.25\linewidth]{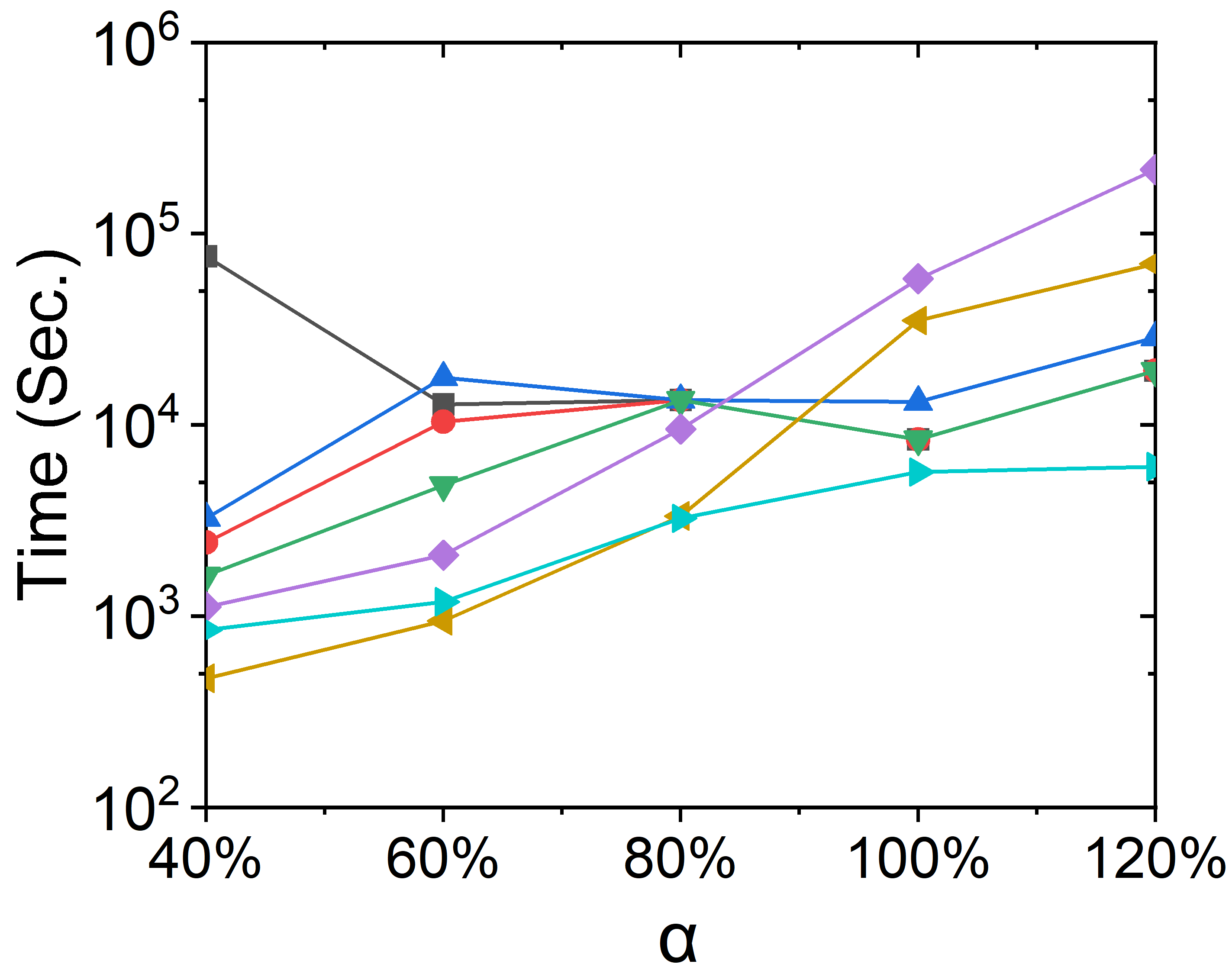}\\
        {\tiny (a) $\mathcal{I}^{ID} = 1\%$, $|\mathcal{A}| = 100$} &
        {\tiny (b) $\mathcal{I}^{ID} = 2\%$, $|\mathcal{A}| = 50$} &
        {\tiny (c) $\mathcal{I}^{ID} = 5\%$, $|\mathcal{A}| = 20$} &
        {\tiny (d) $\mathcal{I}^{ID} = 10\%$, $|\mathcal{A}| = 10$}\\
    \end{tabular}
    \caption{Efficiency Study on Bank location type}
    \label{Fig:Bank_Time}
\end{figure}

\begin{figure}[h!]
    \centering
    \begin{tabular}{lclclclclclclc}
        EBTE & \includegraphics[width=0.11\linewidth]{EBTE.png} & EBOE & \includegraphics[width=0.068\linewidth]{EBOE.png} & EAOE & \includegraphics[width=0.068\linewidth]{EAOE.png} & EA & \includegraphics[width=0.068\linewidth]{EA.png}& ALS & \includegraphics[width=0.068\linewidth]{ALS.png}& SG & \includegraphics[width=0.068\linewidth]{SG.png}& BG & \includegraphics[width=0.068\linewidth]{BG.png}\\
    \end{tabular}
    \begin{tabular}{cccc}     
        \includegraphics[width=0.25\linewidth]{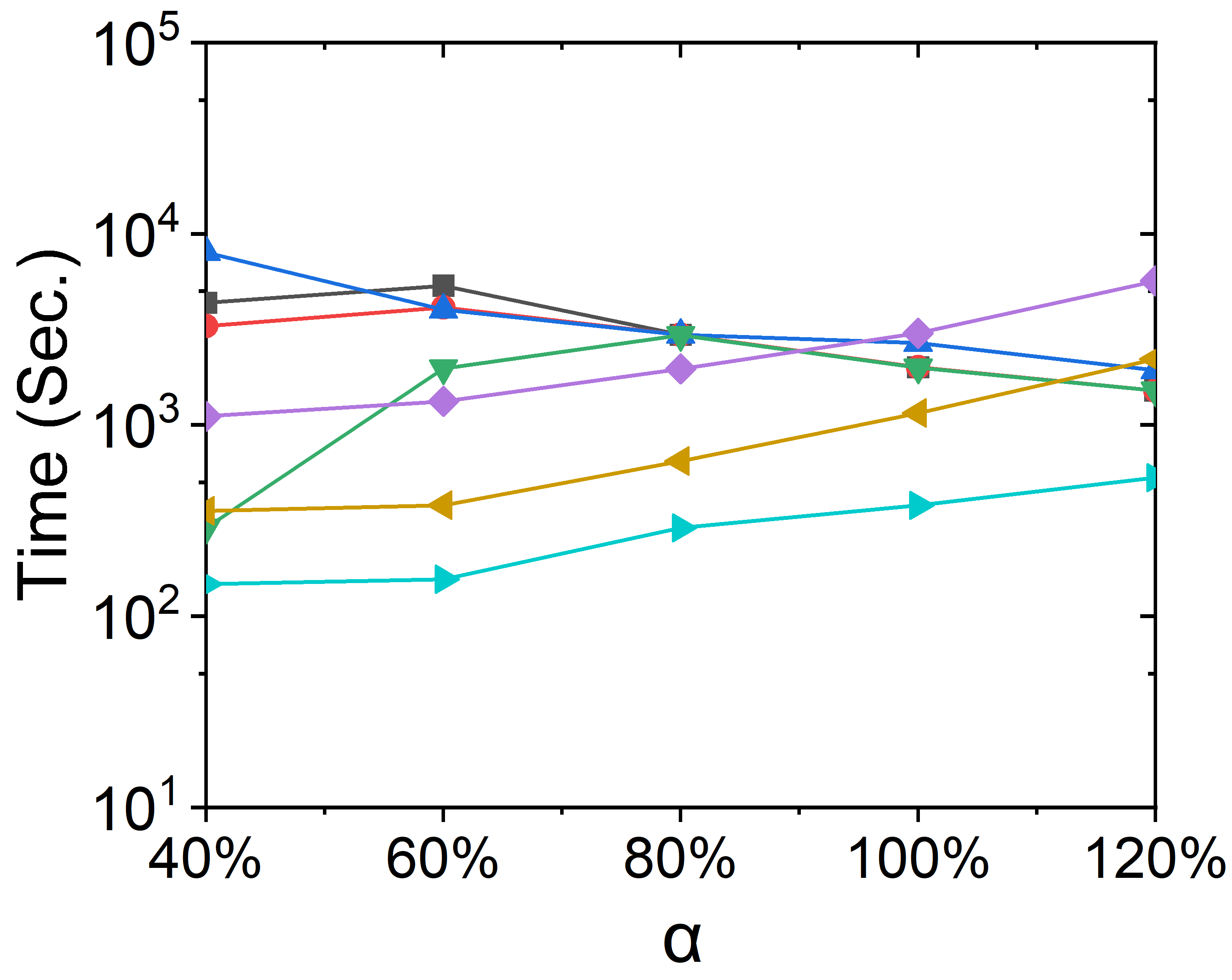} &
        \includegraphics[width=0.25\linewidth]{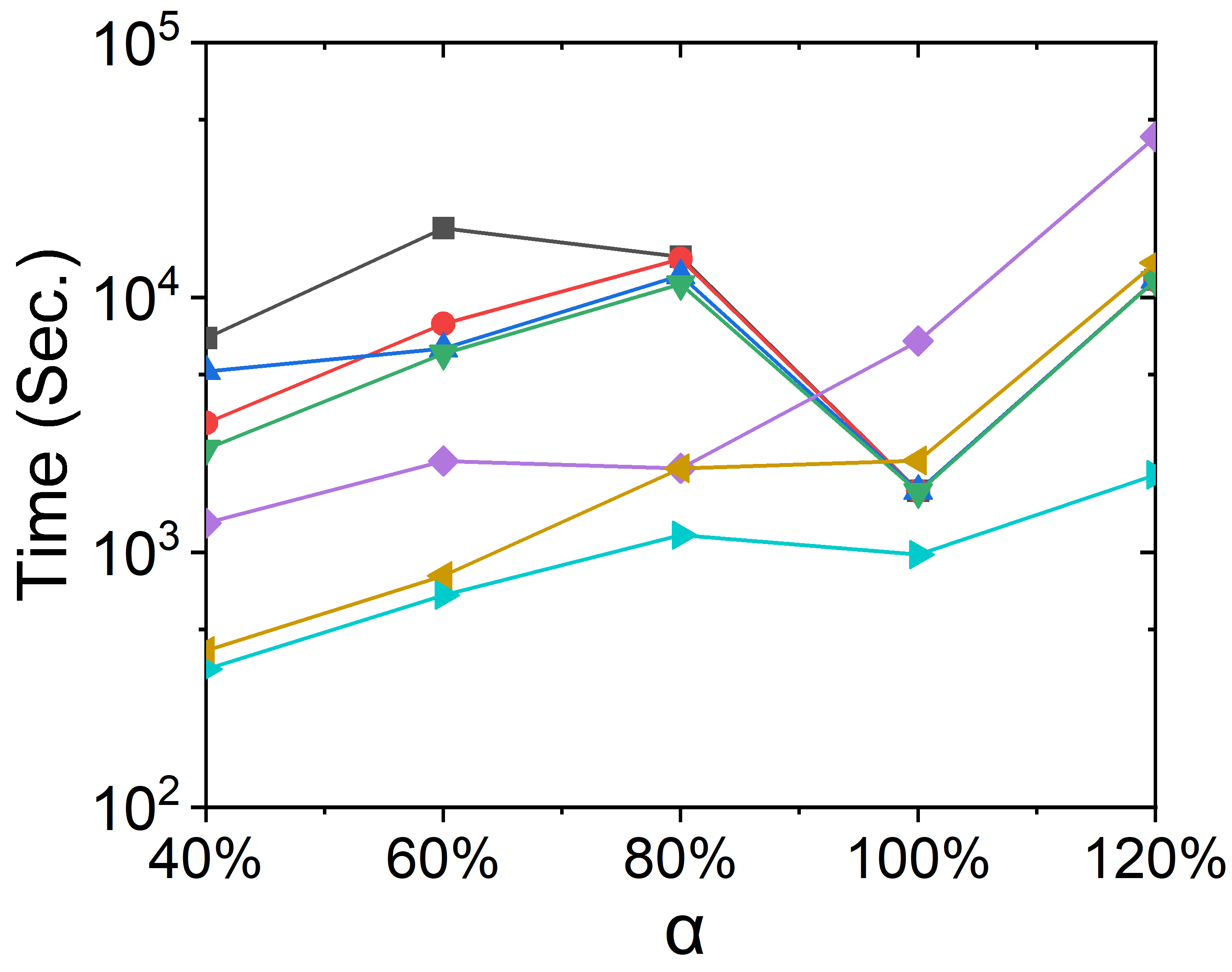} &
        \includegraphics[width=0.25\linewidth]{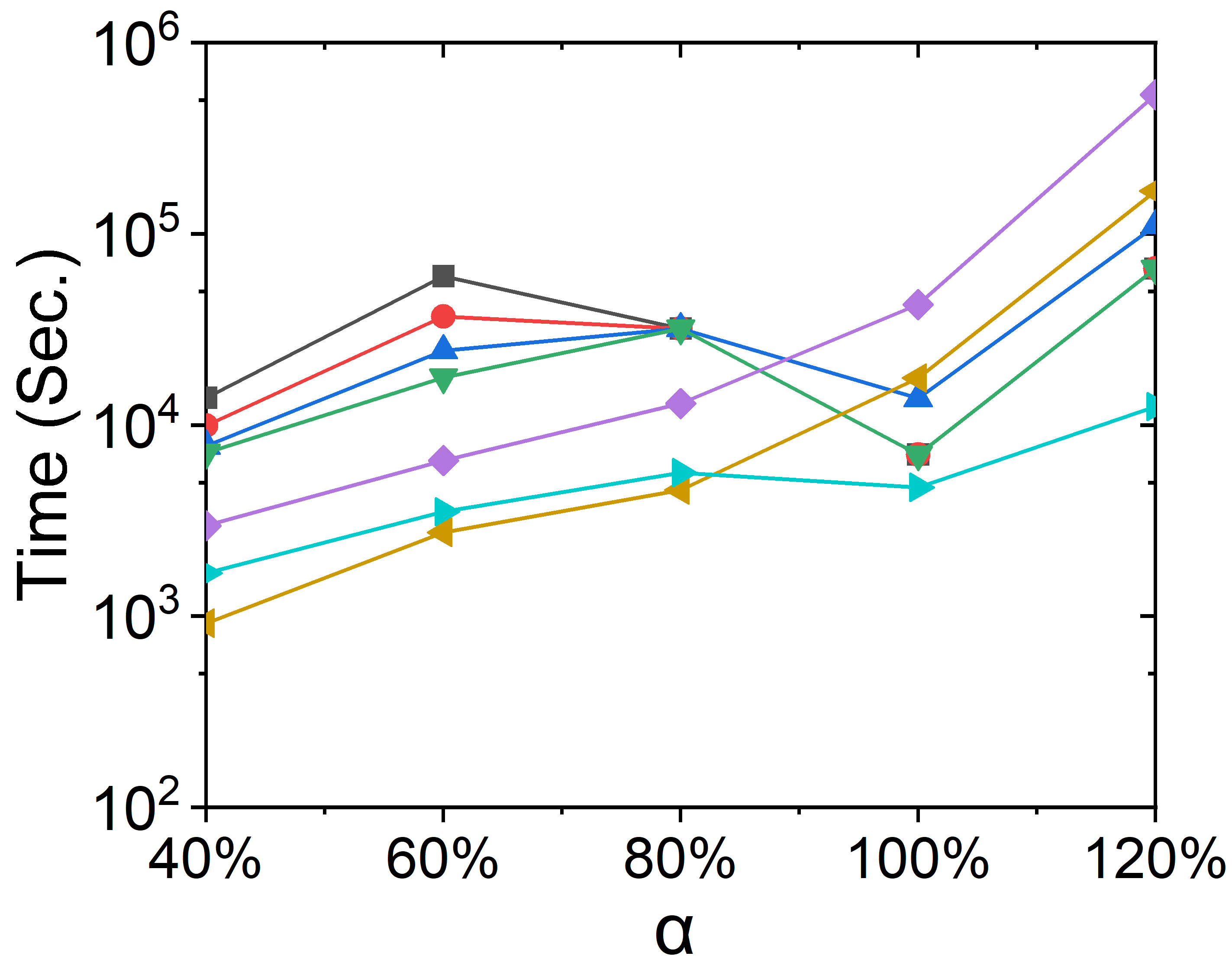} &
        \includegraphics[width=0.25\linewidth]{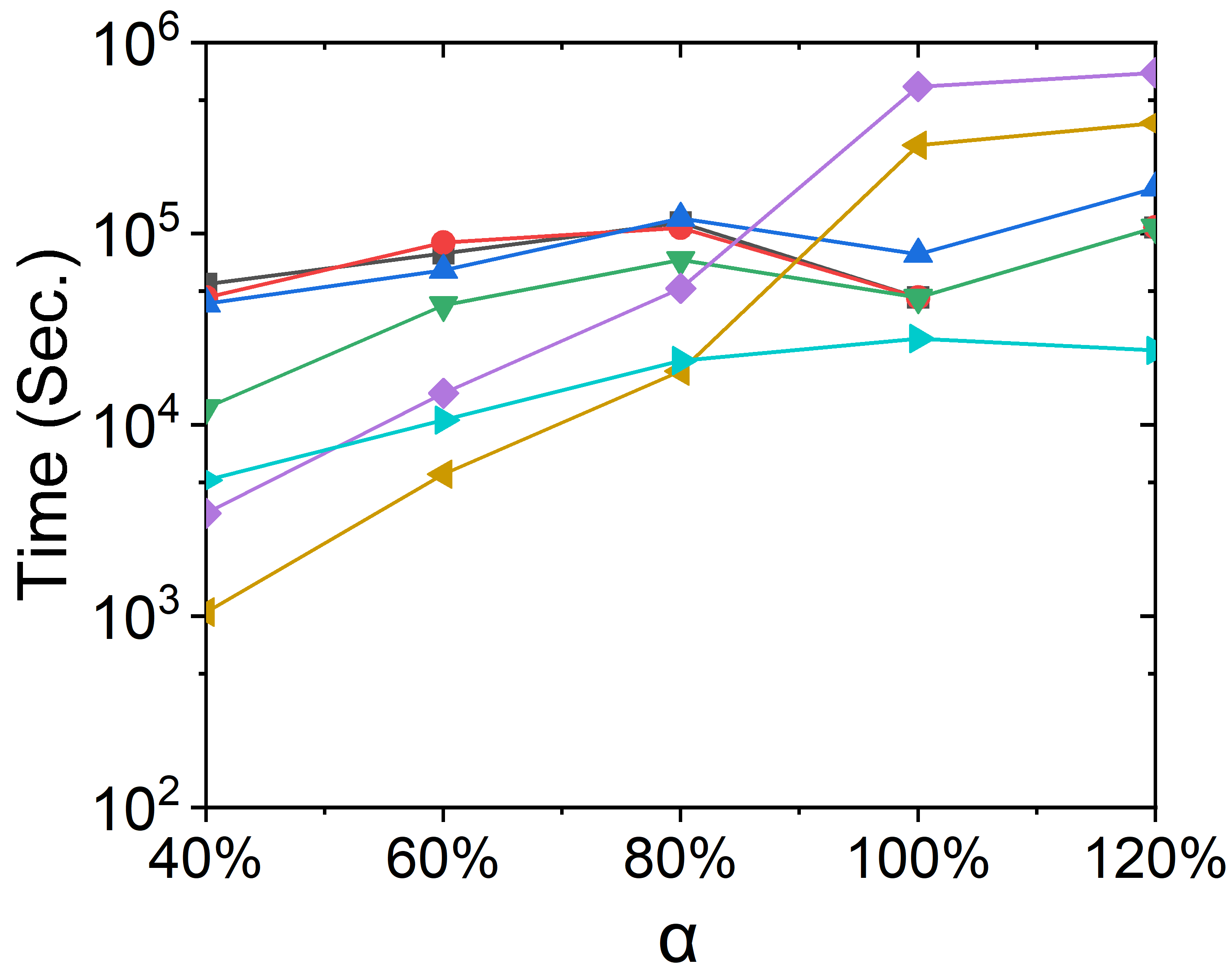}\\
        {\tiny (a) $\mathcal{I}^{ID} = 1\%$, $|\mathcal{A}| = 100$} &
        {\tiny (b) $\mathcal{I}^{ID} = 2\%$, $|\mathcal{A}| = 50$} &
        {\tiny (c) $\mathcal{I}^{ID} = 5\%$, $|\mathcal{A}| = 20$} &
        {\tiny (d) $\mathcal{I}^{ID} = 10\%$, $|\mathcal{A}| = 10$}\\
    \end{tabular}
    \caption{Efficiency Study on Park location type}
    \label{Fig:Park_Time}
\end{figure}

\begin{figure}[h!]
    \centering
    \begin{tabular}{lclclclclclclc}
        EBTE & \includegraphics[width=0.11\linewidth]{EBTE.png} & EBOE & \includegraphics[width=0.068\linewidth]{EBOE.png} & EAOE & \includegraphics[width=0.068\linewidth]{EAOE.png} & EA & \includegraphics[width=0.068\linewidth]{EA.png}& ALS & \includegraphics[width=0.068\linewidth]{ALS.png}& SG & \includegraphics[width=0.068\linewidth]{SG.png}& BG & \includegraphics[width=0.068\linewidth]{BG.png}\\
    \end{tabular}
    \begin{tabular}{cccc}     
        \includegraphics[width=0.25\linewidth]{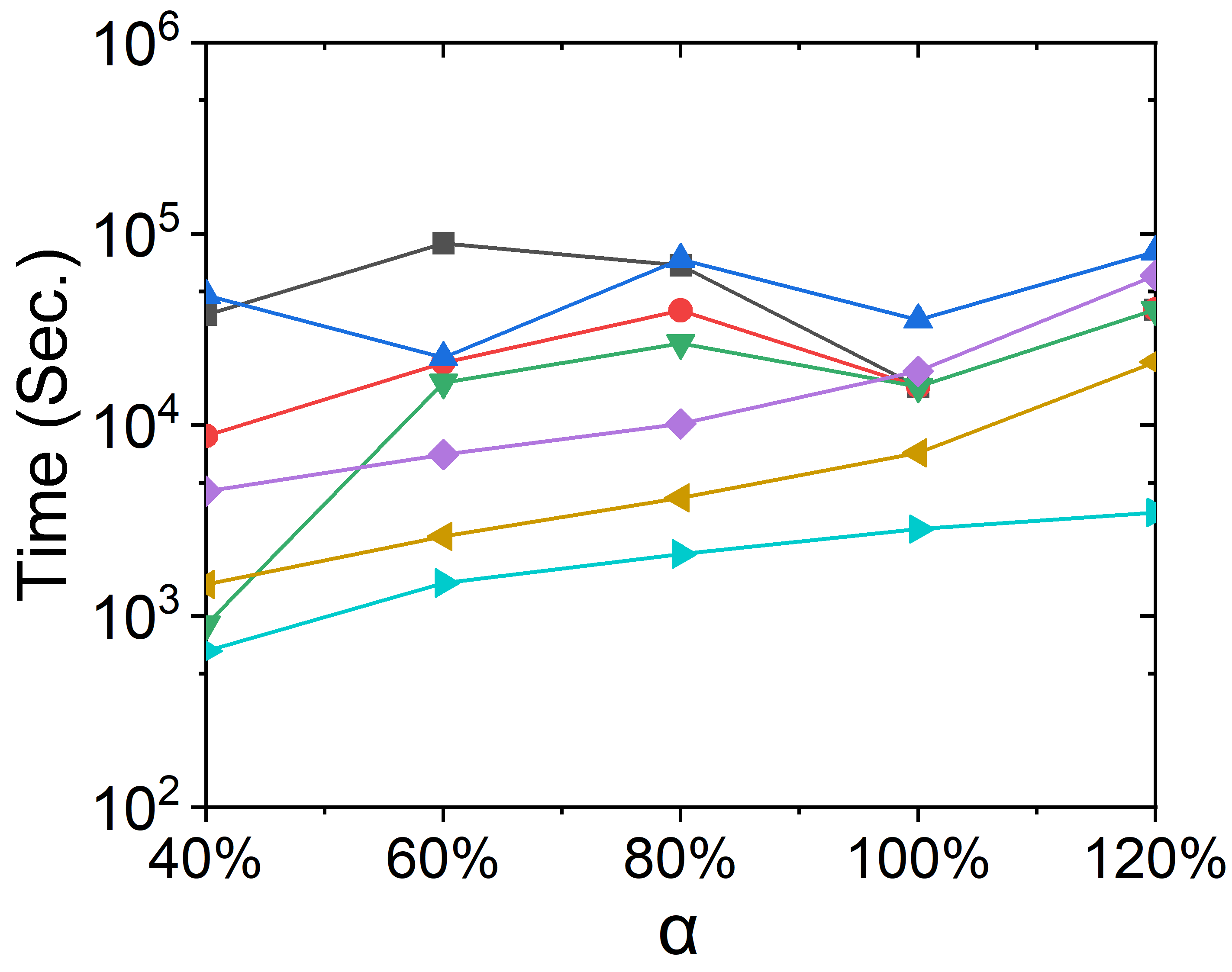} &
        \includegraphics[width=0.25\linewidth]{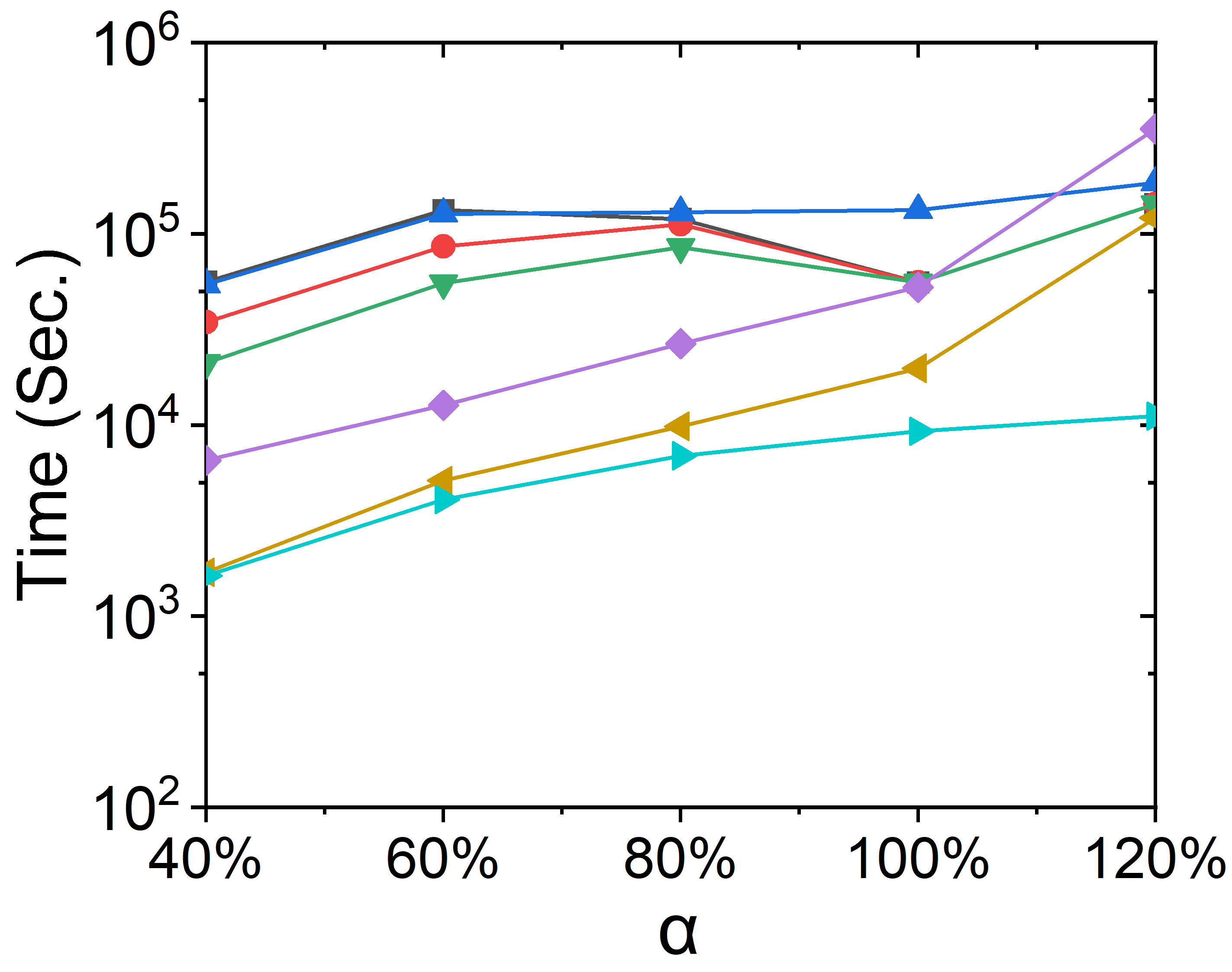} &
        \includegraphics[width=0.25\linewidth]{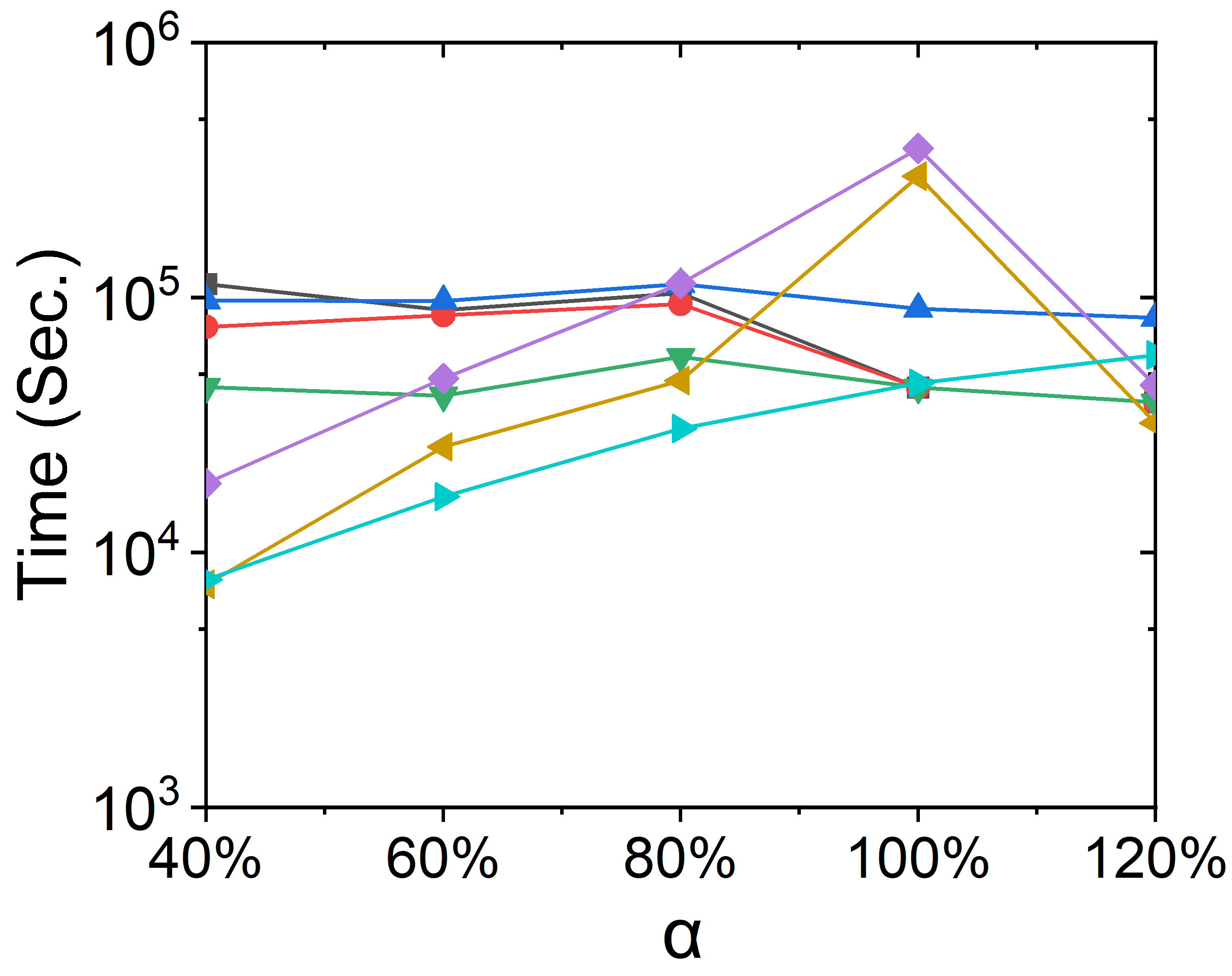} &
        \includegraphics[width=0.25\linewidth]{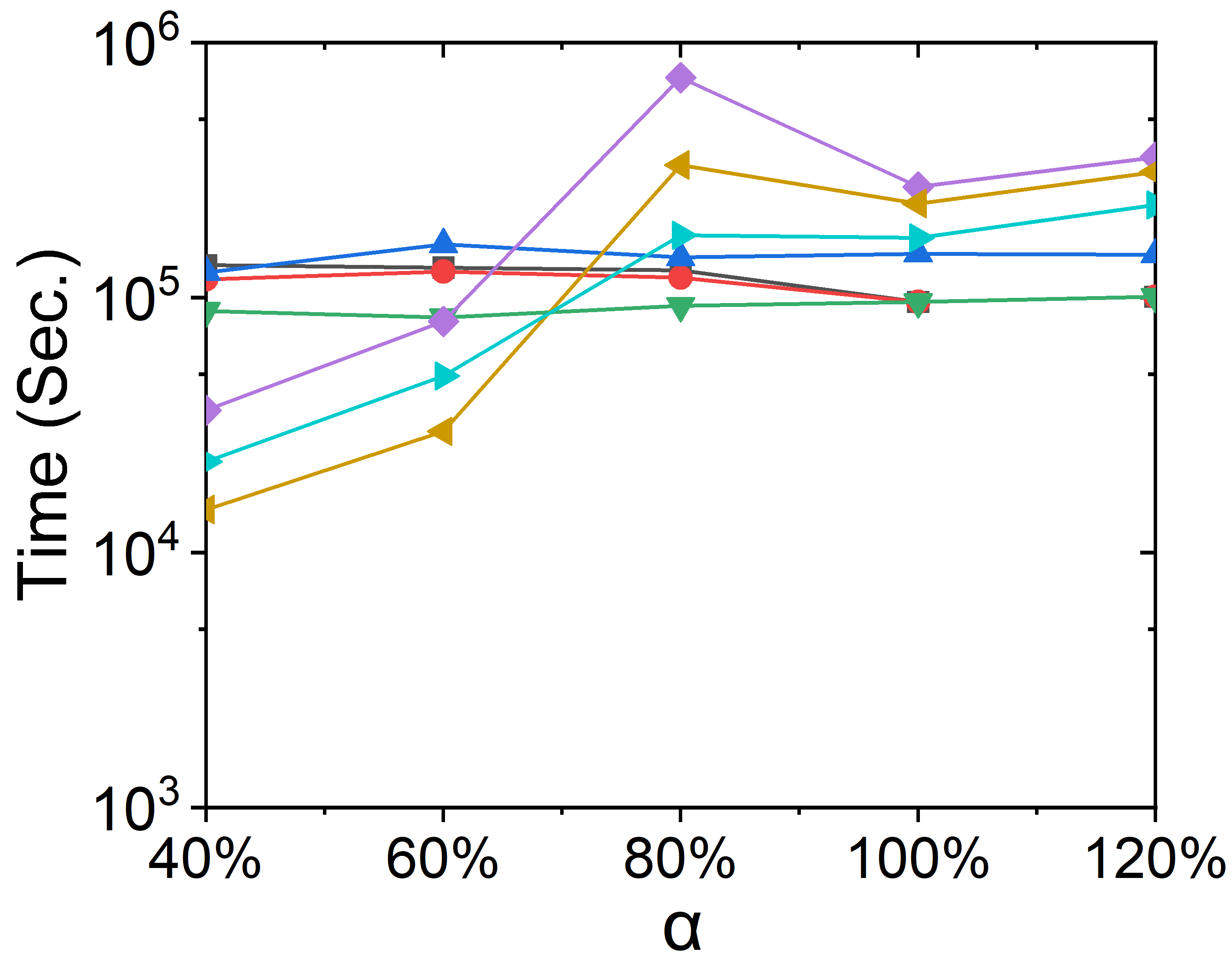}\\
        {\tiny (a) $\mathcal{I}^{ID} = 1\%$, $|\mathcal{A}| = 100$} &
        {\tiny (b) $\mathcal{I}^{ID} = 2\%$, $|\mathcal{A}| = 50$} &
        {\tiny (c) $\mathcal{I}^{ID} = 5\%$, $|\mathcal{A}| = 20$} &
        {\tiny (d) $\mathcal{I}^{ID} = 10\%$, $|\mathcal{A}| = 10$}\\
    \end{tabular}
    \caption{Efficiency Study on Airport location type}
    \label{Fig:Airport_Time}
\end{figure}

Efficiency study is an important parameter in this regret minimization problem as every day, more than a thousand advertisers come to the influence provider with the required influence demand. Hence, we conduct efficiency evaluation to find out computational cost varying different $\alpha$, $\mathcal{I}^{ID}$ and the number of advertisers, $|\mathcal{A}|$. The results are reported in Figure \ref{Fig:Beach_Time},\ref{Fig:Mall_Time},\ref{Fig:Bank_Time},\ref{Fig:Park_Time},\ref{Fig:Airport_Time} over different datasets. We have some observations over different datasets.

First, among baseline methods, both `ALS' and `SG' take more computational time compared to the `BG' approach as they use `BG' as their initial allotment plan of billboard slots and try to improve the allocation further over it by exchanging billboard slots among the advertisers. Out of all the proposed approaches, the `EA'  takes less time compared to other ones because they use the `EA' as their initial allocation plan for slots allocation like the `ALS' and `SG' and further exchange billboard slots among advertisers to get an effective allotment plan requires more computational search time.

Second, when the $\alpha$ value increases from $40\%$ to $120\%$, all the algorithms take more computational search time to explore all the billboard slots. In this case, the influence provider must allocate more billboard slots to the advertisers. Next, when $\mathcal{I}^{ID}$ value increases from $1\%$ to $10\%$, the individual influence demand of all the advertisers increases, and almost all the algorithms require more computational time. Among the proposed approaches, `EBTE',  `EBOE', and `EAOE' need a more significant number of iterations to explore all the billboard slots to find a better alternative allocation over the `EA'.

Third, when $\alpha$ value increases from $40\%$ to $60\%$ and the $\mathcal{I}^{ID}$ value from $1\%$ to $2\%$, the computational cost for all the algorithms increases except the `EBTE' because of increasing individual demand value, influence provider needs more billboard slots to satisfy the advertisers. When the $\alpha $ value increases from $100\%$ to $120\%$, the influence demand of all the advertisers is equal to or more than the influence supply from the influence provider. Hence, very few billboard slots remain to allocate further using the `EBTE' or `EBOE' methods. 

In the case of `Bank', `Park', and `Airport' datasets with the increasing value of $\alpha$ and $\mathcal{I}^{ID}$, the computational time for all the algorithms also increases. But, in the case of the `Beach' and `Mall' dataset, when $\alpha$ value rises from $80\%$ to $120\%$, the computational time of `EA' increases because there is a huge number of billboard slots to allocate, and it takes more computational search time. After effectively allocating slots to the advertisers, very few slots remain in the influence provider's hand to use in other proposed approaches. Therefore, in most cases, the proposed methods dominate baselines regarding efficiency.

\subsection{Additional Discussion}
This section discusses the additional parameters used in our experiments, e.g., $\gamma$ and $\lambda$. We have two main observations. 

First, we observe that the total regret for all the proposed and baseline methods increases with the distance $(\lambda)$ increase. As one billboard slot can influence more trajectories, the influence provider influence supply, $\mathcal{I}^{h}$, increases. While increasing $\mathcal{I}^{h}$ but fixing the value of $\alpha$ and $\lambda$, the advertiser's demand increases, and consequently, regret rises proportionally. 

\begin{figure}[h!]
\centering
 \begin{tabular}{lclc}
       Unsatisfied Regret & \includegraphics[width=0.11\linewidth]{Unsatisfied.png} \  & \ Excessive Regret & \includegraphics[width=0.11\linewidth]{Excessive.png} \\
    \end{tabular}

\begin{tabular}{ccccc}
\includegraphics[scale=0.09]{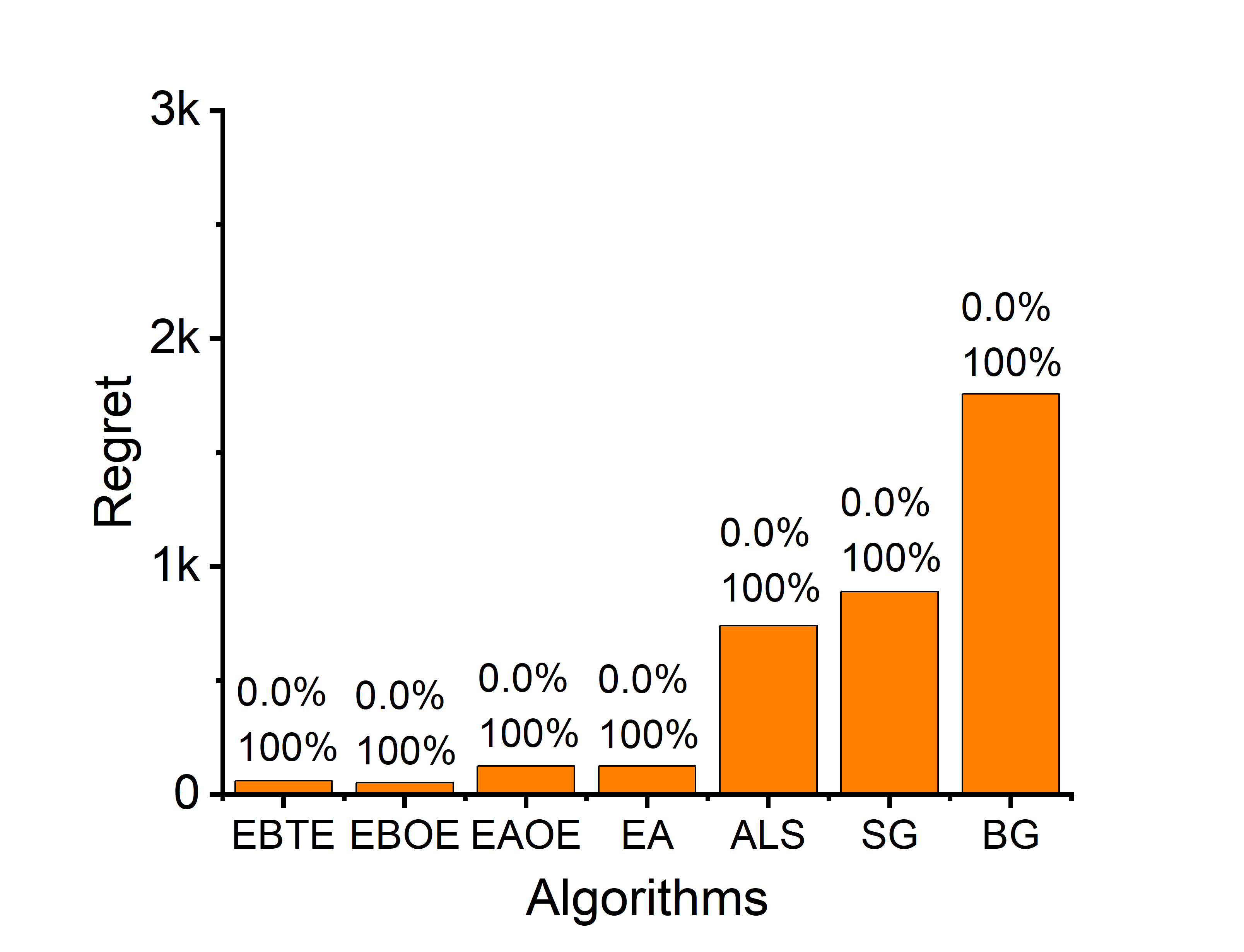} & \includegraphics[scale=0.09]{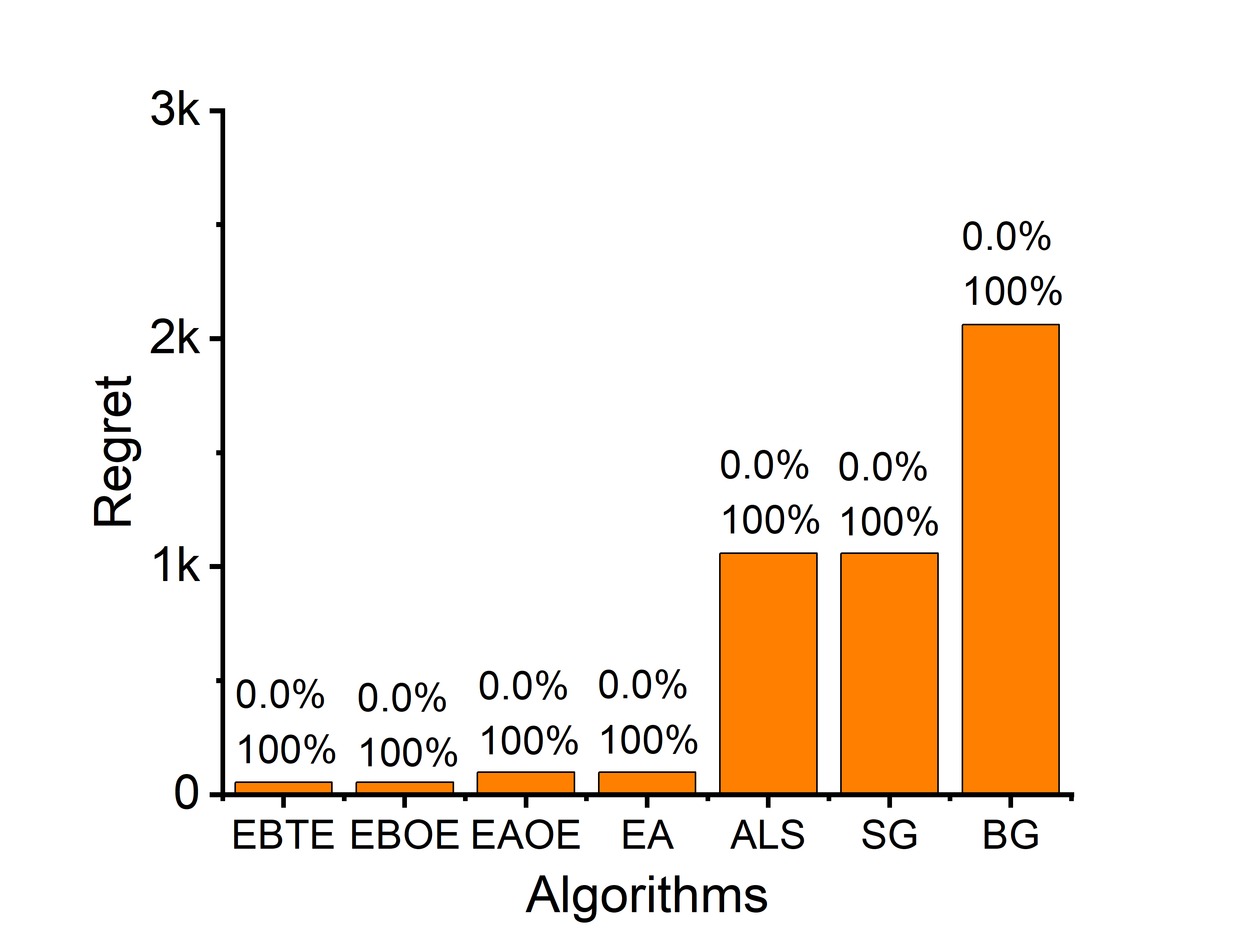}  &\includegraphics[scale=0.09]{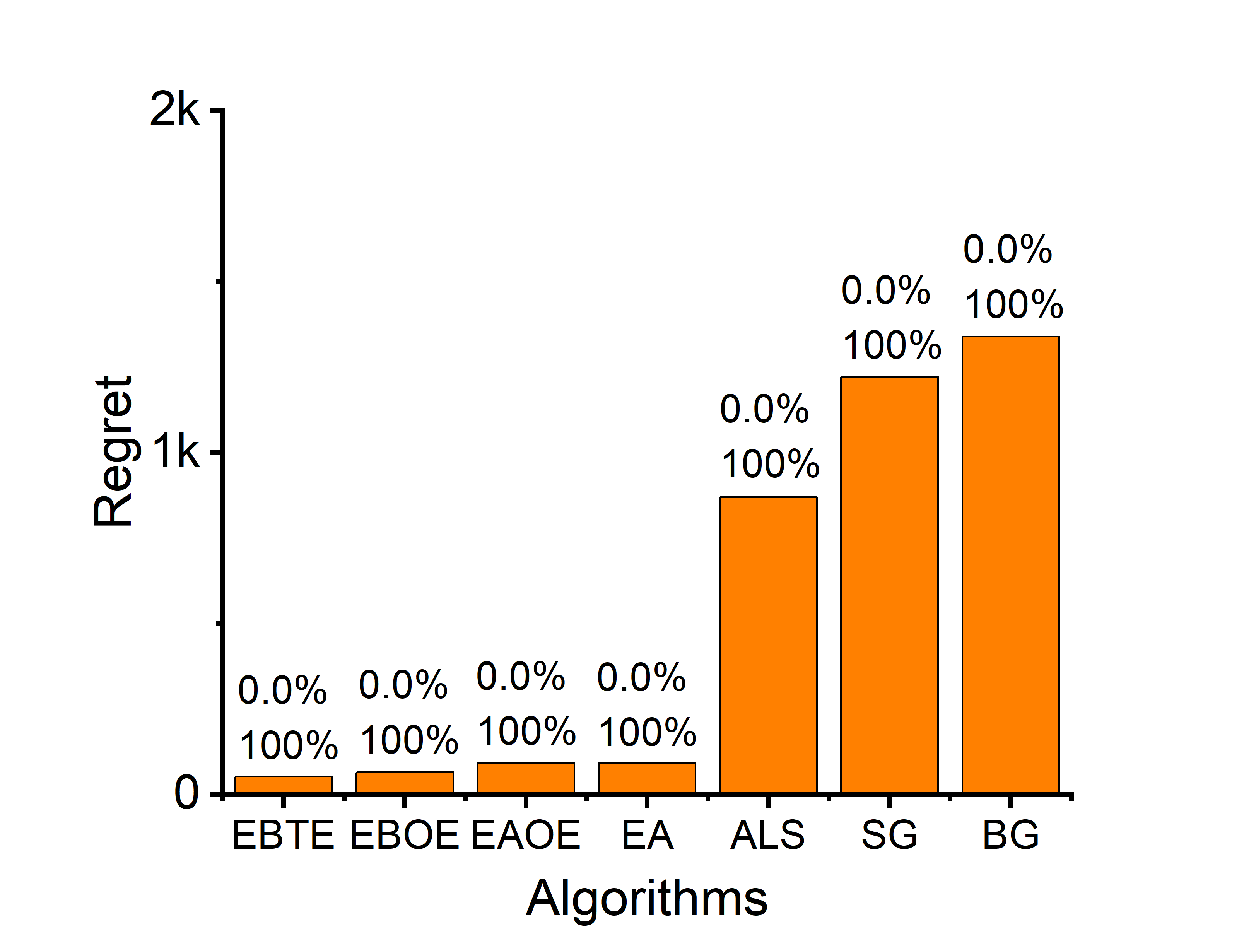} & \includegraphics[scale=0.09]{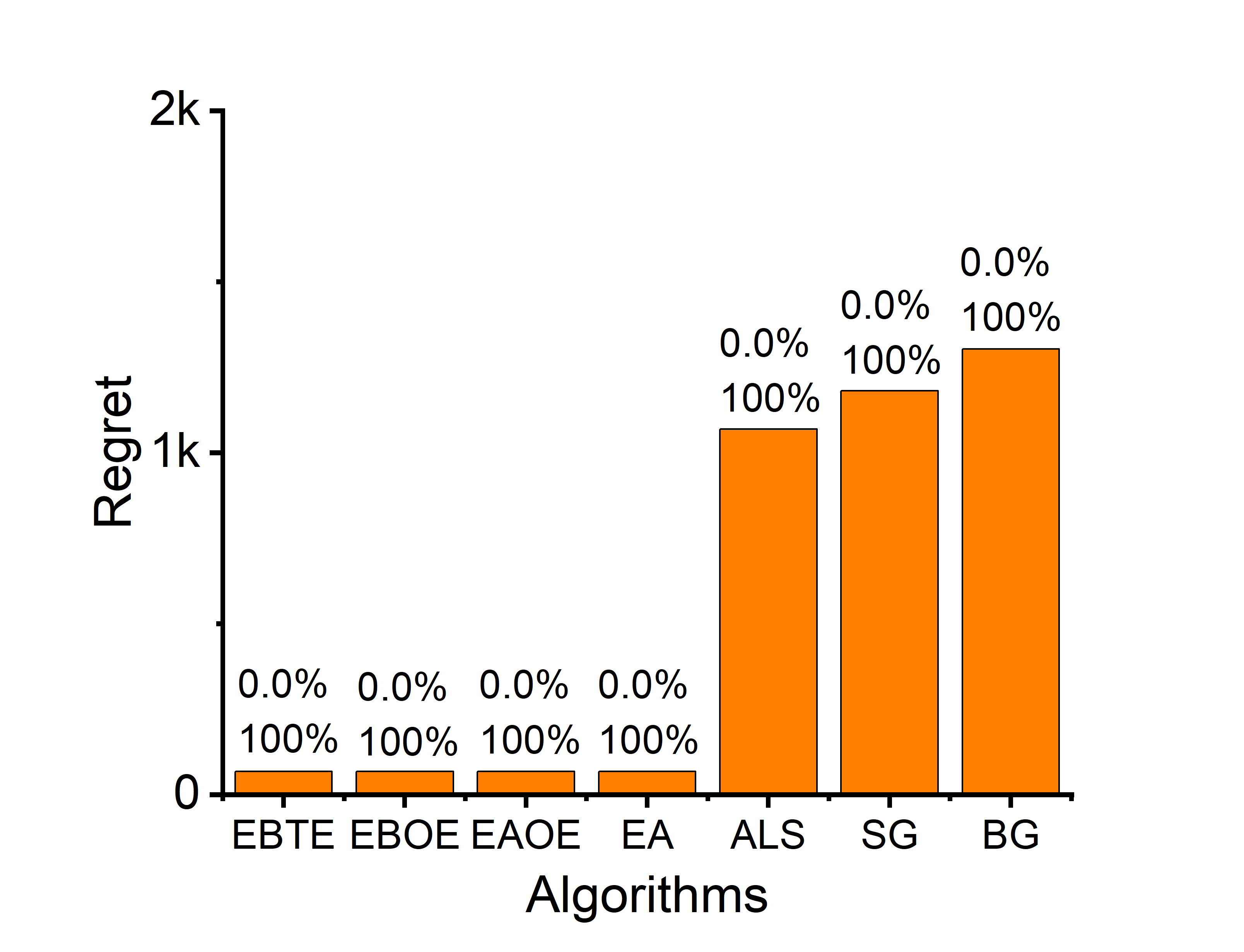} & \includegraphics[scale=0.09]{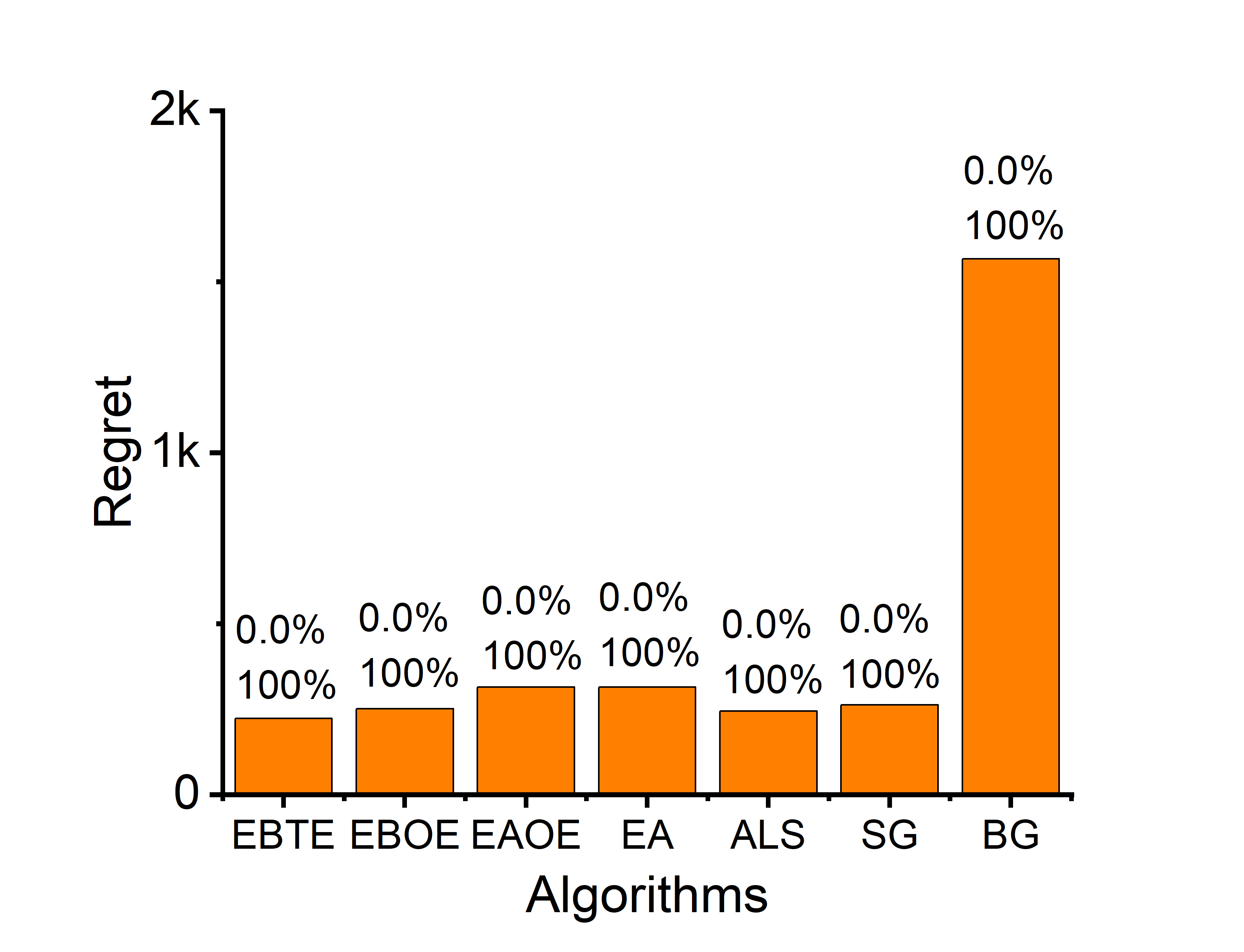}\\
{\tiny (a) $\gamma = 0$} & {\tiny (b) $\gamma = 0.25$} & {\tiny (c) $\gamma = 0.5$} & {\tiny (d) $\gamma = 0.75$} &{\tiny (e) $\gamma = 1$}\\

\end{tabular}
\caption{Regret of varying $\gamma$ when $\mathcal{I}^{ID} = 5\%, \mathcal{|A|} = 20$ (a, b, c, d, e) for Beach location }
\label{Fig:Beach_Gamma}
\end{figure}

\begin{figure}[h!]
\centering
    \begin{tabular}{lclc}
       Unsatisfied Regret & \includegraphics[width=0.11\linewidth]{Unsatisfied.png} \  & \ Excessive Regret & \includegraphics[width=0.11\linewidth]{Excessive.png} \\
    \end{tabular}

\begin{tabular}{ccccc}
\includegraphics[scale=0.09]{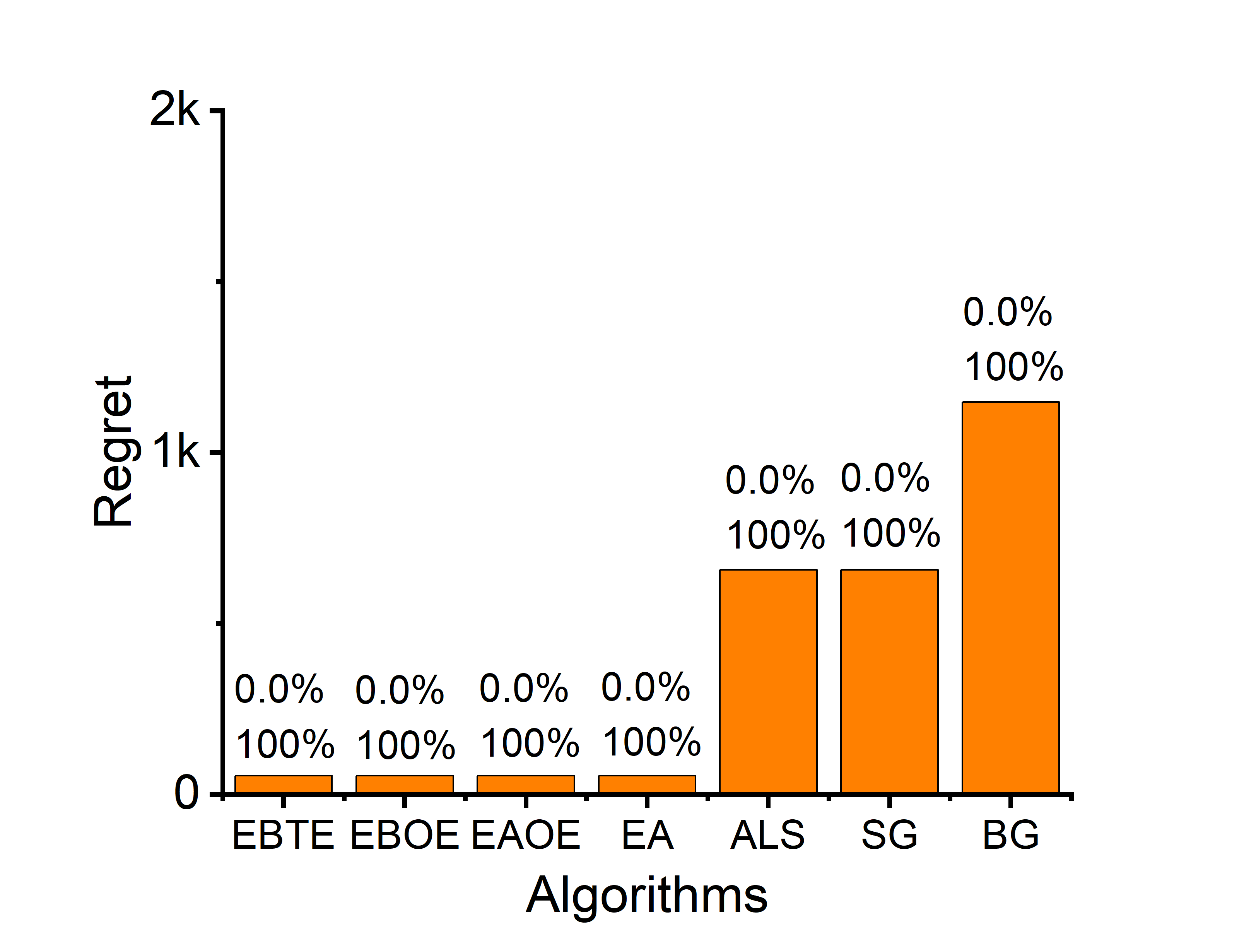} & \includegraphics[scale=0.09]{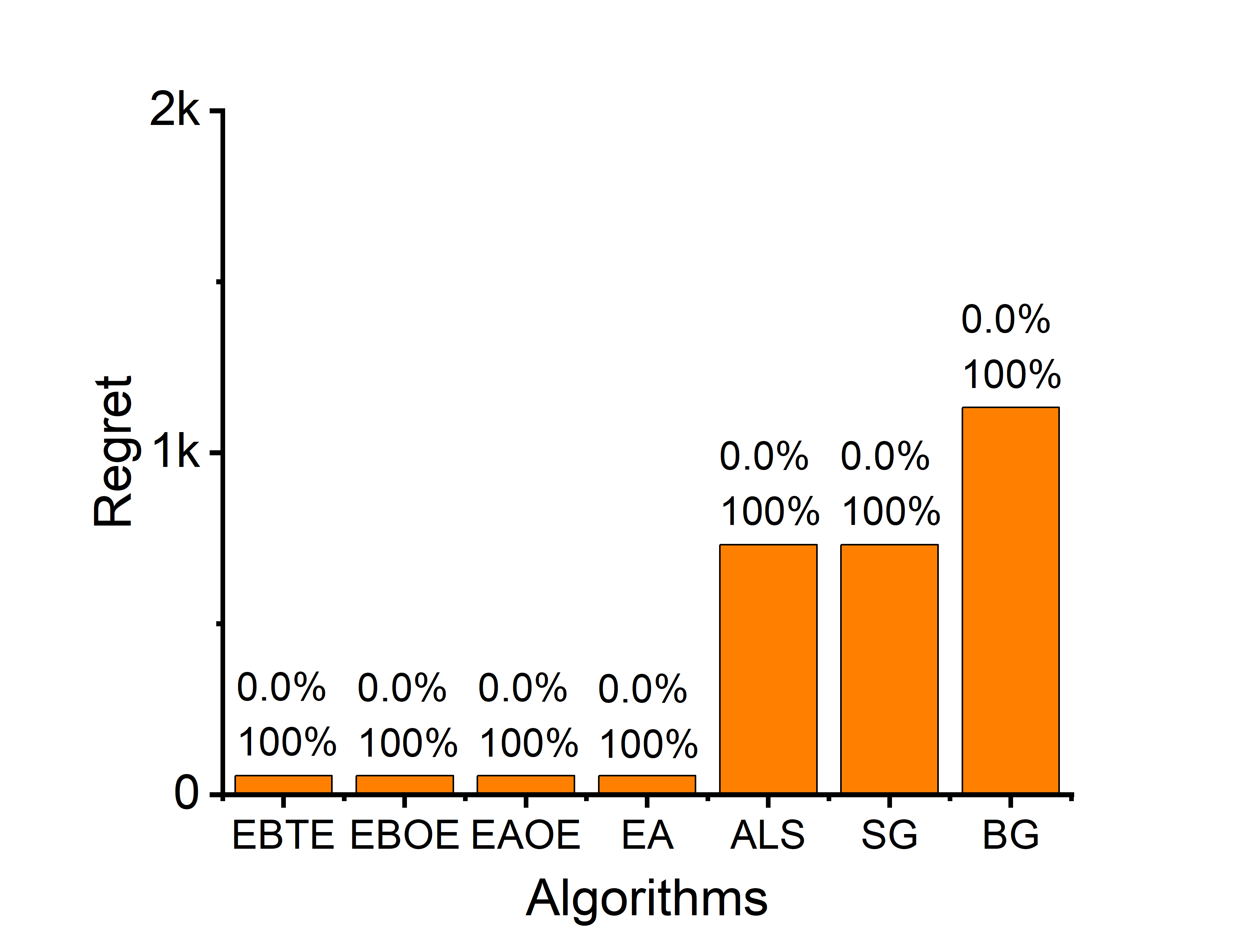}  &\includegraphics[scale=0.09]{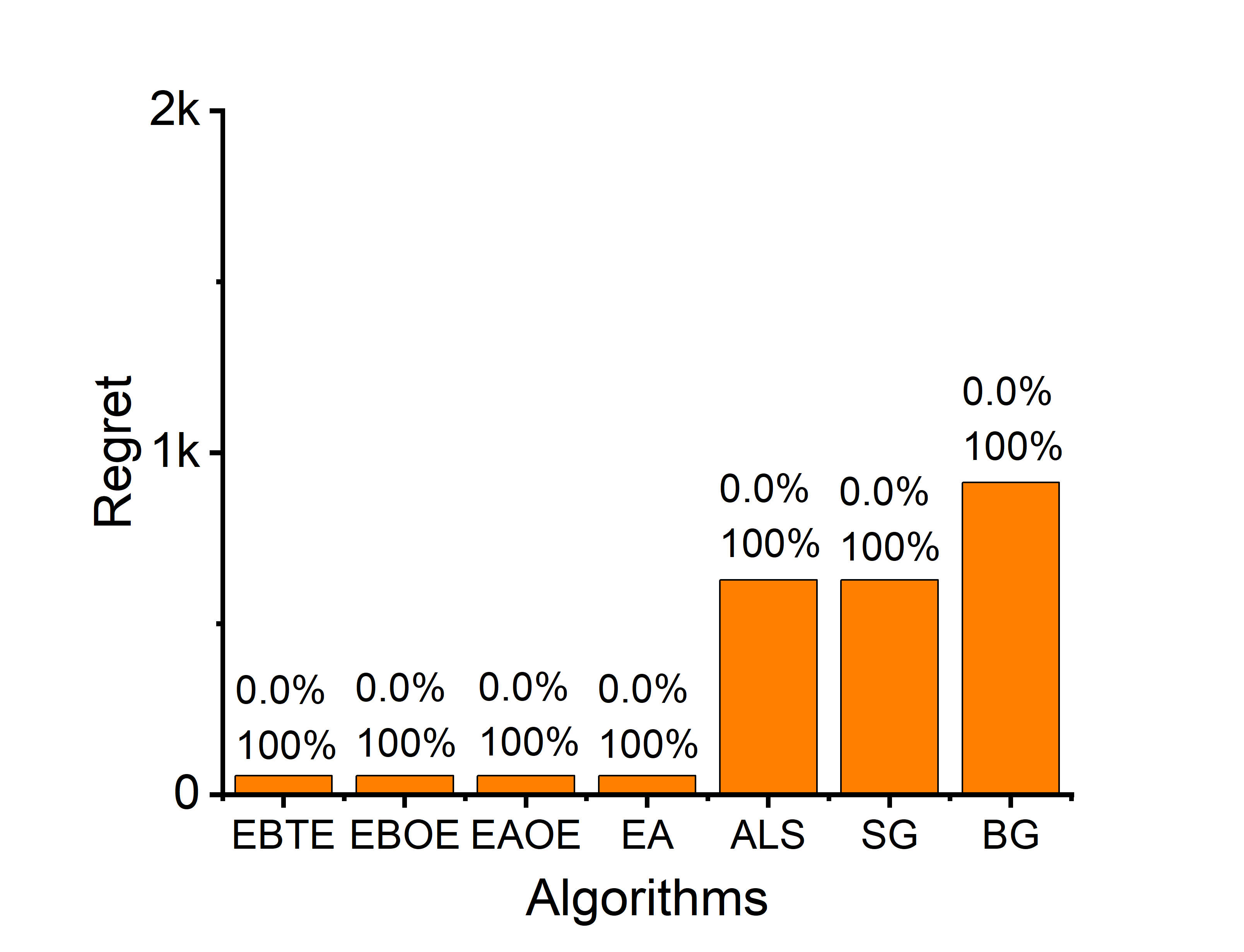} & \includegraphics[scale=0.09]{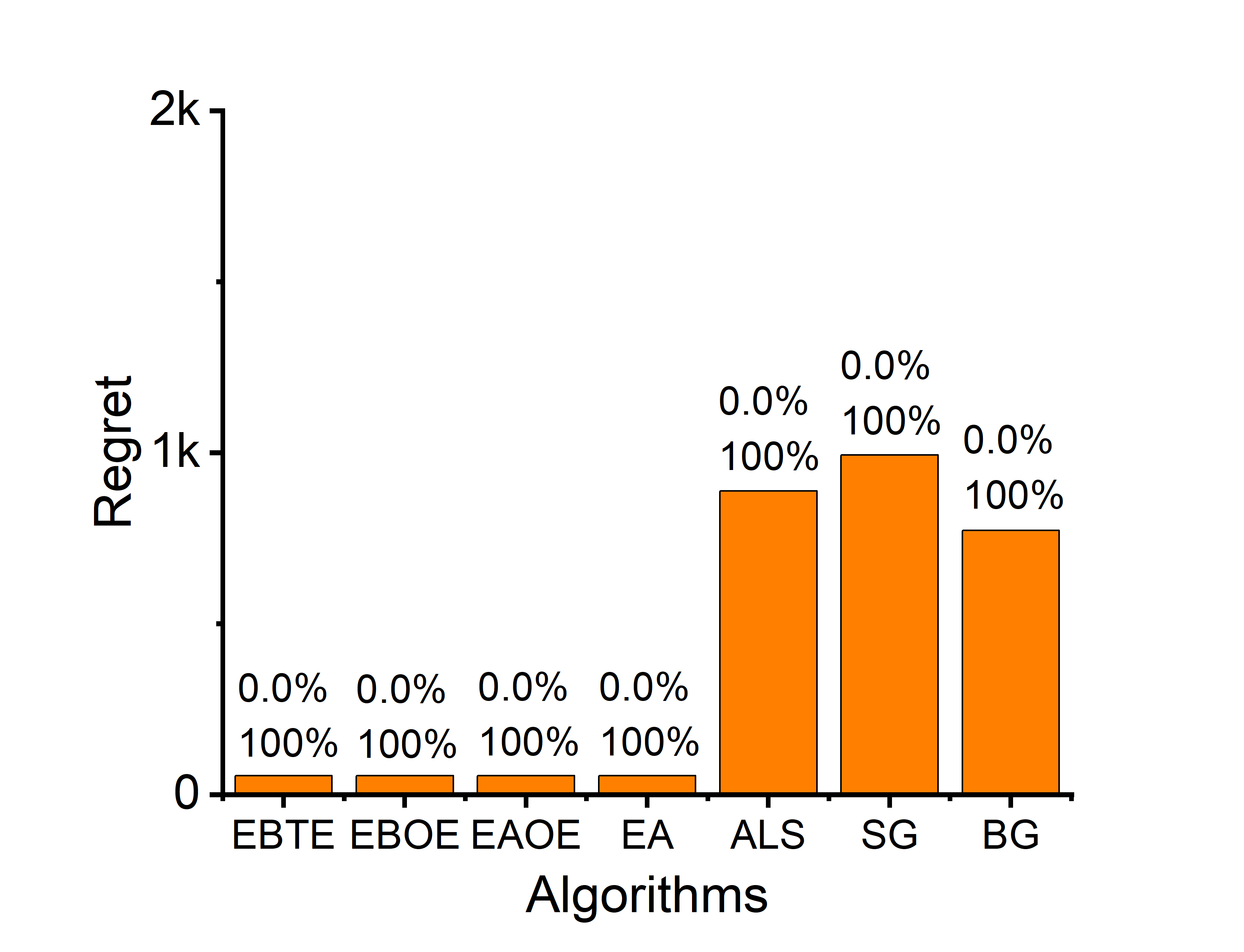} & \includegraphics[scale=0.09]{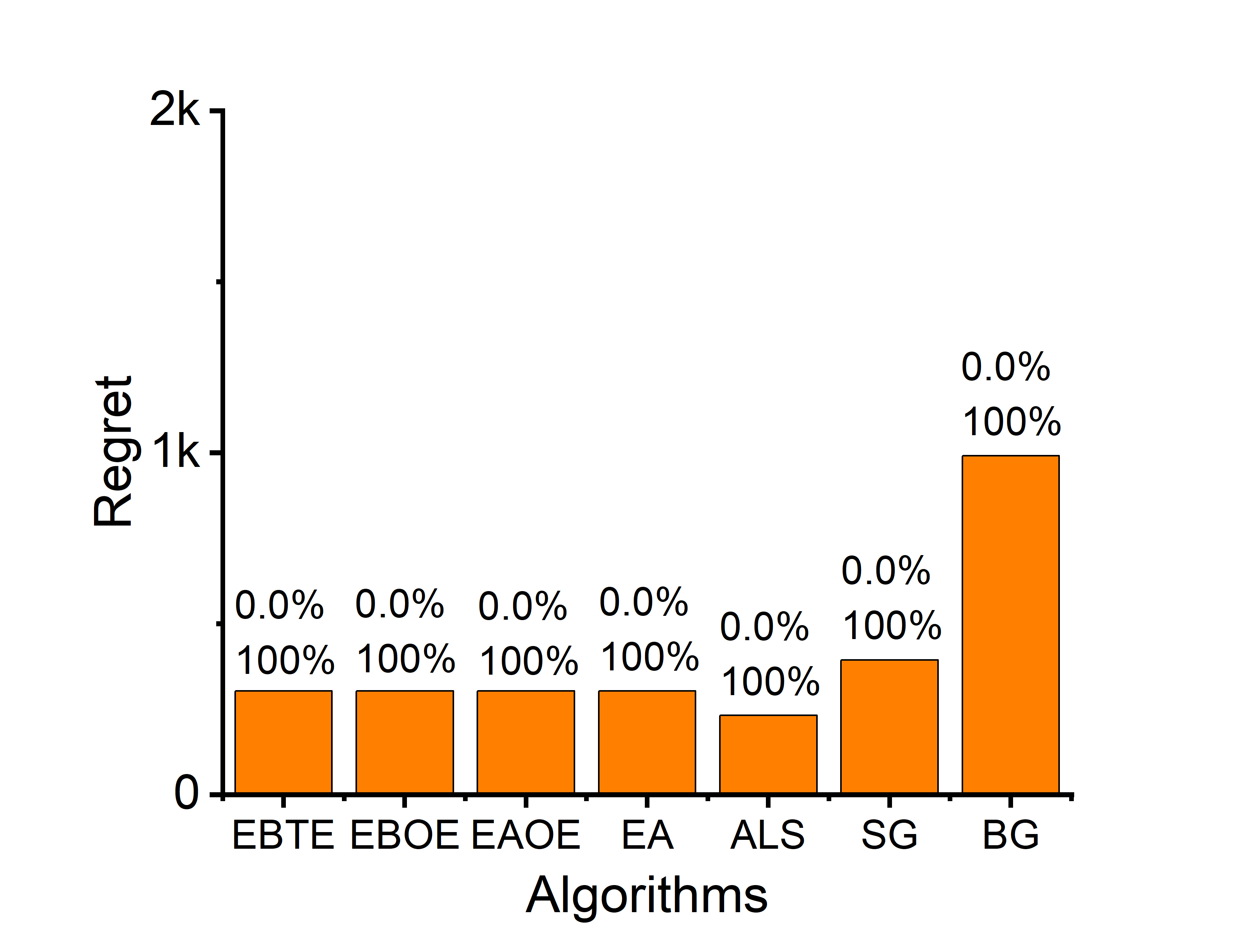}\\
{\tiny (a) $\gamma = 0$} & {\tiny (b) $\gamma = 0.25$} & {\tiny (c) $\gamma = 0.5$} & {\tiny (d) $\gamma = 0.75$} &{\tiny (e) $\gamma = 1$}\\

\end{tabular}
\caption{Regret of varying $\gamma$ when $\mathcal{I}^{ID} = 5\%, \mathcal{|A|} = 20$ (a, b, c, d, e) for Mall location }
\label{Fig:Mall_Gamma}
\end{figure}

\begin{figure}[h!]
\centering
  \begin{tabular}{lclc}
       Unsatisfied Regret & \includegraphics[width=0.11\linewidth]{Unsatisfied.png} \  & \ Excessive Regret & \includegraphics[width=0.11\linewidth]{Excessive.png} \\
    \end{tabular}

\begin{tabular}{ccccc}
\includegraphics[scale=0.09]{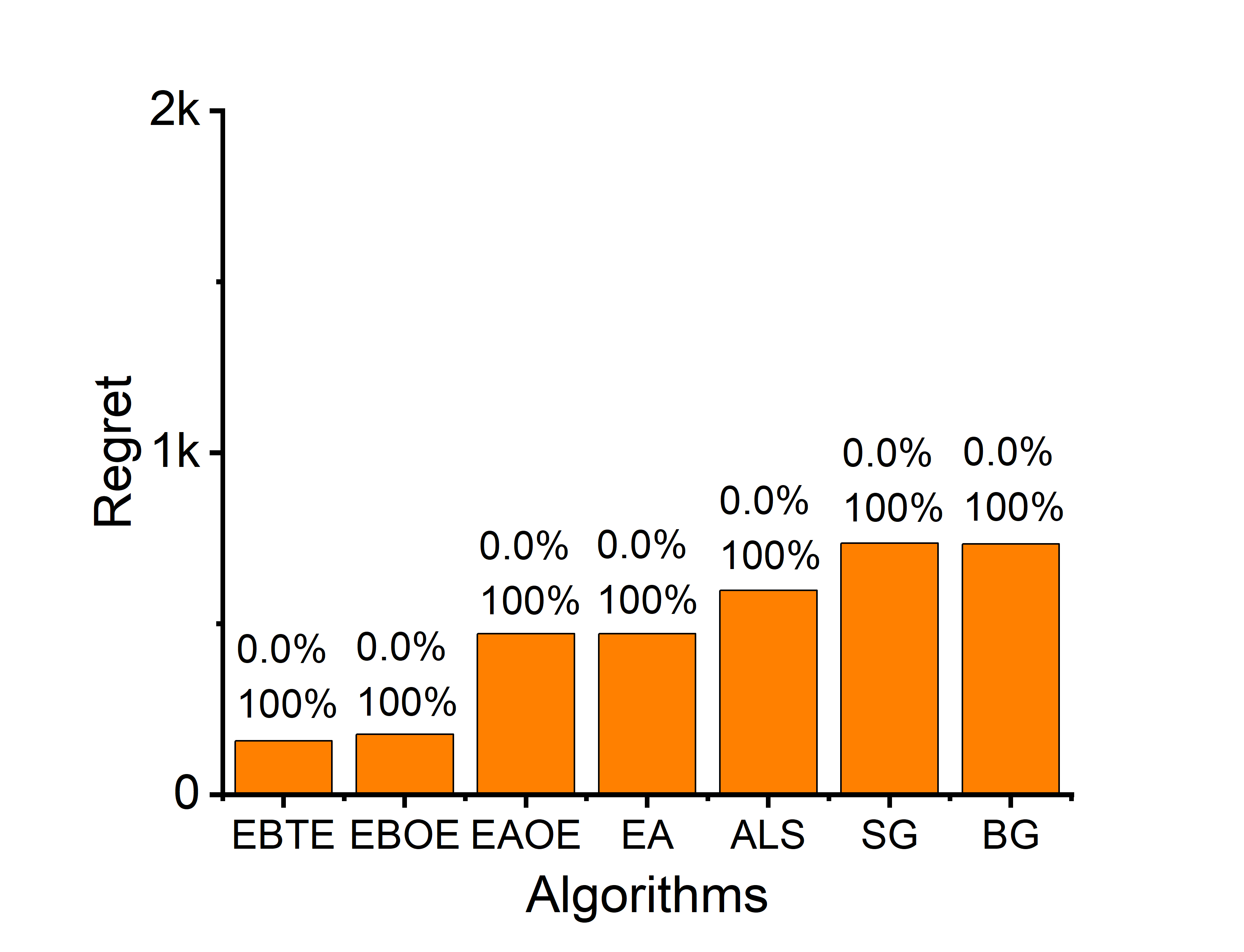} & \includegraphics[scale=0.09]{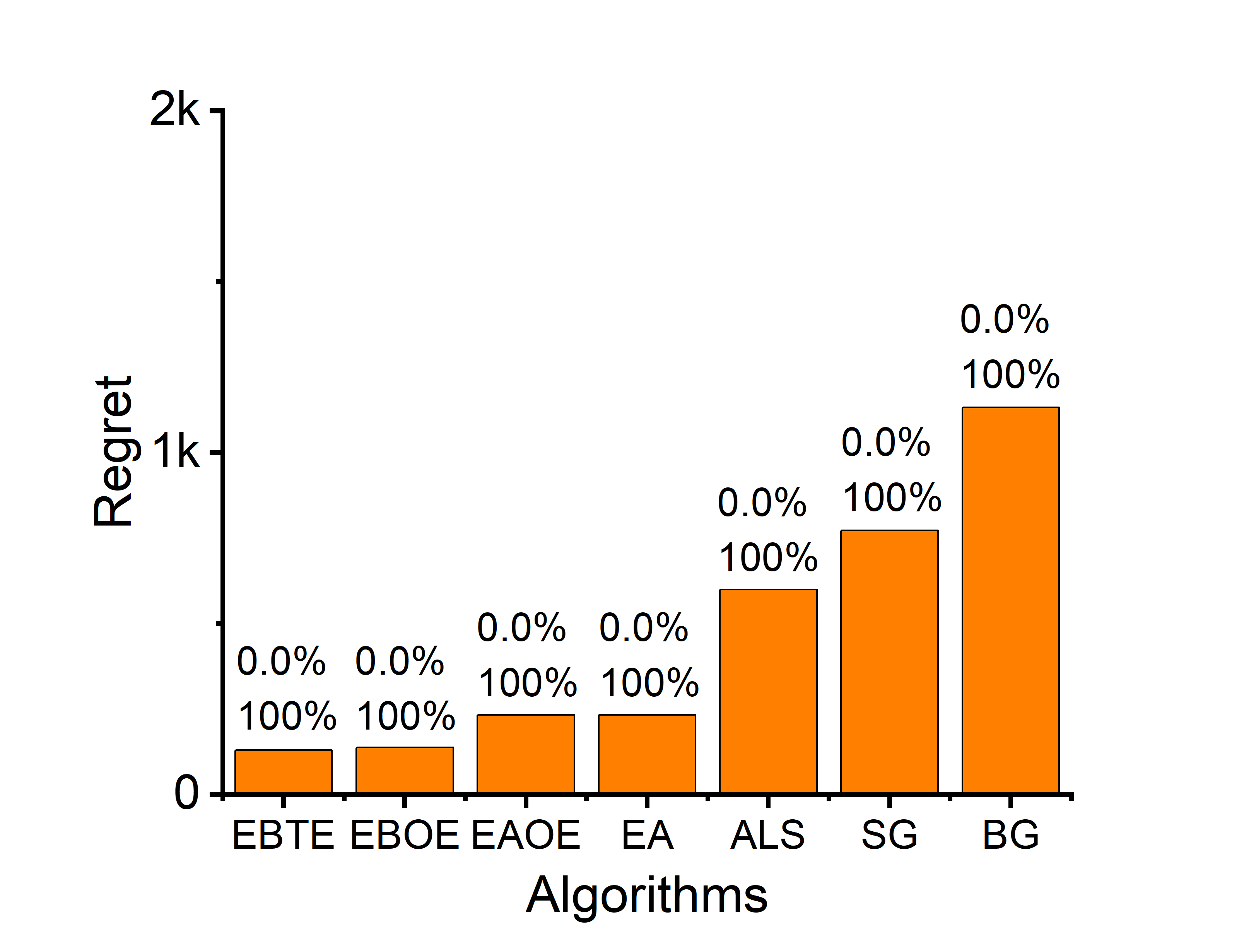}  &\includegraphics[scale=0.09]{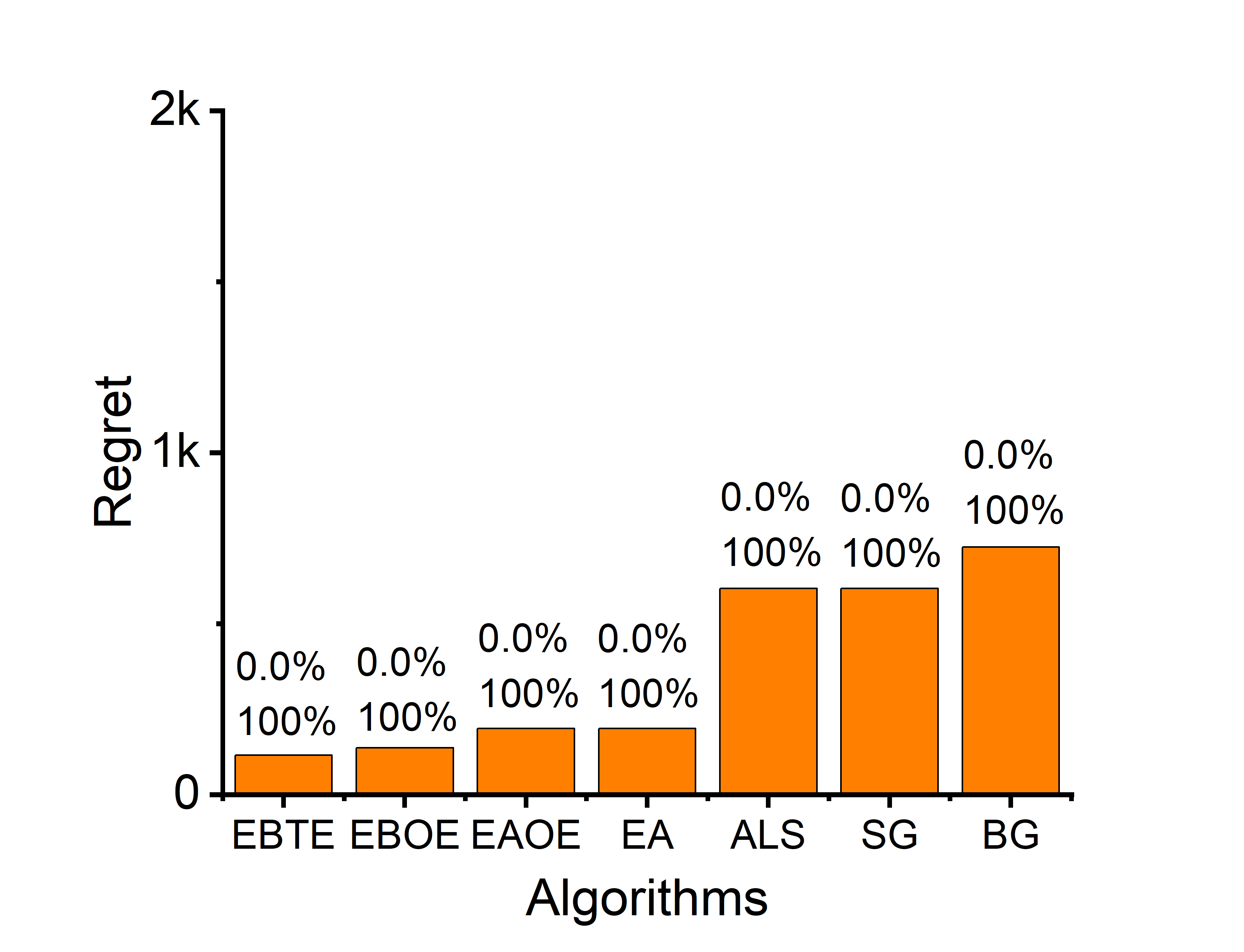} & \includegraphics[scale=0.09]{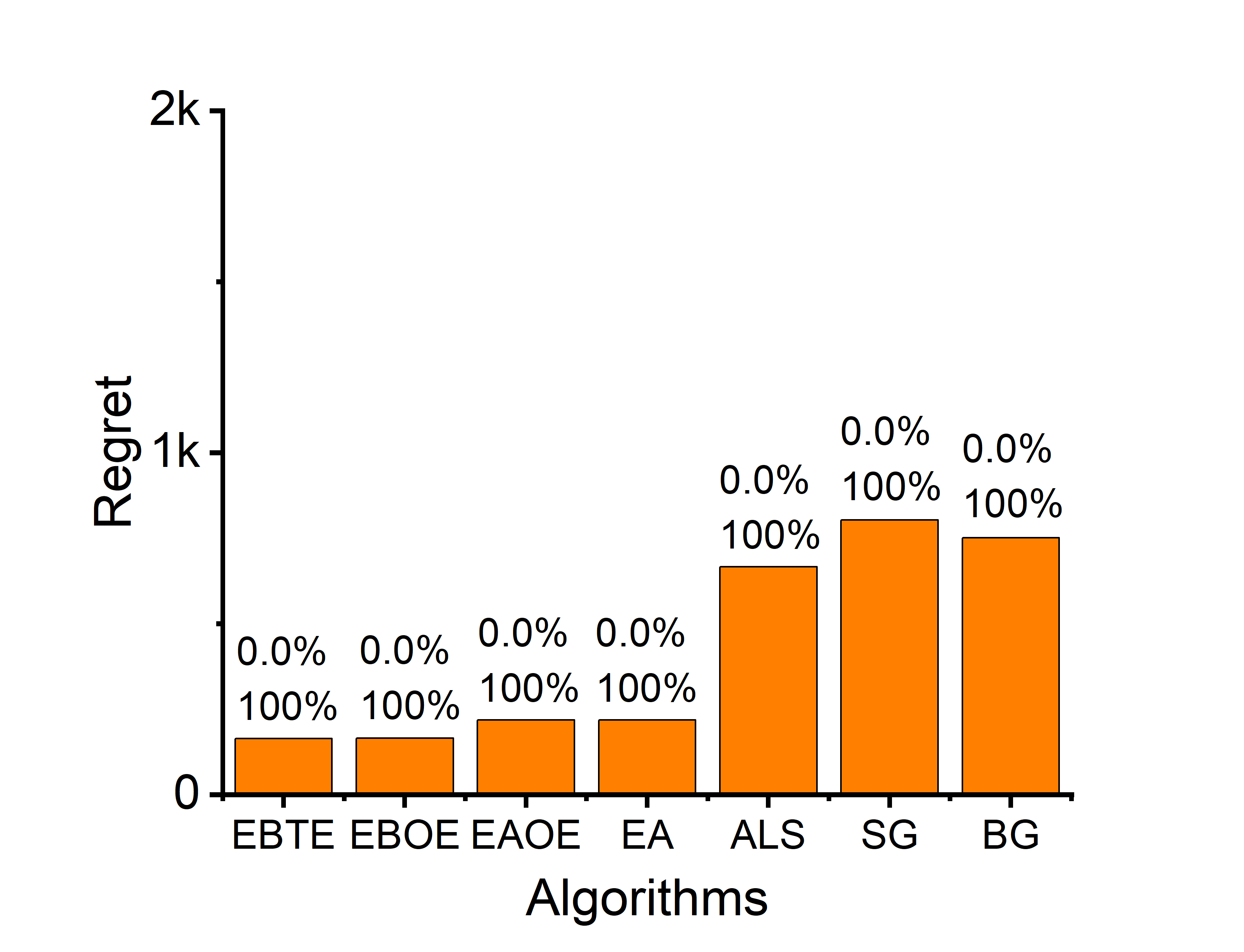} & \includegraphics[scale=0.09]{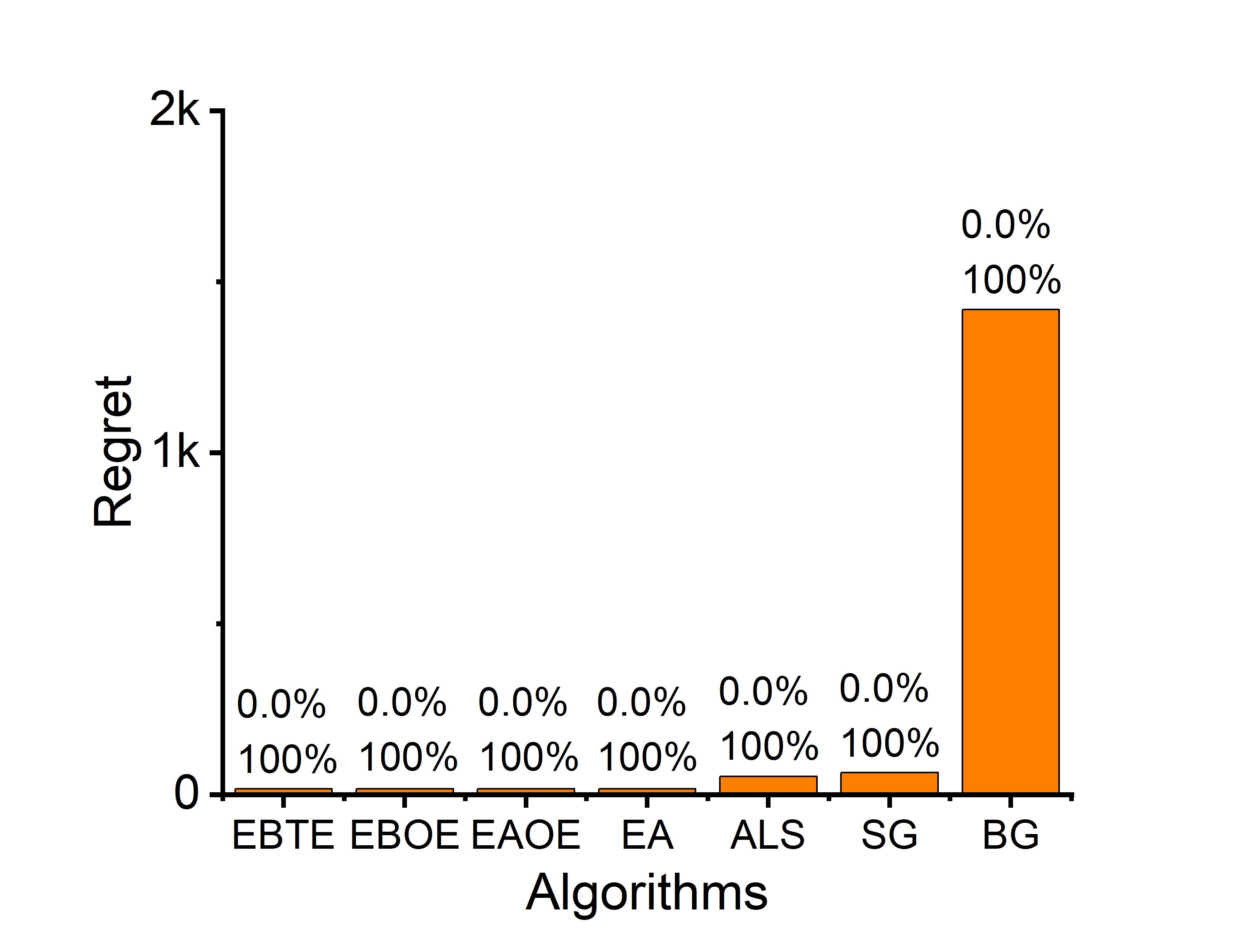}\\
{\tiny (a) $\gamma = 0$} & {\tiny (b) $\gamma = 0.25$} & {\tiny (c) $\gamma = 0.5$} & {\tiny (d) $\gamma = 0.75$} &{\tiny (e) $\gamma = 1$}\\

\end{tabular}
\caption{Regret of varying $\gamma$ when $\mathcal{I}^{ID} = 5\%, \mathcal{|A|} = 20$ (a, b, c, d, e) for Bank location }
\label{Fig:Bank_Gamma}
\end{figure}

\begin{figure}[h!]
\centering
    \begin{tabular}{lclc}
       Unsatisfied Regret & \includegraphics[width=0.11\linewidth]{Unsatisfied.png} \  & \ Excessive Regret & \includegraphics[width=0.11\linewidth]{Excessive.png} \\
    \end{tabular}

\begin{tabular}{ccccc}
\includegraphics[scale=0.09]{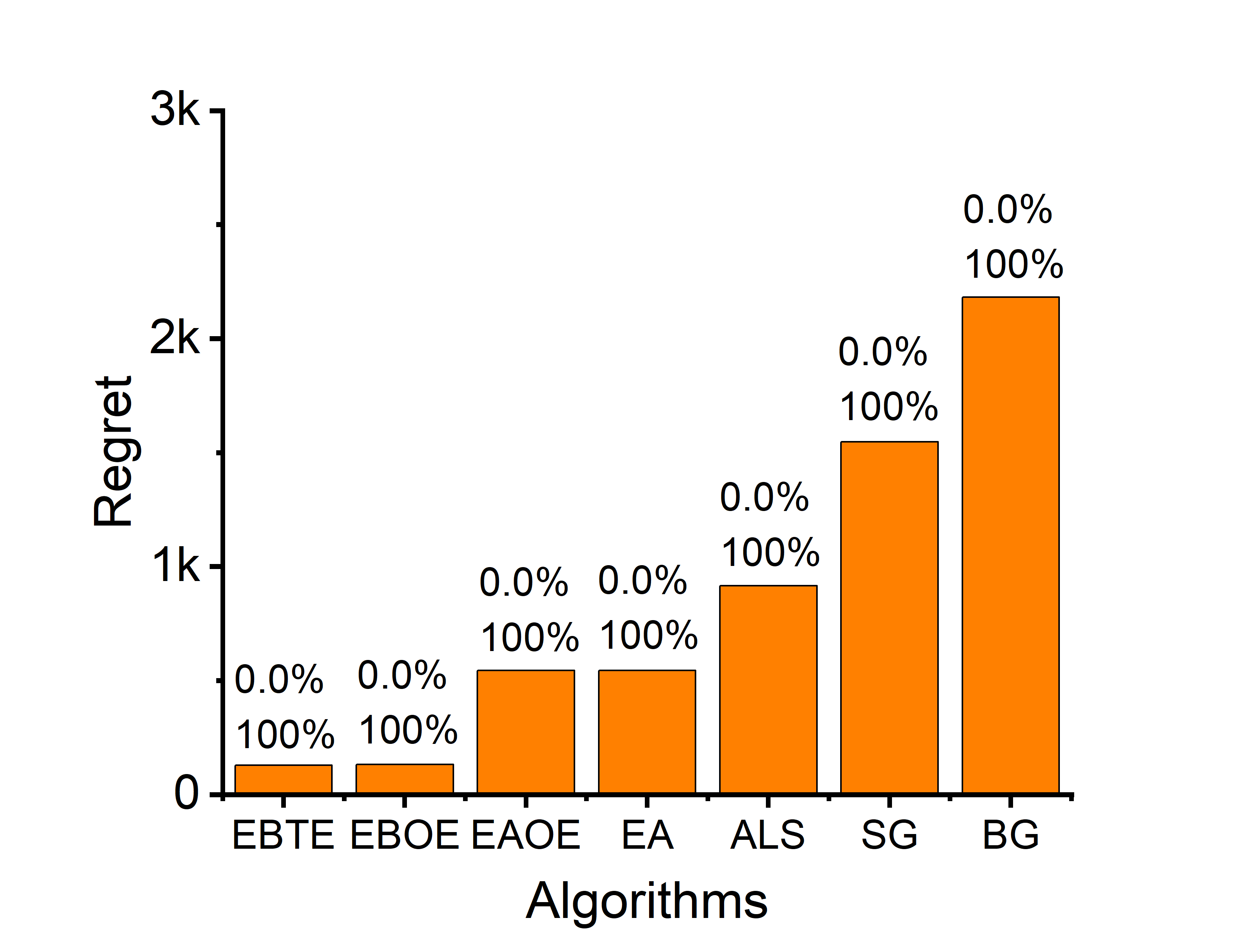} & \includegraphics[scale=0.09]{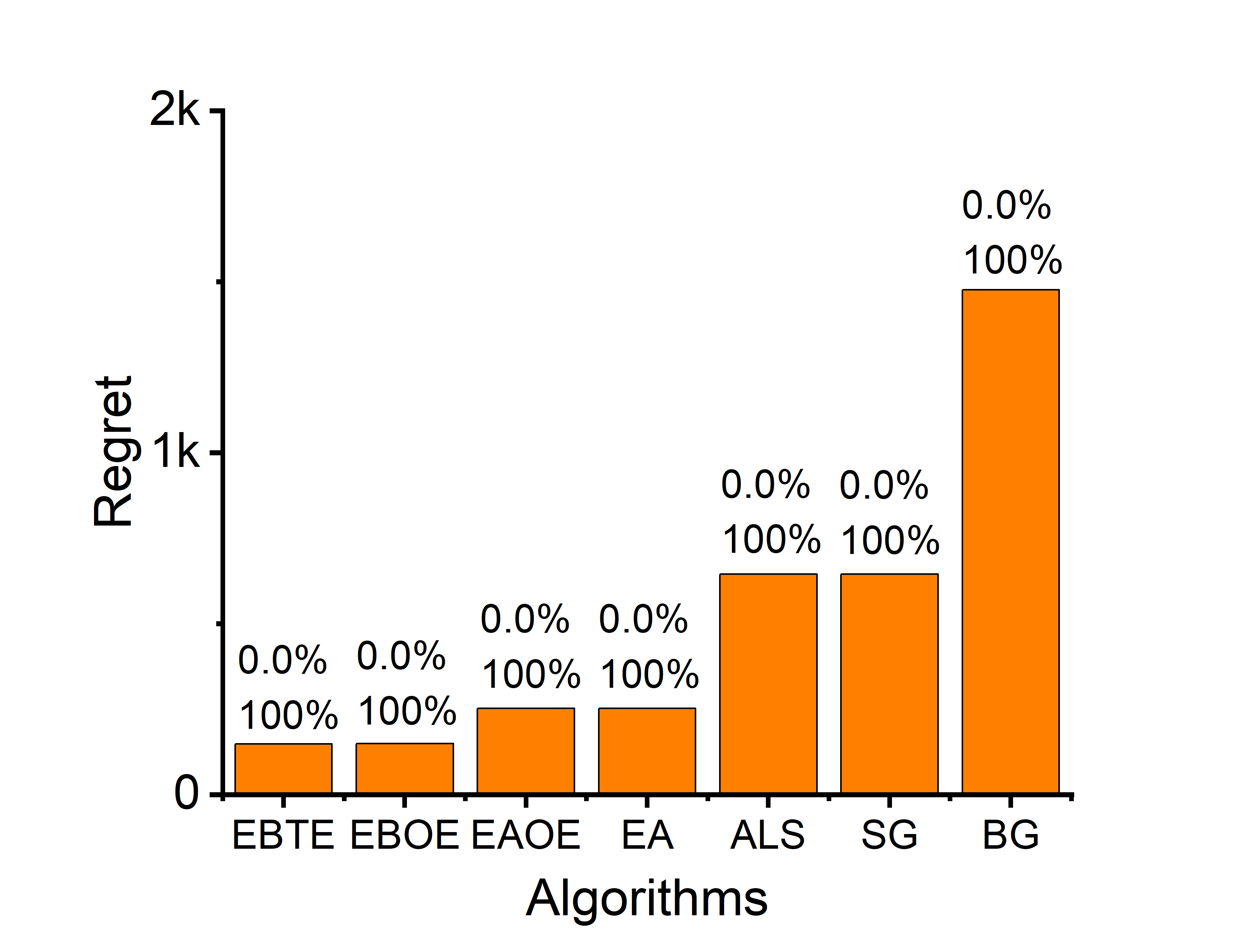}  &\includegraphics[scale=0.09]{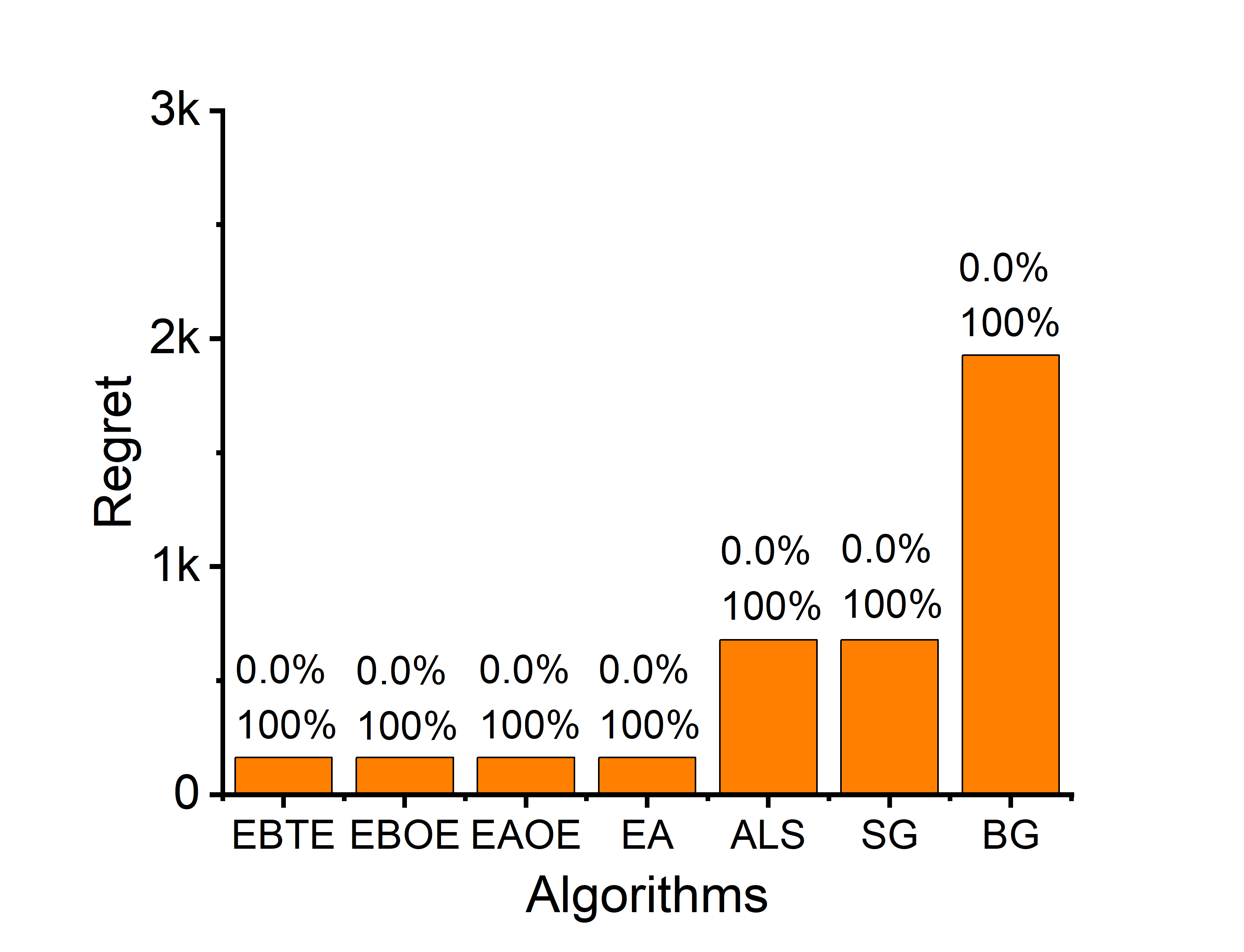} & \includegraphics[scale=0.09]{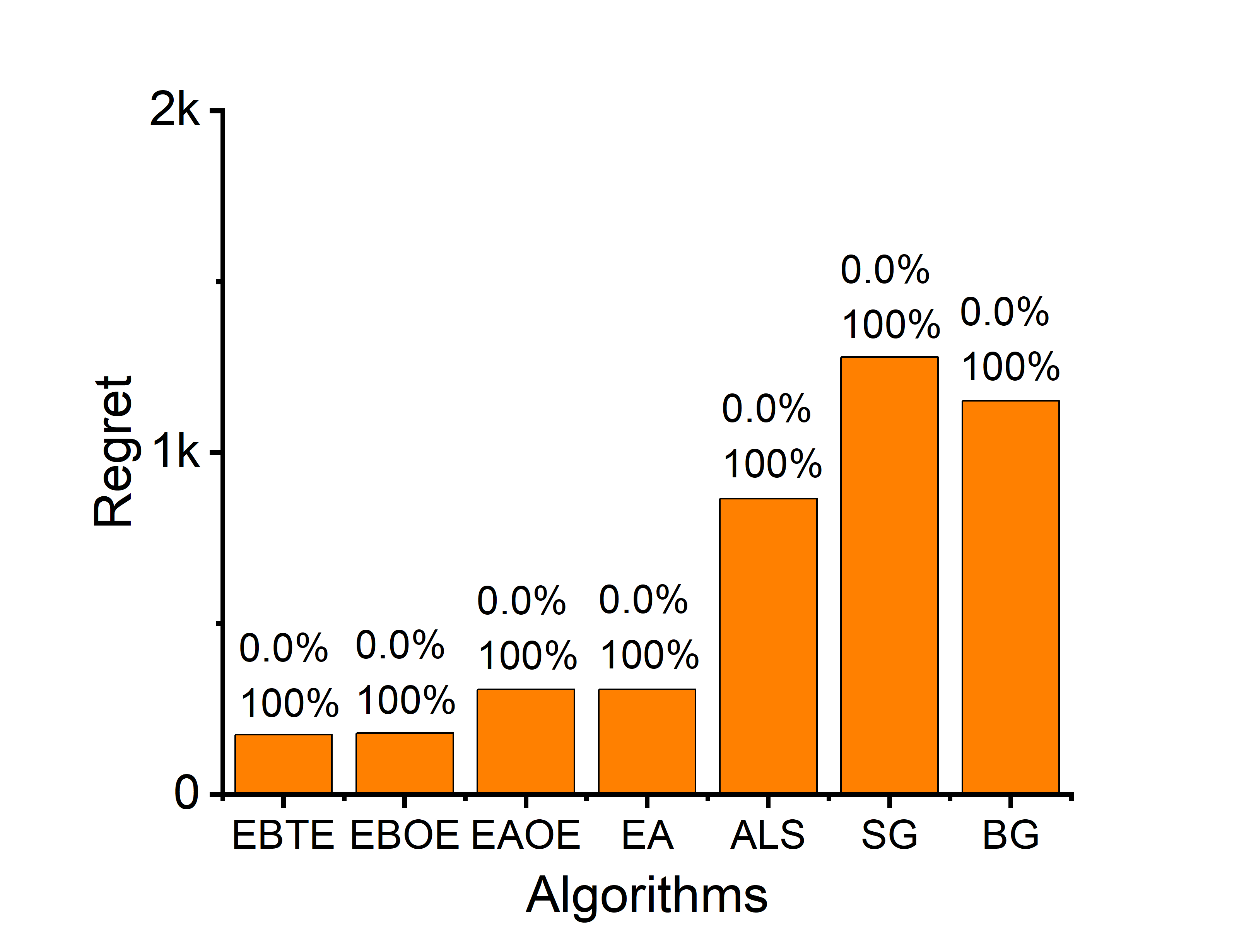} & \includegraphics[scale=0.09]{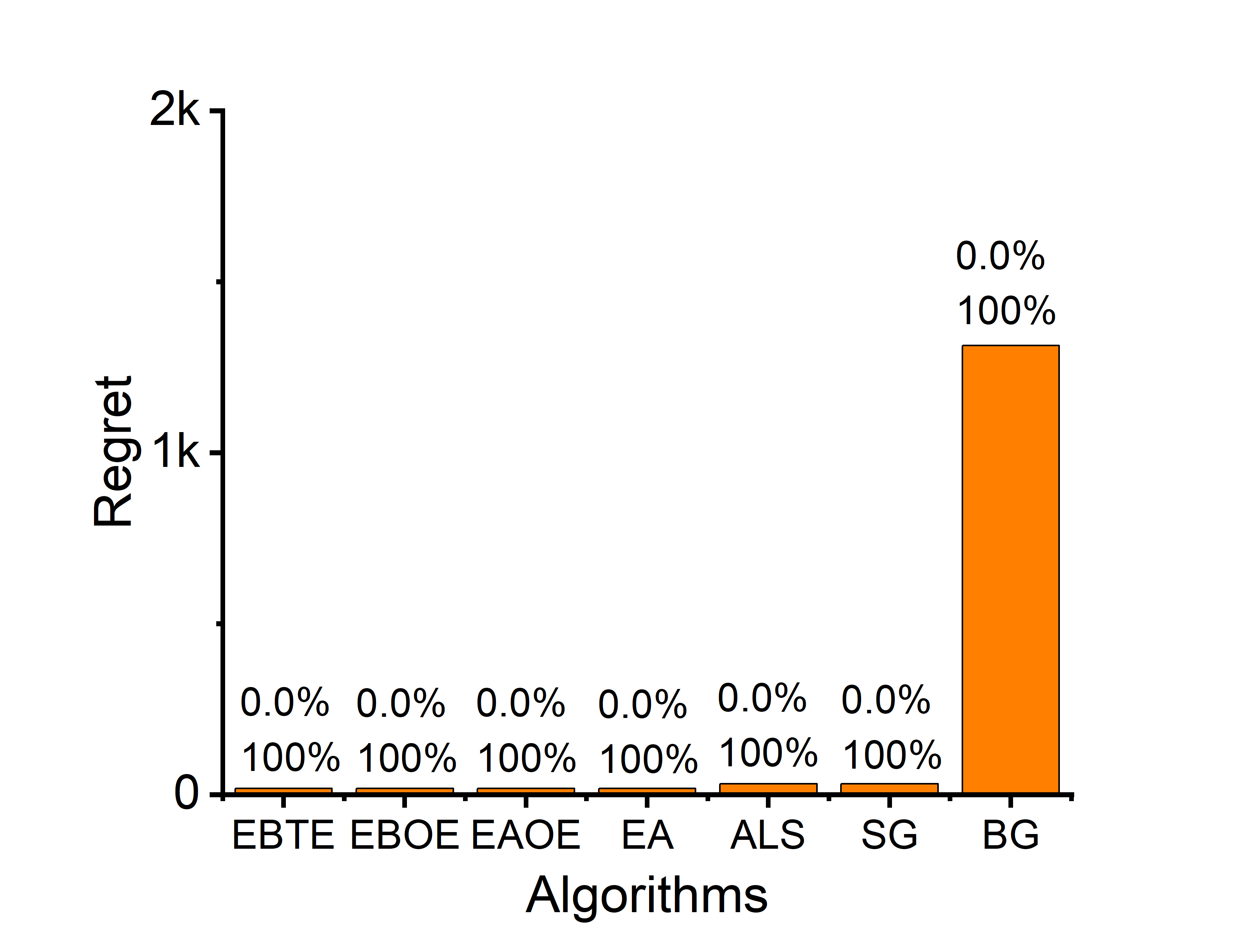}\\
{\tiny (a) $\gamma = 0$} & {\tiny (b) $\gamma = 0.25$} & {\tiny (c) $\gamma = 0.5$} & {\tiny (d) $\gamma = 0.75$} &{\tiny (e) $\gamma = 1$}\\

\end{tabular}
\caption{Regret of varying $\gamma$ when $\mathcal{I}^{ID} = 5\%, \mathcal{|A|} = 20$ (a, b, c, d, e) for Park location }
\label{Fig:Park_Gamma}
\end{figure}

Second, we observed that $\gamma$ plays a key role in the increase of unsatisfied regret. As we previously discussed in Definition \ref{Def:5}, $\gamma$ controls the unsatisfied regret for the unsatisfied advertisers. When the $\gamma$ value is very small, influence providers suffer from higher regret. However, when the $\gamma$ value increases, the unsatisfied regret decreases, as represented in Figure \ref{Fig:Beach_Gamma}, \ref{Fig:Mall_Gamma},\ref{Fig:Bank_Gamma}, and \ref{Fig:Park_Gamma}.


\section{Concluding Remarks and Future Research Directions} \label{Sec:CFD}
In this paper, we have studied the problem of regret minimization in billboard advertisements and proposed a number of heuristic solution approaches. All the algorithms have been analyzed to understand their time and space requirements. Several experiments have been conducted on real-world datasets showing the effectiveness and efficiency of the proposed solution approaches. Now, this work can be extended into several directions. First, we have considered that both the billboard slots and advertiser information are available well before the algorithm starts its execution. However, in practice, the `online' setting, where both the billboard slots and advertisers appear dynamically, is more appealing. Developing more efficient pruning techniques will remain an active area of research in the near future.

\paragraph{\textbf{Acknowledgements}.} The authors would like to acknowledge Indian Institute of Technology (IIT) Jammu, India, with grant number SG100047, for providing the funds required for the project.

 \bibliographystyle{splncs04}
 \bibliography{Paper}
\end{document}